\documentclass[aps,prb,reprint,twocolumn,superscriptaddress]{revtex4-2}

\usepackage{graphicx}
\usepackage{bm}
\usepackage{hyperref}
\usepackage{amsmath}
\usepackage{amssymb}

\usepackage{array}
\usepackage{multirow}
\usepackage{physics}
\usepackage{url}

\def\bk{{\bm{k}}}
\def\bg{{\bm{g}}}
\def\br{{\bm{r}}}
\def\bq{{\bm{q}}}
\def\bR{{\bm{R}}}

\begin{document}
\title{
 { The ideal limit of rhombohedral graphene:\\ Dynamical skyrmion lattices and their excitations }
}
\date{\today}
\author{Tixuan Tan}
\affiliation{Department of Physics, Stanford University, Stanford, CA 94305, USA}
\author{Patrick J. Ledwith}
\affiliation{Department of Physics, Massachusetts Institute of Technology, Cambridge, MA 02139, USA}
\author{Trithep Devakul}
\email{tdevakul@stanford.edu}
\affiliation{Department of Physics, Stanford University, Stanford, CA 94305, USA}
\begin{abstract}
{
We introduce an interacting ideal limit of rhombohedral graphene in which interactions spontaneously generate a correlated Chern insulator.
}
{{ Our analysis reveals that this
Chern insulator  can be} understood as a layer-pseudospin skyrmion lattice stabilized by electronic interactions, and that it gives rise to genuine skyrmion lattice physics{, complete with dynamics.}
This { dynamical} skyrmion lattice naturally leads to experimentally measurable consequences.
In particular, we identify new collective excitations corresponding to chiral shape deformations of the skyrmions, and propose a new positively charged quasiparticle, the skyrmion vacancy.
{ These results establish a qualitatively new form of moir\'e Chern physics in rhombohedral graphene: an interaction-generated skyrmion lattice whose skyrmions are dynamical low-energy degrees of freedom.}
}
\end{abstract}
\maketitle

{
\section{Introduction}

Progress in theoretical understanding of complex systems often comes from identifying unifying structures that are obscured by microscopic detail.
This perspective has proven especially fruitful in two-dimensional moir\'e materials, where Chern physics has now been realized across a broad range of systems~\cite{cao2025fractional}, from twisted graphene~\cite{andrei2020graphene} to transition metal dichalcogenides~\cite{mak2022semiconductor,li2025review}.
Despite their disparate microscopic origins, the Chern bands in these systems often exhibit deep similarities to Landau levels (LLs) at strong magnetic fields.

The origin of these similarities can be naturally explained in terms of effective magnetic fields generated by pseudospin skyrmion lattices~\cite{wu2019topological,yu2020giant,zhai2020theory,morales2024magic,guerci2025layer,kolavr2025robustness,zhang2025experimental,thompson2025microscopic}.
At low energies, many moir\'e systems  are described by electrons carrying an internal pseudospin degree of freedom (e.g. layer index in twisted MoTe$_2$~\cite{wu2019topological}), whose real-space orientation $\bm{n}(\bm{r})$ is influenced by a moir\'e superlattice potential.
A smooth spatial variation of this pseudospin texture causes electrons adiabatically following it to experience an effective magnetic field proportional to the local skyrmion density,
$\rho_{\text{sk}}=\frac{1}{4\pi}\bm{n}\cdot(\partial_x\bm{n}\times\partial_y\bm{n})$.
The same principle underlies the `topological Hall effect' in magnetic skyrmion materials~\cite{nagaosa2013topological} and the electric charge of spin skyrmions in quantum Hall ferromagnets~\cite{girvin2002quantum}.
When the pseudospin texture forms a skyrmion lattice (corresponding to one skyrmion per moir\'e unit cell), the resulting effective magnetic field averages to one flux quantum per moir\'e unit cell, thus explaining the emergence of LL-like Chern bands.

This picture becomes most transparent in certain idealized limits of moir\'e systems, where the analogy can be made exactly~\cite{ledwith2020fractional,wang2021exact,dongCompositeFermiLiquid2023,guerci2025layer}.
A notable example is the chiral limit~\cite{tarnopolsky2019origin,ledwith2020fractional} of twisted bilayer graphene~\cite{bistritzer2011moire},
though similar limits have also emerged in twisted graphene multilayers~\cite{khalaf2019magic,popov2023magic,ledwith2022family,wang2022hierarchy,popov2023magic2,devakul2023magic,gaountwisting2023,guerci2024chern,guerci2024nature,foo2024extended,kwan2024strong,guineachiralmultilayer}, transition metal dichalcogenides~\cite{crepel2024chiral,shi2024adiabatic}, and other related structures~\cite{li2025variational,crepelrealizedinmonolayer,sarkar2025ideal,wan2023topological,sommer2025idealopticalfluxlattices,paul2025emergenttopologyflatbands,unconventionalfractional2025}.
In all these examples, flat Chern bands emerge that admit an exact description in terms of electrons moving in a periodic magnetic field~\cite{PhysRevA.19.2461,DubrovinNovikov1980,wang2021exact,dongCompositeFermiLiquid2023,guerci2025layer} generated by a real-space pseudospin skyrmion lattice.

A qualitatively distinct regime of Chern physics is realized in
moir\'e heterostructures based on rhombohedral $N$-layer graphene (R$N$G).
Recent experiments on R$N$G ($N=4,5,6$) aligned with hBN (R$N$G/hBN) have revealed some of the sharpest signatures to date of integer and fractional Chern insulators~\cite{lu2024fractional,waters2025chern,aronson2025displacement,choi2025superconductivity,xie2025tunable,zheng2025switchable,lu2025extended,chen2020tunable}.
Unlike other moir\'e platforms where similar phenomena originate from single-particle Chern bands (such as {{{hBN-aligned graphene~\cite{spantonObservationFractionalChern2018}, hBN-aligned twisted bilayer graphene~\cite{xieFractionalChernInsulators2021,finneyExtendedFractionalChern2025a,finneyExtendedFractionalChern2025a,abouelkomsanParticleHoleDualityEmergent2020a,ledwith2020fractional,repellinChernBandsTwisted2020,wilhelmInterplayFractionalChern2021,shefferChiralMagicangleTwisted2021a,parkerFieldtunedZerofieldFractional2021}}}}, and twisted bilayer MoTe$_2$~\cite{mak2022semiconductor,li2025review,wu2019topological,cai2023signatures,zeng2023thermodynamic,park2023observation,xu2023observation,anderson2024trion}), the Chern band in R$N$G/hBN emerges fundamentally from electronic interactions~\cite{dong2024theory,zhou2024fractional,dong2024anomalous,guo2024fractional,kwan2025moire,herzog2024moire,guo2024correlation,zhou2025new,huang2024self,yu2025moire,huang2025fractional,li2025multiband,uchida2025non,wei2025edge,kudo2024quantum,xie2024integer,bernevig2025berry}.
This has motivated extensive efforts to understand how interactions generate Chern bands in this general setting~\cite{desrochers2025elastic,tan2024parent,dong2024stability,lu2025general,tan2025variational,soejima2025topological,dong2025phonons,zeng2025berry,xiang2025continuously,perea2025quantum,lu2025generic,tan2024parent,desrochers2025electronic,soejima2025jellium,soejima2024anomalous,crepel2025efficient}, much of which is based on numerical Hartree-Fock (HF) or perturbative analyses in momentum-space.
At the same time, recent numerical studies show that the correlated phase diagram of R$N$G/hBN can depend sensitively on microscopic modeling details~\cite{huo2025moire,yu2025moire,li2025multiband}.
Moreover, obtaining an accurate microscopic model that quantitatively describes experiments remains an active area of research~\cite{klein2024imaging,uzan2025hbn,seewald2025mapping,zhang2025moir}.
Taken together, these observations underscore the need for a more transparent physical picture for R$N$G/hBN, akin to those that have proven fruitful in other moir\'e materials, which can lead to physical predictions that are less sensitive to microscopic details.%

In this work, we demonstrate that the Chern insulator in R$N$G/hBN realizes a \emph{dynamical} skyrmion lattice.
Unlike in other moir\'e Chern systems, this texture is not imposed by the single-particle Hamiltonian, but emerges spontaneously from electron interactions.
This distinction underlies both the challenge of establishing the skyrmion lattice picture and the new physical consequences that follow from it.
{

In other moir\'e Chern systems, skyrmion textures can often be inferred directly from the structure of the single-particle Hamiltonian~\cite{wu2019topological,yu2020giant,zhai2020theory,morales2024magic,guerci2025layer,kolavr2025robustness,zhang2025experimental,thompson2025microscopic}.
In R$N$G/hBN, by contrast, the microscopic Hamiltonian contains no preexisting texture of this kind.
This makes the origin of the skyrmion lattice substantially more challenging to establish.
{We address this challenge by introducing an interacting
\emph{ideal limit} of rhombohedral graphene, together with a real-space framework
that renders its many-body ground states analytically tractable.} This framework identifies the microscopic mechanism driving skyrmion lattice formation { in R$N$G/hBN}, which we show remains operative in the experimentally realistic setting.

The interaction-driven origin of the skyrmion lattice has important physical consequences beyond establishing the texture itself.
Because the texture is generated spontaneously, its deformations become intrinsic low-energy degrees of freedom.
We show that these dynamics give rise to collective excitations closely paralleling the gyrotropic, breathing, and shape modes familiar from magnetic skyrmion crystals~\cite{garst2017collective,lonsky2020dynamic,nagaosa2013topological}.
It also changes the nature of charged excitations: we show that the lowest-energy positively charged excitation is a skyrmion vacancy, corresponding to a local defect of the skyrmion texture.

Together, these results establish a qualitatively new form of moir\'e Chern physics in R$N$G/hBN, where an interaction-generated skyrmion texture acquires intrinsic dynamics, giving rise to distinctive collective modes and quasiparticle excitations.
{ We further expect the ideal limit framework developed in our work to be useful more broadly in the study of correlated phases in rhombohedral graphene.}
}

\section{Summary of main results}
{ \noindent { \textbf{An interacting ideal limit of rhombohedral graphene.}}}
{ The principal theoretical advance of this work} is:
(i) the introduction of an idealized limit of the { R$N$G} Hamiltonian, together with (ii) the development of a real-space formalism that { makes its interacting physics analytically tractable.}
Several noteworthy aspects { of our analysis} are highlighted here:
\begin{itemize}
\item It directly connects the physics of the Chern insulator to that of electrons in an effective magnetic field generated by a skyrmion lattice of the $N$-component layer pseudospin.

\item { This ideal limit} includes interactions.
It describes an interaction-induced Chern band, unlike all previously known ideal limits which are single-particle.

{
\item It identifies the microscopic mechanism for skyrmion lattice formation.
The ideal limit shows that the essential ingredients are the holomorphic layer-pseudospin structure of R$N$G near the band edge and repulsive short-range interactions.
Together, these ingredients favor the chiral two-particle correlations associated with LLL physics, leading to the spontaneous formation of a skyrmion lattice.
}

{
\item This mechanism remains operative in realistic R$5$G/hBN.
We show that the skyrmion lattice persists to the experimentally relevant regime, which includes realistic band dispersion, longer-range Coulomb interactions, hBN-induced moir\'e potentials, and spinors that are only approximately holomorphic.
}
\end{itemize}

\noindent \textbf{Skyrmion lattice picture for the correlated Chern insulator.}
{ The fact that the Chern insulator in R$N$G/hBN can be understood as an interaction-induced skyrmion lattice}
 provides a natural explanation for several aspects of the ground state:
\begin{itemize}
\item It provides an intuitive real-space picture for the origin of topology.
Previous analyses have offered intuition primarily in momentum space, or by analogy with pinned Wigner or Hall crystals~\cite{dong2024anomalous,soejima2024anomalous,tan2024parent}.
By contrast, the skyrmion lattice and its effective magnetic field provide a far more direct and intuitive explanation for the Chern band and its connection to quantum Hall physics.

\item The skyrmion lattice exhibits a signature charge density distribution.
Because the skyrmion texture resides in the layer pseudospin, it is clearly visible in the layer-resolved charge density.
In particular, sharp minima in the top-layer charge density  (furthest from hBN) mark the ``skyrmion cores'', where the local layer-pseudospin texture points in a direction orthogonal to the top layer.
Because most of the wavefunction weight lies in the top layer, this structure is also apparent in the total charge density (summed over all layers) though with shallower minima, explaining the honeycomb-like charge distribution observed in previous HF studies~\cite{dong2024anomalous,kwan2025moire,zeng2024sublattice}.
Importantly, this makes it possible to infer the presence and shape of skyrmions from the top-layer charge density.
\end{itemize}

\noindent \textbf{The skyrmions are dynamical objects.}
{ A key physical consequence of the interaction-generated texture is that the
skyrmions become dynamical low-energy degrees of freedom.}
We demonstrate this with two cases that clearly illustrate the important role of skyrmions in the low-energy physics:
\begin{itemize}
\item The skyrmions have internal shape excitation modes.
Beyond the phonon-like modes associated with sliding motion of the skyrmion lattice~\cite{kwan2025moire,dong2025phonons}, we identify a hierarchy of new collective excitations corresponding to chiral multipolar shape deformations of the skyrmions,
which closely parallel those found in magnetic skyrmion materials~\cite{garst2017collective,lonsky2020dynamic,nagaosa2013topological}.
Specifically, we predict quadrupolar, sextupolar, and breathing modes, and discuss routes to experimentally probing them.
The quadrupolar mode, in particular, is reminiscent of the ``chiral graviton'' mode in fractional quantum Hall systems~\cite{haldaneGeometricalDescriptionFractional2011,yangModelWaveFunctions2012,maciejkoFieldTheoryQuantum2013a,gromovBimetricTheoryFractional2017a,nguyenProbingSpinStructure2021,duChiralGravitonTheory2025}, which was recently observed experimentally~\cite{liangEvidenceChiralGraviton2024}.
\item The skyrmion lattice supports a new low-energy positively charged quasiparticle: the skyrmion-vacancy.
The absence of a skyrmion at a site of an otherwise regular lattice corresponds to a missing magnetic flux quantum for the electrons, and therefore carries a net positive charge.
We show that such vacancies can be pinned by potential disorder in real samples (e.g. a charged point defect in the substrate).
Its experimental observation would provide direct evidence for the dynamical skyrmion lattice in R$N$G/hBN.
\end{itemize}
{\noindent  These predictions are direct consequences of the dynamical nature of the skyrmion lattice in R$N$G/hBN, and have no analogue in other moir\'e systems with static skyrmion textures.

Taken together, these results reveal R$N$G/hBN to be a remarkably rich setting in which skyrmion lattice dynamics are intertwined with quantum Hall-like Chern physics, driven by the interplay of Coulomb interactions and the quantum geometry of R$N$G.
}

\section{Structure of this paper}
{
}

We begin with Sec~\ref{sec:idealsection}, which introduces the ideal limit of rhombohedral graphene and develops a real-space formalism for {explaining the emergence of skyrmion lattices}.
In Sec~\ref{sec:idealrng}, we introduce the ideal limit of R$N$G.
In Sec~\ref{sec:exactgroundstates}, we characterize its exact many-body ground states, making the direct connection to a Dirac electron in an effective magnetic field.
In Sec~\ref{sec:layerskyrmionlattices}, the structure of periodic ground states is { shown to be} a skyrmion lattice.
In Sec~\ref{sec:ansatz}, we introduce an ansatz wavefunction for the { skyrmion lattice} ground state in the presence of dispersion, which is then verified by numerical analysis in Sec~\ref{sec:HFdispideal}.

{In Sec~\ref{sec:connecting}, we show that the mechanism for skyrmion lattice formation identified in the ideal limit remains operative in the experimentally-relevant regime of realistic R$5$G/hBN.
We demonstrate this through an interpolation connecting the ideal and realistic limits, together with direct signatures of the skyrmion lattice in the local density matrix and layer-resolved charge density.
}

We then turn to the dynamical nature of the skyrmion lattice {and its experimental consequences}.
In Sec~\ref{sec:collective}, we perform a detailed analysis of the collective modes, identifying new low-energy excitations corresponding to skyrmion shape modes.
In Sec~\ref{sec:vacancy}, we discuss the role of skyrmion vacancies as {positively} charged excitations, and show that such vacancies can be pinned by potential disorder.
Finally, we conclude in Sec~\ref{sec:discussion} with a discussion of experimental and theoretical implications, as well as future directions.
}

{
{\section{ The ideal limit of rhombohedral graphene}
\label{sec:idealsection}}
In this section, we introduce an interacting ideal limit of R$N$G and develop a real-space
formalism that permits its many-body ground states to be characterized analytically.
The resulting construction explains how repulsive short-range interactions generate a Chern band, identifying a layer-pseudospin skyrmion lattice as its microscopic origin.
Later, in Sec.~\ref{sec:connecting}, we will show that this same mechanism remains
operative in realistic R$5$G/hBN.\\
}

\subsection{ Ideal rhombohedral graphene}
\label{sec:idealrng}
\begin{figure}
    \centering
    \includegraphics[width=1\linewidth]{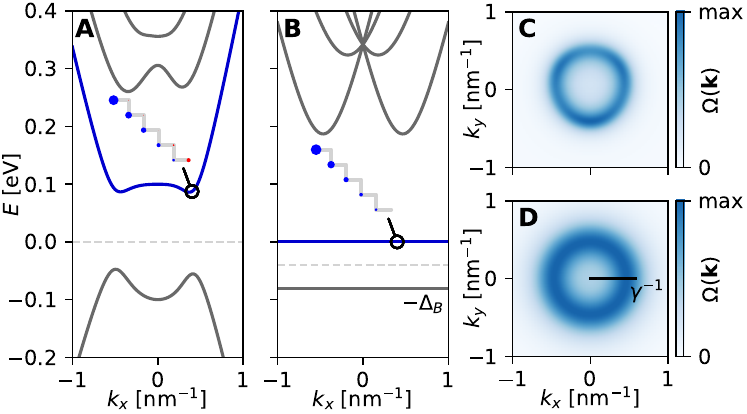}
    \caption{The single particle bands of realistic and ideal R$N$G.  Band structures of the (\textbf{A}) realistic ($h_{\text{real}}$) and (\textbf{B}) ideal ($h_{\text{flat}}$) single-particle models for R$5$G.
    The first conduction band is highlighted in blue.
    Insets illustrate the layer- and sublattice-resolved densities of a representative state.
    Berry curvature $\Omega(\bk)$ of the first conduction band of (\textbf{C}) $h_{\text{real}}$ and (\textbf{D}) $h_{\text{flat}}$.
    }
    \label{fig:model}
\end{figure}

To set the stage, Fig.~\ref{fig:model}A shows the band structure of $h_{\text{real}}$, the full single-particle model for R5G near the $K$ valley, incorporating all microscopic terms at an experimentally relevant displacement field (for full model details, see supplemental material (SM)~\cite{supp}).
For concreteness, we focus primarily on $N=5$ layers in this work, but our analytical framework can also be applied to other $N$.
We are interested in physics at small electron doping of R$5$G/hBN~\cite{lu2024fractional},
where only the first R5G conduction band (highlighted in blue) is partially occupied.
This band consists mainly of states on the top layer (furthest from hBN) and $A$ sublattice, as illustrated in the inset.
In the regime of interest, the flat band bottom induces a spin-valley-polarized phase which can be understood via Stoner flavor-ferromagnetism.
We take this to be the starting point for our theoretical analysis.
Time reversal symmetry is thus broken from the outset, and
 we henceforth focus on a single spin and the $K$ valley.

The ideal limit of R$N$G consists of two ingredients:
a simplified single-particle Hamiltonian obtained by neglecting certain terms in the full R$N$G model, and a short-range interaction term.

The single-particle part of the Hamiltonian is given by
\begin{equation}
h_{\text{flat}} =
\begin{pmatrix}
0 & \mathcal{D}^\dagger \\
\mathcal{D} & -\Delta_B\mathbb{I}
\end{pmatrix}
,\;\;
\begin{array}{l}
\mathcal{D}_{\ell,\ell+1} = -t_\perp\\
\mathcal{D}_{\ell,\ell} = -2iv\partial_{\bar{z}} \text{ for } \ell\neq N-1\\
\mathcal{D}_{N-1,N-1} = 0
\end{array}
\end{equation}
where $h_{\text{flat}}$ is written in block form with respect to the sublattice.
Here, $\mathcal{D}$ is an $N\times N$ matrix acting in layer space, $\ell\in[0,N-1]$ is the layer index (with $\ell=0$ the top-most layer), $\Delta_B>0$, $\mathbb{I}$ the identity matrix, $v=10^{6}$m/s and $t_\perp=0.38$eV.
All other unspecified matrix elements of $\mathcal{D}$ are zero.
Throughout, we often use holomorphic coordinates $z= x+iy$, $\bar{z}=z^*$, $\partial_{\bar{z}}=\frac{1}{2}(\partial_{x}+i\partial_y)$, where $(x,y)$ is the continuous 2D position of the electron.

Notice that $\mathcal{D}$ is ``missing'' the final diagonal element, $\mathcal{D}_{N-1,N-1}=0$.
Had this element instead been $-2iv\partial_{\bar{z}}$, then $h_{\text{flat}}$ would describe the standard nearest-neighbor model for R$N$G expanded near the $K$ point, with sublattice potential $\Delta_B$ playing a similar role as a displacement field in breaking inversion symmetry.
Setting $\mathcal{D}_{N-1,N-1}=0$ amounts to decoupling the $A$ and $B$ sublattices on the bottom layer.
In this work, we take this limit because it enables a dramatic level of analytic progress in understanding the interacting ground states.

With $\mathcal{D}_{N-1,N-1}=0$, the operator $\mathcal{D}$ supports a band of exact zero modes, $\mathcal{D}|\varphi_{\bk}\rangle=0$.
They take the form of plane waves
\begin{equation}
\langle\br,\ell|\varphi_{\bk}\rangle\equiv\varphi_{\bk,\ell}(\br) = e^{i\bk\cdot\br} s_{\bk,\ell}
\label{eq:varphik_def}
\end{equation}
with layer pseudospinor components
\begin{equation}
s_{\bk,\ell} = N_{\bk}\,[\gamma(k_x+ik_y)]^\ell,\;\;\;
N_\bk = \left[\textstyle\sum_{\ell=0}^{N-1}|\gamma \bk|^{2\ell}\right]^{-1/2}
\label{eq:skl}
\end{equation}
where $\gamma\equiv v/t_\perp\approx 1.73$nm.
As a result, $h_{\text{flat}}$ admits two exactly flat sublattice-polarized bands, $|\varphi_\bk^A\rangle = (|\varphi_\bk\rangle,0)^{\mathsf{T}}$ and $|\varphi_\bk^B\rangle = (0,|\tilde{\varphi}_\bk\rangle)^{\mathsf{T}}$, where $\mathcal{D}^\dagger|\tilde{\varphi}_\bk\rangle=0$ is the adjoint zero mode.
The $-\Delta_B$ term is introduced to shift the $\{|\varphi^B_\bk\rangle\}$ band down in energy, rendering it fully filled and inert at charge neutrality.  This allows us to focus on the physics at small electron filling of the zero-energy $\{|\varphi_\bk^A\rangle\}$ band.

Fig.~\ref{fig:model}B shows the band structure of $h_{\text{flat}}$, featuring the exactly flat band at $E=0$.
We also compute the Berry curvature,
 $\Omega(\bk) = 2\textrm{Im} \langle\partial_{k_{x}}u_{\bm{k}}|\partial_{k_{y}}u_{\bm{k}}\rangle$,
where $|u_{\bk}\rangle\equiv e^{-i\bk\cdot\br}|\varphi_\bk\rangle$.
Fig.~\ref{fig:model}(C,D) show $\Omega(\bk)$ for the first conduction band of $h_{\text{real}}$ and $h_{\text{flat}}$, which both exhibit qualitatively similar ring-like structures.

In fact, the pseudospinor in Eq~\ref{eq:skl} has been previously proposed as an approximate description of R$N$G at $|\bk|\lesssim\gamma^{-1}$~\cite{soejima2024anomalous,tan2024parent,bernevig2025berry,han2025exact,desrochers2025electronic} (taking $\mathcal{D}^\dagger\mathcal{D}$ recovers a model derived in the SM of Ref~\cite{han2025exact}, which is equivalent for $N=2$ to the model of Ref~\cite{soejima2025jellium} up to dispersion).
In $h_{\text{flat}}$, this pseudospinor is exact for all $\bk$ and the band is perfectly flat.
The insets of Fig.~\ref{fig:model}(A,B) confirm that near the band edge of $h_{\text{real}}$, the pseudospinor structures of the two are similar: the primary difference being that $h_{\text{real}}$ is only approximately polarized to the $A$ sublattice, whereas $h_{\text{flat}}$ is fully polarized.
{ In the SM~\cite{supp}, we quantify this by computing the overlap between the real and ideal pseudospinors.}

We caution that this flat band is \emph{not} to be regarded as a conventional flat Chern band, as $0\leq |\bk|<\infty$ is unbounded.  To explain the emergence of $C=1$ physics at small densities in R$N$G/hBN, Chern \emph{mini}bands must be formed from this parent flat band.
Previous studies on R$5$G/hBN have shown numerically that electron interactions are essential for this reconstruction, yielding a $C=1$ self-consistent HF band~\cite{dong2024theory,zhou2024fractional,dong2024anomalous,guo2024fractional,kwan2025moire}.
Our goal { in this section} is to derive
{ the microscopic mechanism driving this process analytically, thereby clarifying the origin of the Chern band.}

To this end, the final essential ingredient of the model is a short-range interaction term.
The full ideal Hamiltonian is
\begin{equation}
H_{\text{ideal}}=H_{\text{flat}}+H_{V_0},\qquad H_{V_0}=V_0\sum_{i<j}\delta(\hat{\br}_i-\hat{\br}_j)
\end{equation}
written in first-quantized form, where $\hat{\br}_i$ is the continuous $(x,y)$ position operator of particle $i$, $V_0>0$, $\delta(\br)$ is the 2D Dirac delta, and $H_{\text{flat}}$ is the many-body version of $h_{\text{flat}}$.
Note that contact interaction is nontrivial because the band carries a layer pseudospin, allowing two electrons with different layer indices to coincide in their $(x,y)$ coordinates.
We focus on small electron density above charge neutrality
and assume a large single-particle band gap of $h_{\text{flat}}$ relative to interactions,
so that only the $\{|\varphi_{\bk}^A\rangle\}$ band is relevant.

{
We may also incorporate dispersion to this model.
Specifically, we will later consider
\begin{equation}
H_{\text{disp-ideal}} = H_{\text{ideal}}+H_{\text{disp}}
\label{eq:Hdispideal}
\end{equation}
where $H_{\text{disp}}$ is the many-body version of
\begin{equation}
h_{\text{disp}}=\begin{pmatrix}
\mathcal{E}(-i\bm{\nabla})\mathbb{I} & 0 \\
0 & -\mathcal{E}(-i\bm{\nabla})\mathbb{I}
\end{pmatrix}
\label{eq:hdisp}
\end{equation}
which preserves the sublattice-polarized zero-mode subspace $\{|\varphi^A_{\bk}\rangle\}$, while assigning it a positive kinetic energy $\mathcal{E}(\bk)>0$.
We leave the precise form of $\mathcal{E}(\bk)$ unspecified for now.
This dispersion will be crucial for regularizing the model, enabling numerical approaches and connecting with realistic R$N$G/hBN.

We first develop a real-space formalism for exact ground states of $H_{\text{ideal}}$.
As we will show, $H_{\text{ideal}}$ hosts an infinite manifold of exact zero-energy ground states with rich analytic structure.
Introducing dispersion via $H_{\text{disp}}$ lifts this degeneracy.
While exact zero-energy ground states no longer exist with dispersion, the formalism developed for $H_{\text{ideal}}$ provides a natural framework for understanding numerical HF results of $H_{\text{disp-ideal}}$, which we will show connects smoothly to those of real R$5$G/hBN.
}

In what follows, we start from a completely general many-body wavefunction and successively impose physical constraints, ultimately clarifying the origin of $C=1$ physics.
{
Unlike most prior treatments of ideal Chern bands which work primarily in momentum space~\cite{Liu_review_2023,parameswaran2013fractional,roy2014band,ledwith2020fractional,wang2021exact,ledwith2022family,wang2022hierarchy,dong2023many,wang2023origin} (though see \cite{ledwith2023vortexability,Okuma,Estienne2023,fujimotohigher}), our analysis is carried out entirely in real space.
This approach offers direct physical intuition and makes the role of interactions transparent.
}

\subsection{ Exact ground states}
\label{sec:exactgroundstates}

To begin, consider a general single-particle state in the zero-energy manifold of $h_{\text{flat}}$.
Such states are of the form $(|\psi\rangle,0)^{\mathsf{T}}$ satisfying the zero-mode condition $\mathcal{D}|\psi\rangle=0$.
We henceforth ignore the vanishing $B$-sublattice component and work only with $|\psi\rangle$.
In real space, with $\psi_{\ell}(\br)\equiv\langle \br,\ell|\psi\rangle$, $\mathcal{D}|\psi\rangle=0$ implies the recurrence relation $\psi_{\ell+1}(\br)=-2i\gamma\partial_{\bar z}\psi_{\ell}(\br)$.
Iterating this gives
\begin{equation}
\psi_{\ell}(\br)=(-2i\gamma\partial_{\bar z})^{\ell}\psi_0(\br)
\label{eq:psil_iterated}
\end{equation}
so that all $\ell>0$ layer components are uniquely determined from $\psi_0(\br)$ by successive antiholomorphic derivatives $\partial_{\bar{z}}$.
The zeroth component $\psi_0(\br)$ thus serves as a ``generator'' of the full layer pseudospinor structure.  It may be chosen freely, provided the resulting $\psi_{\ell}(\br)$ defines a normalizable wavefunction.
Unless stated otherwise, overall normalization factors in wavefunctions are omitted for brevity.

Extending to the many-body case, let $\Psi_{\{\ell\}}(\{\br\})=\langle\{\br\},\{\ell\}|\Psi\rangle$ denote the $N_e$-electron wavefunction, with electron coordinates $\{\br\}\equiv (\br_1,\dots,\br_{N_e})$ and layer indices $\{\ell\}\equiv (\ell_1,\dots,\ell_{N_e})$.
The zero-mode condition ($\mathcal{D}_j|\Psi\rangle=0$ for all particles $j$) then implies
\begin{equation}
\Psi_{\{\ell\}}(\{\br\})=\left[\textstyle\prod_{j=1}^{N_e}(-2i\gamma\partial_{\bar{z}_j})^{\ell_j}\right]\Psi_{\{0\}}(\{\br\})
\label{eq:manybody_iteration}
\end{equation}
so the full wavefunction is determined entirely by $\Psi_{\{0\}}$, the component with all electrons on layer $0$.

While all such states are zero-energy eigenstates of $H_{\text{flat}}$ by construction, they generically acquire positive energy under $H_{V_0}$.
To see this, consider a two-particle wavefunction $\Psi_{\ell_1\ell_2}(\br_1,\br_2)$.
By antisymmetry, the zeroth component has a node as $\delta\br\equiv \br_1-\br_2\rightarrow 0$, and can be expanded as
\begin{equation}
\Psi_{00}(\br+\delta\br,\br)=c_1(\br)\delta z+c_2(\br)\delta\bar{z} + O(\delta\br^2)
\end{equation}
for some smooth $c_{1,2}(\br)$.
The $\Psi_{10}$ component is then
\begin{equation}
\Psi_{10}(\br+\delta\br,\br)\propto \partial_{\bar{z}_1}\Psi_{00}= c_2(\br)+O(\delta\br)
\end{equation}
which remains finite as $\delta\br\rightarrow 0$, thereby producing an energy cost under $V_0$ unless $c_2(\br)=0$.
Thus, short-range interactions penalize the presence of antichiral $(\delta\bar{z})$ nodes in $\Psi_{00}$.
More generally, the $\Psi_{\ell_1\ell_2}$ components penalize terms in $\Psi_{00}$ of the form $(\delta\bar{z})^{\ell_1+\ell_2}$.
This conclusion generalizes to every two-particle node in the full $N_e$-body wavefunction $\Psi_{\{0\}}$.

Therefore, a sufficient condition for zero contact energy across all layer configurations is that all two-particle zeros are chiral.
Equivalently, the many-body wavefunction (on the infinite plane) factors as
\begin{equation}
\Psi_{\{\ell\}}(\{\br\})=\left[\textstyle\prod_{i<j}(z_i-z_j)\right]\Xi_{\{\ell\}}(\{\br\})
\label{eq:psijastrow}
\end{equation}
where $\Xi$ is smooth, bosonic (symmetric under interchange), and satisfies the zero-mode condition $\mathcal{D}_j|\Xi\rangle=0$ (i.e. satisfies Eq~\ref{eq:manybody_iteration}).
Because $\mathcal{D}$ contains only $\partial_{\bar{z}}$, it commutes with the Jastrow factor $\prod(z_i-z_j)$, ensuring $\mathcal{D}_j|\Psi\rangle=0$.
As all layer components share the same Jastrow factor, every pair of particles has a guaranteed zero.
Since $\Xi_{\{0\}}$ can be chosen freely, this construction yields an infinite family of exact zero-energy ground states of $H_{\text{ideal}}$.
In ideal R$N$G, this factorization is only violated by the appearance of $(\delta\bar{z})^{2N-1}$ or higher-order antichiral terms in the two-particle expansion of $\Psi_{\{0\}}$.

To further investigate the structure of the ground state manifold and connect with HF analyses, we look for solutions in the form of Slater determinants,
\begin{equation}
\Psi^{}_{\{\ell\}}(\{\br\})=\det_{jm}[\langle\br_j\ell_j|\psi_m\rangle]=\det_{jm}[\psi_{m,\ell_j}(\br_j)]
\end{equation}
where $\{|\psi_m\rangle\}_{m\in[1,N_e]}$ are the occupied single-particle orbitals.
Focusing on the zeroth component, let
$\bm{C}(\br)\equiv(\psi_{0,0}(\br),\dots,\psi_{N_e,0}(\br))^{\mathsf{T}}$
denote the column vector of orbitals evaluated at $\br$, such that $\Psi_{\{0\}}(\{\br\})=\det[\bm{C}(\br_1),\cdots,\bm{C}(\br_{N_e})]$.
Setting $\br_1=\br+\delta\br$ and $\br_2=\br$, the Slater determinant expands as $\delta\br\rightarrow 0$,
\begin{equation}
\begin{split}
\Psi^{}_{\{0\}}(\{\br\})&\sim \delta z\det[\partial_z \bm{C}(\br),\bm{C}(\br),\bm{C}(\br_3),\dots] \\
&\;\;\;\;+ \delta\bar z\det[\partial_{\bar z}\bm{C}(\br),\bm{C}(\br),\bm{C}(\br_3),\dots].
\end{split}
\end{equation}
For the state to have zero contact energy, this node must be chiral: the coefficient of $\delta\bar{z}$ must vanish for all $\br$ and $\br_{3\dots N_e}$.
This occurs if and only if $\partial_{\bar z}\bm{C}(\br)$ is parallel to $\bm{C}(\br)$ for all $\br$~\cite{supp}, so that the second determinant vanishes.
In terms of $\psi_{m,0}(\br)$, this condition implies $\partial_{\bar z}\psi_{m,0}(\br)=\lambda(\br)\psi_{m,0}(\br)$ with an $m$-independent function $\lambda(\br)$.
From this, it follows that all orbitals must factorize as $\psi_{m,0}(\br)=f_m(z)\xi_0(\br)$, where $f_m(z)$ is holomorphic in $z$, such that
$\partial_{\bar{z}}\psi_{m,0}(\br)=[\partial_{\bar{z}}\log \xi_0(\br)]\psi_{m,0}(\br)$.
Notice that this also leads to all higher-order two-body antichiralities $(\delta\bar{z})^{n>1}$ vanishing, since all $\partial_{\bar z}^n\bm{C}(\br)$ remain parallel to $\bm{C}(\br)$.
The $\ell>0$ components follow from the zero-mode equation, giving
\begin{equation}
\begin{split}
\psi_{m,\ell}(\br)&=f_m(z)\xi_{\ell}(\br)\\
\xi_{\ell}(\br)&=(-2i\gamma\partial_{\bar z})^{\ell}\xi_0(\br)
\end{split}
\label{eq:single_psim_factor}
\end{equation}
where $f_m(z)$ and $\xi_0(\br)$ can be chosen freely.
Because $\xi_{\ell}(\br)$ is $m$-independent, it factors out of the determinant, resulting in
\begin{equation}
\Psi^{}_{\{\ell\}}(\{\br\})=\det_{jm}\left[f_m(z_j)\right]\prod_{i=1}^{N_e}\xi_{\ell_i}(\br_i)
\label{eq:psidet_factor}
\end{equation}
which holds for \emph{all} Slater determinant ground states.

{
This Slater determinant has a natural physical interpretation.
The holomorphic determinant is reminiscent of electrons in the lowest Landau level (LLL), while the accompanying single-particle factor aligns each electron's layer pseudospin with the local texture $\bm{\xi}(\br)=(\xi_0(\br),\dots,\xi_{N-1}(\br))^{\mathsf{T}}$.
Indeed, we can perform a rotation to the $\hat{\bm{\xi}}(\br)\equiv\bm{\xi}(\br)/|\bm{\xi}(\br)|$ local frame (\`a la the ``adiabatic approximation''~\cite{morales2024magic,li2025variational,shi2024adiabatic,guerci2025layer,sommer2025idealopticalfluxlattices}) by writing $\psi_{\ell}(\br)=\phi(\br)\hat{\xi}_{\ell}(\br)$.
The zero-mode equation $\mathcal{D}|\psi\rangle=0$ then leads to
\begin{equation}
\begin{split}
\left[\hat{\xi}_{\ell}(\br)\partial_{\bar{z}}+\partial_{\bar{z}}\hat{\xi}_{\ell}(\br)+\frac{1}{2i\gamma}\hat{\xi}_{\ell+1}(\br)\right]\phi(\br)=0
\end{split}
\end{equation}
which must vanish for all $\ell\in[0,N-2]$.
For $\xi_{\ell}(\br)$ satisfying Eq~\ref{eq:single_psim_factor}, this equation becomes $\ell$-independent,
\begin{equation}
\begin{split}
(\partial_{\bar{z}}-\partial_{\bar{z}}\log|\bm{\xi}(\br)|)\phi(\br)=0.
\end{split}
\end{equation}
Interpreting $\partial_{\bar{z}}\log|\bm{\xi}(\br)|$ as an emergent gauge field,
this precisely describes a Dirac electron in the LLL of an inhomogeneous magnetic field $-B_{\text{eff}}(\br)\hat{z}$~\cite{supp},
\begin{equation}
B_{\text{eff}}(\br)=-\nabla^2\log|\bm{\xi}(\br)|,
\label{eq:Beff}
\end{equation}
whose exact eigenstates are given by Aharonov and Casher~\cite{aharonov1979ground}: $\phi_{\text{AC}}(\br)=f(z)|\bm{\xi}(\br)|$ for holomorphic $f(z)$.
In the unrotated frame, $\psi_{\ell}(\br)=\phi_{\text{AC}}(\br)\hat{{\xi}}_{\ell}(\br)=f(z)\xi_{\ell}(\br)$ recovers Eq~\ref{eq:single_psim_factor}.
Thus, Eq~\ref{eq:psidet_factor} precisely describes a determinant of Dirac electrons in the LLL of $B_{\text{eff}}(\br)$, with layer pseudospin locked to $\hat{\bm{\xi}}(\br)$.

Intuitively, the origin of $B_{\text{eff}}$ is the real-space Berry phase of the layer pseudospin texture, which mimics the Aharonov-Bohm phase of a real magnetic field.
By creating an emergent magnetic field through this pseudospin texture, electrons can inherit the chiral nodes from the LLL, which are energetically favorable with short-range interactions.

Note that this connection to the LLL follows purely as a consequence of the Slater determinant and chiral nodal structure imposed by $V_0$ interactions.
No reference to any momentum space structure or $C=1$ band was made,
and the magnetic field $B_{\text{eff}}(\br)$ need not be periodic for this to be true.
}

\subsection{ Layer pseudospin skyrmion lattices}
\label{sec:layerskyrmionlattices}
We now further restrict the set of relevant ground states by requiring translation symmetry.
For any $N_e\geq 2$, we can prove that no Slater determinant ground state retains continuous translation invariance~\cite{supp}.
They can, however, be chosen to respect discrete lattice translations generated by a pair of primitive vectors $\{\bm{a}_1,\bm{a}_2\}$.
This requires a spatially periodic layer pseudospin texture, generating a $B_\text{eff}(\br)=B+\delta B(\br)$ with a positive average component $B>0$ to accommodate a finite density of LLL electrons, and a periodic modulation $\delta B(\br)$ that averages to zero.
As we shall see, such configurations correspond precisely to skyrmion lattices of the $N$-component layer-pseudospin texture.

To construct periodic solutions, let us write
\begin{equation}
\xi_\ell(\br)=e^{-\frac{1}{4}l_B^{-2}|\br|^2}\chi_\ell(\br)
\label{eq:chidef}
\end{equation}
with $l_B\equiv1/\sqrt{B}$, so that Eq~\ref{eq:psidet_factor} becomes
\begin{equation}
\Psi^{}_{\{\ell\}}(\{\br\})=
\Phi_{\text{LLL}}^{}(\{\br\}) \prod_i\chi_{\ell_i}(\br_i)
\label{eq:LLLchi}
\end{equation}
where
\begin{equation}
\Phi^{}_{\text{LLL}}(\{\br\})=\det_{jm}[f_m(z_j)] e^{-\frac{1}{4}l_B^{-2}\sum_i|\br_i|^2}
\end{equation}
describes a symmetric-gauge LLL Slater determinant in a \emph{uniform} magnetic field $B>0$.
For example, choosing $f_m(z)=z^m$ gives $\det_{jm}[z_j^m]=\prod_{i<j}(z_i-z_j)$, the Vandermonde determinant, which yields the filled LLL ($\nu=1$ quantum Hall state); more generally, other choices yield determinants at partial filling of the LLL.

Note that because of the Gaussian factor in Eq~\ref{eq:chidef}, the zero-mode condition for $\chi$ becomes
\begin{equation}
\begin{split}
\chi_{\ell+1}(\br)&=-2i\gamma (\partial_{\bar{z}}-z/4l_B^2) \chi_{\ell}(\br) \\
&\equiv (-\sqrt{2}\gamma /l_B)\bar{A}^\dagger \chi_{\ell}(\br)
\end{split}
\label{eq:chizeromode}
\end{equation}
where we have defined
\begin{equation}
 \bar{A}^\dagger = i(2l_B\partial_{\bar{z}}-z/2l_B)/\sqrt{2},
\end{equation}
the conjugate LL raising operator.
Thus, all $\chi_{\ell>0}(\br)$ are fully determined from $\chi_0(\br)$ via applications of $\bar{A}^\dagger$.

A periodic $|\Psi\rangle$ can then be constructed for any $|\Phi^{}_{\text{LLL}}\rangle$ that is \emph{magnetic} periodic,
\begin{equation}
\Phi^{}_{\text{LLL}}(\{\br+\bm{a}\})=e^{i\bm{K}\cdot\bm{a}}\Phi^{}_{\text{LLL}}(\{\br\})\textstyle\prod_j e^{\frac{i}{2}l_B^{-2}\bm{a}\times\br_j}
\end{equation}
provided that
\begin{equation}
\chi_\ell(\br+\bm{a})=e^{i\bm{q}\cdot\bm{a}}\chi_\ell(\br)e^{-\frac{i}{2}l_B^{-2}\bm{a}\times\br}
 \label{eq:magtrans}
\end{equation}
satisfies the inverse magnetic periodicity, for some $\bm{K}$ and $\bm{q}$.
Their product is then periodic under ordinary (non-magnetic) lattice translations with total Bloch momentum $\bm{K}+N_e\bm{q}$.

Let us now determine the valid solutions for $\chi_{\ell}(\br)$.
The only degree of freedom is the zeroth component $\chi_0(\br)$.
A useful analogy is to view $\prod_i\chi_{0}(\br_i)\equiv X_{\{0\}}(\{\br\})$ as a fictional wavefunction of bosons, which have condensed into a single state $\chi_0(\br)$.
Eq~\ref{eq:magtrans} then implies these bosons see an opposite uniform magnetic field $-B$.
Magnetic periodic solutions exist if and only if the unit cell encloses an integer number of flux quanta, $l_B^{-2}|\bm{a}_1\times\bm{a}_2|=2\pi N_{\phi}$, with $N_{\phi}$ a positive integer.
They correspond to $\chi_0$ being Abrikosov \emph{anti}vortex lattices with net $N_{\phi}$ antivortices per unit cell.

{
\begin{figure}[t]
    \centering
    \includegraphics[width=0.8\linewidth]{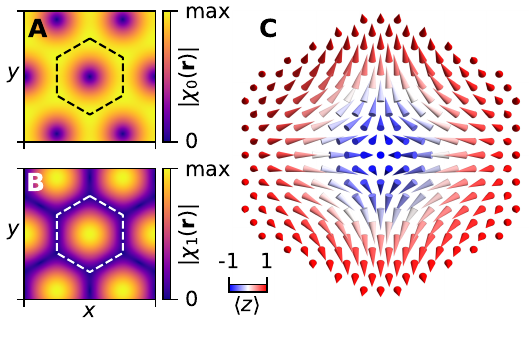}
    \caption{
    Visualizing the layer pseudospin skyrmion lattice in ideal R$2$G.
    The amplitude of (\textbf{A}) $|\chi_0(\br)|$ and (\textbf{B}) $|\chi_1(\br)|$, for the example $\chi_0=w_0$ defined in the main text.
    As discussed in the main text, $\chi_0(\br)$ can be viewed as an Abrikosov antivortex lattice with a node at each skyrmion core.
    (\textbf{C}) The direction of the normalized pseudospinor $\hat{\bm{\chi}}(\br)\equiv\bm{\chi}(\br)/|\bm{\chi}(\br)|$ on the Bloch sphere, with the color indicating the $z$-component.
    Because $\hat{\bm{\chi}}(\br)$ wraps the Bloch sphere once per unit cell, this texture realizes a skyrmion lattice with winding number $N_\phi=1$.
    }
    \label{fig:skyrmion}
\end{figure}
Consider the simplest case $N_{\phi}=1$.
In this case, a complete basis for functions satisfying Eq~\ref{eq:magtrans} is given by (the complex conjugate of) the Landau level (LL) magnetic Bloch states, which we denote by $\{w_n^{(\bq)}(\br)\}$, where $n=0,1,\dots$ is the LL index.
We first define
\begin{equation}
    w_0^{(\bq)}(\br) = \left[e^{-\frac{i}{2} \bar{q}' z} {\sigma(z-il_B^2 q')} e^{-\frac{1}{4}l_B^{-2}|\br|^2}\right]^*
\end{equation}
 where $q'=q-(b_1+b_2)/2$, $\bm{b}_{i}$ are reciprocal lattice basis vectors ($\bm{b}_i\cdot\bm{a}_j=2\pi\delta_{ij}$), $q = q_x+i q_y$, $b_i = b_{ix}+ib_{iy}$.
 Here, $\sigma(z)$ is the (modified~\cite{haldaneModularinvariantModifiedWeierstrass2018}) Weierstrass sigma function~\cite{supp} which satisfies
 $\sigma(z+a) = \eta_{\bm a}e^{\frac{1}{2}l_B^{-2}\overline{a}(z + a/2)}\sigma(z)$, where $\bm{a}=n_1\bm{a}_1+n_2\bm{a}_2$ ($n_{1,2}$ integers) is a lattice vector, $a = a_x + i a_y$, and $\eta_{\bm a} = 1$ $(-1)$ if $\bm{a}/2$ is (not) a lattice vector.
 From this property, one can verify
 magnetic Bloch translation symmetry (Eq~\ref{eq:magtrans}).
We can then define higher LLs as $w_n^{(\bm{q})}(\br)=(n!)^{-\frac{1}{2}}(\bar{A}^\dagger)^nw_0^{(\bm{q})}(\br)$.
The most general solution to Eq~\ref{eq:magtrans} is a linear combination $\sum_n c_n w_n^{(\bq)}(\br)$.
These different solutions correspond to different antivortex lattices and will result in different periodic magnetic field profiles $\delta B(\br)=-\nabla^2\log |\bm{\chi}(\br)|$.
Here, $\bm{q}$ determines both the magnetic Bloch momentum and the overall translational degree of freedom of the antivortex lattice.  {For example, since $\sigma(a)=0$, the antivortices of $w_0^{(\bq)}$ are located at the points
$z = il_B^2q'+a$.
}
We will often suppress the $\bm{q}$ superscript when it is unimportant.

We illustrate the pseudospin texture corresponding to the simplest solution, $\chi_0(\br)=w_0(\br)$ on a hexagonal lattice, in Fig~\ref{fig:skyrmion}.
Note that this choice $\chi_0=w_0$ corresponds to the simplest ``textbook'' Abrikosov lattice wavefunction~\cite{tinkham2004introduction}.
The zero-mode condition (Eq~\ref{eq:chizeromode}) then fixes all higher components as $\chi_{\ell}(\br)\propto w_\ell(\br)$.
For visualization purposes, we consider R$2$G here so that
the normalized layer pseudospinor $\hat{\bm{\chi}}(\br)=[\chi_0(\br),\chi_1(\br)]^{\mathsf{T}}/|\bm{\chi}(\br)|$ maps each real-space position to a direction on the Bloch sphere.
Fig~\ref{fig:skyrmion}A shows $|\chi_0(\br)|$, the Abrikosov lattice wavefunction, and Fig~\ref{fig:skyrmion}B shows $|\chi_1(\br)|$.
Fig~\ref{fig:skyrmion}C shows the direction in the Bloch sphere of $\hat{\bm{\chi}}(\br)$ over a unit cell.
This pseudospin texture covers the Bloch sphere exactly once per unit cell, and corresponds to a topologically nontrivial skyrmion lattice with $N_{\phi}=1$.
To see this, notice that $\chi_0(\br)=w_0(\br)$ has a single zero per unit cell at the vortex core, corresponding to where $\hat{\bm{\chi}}\propto(0,1)^\mathsf{T}$ points to the south pole, so it covers the sphere once.

More generally, it can be proven that any $\chi_\ell(\br)$ satisfying Eq~\ref{eq:magtrans} describes a topologically non-trivial skyrmion texture in $\mathbb{C}P^{N-1}$ with winding number~\cite{supp}
\begin{equation}
\frac{1}{2\pi i} \int_{\text{unit cell}} \Tr P [\partial_x P, \partial_y P] d^2 \br = N_{\phi},
\label{eq:windingmaintext}
\end{equation}
where $P(\br) = \hat{\chi}(\br) \hat{\chi}(\br)^\dagger$ is the projector onto the direction of $\hat{\chi}(\br)$. For $N=2$, $P(\br) = \frac{1}{2}(1+\bm{n}(\br) \cdot \vec{\sigma})$ with $\vec{\sigma}=(\sigma_x,\sigma_y,\sigma_z)$ Pauli matrices, defines a map to the Bloch sphere and Eq.~\ref{eq:windingmaintext} reduces to the integral of the familiar Pontryagin density $\frac{1}{4\pi} \bm{n} \cdot (\partial_x \bm{n} \times \partial_y \bm{n})$.
For ideal R$N$G with $N>2$, $\mathbb{C}P^{N-1}$ describes a $2N-2$ dimensional ``Bloch sphere'' manifold that can no longer be as easily visualized.
Nevertheless, it will be instructive to view $|\chi_0|$ as a measure of the skyrmion profile, with its zeros marking the skyrmion ``cores'' at which the local pseudospinor is orthogonal to the top layer (the higher-dimensional analogue of the ``south pole'' in Fig.~\ref{fig:skyrmion}C), though the full texture in $\mathbb{C}P^{N-1}$ is obviously more intricate.
This
provides a clear physical picture for the potential emergence of LLL physics: it arises from an emergent periodic magnetic field generated by a pseudospin skyrmion lattice.

}

{
\subsection{ Ansatz wavefunction with dispersion}
\label{sec:ansatz}
Thus far, we have identified an infinite class of periodic Slater determinants that are all exact zero-energy ground states of $H_{\text{ideal}}$.
The underlying source of this infinite degeneracy is that the single-particle band (Fig~\ref{fig:model}B) remains flat for all $|\bk|$.
We now consider the effect of band dispersion, incorporated by $h_{\text{disp}}$ (Eq~\ref{eq:hdisp}), which assigns a kinetic energy $\mathcal{E}(\bk)$ while preserving the zero-mode subspace $\{|\varphi^A_{\bk}\rangle\}$.
A physically realistic $\mathcal{E}(\bk)$ will energetically penalize wavefunctions with large $|\bk|$ components, i.e. those that vary quickly in real space.

In the presence of dispersion, the Slater determinants identified earlier are no longer exact eigenstates.
However, we can still seek the lowest-energy state within the variational subspace of periodic Slater determinant ground states of $H_{\text{ideal}}$.
In this section, we introduce a particular ansatz wavefunction at filling of one electron per unit cell (density $\rho=1/|\bm{a}_1\times\bm{a}_2|$), which will be shown to be nearly optimal with dispersion.
It will provide the basis for our understanding of the numerical HF results and collective excitations in later sections.

The ansatz wavefunction is, for a given $\{\bm{a}_1,\bm{a}_2\}$, \begin{equation}
\Psi_{\{\ell_j\}}^{\text{ansatz}}(\{\br\})=\Phi_{\text{LLL}}^{\nu=1}(\{\br\})\prod_j \chi_{\ell_j}(\br_j)\\
\label{eq:final_psi_nu1}
\end{equation}
where
\begin{equation}
\Phi_{\text{LLL}}^{\nu=1}(\{\br\}) = \prod_{i<j}(z_i-z_j)\prod_ie^{-\frac{1}{4}l_B^{-2}|\br_i|^2}
\end{equation}
and with the choice $\chi_{\ell}(\br)=\chi_\ell^{\text{ansatz}}(\br)$,
\begin{equation}
\chi_\ell^{\text{ansatz}}(\br)= \sqrt{\ell!}\left(-\sqrt{2}\gamma/l_B\right)^{\ell} w_{\ell}^{(\bq_0)}(\br).
\end{equation}
with $l_B^2=|\bm{a}_1\times\bm{a}_2|/2\pi$, and for any $\bm{q}_0$.
This simply corresponds to the fully filled $\nu=1$ LLL combined with the simplest antivortex lattice $\chi_0^{}(\br)=w_0(\br)$ with $N_{\phi}=1$ (as in Fig~\ref{fig:skyrmion}), and $\bm{q}_0$ representing an overall translation.
This corresponds to the fictional wavefunction of bosons $X_{\{0\}}(\{\br\})=\prod_i w_0(\br_i)$ all condensing into the conjugate LLL.
This wavefunction describes a fully filled $C=1$ Chern band (inherited from $\Phi^{\nu=1}_{\text{LLL}}$, since the single-particle factor $\prod \chi$ does not affect band topology), and is an exact zero-energy eigenstate of $H_{\text{ideal}}$.

We now argue that, within the space of $\bm{a}_{1,2}$-periodic Slater determinants at this density, this ansatz nearly minimizes the kinetic energy $H_{\text{disp}}$.
In the SM~\cite{supp}, we derive the momentum-space representation of $|\Psi^{\text{ansatz}}\rangle$.
The result is that the kinetic energy is given by $E_{\text{kin}}=\sum_{\bk}\mathcal{E}(\bk)n(\bk)$, with a momentum distribution function
\begin{equation}
\begin{split}
n(\bk)=\frac{P(\bk)}{\textstyle\sum_{\bg} P(\bk+\bg)},\qquad
P(\bk)=e^{-\frac{1}{2}l_B^2|\bk|^2}N_{\bk}^{-2}
\label{eq:nk}
\end{split}
\end{equation}
where $\bk$ runs over all momenta, $N_{\bk}$ is given in Eq~\ref{eq:skl}, and the sum over
$\bg=n_1\bm{b}_1+n_2\bm{b}_2$ ($n_{1,2}$ integers) spans the reciprocal lattice.

Notice that $P(\bk)$ is simply a Gaussian multiplied by a $2(N-1)$-th order polynomial, and that $n(\bk)$ is just $P(\bk)$ divided by a $\bm{g}$-periodic normalization factor $\sum_{\bg}P(\bk+\bg)$.
The Gaussian factor implies that this state costs finite kinetic energy and can be well captured numerically with a momentum cutoff $|\bk|< \Lambda \sim O(l_B^{-1})$.
We remark that since $l_B^2=1/(2\pi\rho)$, for small $\rho$, $P(\bk)$ is dominated by the Gaussian and is a monotonically decreasing function of $|\bk|$; at larger $\rho$, the polynomial factor means that $P(\bk)$ can be a non-monotonic function of $|\bk|$.

To see why this nearly minimizes the kinetic energy, notice that at this filling and $N_{\phi}=1$, the LLL factor $\Phi^{\nu=1}_{\text{LLL}}$ is fixed;  the only variational degree of freedom is the antivortex lattice $\chi_0(\br)$.
Since $\{w_n(\br)\}$ form a complete basis for solutions to Eq~\ref{eq:magtrans}, we may expand a generic $\chi_0(\br)=\sum_{n=0}^{\infty} c_n w_n(\br)$.
This yields a state with the same kinetic energy form but with $P(\bk)\rightarrow P_f(\bk)=|f(k)|^2 P(\bk)$, where $f(k)=\sum_{n=0}^{\infty} f_n (k_x+ik_y)^n$ (with $f_n= c_n(-l_B/\sqrt{2})^n/\sqrt{n!}$)~\cite{supp}.
Since any non-constant holomorphic $f$ is unbounded, this will tend to shift the weight of $P_f(\bk)$ towards larger $|\bk|$.
The choice $f=1$ ($\chi_0=w_0$) thus yields a $P_f(\bk)$ that is most concentrated at small $|\bk|$.
Intuitively, $\chi_0=w_0$ is the ``smoothest'' choice of antivortex function, and hence kinetically favorable.
However, in general the \emph{true} minimum, obtained by minimizing $E_{\text{kin}}$ over all possible $f$, will contain small symmetry-allowed admixtures of higher-order terms, though we shall find they are negligibly small in the cases of interest.

This expression also \emph{a posteriori} justifies $N_{\phi}=1$, which maximizes $l_B^2$ so that the Gaussian factor in $P(\bk)$ concentrates weight at small $|\bk|$.
Physically, $l_B$ sets the characteristic lengthscale of variations in the wavefunction, and so a larger $l_B$ (smaller $B$) is kinetically preferable.

We note that this argument always yields (independent of choice of $\chi_\ell(\br)$) a $C=1$ Chern insulator.
Interestingly, this conclusion is independent of detailed features of R$N$G, such as its Berry curvature distribution.
In this regard, $C=1$ is special (as opposed to, say, any other $C>1$).
This reasoning applies to other models with similar holomorphic structure~\cite{tan2024parent,soejima2025jellium,desrochers2025electronic},
predicting that large $V_0$ (compared to band dispersion) should always produce a $C=1$ HF ground state.
This can be viewed as a generalization of the result that an ideal flat Chern $C$ band at filling $\nu=1/C$ hosts exact Chern $1$ ground states under contact interactions~\cite{dong2023many,wang2023origin}: extended to unbounded bands, our result shows that $C=1$ states can arise at \emph{any} filling.

}

\subsection{ Numerical phase diagram}
\label{sec:HFdispideal}

We now confirm our analysis numerically with self-consistent HF theory.
We consider dispersive ideal R$5$G,
$H_{\text{disp-ideal}}$, defined in Eq~\ref{eq:Hdispideal},
with a parabolic dispersion $\mathcal{E}(\bk)=\alpha|\bk|^2$ for simplicity.
Projected to the $\{|\varphi^A_\bk\rangle\}$ band, this model has only two dimensionless parameters:
(i) $\rho/\rho_{\gamma}$, the total density $\rho$ relative to $\rho_{\gamma}\equiv1/(4\pi\gamma^2)$ (the density of a circular Fermi surface with radius $k_F=\gamma^{-1}$) and (ii) $\alpha/V_0$, the ratio of the kinetic to the interaction energy.

For each density $\rho$, we assume the ground state is translation-invariant with respect to a hexagonal lattice $\bm{a}_1=a(1,0),\bm{a}_2=a(\frac{1}{2},\frac{\sqrt{3}}{2})$ with one electron per unit cell, so $a^2=2/(\sqrt{3}\rho)$.
We use self-consistent HF theory to find the lowest energy periodic Slater determinant.
Technical details are described in the SM~\cite{supp}.

We argued above that as $\alpha/V_0\rightarrow 0$, the HF ground state at all densities corresponds to a filled $C=1$ Chern band, approximately described by the ansatz wavefunction Eq~\ref{eq:final_psi_nu1} with $\chi_0(\br)=w_0^{(\bm{q}_0)}(\br)$, degenerate with respect to $\bm{q}_0$ (an overall translation).
A key feature of this ansatz is its distinctive momentum distribution function $n(\bk)$, Eq~\ref{eq:nk}, which can be directly compared with HF results.

\begin{figure}[t]
    \centering
    \includegraphics[width=1\linewidth]{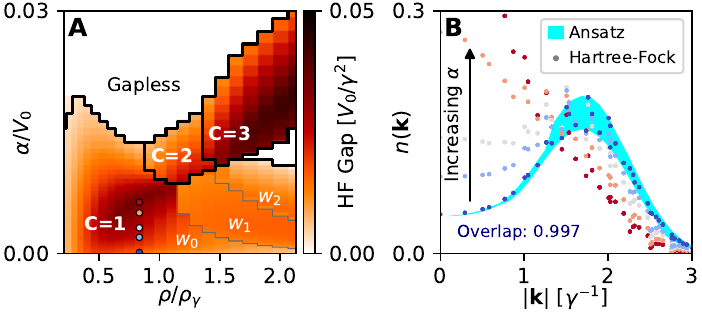}
    \caption{
    Hartree-Fock (HF) phase diagram of dispersive ideal R$5$G with
    parabolic dispersion $\mathcal{E}(\bk)=\alpha|\bk|^2$.
    (\textbf{A}) The phase diagram as a function of $\alpha/V_0$ and density $\rho/\rho_\gamma$.
    Here, $\rho_{\gamma}=1/(4\pi\gamma^2)$ is the density of a circular Fermi surface with radius $k_F=\gamma^{-1}$.
    (\textbf{B}) The momentum distribution function $n(\bk)$ at several points in the phase diagram along $\rho=0.833\rho_\gamma$ [indicated by the colored points in (A)].
    The shaded blue region shows $n(\bk)$ of the ansatz, Eq~\ref{eq:nk}, with the shading width indicating the angular variation.
    As $\alpha\rightarrow 0$, the HF state approaches the ansatz wavefunction Eq~\ref{eq:final_psi_nu1}, with a near-unity per-particle overlap $(|\langle\Psi^{\text{HF}}|\Psi^{\text{ansatz}}\rangle|^2)^{\frac{1}{N_e}}\approx 0.997$.
    }
    \label{fig:ideal}
\end{figure}

Fig~\ref{fig:ideal}A shows the self-consistent HF phase diagram calculated as a function of $\alpha/V_0$ and $\rho/\rho_\gamma$.
At small $\alpha/V_0$, we universally find $C=1$ phases, in agreement with our analysis.
We identify three distinct $C=1$ phases, all described by Eq~\ref{eq:final_psi_nu1} but distinguished by the choice of antivortex lattices $\chi_0$:
the phase as $\alpha/V_0\rightarrow 0$ corresponds to the proposed ansatz $\chi_0=w_0$, while small $\alpha$ also stabilizes phases smoothly connected to $\chi_0=w_1$ and $\chi_0=w_2$~\cite{supp}.
These phases are {all skyrmion lattices, which can be} distinguished by their $C_{3z}$ rotation eigenvalue (we note that the $w_{1}$ and $w_{2}$ phases are ``halo'' anomalous Hall crystals~\cite{desrochers2025electronic}).

{ To get a sense of scale for $\alpha/V_0$ in Fig~\ref{fig:ideal}, we note that the gate-screened Coulomb interaction admits a short-range pseudopotential expansion\cite{TrugmanKivelson1985,pokrovskySimpleModelFractional1985}, $V_{\text{Coul}}(\bm{q})=\frac{e^2 \tanh(|\bm{q}|d)}{2\epsilon_0\epsilon_r|\bm{q}|}\approx V_0+O(d^3)$,
with $V_0=\frac{e^2d}{2\epsilon_0\epsilon_r} \approx 45.2\text{eVnm$^2$}$, where we have used a gate distance $d=25$nm and dielectric constant $\epsilon_r=5$.
A rough parabolic approximation to the realistic R5G dispersion yields $\alpha\approx0.2$eVnm$^2$, giving $\alpha/V_0\approx 0.004$: deep inside the $C=1$ phase in Fig.\ref{fig:ideal}.
However, we emphasize that this comparison is used merely to illustrate the scale of the axes in Fig.~\ref{fig:ideal}, and
should not be taken quantitatively.
}

As $\alpha$ increases further, phases with $C>1$ appear and, in the limit $\alpha\rightarrow\infty$, the system becomes metallic.
These intermediate $C>1$ phases arise from competition between kinetic and interaction energy and do not have representatives in the zero-energy manifold of $H_{\text{ideal}}$.

These $C>1$ phases may be intuitively interpreted in terms of ``Berry curvature rounding'' of the parent Berry curvature (Fig~\ref{fig:model}D) to the first Brillouin zone.
The $C=1$ phase as $\alpha/V_0\rightarrow 0$, interestingly, violates this notion of Berry curvature rounding.
In this limit, $V_0$ is large compared to the kinetic energy, so perturbative arguments based on weakly-gapped Fermi surfaces or high-symmetry points~\cite{dong2024stability,soejima2024anomalous,crepel2025efficient} are not applicable.
From our perspective of a real-space effective magnetic field, however, the prevalence of $C=1$ is perfectly natural since the LLL always carries $C=1$.
{
We note that the inclusion of longer-ranged interactions, such as Coulomb, will favor competing phases such as a trivial Wigner crystal in the low-density limit (by analogy with the two-dimensional electron gas).
}

Finally, Fig~\ref{fig:ideal}B shows the HF momentum distribution function $n(\bk)$ at several points in the dominant $C=1$ phase.
As $\alpha/V_0\rightarrow 0$, the HF $n(\bk)$ approaches that of the ansatz in Eq~\ref{eq:nk}.
We have chosen a value of $\rho$ here such that Eq~\ref{eq:nk} exhibits a non-monotonic dependence on $|\bk|$, which is perfectly reproduced by HF.
Furthermore, for our smallest $\alpha$, the HF ground state has a near-unity overlap with Eq~\ref{eq:final_psi_nu1} (optimized over $\bm{q}_0$), confirming the accuracy of the ansatz wavefunction beyond just $n(\bk)$.
Further technical details are presented in the SM~\cite{supp}.

{
\subsection{Generalizations}
Before continuing, we briefly discuss generalizations of our formalism beyond the specific case studied thus far, as we believe the techniques we have developed will be useful for future theoretical studies.

First, our formalism can be broadly generalized to any continuum Hamiltonian with factorizable spinor structure generalizing Eq~\ref{eq:skl}: $s_{\bk,\ell}=N_{\bk}\tilde{s}_\ell(k)$ where $\tilde{s}_\ell(k)$ is an entire function of $k$.
Our derivation above largely carries over, with the main difference being  Eq~\ref{eq:psil_iterated}, replacing $(-2i\gamma\partial_{\bar{z}})^\ell$ with $\tilde{s}_{\ell}(k\rightarrow-2i\partial_{\bar{z}})$.
This includes several recently proposed jellium-like systems with Berry curvature~\cite{tan2024parent,soejima2025jellium,desrochers2025electronic}.

Second, other types of states can also be addressed.
Our discussion so far has focused primarily on the $N_{\phi}=1$ states, with one skyrmion per unit cell.
More generally, however, skyrmions can rearrange if it is energetically favorable to do so.
For example, Ref~\cite{zhou2025new} discovered $C=1$ phases in R$N$G/hBN, referred to as ``QAHC-$z$'' ($z=2,3$), with $z$-fold enlarged unit cells .
These are naturally described within our formalism by allowing $\chi_0$ to describe LLL Bloch states in the enlarged unit cell with $N_{\phi}=z$.
Since these states have one electron per skyrmion, the total Chern number remains $C=1$.

}

{
\section{Skyrmion lattice in real R$5$G/hBN}
\label{sec:connecting}
We now consider real R5G/hBN { which contains realistic band dispersion, longer-range Coulomb interactions, hBN-induced moir\'e potentials, and deviations from exact holomorphic spinors.}
{ We demonstrate that the microscopic mechanism driving skyrmion lattice formation developed in Sec~\ref{sec:idealsection} remains operative in this  realistic setting}.
We do this by (i) establishing adiabatic continuity to the exact skyrmion lattice of the ideal limit, and (ii) identifying its characteristic signatures in the local density matrix and the layer-resolved charge density.

}

{To demonstrate adiabatic continuity between the ideal limit and fully realistic R$5$G/hBN, we introduce a single-particle Hamiltonian that smoothly interpolates between the two,
\begin{equation}
h_{\lambda}=(1-\lambda) (h_{\text{flat}}+h_{\text{disp}}) + \lambda h_{\text{real}}
\end{equation}
with $0\leq\lambda\leq 1$.
}
In $h_{\text{disp}}$, we set $\mathcal{E}(\bk)=\mathcal{E}_{\text{real}}(\bk)$ to be the dispersion of $h_{\text{real}}$, so that $\lambda=0$ and $\lambda=1$ describe R$5$G bands with identical dispersion, only differing in their pseudospinor structure.
For $0<\lambda<1$, we have confirmed that the first conduction single-particle band evolves continuously with no gap closing~\cite{supp}.

We consider the many-body Hamiltonian
\begin{equation}
H=H_{\lambda}+H_{\text{moir\'e}}+H_{\text{Coul}}+H_{V_0}
\label{eq:Hinterp}
\end{equation}
projected to the first conduction band of $H_{\lambda}$.
{
Here, $H_{\text{moir\'e}}$ contains the moir\'e potential effect from aligned hBN, which includes the ``moir\'e capacitor effect''~\cite{moirecapacitor} --- an electrostatic (Hartree) contribution from the periodic background charge density of the filled valence bands~\cite{supp}.
We also take the realistic gate-screened Coulomb interaction $H_{\text{Coul}}$.
Together, $H_{\text{Coul}}+H_{V_0}$ describes a density-density interaction with the Fourier transform $V(\bm{q})=\frac{e^2\tanh(|\bq|d)}{2\epsilon_0\epsilon_r|\bq|}+V_0$, with $d=25$nm and $\epsilon_r=5$.
The realistic model corresponds to $\lambda=1$ and $V_0=0$, while
at $\lambda= 0$ and large $V_0$ we recover the ideal limit with dispersion, moir\'e, and Coulomb as perturbations.

Unless stated otherwise, we consider device parameters relevant to the experiment in Ref~\cite{lu2024fractional}.
We take the graphene-hBN twist angle to be $\theta=0.77^\circ$ and a displacement field modeled by a layer potential in $h_{\text{real}}$ of $u_D=50$meV  (see SM~\cite{supp} for full model details).
Note that our results do not depend sensitively on this choice of parameters.
We consider filling one electron per moir\'e unit cell, corresponding to $\rho\approx0.35\rho_{\gamma}$.
}

\begin{figure}[t]
    \centering
    \includegraphics[width=1\linewidth]{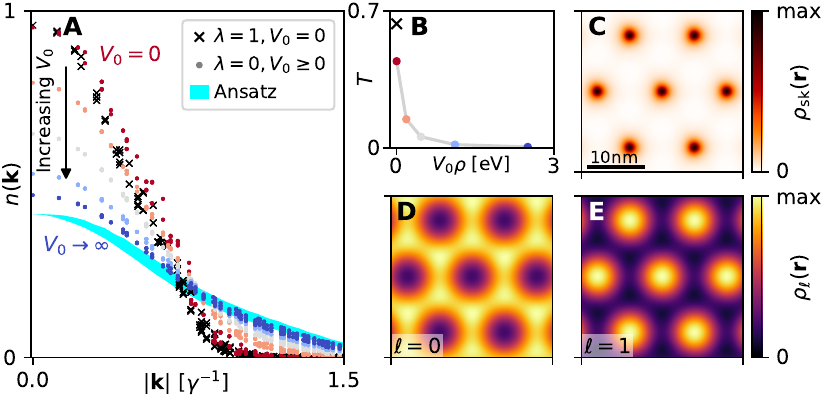}
    \caption{
    {The skyrmion lattice in real R$5$G/hBN.
    Adiabatic continuity between the ideal and real limits of R$5$G/hBN is demonstrated through the smooth evolution of (\textbf{A}) the momentum distribution function $n(\bk)$ interpolating from real ($\lambda=1,V_0=0$) to ideal ($\lambda=0,V_0$ large) and (\textbf{B}) the trace condition violation $T$ along this interpolation.
    The skyrmion texture in the realistic limit is visualized through (\textbf{C}) the skyrmion density $\rho_{\text{sk}}(\br)$ calculated from the dominant eigenvector of the local density matrix, and the layer-resolved charge density in the top two layers, (\textbf{D}) $\rho_{\ell=0}(\br)$ and (\textbf{E}) $\rho_{\ell=1}(\br)$.
    }
    }
    \label{fig:real}
\end{figure}

{
We calculate the HF ground state along this interpolation.
Fig~\ref{fig:real}A shows the momentum distribution function, $n(\bk)$, which evolves smoothly without any sharp transition.
Starting at $V_0=0$, $n(\bk)$ remains nearly unchanged between $\lambda=1$ and $\lambda=0$.
}
This is because the ideal pseudospinor (Eq~\ref{eq:skl}) is an accurate approximation to the real pseudospinor at $|\bk|<\gamma^{-1}$, the region where the majority of the wavefunction weight is concentrated.
As $V_0$ is then increased, $n(\bk)$ approaches that of the ansatz, Eq~\ref{eq:nk}.
Note that for this value of $\rho$, $n(\bk)$ is a monotonically decreasing function of $|\bk|$.

We also consider the ``trace condition violation'' $T$,
\begin{equation}
T=\frac{1}{2\pi}\int_{\text{BZ}} d\bk(\text{Tr}[g(\bk)]-|\Omega(\bk)|)
\end{equation}
where $g_{\mu\nu}(\bk)$ is the Fubini-Study metric~\cite{Liu_review_2023,parameswaran2013fractional} and $\Omega(\bk)$ is the Berry curvature, both calculated for the filled HF band, and $T \geq 0$ by the trace inequality \cite{Liu_review_2023,parameswaran2013fractional,roy2014band}.
Flat Chern bands with $T=0$ have ``ideal" quantum geometry since, regardless of the Berry curvature distribution, $T=0$ enables the construction of exact fractional Chern insulating ground states for short-range repulsive interactions~\cite{ledwith2020fractional,wang2021exact,ledwith2023vortexability,TrugmanKivelson1985,pokrovskySimpleModelFractional1985}.
{A $C=1$ band is ideal if and only if the filled band wavefunction can be factorized as Eq. \ref{eq:final_psi_nu1}~\cite{wang2021exact,dongCompositeFermiLiquid2023,guerci2025layer}}.
{Fig~\ref{fig:real}B shows that $T$ quickly goes to zero as the ideal limit is approached at large $V_0$, thus establishing a limit in which R$5$G/hBN realizes an ideal $C=1$ HF band. }
{
To get a sense of the scale of $V_0$ in Fig~\ref{fig:real}(A,B), we note that the $V_0\approx 45.2$eVnm$^2$ in Sec.~\ref{sec:HFdispideal} corresponds to, at $\theta=0.77^\circ$, $V_0 \rho\approx0.42\text{eV}$.}

{
We now demonstrate that all the signature features of the skyrmion lattice, identified in the ideal limit, remain present in real R$5$G/hBN.
We first identify the presence of the skyrmion texture in the local density matrix in layer and sublattice ($\tau\in\{A,B\}$) space,
$D_{\ell'\tau',\ell\tau}(\br)=\langle c_{\ell^\prime \tau^\prime}^\dagger(\br) c_{\ell\tau}(\br)\rangle$ (neglecting contributions from the filled remote valence bands).
In the ideal limit, $D$ has a single non-zero eigenvalue, and its eigenvector traces out the skyrmion texture with $\br$.
The topology of this real-space texture in the dominant eigenvector is robust under small perturbations away from the ideal limit, protected by the spectral gap of $D$~\cite{kolavr2025robustness}.
However, this real-space topology need not, in general, match the momentum-space Chern number.
To show that this texture survives at the realistic R$5$G/hBN parameter, we compute $\hat{x}_{\ell\tau}(\br)$, the normalized eigenvector of $D$ with the largest eigenvalue.
From this, we compute the skyrmion density
$\rho_{\text{sk}}(\br)=\frac{1}{2\pi i}\text{Tr}P[\partial_x P,\partial_y P]$ with $P=\hat{x}(\br)\hat{x}^\dagger(\br)$,
shown in
 Fig~\ref{fig:real}C.
 Within each unit cell, $\int_{\text{u.c.}}\rho_{\text{sk}}(\br)d^2\br=1$, confirming the presence of a non-trivial real-space texture.
 The skyrmion density is largely concentrated at the sites of a triangular lattice, which marks the skyrmion cores.
}

{
This skyrmion texture leaves clear signatures in the layer-resolved density, $\rho_{\ell}(\br)$.
Fig~\ref{fig:real}(D,E) shows $\rho_{\ell}(\br)$ for the top two layers, $\ell=0,1$, where the majority of the density is located.
Fig~\ref{fig:real}D shows that $\rho_{\ell=0}(\br)$ exhibits a honeycomb-like structure with deep minima at the skyrmion cores (which are exact zeros in the ideal limit, where $|\chi_0(\br)|\rightarrow 0$) at which the texture points in a direction orthogonal to the top-layer.
Fig~\ref{fig:real}E shows that at the skyrmion cores, $\rho_{\ell=1}(\br)$ exhibits a maximum, consistent with this picture (and with expectations based on $|\chi_0(\br)|^2$ and $|\chi_1(\br)|^2$ in Fig~\ref{fig:skyrmion}(A,B)).
Since the majority of the charge density is located on just the top layer, the total charge density, $\rho_{\text{tot}}(\br)=\sum_{\ell}\rho_{\ell}(\br)$, also exhibits a honeycomb-like structure observed in previous HF studies~\cite{dong2024anomalous,kwan2025moire,zeng2024sublattice}, which can be understood as being inherited from $\rho_{\ell=0}(\br)$.
The minima in $\rho_{\ell=0}$ (and $\rho_{\text{tot}}$) can therefore be regarded as a signature of the skyrmion core.

{
We have therefore demonstrated that the microscopic mechanism driving skyrmion lattice formation, derived in the ideal limit, fully applies in realistic R$5$G/hBN.
In the remainder of this manuscript, we consider the dynamical properties of this interaction-generated skyrmion lattice, which distinguishes R$N$G/hBN from other moir\'e Chern systems with static skyrmion textures.
}
}

\section{Collective excitations of the skyrmion lattice}
\label{sec:collective}

{
In this section, we turn to the characteristic collective excitations of the skyrmion lattice in R$N$G/hBN.
The interaction-driven origin of the skyrmion lattice makes R$N$G/hBN conceptually much more similar to magnetic skyrmion crystals~\cite{nagaosa2013topological}, compared to other moir\'e platforms.
In the context of magnetic skyrmion materials, such textures} are known to exhibit rich collective dynamics~\cite{garst2017collective,lonsky2020dynamic}.
In magnetic skyrmion materials, collective excitations (magnons) are well characterized both theoretically and experimentally~\cite{garst2017collective,lonsky2020dynamic,nagaosa2013topological}.
The most studied are the clockwise (CW) and counterclockwise (CCW) gyrotropic modes, in which skyrmions orbit about their equilibrium positions, and the breathing mode, in which the skyrmion radius oscillates in time~\cite{mochizuki2012spin,onose2012observation,okamura2013microwave,schwarze2015universal}.
Higher-order internal modes associated with multipolar deformations of the skyrmion shape have also been identified~\cite{diaz2020chiral,makhfudz2012inertia,schutte2014magnon,lin2014internal,takagi2021hybridized}.
As we show below, all these collective modes have direct analogs in  R$N$G/hBN.
Throughout this section, we restrict attention to collective excitations within a single spin and valley flavor.
{
}

\subsection{ Analytical understanding of skyrmion shape modes in the ideal limit}
\begin{figure}[t]
    \centering
    \includegraphics[width=1\linewidth]{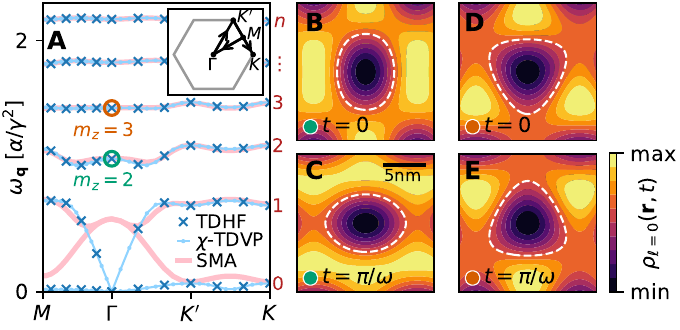}
    \caption{
    Collective excitations of dispersive ideal R$5$G at $\rho=0.3\rho_\gamma$ in the limit $\alpha\rightarrow 0$.
    (\textbf{A}) The TDHF spectrum shows two gapless phonon branches and a ladder of weakly dispersing excitations, labeled by $n$, which we identify as $n$-fold shape distortions of the skyrmion texture.
    Also shown are spectra calculated from $\chi$-TDVP and the single-mode approximation (SMA) as described in the main text.
    The top-layer charge density $\rho_{\ell=0}(\br,t)$ at two time instances for a coherent excitation of the (\textbf{B},\textbf{C}) quadrupolar ($n=2$) and (\textbf{D},\textbf{E}) sextupolar ($n=3$) shape modes.
    }
    \label{fig:collectiveideal}
\end{figure}

We begin with the ideal limit, where we can obtain an analytic understanding of all the collective modes.
We consider the dispersive ideal R$5$G Hamiltonian $H_{\text{disp-ideal}}$ (Eq~\ref{eq:Hdispideal}), with parabolic dispersion $\mathcal{E}(\bk)=\alpha|\bk|^2$, in the limit $\alpha/V_0\rightarrow 0$.
The full collective mode spectrum can be obtained using the time-dependent HF (TDHF) formalism~\cite{thoulessStabilityConditionsNuclear1960,thoulessVibrationalStatesNuclei1961,RevModPhys.54.913}.
The TDHF approach can be viewed as an application of the time-dependent variational principle (TDVP)~\cite{kramerGeometryTimeDependentVariational1981}, $\delta S=0$ with the action $S=\int\langle\Psi|i\partial_t-H|\Psi\rangle$, restricted to the variational manifold of Slater determinants.
By expanding about the self-consistent HF state, the frequencies $\omega_{\bm{q}}$ of small harmonic oscillations about it are obtained as functions of their momentum transfer $\bm{q}$.
The resulting TDHF spectrum is shown in
 Fig~\ref{fig:collectiveideal}A (labeled TDHF).
At low frequencies near $\bm{q}=\bm{\Gamma}=0$, two gapless Goldstone branches are visible.
These are the acoustic phonon modes associated with a sliding motion of the skyrmion lattice.
The upper branch disperses rapidly away from $\bm{\Gamma}$ and remains relatively flat across most of the Brillouin zone.
Above these, we find an almost evenly spaced ladder of weakly dispersing gapped modes.
This trend continues to higher frequencies beyond the range shown here.

We now develop an analytic understanding of this spectrum.
As shown in Sec~\ref{sec:HFdispideal}, the HF ground state in the limit $\alpha/V_0\rightarrow 0$ is exceedingly well approximated by the analytic ansatz wavefunction $|\Psi^{\text{ansatz}}\rangle$ (Eq~\ref{eq:final_psi_nu1}).
This state is characterized by $\chi_0(\bm{r}) = w_0^{(\bm{q}_0)}(\bm{r})$ for some $\bm{q}_0$, with all higher components $\chi_{\ell>0}$ determined by the zero-mode equation (Eq.~\eqref{eq:chizeromode}).
Deformations of $\chi_0$ away from $w_0$, while maintaining the zero-mode constraint for $\chi_{\ell>0}$,
 remain entirely within the zero-energy manifold of $H_{\text{ideal}}$, so their energy cost arises solely from dispersion and is proportional to $\alpha$.
We therefore expect that all low-energy excitations in the $\alpha/V_0\rightarrow 0$ limit correspond to such deformations.

To confirm this, we apply the TDVP directly within the variational subspace spanned by Eq~\ref{eq:final_psi_nu1} allowing for arbitrary time-dependent $\chi_0$, a procedure we refer to as $\chi$-TDVP.
Because all such states are exact zero-energy eigenstates of $H_{\text{ideal}}$, $\chi$-TDVP reduces to extremizing only the kinetic part of the action $S_{\text{disp}}=\int\langle\Psi|i\partial_t-H_{\text{disp}}|\Psi\rangle$ with respect to $\chi_0$.
The frequencies of harmonic oscillations about the ansatz $\chi_0=w_0$ can be obtained by solving the resulting linearized equations of motion, as detailed in the SM~\cite{supp}.
The resulting spectrum, shown in Fig~\ref{fig:collectiveideal}A (labeled $\chi$-TDVP), is indistinguishable from the full TDHF calculation. This confirms that the entire spectrum, the tower of gapped modes as well as the small-$\bq$ acoustic phonons, can be understood through deformations of $\chi_0$ away from $w_0$ alone.

We now show that the gapped modes correspond to relatively simple excitations of $\chi_0$ which can be understood in terms of a generalized single-mode approximation (SMA)\cite{feynmanAtomicTheoryTwoFluid1954,bijlPropertiesLiquidHelium1941,girvinMagnetorotonTheoryCollective1986,girvinQuantumHallEffect1999}.
We define an operator $M_{n\bm{q}}$ by its action on $\Psi^{\text{ansatz}}_{\{0\}}(\br)$~\cite{supp},
\begin{equation}
e^{\eta(t) M_{n\bq}}\Psi^{\text{ansatz}}_{\{\ell\}}(\br)=\Phi^{\nu=1}_{\text{LLL}}(\{\br\})\prod_j\chi^{\text{exc}}_{\ell_j}(\br_j,t)
\end{equation}
where
\begin{equation}
\chi^{\text{exc}}_{0}(\br,t)\propto w_0^{(\bq_0)}(\br)+\eta(t) w_n^{(\bq_0+\bq)}(\br)
\label{eq:chi0exc}
\end{equation}
with $\chi^{\text{exc}}_{\ell>0}$ determined by the zero-mode condition Eq~\ref{eq:chizeromode}.
Within the SMA, this state evolves in time according to $\eta(t)=\eta e^{-i\omega t}$, where $\omega=\omega_{n\bq}^{\text{SMA}}$ is the energy cost of creating a single such excitation.
It can be computed by defining $|\Psi^{\text{SMA}}_{n\bq}\rangle \equiv M_{n\bq}|\Psi^{\text{ansatz}}\rangle$, which describes an excited state in which a single electron has its $\chi_0$ promoted from the conjugate LLL ($w_0$) to the $n$th conjugate LL ($w_n$) together with a momentum boost $\bq$.
This excitation has a purely kinetic energy cost given by
\begin{equation}
\omega_{n\bq}^{\text{SMA}}=\frac{\langle\Psi^{\text{SMA}}_{n\bq}| (H_{\text{disp}}-E_0)|\Psi^{\text{SMA}}_{n\bq}\rangle}{\langle\Psi^{\text{SMA}}_{n\bq}|\Psi^{\text{SMA}}_{n\bq}\rangle}
\label{eq:omegaSMA}
\end{equation}
where $E_0$ is the kinetic energy of $|\Psi^{\text{ansatz}}\rangle$.
The resulting spectrum is plotted in Fig~\ref{fig:collectiveideal}A, and exhibits excellent agreement with both TDHF and $\chi$-TDVP everywhere except for the $n=0,1$ modes as $\bq\rightarrow\bm{\Gamma}$, where the SMA incorrectly predicts a finite gap.
This deviation for the acoustic phonon modes will be explained shortly.

Taken together, Fig~\ref{fig:collectiveideal}A demonstrates that the majority of the collective mode spectrum can be understood through simple time-dependent ansatzes for $\chi_0$ involving excitations of a single $w_n$ mode.

We now show that these excitations correspond physically to shape modes of the skyrmions.
We focus on the $n=2$ and $n=3$ modes at $\bm{q}=\bm{\Gamma}$.
A coherent excitation of each of these modes can be described by a time-dependent $\chi_0^{\text{exc}}(\br,t)$ given by Eq~\ref{eq:chi0exc} with $\eta(t) = \eta e^{-i\omega t}$.
{
Expanding near a zero of $\chi_0$ (the skyrmion core, taken at the origin),
 the basis functions behave as
$w_0 \propto \bar{z}$ and $w_n\propto z^{n-1}$ for $n>0$.
The corresponding excited state therefore behaves as $\chi_0^{\text{exc}}\propto  \bar{z}+\alpha e^{-i\omega t}z^{n-1}$ for some small $\alpha$ (taken real), which leads to
\begin{equation}
|\chi_0^{\text{exc}}(\br,t)|^2 \propto |{\br}|^2 + 2\alpha |\br|^{n} \cos (n\theta-\omega t)+O(\alpha^2)
\label{eq:chi0shape}
\end{equation}
where $\theta$ is the polar angle defined by $z = |z|e^{i\theta}$.
Interpreting $|\chi_0|^2$ near its zero as a measure of the skyrmion shape, we thus see that this excitation corresponds an $n$-fold deformation of the skyrmion profile that rotates counterclockwise in time.
}

This picture is confirmed by direct examination of the TDHF collective modes.
We visualize the shape of the skyrmions through the top-layer charge density $\rho_{\ell=0}(\br,t)$ in a coherent excitation of the $n=2$ and $n=3$ modes at $\bm{\Gamma}$, obtained from TDHF.
Near its minima, $\rho_{\ell=0}$ should follow that of $|\chi_0^{\text{exc}}(\br,t)|^2$ in Eq~\ref{eq:chi0shape}.
Indeed, Fig~\ref{fig:collectiveideal}(B,C) illustrates the $n=2$ mode at two time instances within its period, revealing a quadrupolar distortion of the skyrmion profile (an elliptical deformation).
Fig~\ref{fig:collectiveideal}(D,E) similarly show the $n=3$ mode, which exhibits a sextupolar distortion (a triangular deformation).
For visual clarity, the amplitude of the collective motion has been exaggerated in these images (see SM for the full time dependence and further details~\cite{supp}).

For the $K$ valley considered here, each mode carries positive azimuthal angular momentum quantum number $m_z = n$ (defined modulo six) and rotates counterclockwise in time.
Since they only carry one sign of angular momentum and rotate in a single direction, we refer to them as chiral shape modes.
This chirality arises from the system's broken time-reversal symmetry and is determined by the overall valley polarization.
{
We remark that the quadrupolar mode is reminiscent of the ``chiral graviton" mode in fractional quantum Hall systems~\cite{haldaneGeometricalDescriptionFractional2011,yangModelWaveFunctions2012,maciejkoFieldTheoryQuantum2013a,gromovBimetricTheoryFractional2017a,nguyenProbingSpinStructure2021,duChiralGravitonTheory2025},
in that both are gapped excitations with angular momentum $m_z=2$.
In the fractional quantum Hall case, the chiral graviton corresponds to the small-$\bq$ limit of the magnetoroton~\cite{girvinMagnetorotonTheoryCollective1986} and was recently observed via circularly polarized resonant Raman scattering~\cite{liangEvidenceChiralGraviton2024,nguyenProbingSpinStructure2021}.
Higher $m_z>2$ modes have also been proposed in fractional quantum Hall states near half filling ($\nu=\frac{1}{2}$), where they correspond to shape modes of the composite Fermi sea~\cite{golkarHigherSpinTheoryMagnetorotons2016}.
}

{
We now discuss how the low energy acoustic phonons emerge from $\chi$-TDVP.
These gapless modes as $\bq\rightarrow 0$ cannot be captured by a single $w_n^{(\bq_0+\bq)}$ as in the SMA, but instead requires a superposition of $w_0^{(\bq_0\pm\bq)}$ and $w_1^{(\bq_0\pm\bq)}$.
To see this, let $\eta_{n,\pm\bq}(t)$ denote the time-dependent coefficient of $w_n^{(\bq_0\pm \bq)}(\br)$,
and define $\bm{\eta}\equiv (\eta_{0,\bq},\bar{\eta}_{0,-\bq},\eta_{1,\bq},\bar{\eta}_{1,-\bq})^{\mathsf{T}}$.
The $\chi$-TDVP linearized equation of motion to leading nonsingular order in $q=q_x+iq_y$ is derived in the SM~\cite{supp}
\begin{equation}
i\partial_t\bm{\eta}=\omega_c\begin{pmatrix}
1 & 1 & 0 & -\sqrt{2}/(l_B q) \\
-1 & -1 & -\sqrt{2}/(l_B \bar{q}) & 0\\
0 & 0 & 1 & \bar{q}/q \\
0 & 0 & -q/\bar{q} & -1
\end{pmatrix}
\bm{\eta}
\end{equation}
where $\omega_c>0$ is a constant.
Restricting to a single $w_n^{(\bq_0+\bq)}$ would thus erroneously result in a gapped mode with frequency $\omega_c$.
From the full coupled equations of motion, however, we can read off two gapless solutions satisfying $i\partial_t\bm{\eta}=0+O(|\bq|)$:
$\bm{\eta}^{\text{shear}} = (1,-1,0,0)^{\mathsf{T}}$ and $\bm{\eta}^{\text{long}} = i(1,1,-\sqrt{2}l_B \bar{q}, \sqrt{2}l_B q)^{\mathsf{T}}$.
These correspond to the shear and longitudinal acoustic phonon modes of the skyrmion lattice, which can be verified by expanding $\chi_0^{\text{exc}}$ near its zeros~\cite{supp}.
}

{

}

\begin{figure}[t]
    \centering
    \includegraphics[width=1\linewidth]{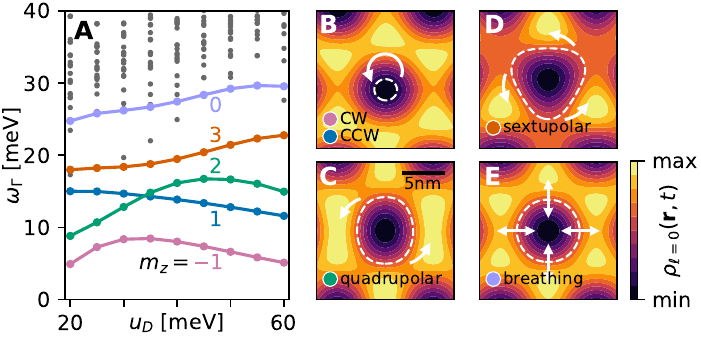}
    \caption{
    Collective excitations of real R$5$G/hBN.
    (\textbf{A}) The $\bm{q}=\bm{\Gamma}$ point spectrum as a function of displacement field, modeled as a layer potential $u_D$.
    Several modes with clear physical meaning are highlighted and labeled by their azimuthal angular momentum quantum number $m_z$:
    the {  clockwise (CW) ($m_z=-1$), counterclockwise (CCW) ($m_z=1)$,} quadrupolar ($m_z=2$), sextupolar ($m_z=3$), and breathing ($m_z=0$) modes.
    Note that $m_z$ is only strictly well-defined mod 3.
    (\textbf{B-E}) The top-layer charge density at { $u_D=50$meV} at a time instance for each mode, illustrating their nature.
    }
    \label{fig:collectivereal}
\end{figure}

\subsection{ Skyrmion shape modes in real R5G/hBN}
Having developed a complete understanding of the collective excitation spectrum in the ideal limit, we now turn to the realistic R$5$G/hBN model.
Fig~\ref{fig:collectivereal}A shows the collective mode spectrum at $\bm{q}=\bm{\Gamma}$, computed with TDHF, for a broad range of realistic displacement fields $u_D$.
A common feature across all $u_D$ is the presence of several isolated modes at low energies, well-separated from the particle-hole continuum at higher energies.
Selected low-energy modes are highlighted in Fig~\ref{fig:collectivereal}A and discussed below.

{ The modes labeled $m_z=\pm 1$ are the (formerly) shear and longitudinal phonon modes, which are gapped by the hBN moir\'e potential.}
These correspond to CW and CCW chiral phonon-like modes of the skyrmion lattice, as illustrated in Fig~\ref{fig:collectivereal}B.

{
We further identify two additional modes, labeled $m_z=2,3$, as the quadrupolar and sextupolar chiral shape modes.}
They carry the expected angular momenta $m_z=n$ (modulo three, due to the threefold rotation symmetry of the hBN moir\'e potential) and connect smoothly to their counterparts in the ideal limit~\cite{supp}.
The top-layer charge densities, shown in Fig~\ref{fig:collectivereal}(C,D), closely resemble those of the ideal limit [Fig~\ref{fig:collectivereal}(B-E)], confirming their nature.

Finally, we highlight an additional gapped mode { labeled $m_z=0$}.
Inspection of the top-layer charge density identifies it as a breathing mode, in which the skyrmion radius oscillates in time~\cite{supp}, illustrated in Fig~\ref{fig:collectivereal}E.
The breathing motion periodically enlarges the region where the layer-pseudospin tilts away from the top layer, producing large oscillations in the layer polarization and, consequently, strong coupling to a time-dependent out-of-plane displacement field.
Indeed, we identify this mode in Fig.~\ref{fig:collectivereal}A by its large coupling matrix element with the displacement field operator~\cite{supp}.

These features of the low-lying spectrum are robust in R$5$G/hBN over a broad range of realistic parameters.
Previous TDHF studies~\cite{kwan2025moire,dong2025phonons} have primarily focused on the phonon-like modes, but also report additional gapped (intra-spin/valley) excitations, present across different ``interaction schemes''~\cite{kwan2025moire}, well below the continuum whose nature was not identified.
These are most likely the shape modes discussed above.
Recognizing the ground state of R$5$G/hBN as a layer-pseudospin skyrmion lattice thus provides a unified framework for organizing the collective excitations, endowing them with clear physical meaning.

{
\section{Charged skyrmion vacancies at point defects}
\label{sec:vacancy}
{ A second physical consequence of the dynamical skyrmion lattice is that
its defects can appear as charged low-energy quasiparticles.
We demonstrate this explicitly by showing that}  R$N$G/hBN supports a new low-energy positively charged quasiparticle: the skyrmion vacancy.
It corresponds to the absence of a skyrmion at a site of an otherwise regular skyrmion lattice, accompanied by a cloud of positive charge, yielding a net charge $+e$.

To understand the charge of a skyrmion vacancy, it is instructive to draw an analogy with charged spin-skyrmions in quantum Hall ferromagnets~\cite{girvin2002quantum,sondhi1993skyrmions}.
At filling factor $\nu=1$ of a spin-degenerate LLL, the interacting ground state is a fully-filled spin-polarized LLL.
The lowest energy hole-like excitation of this state is not a simple hole in the spin-polarized LLL, but is instead a spin-skyrmion.
This spin-skyrmion contributes $-2\pi$ Berry flux for the electrons, canceling out $2\pi$ magnetic flux of the magnetic field.
Since the average electron and magnetic flux density are equal at $\nu=1$, such a spin-skyrmion effectively carries a net positive charge $+e$.

A similar mechanism may arise in the R$N$G/hBN Chern insulator, but with a skyrmion vacancy.
Here, the skyrmion lattice generates an effective magnetic field with $2\pi$ magnetic flux per unit cell.
A skyrmion vacancy leads to an effective $-2\pi$ flux relative to the background, and therefore carries a net positive charge $+e$.
A skyrmion vacancy could be more energetically favorable than a conventional hole (e.g. emptying a state in the $\nu=1$ HF band) because the vacancy allows the pseudospin texture to locally vary more slowly in space and spend more time pointing towards the top layer, possibly leading to a lower electronic kinetic energy cost.
If such vacancies are pinned by disorder present in real samples, they could potentially be observed via local probes with moir\'e-scale resolution.

\begin{figure}
    \centering
    \includegraphics[width=1\linewidth]{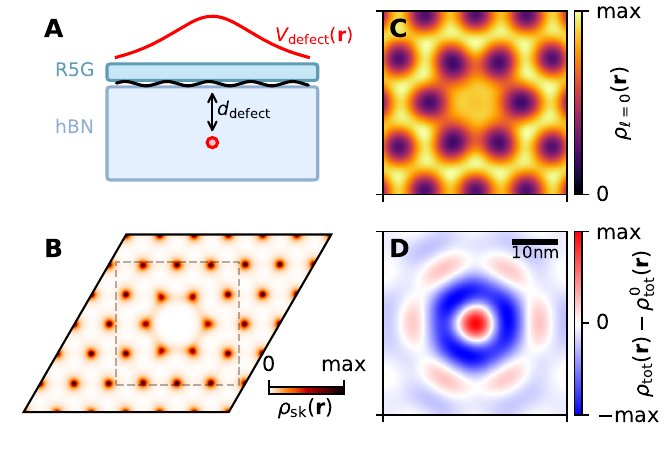}
    \caption{
     Charged skyrmion vacancy pinned by disorder.
    (\textbf{A}) We consider a negatively charged defect in the hBN layer at a distance $d_{\text{defect}}=15$nm away, modeled as a repulsive scalar potential $V_{\text{defect}}(\br)$ for electrons in R$5$G.
    This defect binds a positively charged skyrmion vacancy in the R$5$G layer.
    (\textbf{B}) The skyrmion density, calculated from the local density matrix, is shown for the HF ground state on an $L\times L$ torus with $N_e=L^2-1$ electrons ($L=6$), clearly showing a skyrmion vacancy.
    Zooming in on the dashed region,
    we show (\textbf{C}) the top-layer electron density $\rho_{\ell=0}(\br)$ and
    (\textbf{D}) the change in $\rho_{\text{tot}}=\sum_{\ell}\rho_{\ell}$ relative to that of the $\nu=1$ ground state in the absence of the defect.
    }
    \label{fig:vacancy}
\end{figure}

We now provide numerical evidence for this in realistic R$5$G/hBN.
We consider the scenario shown in Fig~\ref{fig:vacancy}A, in which a negatively charged point defect is present deep in the hBN layer: a specific example of such a defect is the well-studied negatively-charged boron vacancy ($V_B^-$) center~\cite{abdi2018color,gottscholl2020initialization,gottscholl2021room,gottscholl2021spin,udvarhelyi2023planar}.
Such a defect can bind a positively charged excitation in the R$5$G layer to its location.
We model the effect of the defect by a repulsive scalar potential $V_{\text{defect}}(\br)$ on the electrons in the R$5$G layer, as illustrated in Fig~\ref{fig:vacancy}A.
The screening of the gates is taken into account in $V_{\text{defect}}(\br)$, as described in the SM~\cite{supp}.

The presence of the defect breaks the moir\'e translation symmetry of the system.
We therefore take a finite-size system on an $L\times L$ torus (with $L=6$), and consider the HF ground state in which no moir\'e periodicity is assumed.
First, we have confirmed in the SM that the $N_e=L^2$ (i.e. $\nu=1$) electron ground state, in the absence of a defect, indeed corresponds to the periodic $C=1$ Chern insulator found in Sec~\ref{sec:connecting}~\cite{supp}.
We note that, for the experimentally relevant parameter chosen here, the Hartree contribution from the filled valence bands in $H_{\text{moir\'e}}$ is essential to correctly stabilize the $\nu=1$ state when moir\'e periodicity is not enforced~\cite{moirecapacitor} (as noted in Refs~\cite{kwan2025moire,dong2025phonons}).
We then consider the HF ground state with $N_e=L^2-1$ electrons in the presence of a defect located at an out-of-plane distance $d_{\text{defect}}=15$nm away from the R$5$G (roughly half of the distance to the gate, $d=25$nm).
This state should contain one hole-like excitation relative to the $\nu=1$ ground state.

Fig~\ref{fig:vacancy}B shows the skyrmion density, $\rho_{\text{sk}}(\br)$, as defined in Sec~\ref{sec:connecting}, calculated for this $N_e=L^2-1$ electron state.
Clearly, compared to the uniform lattice (Fig~\ref{fig:real}C), one of the peaks is missing at the origin where the defect is located.
Integrating $\rho_{\text{sk}}(\br)$ over the entire system yields in $N_{\text{sk}}\equiv\int d^2\br \rho_{\text{sk}}=L^2-1$, as opposed to $N_{\text{sk}}=L^2$ for a conventional hole excitation that does not alter the pseudospin texture.
Fig~\ref{fig:vacancy}C shows the top-layer electron density, $\rho_{\ell=0}(\br)$, which is also clearly missing one of its minima (which mark the skyrmion cores).
This state therefore describes a skyrmion vacancy pinned to the position of the defect.
In the SM~\cite{supp}, we show that the skyrmion vacancy observed here is quite typical over a range of realistic $u_D$ and $d_{\text{defect}}$.

In Fig~\ref{fig:vacancy}D, we show the change in the total electron density, $\delta\rho_{\text{tot}}(\br) =\rho_{\text{tot}}(\br)-\rho_{\text{tot}}^0(\br)$, where $\rho_{\text{tot}}^0(\br)$ is the density of the $\nu=1$ ground state in the absence of the defect.
Interestingly, the absence of a skyrmion core (which are also minima of $\rho_{\text{tot}}$) means that $\rho_{\text{tot}}$ is actually increased right at the origin where $V_{\text{defect}}(\br)$ is centered.
This is compensated for by a ``halo'' of decreased density in the immediate surrounding area, integrating to a total of $\int\delta\rho_{\text{tot}}d^2\br=-1$.
This counterintuitive density redistribution, in which the density is increased at the origin, is a signature of the skyrmion vacancy that would be difficult to explain without the understanding developed in this work.

We also remark on the importance of ``HF band-mixing'' in capturing the skyrmion vacancy state.
In the SM~\cite{supp}, we consider the skyrmion vacancy state (which is at $\nu=(L^2-1)/L^2\approx0.97$) projected into the basis of the $\nu=1$ HF bands, and show that the skyrmion vacancy state has a significant portion of its weight in the higher $\nu=1$ HF bands (to achieve $>0.9$ overlap with the full skyrmion vacancy state, at least three of the $\nu=1$ HF bands must be kept).
{We have also performed HF at $\nu=(L^2-1)/L^2$ but projected to the lowest $\nu=1$ band, which incorrectly results in a conventional hole localized at the defect and not a skyrmion vacancy~\cite{supp}. }
This can be understood intuitively as follows: the lowest $\nu=1$ HF band describes electrons in the LLL of a fixed skyrmion lattice texture (an exact statement in the ideal limit), thus in order to modify the skyrmion lattice itself, higher HF bands must be involved.
We note that early numerical studies of the fractional Chern insulator in R$5$G/hBN at $\nu=2/3$ performed exact diagonalization projected to the lowest HF band of the $\nu=1$ state~\cite{dong2024theory,zhou2024fractional,dong2024anomalous,guo2024fractional},
the validity of which is questionable~\cite{yu2025moire,li2025multiband}.
Our present analysis suggests that even at infinitesimally small doping below $\nu=1$, and even at mean-field level, projection to the lowest $\nu=1$ HF band can already result in qualitatively incorrect physics.

{
Finally, we remark that skyrmion vacancies are a unique consequence of the interaction-generated skyrmion lattice in R$N$G/hBN.
In other systems, such as twisted MoTe$_2$~\cite{wu2019topological}, where the skyrmion texture is fixed by the single-particle Hamiltonian,  skyrmion vacancies simply cannot emerge as low-energy quasiparticle excitations.
}
}

\section{Discussion}
\label{sec:discussion}

{
In this work, we have {introduced an ideal limit of rhombohedral graphene. }
{ Our analysis reveals } a qualitatively new form of moir\'e Chern physics in R$N$G/hBN, in which an interaction-generated skyrmion lattice acquires intrinsic dynamics.
{ The} skyrmion lattice plays an essential role in the low-energy { excitations,} { leading to sharp experimental signatures} as demonstrated through our analysis of { skyrmion shape modes} and skyrmion vacancies.
We now close by discussing some of the experimental and theoretical implications of our work, and highlighting potential future directions.
}

The skyrmion texture gives rise to experimentally measurable signatures in local probes.
The layer skyrmion texture implies a strongly modulated layer-resolved charge density.
In particular, the top-layer charge density, $\rho_{\ell=0}(\br)$,  is strongly suppressed at the skyrmion cores (Fig~\ref{fig:real}D), a characteristic signature that could be detected through scanning tunneling microscopy (STM), which directly probes the local density of states on the top layer.
Similar layer skyrmion textures have been imaged in moir\'e TMDs~\cite{zhang2025experimental,thompson2025microscopic}, and recently applied to R$3$G/hBN~\cite{seewald2025mapping}, though not yet in the correlated phases.
{ A key experimental prediction of our theory is that charged skyrmion vacancies at fillings just below $\nu=1$ can be pinned by defects in the substrate.
Such a vacancy can be directly identified through the absence of a minimum in $\rho_{\ell=0}(\br)$ (Fig~\ref{fig:vacancy}C) or by its signature $\delta\rho_{\text{tot}}$ (Fig~\ref{fig:vacancy}D), in an otherwise moir\'e-periodic atomic structure.
This decoupling of the skyrmion texture from the underlying moir\'e superlattice can occur only in such an interaction-generated skyrmion lattice, so it would serve as a definitive signature of the interaction-driven origin of the Chern band in R$N$G/hBN.
}

Beyond the ground state,
our theory predicts internal excitation modes of the skyrmions that have characteristic energies in the few tens of meV.
{ The quadrupolar shape mode carries angular momentum $m_z=2$ and may be detectable via circularly polarized Raman scattering, analogous to the chiral graviton mode in fractional quantum Hall systems~\cite{nguyenProbingSpinStructure2021,liangEvidenceChiralGraviton2024}.}
These excitations could also be probed with recently developed on-chip THz spectroscopy techniques~\cite{seo2024chip,stiewe2022spintronic,gallagher2019quantum,zhao2023observation} tailored to microscopic van der Waals heterostructures, although quantitative estimates of their coupling to different probes will need to be performed on a case-by-case basis.
In particular, the breathing mode, characterized by an oscillating layer polarization, should be especially amenable to detection through its coupling to a uniform out-of-plane electric field~\cite{supp}.

{

}

The real-space framework we have developed for ideal R$N$G provides a foundation for systematically generating analytic many-body wavefunctions that extend beyond the Slater determinant states analyzed here.
Any wavefunction $\Psi_{\{\ell\}}(\{\br\})=\Phi(\{\br\})X_{\{\ell\}}(\{\br\})$, with $\Phi$ describing LLL fermions in field $B$, $X_{\{0\}}$ describing bosons in field $-B$, and $X$ satisfying the many-body version of Eq~\ref{eq:chizeromode}, are exact zero-energy eigenstates of $H_{\text{ideal}}$.
This charge-pseudospin separation evokes parton constructions of quantum Hall states: indeed, a parton construction reproduces Eq~\ref{eq:final_psi_nu1} exactly~\cite{supp}
and extends naturally to non-Slater determinant ansatzes.
For instance, when $X=\prod \chi$ condenses to a skyrmion lattice, fractional Chern insulators follow by choosing $\Phi$ to be the corresponding fractional quantum Hall states in the LLL.
More broadly, $X_{\{0\}}$ may represent any correlated bosonic state in field $-B$, generating a broad family of variational wavefunctions.
If $X_{\{0\}}$ preserves continuous magnetic translation symmetry, such wavefunctions describe strongly-correlated continuous translation-invariant states in unaligned R$N$G { (similar in spirit to those of Refs.~ \cite{Kim_2025,kim2026variationalmontecarlooptimization}).}
These states can be interpreted as skyrmion liquids, and the $m_z \geq 2$ shape modes may persist in these states. A similar SMA to Eq.~\ref{eq:omegaSMA} can be applied to study them; for instance, $(\partial_{\bar{z}})^n$ should create a $\bq=0$ shape excitation with $m_z = n$.
Further work is necessary to explore these candidate states and assess their energetic stability.

The intuition and techniques developed here also illuminate physical phenomena more broadly.
A notable example is the chiral superconductor recently reported in R$4$G~\cite{han2025signatures}.
The pairing chirality (within a given valley) was recently shown to be opposite in sign for superconductivity driven by attractive versus repulsive interactions via the Kohn-Luttinger mechanism~\cite{may2025pairing}.
Our analysis provides a simple real-space interpretation: a repulsive $V_0>0$ favors chiral two-particle nodes on the zeroth layer, while for $V_0<0$ the same reasoning implies that paired electrons favor antichiral nodes.
Our work also offers a complementary perspective on anomalous Hall crystals, framing them as spontaneously formed pseudospin skyrmion crystals whose effective magnetic field drives quantum Hall physics.
This real-space perspective makes the microscopic origin of its topology more transparent than the ``topological Wigner crystal'' picture,
 provides physical meaning to its collective excitations, and could serve as a natural foundation for effective low-energy descriptions of these phases.
Finally, we note that much of our analysis can be straightforwardly generalized to other recently proposed models with similar pseudospinor structure~\cite{tan2024parent,soejima2025jellium,desrochers2025electronic}.
We expect that some of the formalism developed in this work will also be useful for developing new beyond-mean-field numerical approaches for this class of systems.

{
The skyrmion vacancies identified in this work may have far broader implications for physics in R$N$G than discussed thus far.
It is possible that skyrmion vacancies could serve as the dominant charge carriers of current below $\nu=1$.
Since skyrmion vacancies likely have a higher effective mass than regular holes, they may be more prone to crystallization and/or localization, potentially explaining the ``extended'' quantum anomalous Hall effect observed over a wide range of fillings below $\nu=1$~\cite{lu2025extended}.
Furthermore, near fractional fillings where fractional Chern insulators are observed,
the skyrmion vacancies could also bind fractional charge.
They therefore compete with  conventional fractionally-charged anyons that do not modify the pseudospin texture.
A proper description of such skyrmion vacancies necessitates incorporating band-mixing effects, which is computationally demanding in exact diagonalization studies, underscoring the need for improved theoretical approaches for modeling these systems.
It is possible that the presence of skyrmion vacancies is, in fact, the underlying physical reason that band-mixing effects are so strong in R$N$G/hBN fractional Chern insulators~\cite{yu2025moire,li2025multiband,moirecapacitor}.
The systematic study of skyrmion vacancies is an important topic for future investigation.
}

Our work draws a surprising connection between R$N$G/hBN Chern insulators and the broader field of magnetic skyrmion crystals.
Although they arise from entirely different microscopic details, the resulting states share remarkable similarities in their ground state and dynamical properties.
This parallel suggests that rhombohedral graphene could serve as a layer-pseudospin analog of magnetic skyrmion materials~\cite{petrovic2025colloquium,tokura2020magnetic,everschor2018perspective}.
Looking ahead, it will be fascinating to explore whether other non-equilibrium skyrmion phenomena, such as current-driven motion or the skyrmion Hall effect~\cite{zangDynamicsSkyrmionCrystals2011,nagaosa2013topological}, also have manifestations in this setting.
{
For example, { since skyrmion vacancies carry positive charge, they can} be directly manipulated via electric fields.
}
This direction could open new avenues for realizing and controlling skyrmion textures and dynamics in quantum materials, potentially enabling a form of ``layer skyrmionics''~\cite{fert2017magnetic,back20202020,petrovic2025colloquium}.

{ More broadly, ideal R$N$G provides an analytically controlled framework for
strongly interacting phases in Jellium-like systems with Berry curvature.
We expect this framework to serve as a foundation for identifying and understanding further correlated phases in rhombohedral graphene and related systems.}

\acknowledgments
We acknowledge useful discussions with Ashvin Vishwanath, Alfred Zong, Yves Kwan, Eslam Khalaf, Steve Kivelson, { and Jonah Herzog-Arbeitman}.
TD acknowledges support from the Sloan Research Fellowship.
TD performed part of this work at the Aspen Center for Physics, which is supported by National Science Foundation grant PHY-2210452.
TT is supported by the Stanford Graduate Fellowship.
PJL is supported by the MIT Pappalardo Fellowship in Physics.
This material is based upon work supported by the Air Force Office of Scientific Research under award number FA9550-25-1-0343.

\bibliography{ref}

\clearpage
\onecolumngrid
\appendix

\begin{center}
\large \textbf{Supplementary Material for ``The ideal limit of rhombohedral graphene: Dynamical skyrmion lattices and their excitations''}
\end{center}
\author{Tixuan Tan}
\affiliation{Department of Physics, Stanford University, Stanford, CA 94305, USA}
\author{Patrick J. Ledwith}
\affiliation{Department of Physics, Massachusetts Institute of Technology, Cambridge, MA 02139, USA}
\author{Trithep Devakul}
\affiliation{Department of Physics, Stanford University, Stanford, CA 94305, USA}
\maketitle

\tableofcontents

\section{Constraints on Slater determinants with strong contact interactions}

In this section we prove two statements that we used in the main text. The first is a technical lemma, which we used in the main text to prove that the orbitals of a $\langle \hat{V}_0 \rangle = 0$ Slater determinant are of the form $f_k(z) \xi_l(\br)$. The second is the claim that no $\langle \hat{V}_0 \rangle = 0$ Slater determinants preserve continuous translation symmetry.

\subsection{Constraining single particle orbitals of Slater determinants}
In the main text, we stated that the vanishing of
\begin{equation}
    \det[\partial_{\bar{z}}\bm{C}(\bm{r}),\bm{C}(\bm{r}),\bm{C}(\bm{r}_3),\dots, \bm{C}(\bm{r_{N_e}})],
    \label{eq:givencondition}
\end{equation}
for all $\bm{r}$ and $\bm{r}_{3},...\bm{r}_{N_{e}}$, implies that $\partial_{\bar{z}}\bm{C}(\bm{r})$ is parallel to $\bm{C}(\bm{r})$ for all $\bm{r}$.  Here, $\bm{C}(\bm{r})$ is a column vector, whose components are $\bm{C}_k(\bm{r})=\braket{\bm{r},0}{\psi_k}$, with $\{\ket{\psi_k}\}_{k\in [1,N_e]}$ being the occupied single-particle orbitals, and $0$ referring to the zeroth-layer component of each single-particle orbital.

The proof is given below.  Let us pick an arbitrary point $\bm{R}_1$, which we will later substitute for $\bm r$. We will assume, without loss of generality, that $\bm{C}(\bm{R}_1)\neq \bm{0}$ and $\partial_{\bar{z}_{\bm{R}_1}}\bm{C}(\bm{R}_1)\neq \bm{0}$. If either is $\bm{0}$, then the statement is trivially true since every vector is parallel to the zero vector. We take $\bm{C}(\bm{R}_1)\neq \bm{0}$ and choose $N_e-1$ coordinate points $\bm{R}_2...\bm{R}_{N_e}$ such that
\begin{equation}
    \det[\bm{C}(\bm{R}_1),\bm{C}(\bm{R}_2),\dots\bm{C}(\bm{R}_{N_e})]\neq 0.
    \label{eq:nonzerodetforCs}
\end{equation}
To see that this is always possible, we compute
\begin{equation}
\begin{split}
&\int |\det[\bm{C}(\bm{R}_1),\bm{C}(\bm{R}_2),\dots\bm{C}(\bm{R}_{N_e})]|^2 d\bm{R}_2d\bm{R}_3..d\bm{R}_{N_e}\\
=&\sum_{p,q}(-1)^{\text{sgn}(p)}(-1)^{\text{sgn}(q)}\bm{C}^*_{p_1}(\bm{R}_1) \bm{C}^*_{p_2}(\bm{R}_2).. \bm{C}_{q_1}(\bm{R}_1) \bm{C}_{q_2}(\bm{R}_2).. d\bm{R}_2d\bm{R}_3..d\bm{R}_{N_e}\\
=&\sum_{i,j=1}^{N_e} \bm{C}^*_i(\bm{R}_1) S^{-1}_{ji} \bm{C}_j(\bm{R}_1) \det(S),
\end{split}
\end{equation}
where the summation is over permutations, $\textrm{sgn}(...)$ is the parity of the permutation, and $S$ is the positive definite overlap matrix (Gram matrix),
\begin{equation}
S_{ij}\equiv \int \bm{C}^*_i(\bm{r})\bm{C}_j(\bm{r}) d\bm{r}.
\end{equation}
Thus, if $\bm{C}(\bm{R}_1)\neq \bm{0}$, then
\begin{equation}
\int |\det[\bm{C}(\bm{R}_1),\bm{C}(\bm{R}_2),\dots\bm{C}(\bm{R}_{N_e})]|^2 d\bm{R}_2d\bm{R}_3..d\bm{R}_{N_e}>0,
\end{equation}
which implies that there must exist $\bm{R}_2...\bm{R}_{N_e}$ such that \eqref{eq:nonzerodetforCs} holds. Equivalently, we have
\begin{equation}
\textrm{span}\{\bm{C}(\bm{R}_1)... \bm{C}(\bm{R}_{N_e})\}=\mathbb{C}^{N_{e}},
\label{eq:spannedbasis}
\end{equation}
where we regard $\bm{C}(\bm{R_i})$ as a complex vector in the $N_e$-dimensional vector space $\mathbb{C}^{N_{e}}$, defined over the complex number field. We can now expand $\partial_{\bar{z}} \bm{C}(\bm{R}_i) = \sum_{i=1}^{N_e} \alpha_i \bm{C}(\bm{R}_i)$.

We now show $\alpha_{i} = 0$ for all $i\in[2,N_e]$. We apply the given condition \eqref{eq:givencondition} to $\bm{r} = \bm{R}_1$ and $\{\bm{r}_3, \ldots, \bm{r}_{N_e} \} = \{\bm{R}_{2}, \ldots, \bm{R}_{N_e} \}_{\mathrm{no }\, \bm{R}_i}$, with $i \in [2,N_e]$. Inserting the basis expansion, we obtain
\begin{equation}
\begin{aligned}
    0 & = \det[\partial_{\bar{z}}\bm{C}(\bm{R_1}),\bm{C}(\bm{R_1}),\ldots,\bm{C}(\bm{R}_j),\ldots]_{\textrm{no}\, \bm{R}_i} \\
    & = \det[\alpha_i C(\bm{R}_i),\bm{C}(\bm{R_1}),\ldots,\bm{C}(\bm{R}_j),\ldots]_{\textrm{no}\, \bm{R}_i} = (-1)^i\alpha_i \det[\bm{C}(\bm{R}_1),\bm{C}(\bm{R}_2),\ldots,\bm{C}(\bm{R}_{N_e})]  .
    \end{aligned}
\end{equation}
Above we used that all other terms $\alpha_j C(\boldsymbol{R}_j)$ in the basis expansion lead to parallel columns and vanishing determinants. We also picked up a factor of $(-1)^i$ upon moving the first column $\bm{C}(\bm{R}_i)$ through to the $i$'th column. Using \eqref{eq:nonzerodetforCs}, we conclude that $\alpha_{i}=0$ for $i \in [2,N_e]$. This implies that $\partial_{\bar{z}}\bm{C}(\bm{R_1}) = \alpha_1 \bm{C}(\bm{R_1})$ for some $\alpha_1$, as we claimed in the main text.

\subsection{No $\langle \hat{V}_0 \rangle =0$ Slater determinant preserves continuous translational symmetry}
In the main text, we mentioned that Slater determinants that are zero modes of contact interactions have single particle orbitals of the form
\begin{equation}
\braket{\bm{r},l}{\psi_k}=\psi_{k,l}(\bm{r})=\xi_{l}(\bm{r})f_{k}(z),\quad \xi_l(\bm{r})=(-i\gamma \bar{\partial}_z)^l \xi_0(\bm{r}).
\end{equation}
Here, $l$ labels the layer components of each single-particle orbital, and $k$ labels different single-particle orbitals. We stated that there can be no $N_e\geq 2$ Slater determinant with such single-particle orbitals that preserves continuous translation symmetry.

Before we proceed, we clarify that there are $N_e>2$ Slater determinants that preserve translation symmetry. For example, the Slater determinant formed by filling different simple plane wave states preserves continuous translation symmetry. However, the single-particle orbitals are not of the required form and such determinants pay nonzero energy to contact interactions, $\langle \hat{V}_0 \rangle \neq 0$.

We first prove that the continuous translation symmetry of the  Slater determinant implies that there exists a basis where each constituent single-particle orbital of the Slater determinant has continuous translation symmetry.  We assume a complete orthogonal basis is $\{c^{\dagger}_1... c^{\dagger}_N\}$. $c^{\dagger}_i$ are electron creation operators that obey canonical anti-commutation relation. The Slater determinant state, written in this basis, is

\begin{equation}
\ket{\Psi}=c^{\dagger}_1 ...c^{\dagger}_{N_e}\ket{0}\quad N_e\leq N.
\end{equation}

The condition that $\ket{\Psi}$ has continuous translation symmetry implies that $\ket{\Psi}$ is an eigenstate of the translation operator $U(\bm{a})$, with the eigenvalue being a pure phase $e^{i\theta(\bm{a})}$. Here, $\bm{a}$ is the amount of translation. Explicitly, $U(\bm{a})$ is of the form

\begin{equation}
U(\bm{a})=\exp(i\sum^{N_e}_{i\in 1} \bm{p}_i \cdot \bm{a}),
\end{equation}
where $\bm{p}_i$ is the momentum operator of particle $i$. Since the basis $\{c^{\dagger}_1... c^{\dagger}_N\}$ is complete, the translation operator acts as
\begin{equation}
U(\bm{a})c^{\dagger}_iU^{\dagger}(\bm{a})=\sum^N_{j=1} M_{ij}c^{\dagger}_{j}\quad i=1...N_e.
\end{equation}
The translation symmetry, $U(\bm{a})\ket{\Psi}=e^{i\theta(\bm{a})}\ket{\Psi}$, has three implications. The first implication is that
\begin{equation}
\det(\tilde{M})=e^{i\theta(\bm{a})},
\end{equation}
where we have defined the submatrix $\tilde{M}$ of the full matrix $M$
\begin{equation}
\tilde{M}_{i,j}=M_{i,j}, \quad i,j\in [1,N_e].
\end{equation}

The second implication is that the coefficients of all terms of the form
\begin{equation}
c^{\dagger}_{a_1}...c^{\dagger}_{a_p} c^{\dagger}_{b_1}...c^{\dagger}_{b_{N_e-p}}\ket{0} \quad a_i \in \{N_e+1...N \} \quad b_i \in \{1...N_e\}
\end{equation}
resulting from $U(\bm{a})\ket{\Psi}$ must vanish. Let us first set $p=1, a_1=N_e+1$. Then there are $N_e$ constraints for the set of coefficients $\{M_{i,N_e+1}\}_{ i\in [1,N_e]}$.
\begin{equation}
U(\bm{a})\ket{\Psi}=\sum_{i=1}^{N_e}C_i c^{\dagger}_{N_e+1}\prod^{j\in [1,N_e]}_{j\neq i} c^{\dagger}_{j}\ket{0}+...
\end{equation}
We require all the coefficients $C_i$ to vanish:

\begin{equation}
\sum^{N_e}_{j=1}(\sum_{\sigma} (-1)^j (-1)^{\text{sgn}(\sigma)}\prod_{p\neq i,n\neq j} M_{\sigma(n),p})M_{j,N_e+1}=0, \quad i\in [1,N_e],
\end{equation}
where the summation over $\sigma$ is over all possible permutations. We notice that
\begin{equation}
\sum_{\sigma} (-1)^{i+j} (-1)^{\text{sgn}(\sigma)}\prod_{p\neq i,n\neq j} M_{\sigma(n),p}=\text{Adj}(\tilde{M})_{ij}.
\end{equation}
Here, $\textrm{Adj}(...)$ is the adjugate matrix. Thus, the $N_e$ constraints, one for each choice of $i$, can be organized into a matrix equation:

\begin{equation}
\textrm{Adj}(\tilde{M})
(M_{1,N_e+1},M_{2,N_e+1}..)^{T}=(0,0,0...)^{T}.
\end{equation}
 Since $\textrm{det}(\tilde{M})\neq 0$, $\textrm{det}(\textrm{Adj}(\tilde{M}))\neq 0$. This implies $M_{i,N_e+1}=0$ for $i\in [1,N_e]$. By similar arguments, we conclude that

\begin{equation}
M_{i,j}=0\quad i=1...N_e, j\geq N_e+1.
\end{equation}

The third implication is that since $U(\bm{a})$ does not change  anticommutation relation, the submatrix $\tilde{M}$ must be unitary.

\begin{equation}
\{U(\bm{a})c_iU^{\dagger}(\bm{a}),U(\bm{a})c_j^{\dagger}U^{\dagger}(\bm{a})\}=\delta_{ij},i,j\in[1,N_e]\to \sum^{N_e}_{k=1}\tilde{M}^{*}_{i,k}\tilde{M}_{j,k}=\delta_{ij}\to \tilde{M}\tilde{M}^{\dagger}=I
\end{equation}

Combining these three implications, we can diagonalize the unitary matrix $\tilde{M}$ to obtain eigenstates $|\tilde{\psi}_k\rangle$ of the translation operator. They are linear combinations of the original basis states $\ket{\psi_k}=c^{\dagger}_k\ket{0}$,

\begin{equation}
\ket{\tilde{\psi}_k}=\sum^{N_e}_{j=1} A_{kj} \ket{\psi_k}.
\end{equation}
The set $\{| \tilde{\psi}_k\rangle \}_{k\in[1,N_e]}$ is also a valid set of basis states for the Slater determinant. The Slater determinant state formed by filling $\{|\tilde{\psi}_k\rangle \}_{k\in[1,N_e]}$ and $\{\ket{\psi_k}\}_{k\in[1,N_e]}$ are the same up to an overall phase. We notice that all the discussion above is for a certain translation $\bm{a}$. However, the translation operators $U(\bm{a})$ commute with each other. Consequently, we can choose $\{| \tilde{\psi}_k\rangle \}_{k\in[1,N_e]}$ to be the common eigenstates of all the translation operators $U(\bm{a})$, i.e. the single-particle orbitals preserve continuous translation symmetry.

We have proved that translationally invariant Slater determinant state implies the existence of translationally invariant single-particle orbitals.  We now prove that such nontrivial Slater determinant does not exist for $N_e\geq 2$. We proceed with a proof by contradiction. By the general argument in the main text, the single particle orbitals are of the form

\begin{equation}
\langle \bm{r},l|\tilde{\psi}_k\rangle =\tilde{\psi}_{k,l}(\bm{r})=\xi_l(\bm{r})\tilde{f}_k(z).
\end{equation}
Since $\tilde{\psi}$ preserves continuous translation symmetry, we must have $\tilde{\psi}_{k,l}(\bm{r}+\bm{a}) = e^{i \bm{\kappa}_k \cdot \bm{a}} \tilde{\psi}_{k,l}(\bm{r})$ for all $\bm{a}$ and some momentum $\bm{\kappa}_k$. Choosing $\bm{a}$ to be arbitrarily small, we find
\begin{equation}
    \tilde{\psi}_{k,l}(\bm{r}) = u_{k,l} e^{i\bm{\kappa}_k \cdot \bm{r}}
\end{equation}
for some position-independent $u_{k,l}$.
We now take the ratio between two orbitals, $k=1,2$ without loss of generality, and fix $l=0$ for concreteness.
\begin{equation}
\frac{\tilde{f}_1(z)}{\tilde{f}_2(z)}= \frac{u_{1,0}}{u_{2,0}}\exp(i(\bm{\kappa}_1-\bm{\kappa}_2)\cdot\bm{r}).
\end{equation}
Differentiating with respect to $\overline{z}$, we find $\bm{\kappa}_1=\bm{\kappa_2}$. Our parent band has a single state for each momentum, so any two orbitals of this form with continuous translation symmetry are proportional to each other. Only one can be occupied, such that there is no Slater determinant with continuous translation symmetry for $N_e \geq 2$ (we assume throughout spin-valley polarization).

\section{Plane wave basis representation, momentum space occupations, and kinetic energy}\label{sec:planewave}
In the main text, we introduced the many-body wavefunction ansatz of the form
\begin{equation}
\Psi_{\{\ell\}}^{\text{ansatz}}(\{\bm{r}\})=\Phi_{\text{LLL}}^{\nu=1}(\{\bm{r}\})\prod_j \sqrt{\ell_j!}(-\sqrt{2}\gamma/ l_B)^{\ell_j}w_{\ell_j}(\bm{r}_j),
\label{eq:supp_final_psi_nu1}
\end{equation}
where $l_B$ is the magnetic length, $w_{\ell}(\bm{r})$, defined in the main text, obeys magnetic translation symmetry. The many-body Slater determinant is obtained by filling a number of single-particle orbitals. This section aims to derive the plane wave expansion of these single-particle orbitals, and from that derive the momentum space occupations and the kinetic energy expression of the Slater determinant.

We will first review the basics of Landau level wavefunctions on the torus (Sec. A,B). In Sec. B we will also review the LLL Bloch states, in terms of the Weierstrass $\sigma$ function, that we quoted in the main text. In Sec. C we compute the plane-wave representation of a product of two wavefunctions, one from a $B>0$ LL and one from a conjugate, $B<0$, LL. We subsequently use this representation in Sec. D to compute the momentum space occupation and kinetic energy of the many-body state, which we quoted in the main text.

\subsection{Review of Landau level wavefunction}
We assume a magnetic field pointing in the $-\hat{z}$ direction, $\bm{B}=-B\hat{z}$, $B>0$. We choose to use the symmetric gauge and set $\hbar=c=1$. The corresponding gauge field is
\begin{equation}
\bm{A}=\frac{B}{2}(y,-x,0)\to \nabla \times \bm{A}=\bm{B}.
\end{equation}
The Hamiltonian describing a free electron in a magnetic field is
\begin{equation}
H=\frac{(\bm{p}+e\bm{A})^2}{2m_e}.
\end{equation}
Here, $m_e$ is electron mass, with $-e<0$ denoting the charge of the electron. $\bm{p}$ is the canonical momentum operator. The guiding center coordinate of the electron is defined as
\begin{equation}
\bm{R}=\bm{r}+l_B^2 \bm{\pi}\times \hat{z},
\end{equation}
where $l_B^2=\frac{1}{eB}$ denotes the magnetic length, $\bm{r}$ represents the position operator of the electrons, $\bm{\pi}=\bm{p}+e\bm{A}$ is the kinetic momentum operator. It can be verified that the following set of commutation relations hold among these operators ($i,j=x,y$):
\begin{equation}
[R_x,R_y]=-il_B^2\quad [R_i,\pi_j]=0\quad [r_i,\pi_j]=i\delta_{ij}\quad [\pi_x,\pi_y]=\frac{i}{l_B^2}.
\end{equation}
In real space, these operators are defined as:
\begin{equation}
R_x=\frac{x}{2}-l_B^2 i\partial_y,\quad R_y=\frac{y}{2}+l_B^2i\partial_x,\quad \pi_x=-i\partial_x+\frac{y}{2l_B^2}, \quad \pi_y=-i\partial_y-\frac{x}{2l_B^2}.
\end{equation}
Furthermore, we can define two pairs of ladder operators that satisfy the canonical commutation relation:
\begin{equation}
a\equiv l_B \frac{\pi_x+i\pi_y}{\sqrt{2}},\quad  a^{\dagger}\equiv l_B \frac{\pi_x-i\pi_y}{\sqrt{2}},\quad [a,a^{\dagger}]=1.
\end{equation}

\begin{equation}
b\equiv \frac{R_x-iR_y}{\sqrt{2}l_B},\quad b^{\dagger}\equiv \frac{R_x+iR_y}{\sqrt{2}l_B} ,\quad [b,b^{\dagger}]=1.
\end{equation}

The Hamiltonian can be expressed using the ladder operators:
\begin{equation}
H=\omega_c a^{\dagger}a+\frac{\omega_c}{2}.
\end{equation}
Here, we have defined the cyclotron frequency $\omega_c=\frac{eB}{m_e}$.
In addition, these two pairs of ladder operators commute with each other, allowing us to introduce the  tensor product Hilbert space $\ket{n,m}\equiv \ket{n}\otimes \ket{m}$. Here, $a^{\dagger}a\ket{n}=n\ket{n}$, $b^{\dagger}b\ket{m}=m\ket{m}$ represent the number basis associated with the two sets of ladder operators. Consequently, the eigenstate of the Hamiltonian can be expressed as:
\begin{equation}
\ket{n,m}=\frac{(a^{\dagger})^n}{\sqrt{n!}} \frac{(b^{\dagger})^m}{\sqrt{m!}}\ket{0,0}=\frac{(a^{\dagger})^n}{\sqrt{n!}} \ket{0}\otimes \frac{(b^{\dagger})^m}{\sqrt{m!}}\ket{0}.
\end{equation}
The eigenvalue of this state is given by $E_n=(n+\frac{1}{2})\omega_c$. The state $\ket{0,0}$ is defined to be annihilated by both $a$ and $b$. It is simple to verify that in real space, the wavefunction is:
\begin{equation}
\braket{\bm{r}}{0,0}=\frac{1}{l_B\sqrt{2\pi}}e^{-\frac{z\bar{z}}{4l_B^2}},\quad z\equiv x+iy,\quad \bar{z}\equiv z^{*}.
\end{equation}
In the subsequent discussion, we will frequently interchange between complex variables $(z,k...)$ and  vector variables $(\bm{r},\bm{k}...)$. Their conversion will consistently adhere to the convention above, unless specified otherwise. For future reference, the normalized wavefunctions of other states in the lowest Landau level (LLL), $n=0$, are

\begin{equation}
\braket{\bm{r}}{n=0,m}=\frac{z^m}{l_B^{m+1}\sqrt{2\pi 2^m m!}}e^{-\frac{z\bar{z}}{4l_B^2}}.
\end{equation}

\subsection{Magnetic translation algebra, Weierstrass $\sigma$ function}
We introduce the magnetic translation operator $t(\bm{d})$ as follows:
\begin{equation}
t(\bm{d})=\exp(i\bm{K}\cdot\bm{d})\quad \bm{K}\equiv \frac{1}{l_B^2}\hat{z}\times \bm{R}.
\end{equation}
This operator is the analog of the ordinary translation operator in the absence of a magnetic field. Since $\bm{R}$ commutes with the Hamiltonian $H$, the magnetic translation is a symmetry of the system. We observe that the magnetic translation operator, when acting on an arbitrary function $f(\bm{r})$, results in both translation and the application of an associated phase factor:
\begin{equation}
t(\bm{d}) f(\bm{r})=\exp(\frac{i(\bm{r}\times \bm{d})}{2l_B^2})\exp(\partial_x d_x+\partial_y d_y) f(\bm{r})=\exp(\frac{i(\bm{r}\times \bm{d})}{2l_B^2}) f(\bm{r}+\bm{d}).
\end{equation}
Using the Baker--Campbell--Hausdorff (BCH) formula, a relationship between the magnetic translation operators can be derived as follows:
\begin{equation}
\begin{split}
t(\bm{d_1})t(\bm{d_2})&=\exp(i(K_x d_{1x}+K_y d_{1y}))\exp(i(K_x d_{2x}+K_y d_{2y}))\\
&=\exp(i\bm{K}\cdot(\bm{d}_1+\bm{d}_2))\exp(\frac{i}{2l_B^2}\hat{z}\cdot(\bm{d_1}\times \bm{d_2})),\\
t(\bm{d_1})t(\bm{d_2})&=t(\bm{d_2})t(\bm{d_1})\exp(\frac{i}{l_B^2}\hat{z}\cdot (\bm{d}_1\times\bm{d}_2)).
\end{split}
\end{equation}
Hence, $t(\bm{d_1})$ and $t(\bm{d_2})$ commute only if the parallelogram formed by $\bm{d_1}$ and $\bm{d_2}$ encloses an integer number of flux quanta, with each flux quantum occupying an area of $2\pi l_B^2$.

We now study Landau level on the torus. The torus is spanned by vector $N_1 \bm{a}_1$ and $N_2 \bm{a}_2$, where $|\bm{a}_1\times \bm{a}_2|=2\pi l_B^2$ is the area of unit cell, and $N_i$ are integers. If we constrain the translation vector $\bm{d}$ to be the lattice points $\bm{d}=n_1\bm{a}_1+n_2\bm{a}_2$, where $n_i=0...N_1-1$, then the set of magnetic translation operators $\{t(\bm{d})\}|_{\bm{d}=n_1\bm{a}_1+n_2\bm{a}_2}$ commute with each other. We can then diagonalize the Hamiltonian and require the wavefunction to be simultaneous eigenstate of the Hamiltonian and the magnetic translation operator.

Furthermore, it is observed that the magnetic translation operators commute with the ladder operators $a$ and $a^{\dagger}$, thereby operating only within the Hilbert space associated with the $b$ and $b^{\dagger}$ degrees of freedom. The magnetic Bloch states $\ket{\bm{k}}$ are defined as the eigenstates of the magnetic translation operator,
\begin{equation}
t(\bm{a}_i)\ket{\bm{k}}=\exp(i\phi_i)\exp(i\bm{k}\cdot \bm{a}_i)\ket{\bm{k}},
\end{equation}
where $i=1,2$, and the phase factors $\phi_i$ are arbitrary; we set both $\phi_{1,2}$ to be 0. The permissible values of $\bm{k}$ in the moir\'e Brillouinze zone (mBZ):
\begin{equation}\label{eq:allowedk_magnetic}
\bm{k}=\frac{n_1}{N_1}\bm{b}_1+\frac{n_2}{N_2}\bm{b}_2,\quad n_i=0...N_i-1.
\end{equation}
Here, $\bm{b}_i$ are the magnetic RL vectors satisfying $\bm{b}_i\cdot \bm{a}_j=2\pi\delta_{ij}$. It can be shown that the state defined below has the correct eigenvalues under the action of magnetic translation operators:
\begin{equation}\label{eq:definemagnetic}
\ket{\bm{k}}=\mathcal{N}_{\bm{k}}\sum^{\infty}_{p,q=-\infty}\exp(-i\bm{k}\cdot \bm{R}_{pq})t(\bm{a}_1)^p t(\bm{a}_2)^{q}\ket{m=0}\to t(\bm{a}_i)\ket{\bm{k}}=\exp(i\bm{k}\cdot \bm{a}_i)\ket{\bm{k}}.
\end{equation}
Here $\bm{R}_{pq}\equiv p\bm{a}_1+q\bm{a}_2$, $\mathcal{N}_{\bm{k}}$ is the normalization constant.

Using the definition provided earlier, the magnetic Bloch states in the lowest Landau level can be expressed as follows:
\begin{equation}\label{eq:mangeticBloch}
\braket{\bm{r}}{n=0,\bm{k}}=\phi_{\bm{k}}^{\textrm{LLL}}(\bm{r})=\mathcal{N}_{\bm{k}}\sum_{m,n}e^{imn\pi}\exp(-i\bm{k}\cdot \bm{R}_{mn})\exp(i\frac{\bm{r}\times \bm{R}_{mn}}{2l_B^2})\exp(-\frac{(z+R_{mn})(\bar{z}+\bar{R}_{mn})}{4l_B^2}).
\end{equation}
In real space, these states have the following boundary condition:
\begin{equation}
\phi_{\bm{k}}^{\textrm{LLL}}(\bm{r}+\bm{a}_i)=\phi_{\bm{k}}^{\textrm{LLL}}(\bm{r}) \exp(i\bm{k}\cdot \bm{a}_i)\exp(-i\frac{\bm{r}\times \bm{a}_i}{2l_B^2}).
\end{equation}
It is easy to verify that the boundary condition is independent of the Landau level index $n$:
\begin{equation}
\phi_{\bm{k}}^{\textrm{nLL}}(\bm{r}+\bm{a}_i)\equiv\frac{(a^{\dagger})^n}{\sqrt{n!}}\phi_{\bm{k}}^{\textrm{LLL}}(\bm{r}+\bm{a}_i)=\phi_{\bm{k}}^{\textrm{nLL}}(\bm{r}) \exp(i\bm{k}\cdot \bm{a}_i)\exp(-i\frac{\bm{r}\times \bm{a}_i}{2l_B^2}).
\end{equation}

For future reference, we now give an analytic form of the magnetic Bloch state, as defined by Eq.\ref{eq:definemagnetic}. We first note that $\phi_{\bm{k}}^{\textrm{LLL}}(\bm{r})$ is holomorphic in $z$ up to a Gaussian factor:

\begin{equation}
\begin{split}
\phi_{\bm{k}}^{\textrm{LLL}}(\bm{r})&=\mathcal{N}_{\bm{k}}\sum_{p,q}e^{ipq\pi}\exp(-i\bm{k}\cdot \bm{R}_{pq})\exp(i\frac{\bm{r}\times \bm{R}_{pq}}{2l_B^2})\exp(-\frac{(z+R_{pq})(\bar{z}+\bar{R}_{pq})}{4l_B^2})\\
&=\mathcal{N}_{\bm{k}}\exp(-\frac{z\bar{z}}{4l_B^2})\sum_{p,q}e^{ipq\pi}\exp(-i\bm{k}\cdot \bm{R}_{pq})\exp(-\frac{2z\bar{R}_{pq}+R_{pq}\bar{R}_{pq}}{4l_B^2})\\
&\equiv  \mathcal{N}_{\bm{k}}\exp(-\frac{z\bar{z}}{4l_B^2}) \tilde{\phi}^{\textrm{LLL}}_{\bm{k}}(z),
\end{split}
\end{equation}
 where we have defined $\tilde{\phi}^{\textrm{LLL}}_{\bm{k}}(z)$ to be the holomorphic function after the summation.
 It has the real-space boundary condition
\begin{equation}
\tilde{\phi}_{\bm{k}}^{\textrm{LLL}}(z+a_i)=\tilde{\phi}_{\bm{k}}^{\textrm{LLL}}(z) \exp(i\bm{k}\cdot \bm{a}_i)\exp(\frac{2z\bar{a}_i+a_i\bar{a}_i}{4l_B^2}),
\label{eq:boundarycondition}
\end{equation}
where $\bm{a}_i$ is a primitive lattice vector. A function that has the same boundary condition is $\varphi_\bk(z) = e^{\frac{i}{2} \overline{k'} z} \sigma(z + i k')$, where $\sigma(z) = \sigma(z|a_1,a_2)$ is the modified Weierstrass sigma function \cite{haldaneModularinvariantModifiedWeierstrass2018} and $k' = k-(b_1+b_2)/2$. It satisfies the properties
\begin{equation}
    \sigma(a) = 0, \qquad \sigma(z + a_i) = - \exp(\frac{2z\bar{a}_i+a_i\bar{a}_i}{4l_B^2}) \sigma(z)
\end{equation}
We note that $\tilde{\phi}_{\bm{k}}^{\textrm{LLL}}(z)/\varphi_\bk(z)$ is a periodic function of $z$ with at most one pole in the unit cell. However, any non-constant periodic function must be an elliptic function with at least two poles per unit cell. We therefore conclude that
\begin{equation}\label{eq:blochtheta}
\tilde{\phi}_{\bm{k}}^{\mathrm{LLL}}(z)\propto e^{\frac{i}{2} \overline{k'} z} \sigma(z + i l_B^2 k').
\end{equation}

We note that in Eq. \ref{eq:definemagnetic} we start with the $\ket{m=0}$ state in the definition of the magnetic Bloch state. This definition is singular at $\bm{k}=\frac{\bm{b_1}+\bm{b}_2}{2}$~\cite{perelomovCompletenessSystemCoherent1971}: a consequence of writing Bloch states as a superposition of exponentially localized states in a topological band~\cite{liConstraintsRealSpace2024}. We can start with a different $\ket{m\neq 0}$, and that definition will become singular at some other point in the BZ.
However, since the real space boundary condition is independent of $m$, starting with a different $m$ will produce  the same magnetic Bloch state, up to an overall normalization constant. We can work in patches of the BZ, and for $m_1\neq m_2$, in the region of the BZ where the definition of magnetic Bloch states using $\ket{m_1}$ and $\ket{m_2}$ are both valid, they produce the same state up to an overall factor. This is due to the uniqueness of holomorphic function with respect to the boundary condition. \footnote{However, the uniqueness argument is true only when there is one flux quantum per magnetic unit cell; more care is required if there are multiple flux quanta per unit cell}.

\subsection{Representation of product of magnetic Bloch states in momentum space}\label{sec:momentumrep}
Since the real-space boundary condition of the mangetic Bloch state is independent of the Landau level index $n$, the product,
\begin{equation}
\psi_{\bm{k}+\bm{q},n}(\bm{r})\equiv \phi_{\bm{k}}^{\textrm{LLL}}(\bm{r})\phi^{\textrm{nLL}*}_{-\bm{q}}(\bm{r}),
\end{equation}
follows Bloch periodicity. This section will provide a plane wave expansion of this Bloch wavefunction. We note that $\psi_{\bm{k}+\bm{q},n}(\bm{r})$ is unnormalized in this definition, which  turns out to be convenient for later sections. In the equation above, $\bm{q}$ is later to be identified with the $\bm{q}$ used in the definition of $\chi_{l}(\bm{r})$ in the main text:
\begin{equation}
\psi_{\bm{k}+\bm{q},n}(\bm{r}+\bm{a}_i)=\exp(i(\bm{k}+\bm{q})\cdot \bm{a}_i) \psi_{\bm{k}+\bm{q},n}(\bm{r}).
\end{equation}

We will focus on the case where $n=0$, leaving $n\neq 0$ case for subsequent discussion. By breaking down $\phi_{\bm{k}}^{\textrm{LLL}}(\bm{r})\phi^{\textrm{LLL}*}_{-\bm{q}}(\bm{r})$ into a lattice sum of a function offset by a shifted origin, we aim to utilize the Poisson resummation formula on it.

\begin{equation}
\begin{split}
&e^{-i(\bm{k}+\bm{q})\cdot\bm{r}}\psi_{\bm{k}+\bm{q},0}(\bm{r})\\
=&e^{-i(\bm{k}+\bm{q})\cdot \bm{r}}\phi^{\textrm{LLL}}_{\bm{k}}(\bm{r})\phi^{\textrm{LLL}*}_{-\bm{q}}(\bm{r})\\
=& \mathcal{N}_{\bm{k}} \mathcal{N}_{-\bm{q}}\sum_{m,n,a,b} e^{imn\pi}e^{-i ab\pi}e^{-i\bm{k}\cdot (\bm{R}_{mn}+\bm{r})}\exp(i\frac{ \bm{r}\times \bm{R}_{mn}}{2l_B^2}) \exp(-i\frac{\bm{r}\times \bm{R}_{ab} }{2l_B^2})\\
&\exp(-\frac{(z+R_{mn})(\bar{z}+\bar{R}_{mn})}{4l_B^2}) \exp(-\frac{(z+R_{ab})(\bar{z}+\bar{R}_{ab})}{4l_B^2})\exp(-i\bm{q}\cdot (\bm{R}_{ab}+\bm{r}))\\
=& \mathcal{N}_{\bm{k}} \mathcal{N}_{-\bm{q}}\sum_{m,n,a,b} e^{imn\pi}e^{-i (a+m)(b+n)\pi}e^{-i(\bm{k}+\bm{q})\cdot (\bm{R}_{mn}+\bm{r})}\exp(i\frac{\bm{r}\times \bm{R}_{mn}}{2l_B^2}) \exp(-i\frac{\bm{r}\times(\bm{R}_{ab}+\bm{R}_{mn})}{2l_B^2})\\
&\exp(-\frac{(z+R_{mn})(\bar{z}+\bar{R}_{mn})}{4l_B^2}) \exp(-\frac{(z+R_{mn}+R_{ab})(\bar{z}+\bar{R}_{mn}+\bar{R}_{ab})}{4l_B^2})\exp(-i\bm{q}\cdot \bm{R}_{ab})\\
=&\mathcal{N}_{\bm{k}} \mathcal{N}_{-\bm{q}}\sum_{m,n,a,b} e^{-i ab\pi}e^{-i(\bm{k}+\bm{q})\cdot (\bm{R}_{mn}+\bm{r})} \exp(-i\frac{ (\bm{r}+\bm{R}_{mn})\times \bm{R}_{ab}}{2l_B^2})\\
&\exp(-\frac{(z+R_{mn})(\bar{z}+\bar{R}_{mn})}{4l_B^2}) \exp(-\frac{(z+R_{mn}+R_{ab})(\bar{z}+\bar{R}_{mn}+\bar{R}_{ab})}{4l_B^2}) \exp(-i\bm{q}\cdot \bm{R}_{ab})\\
\equiv&  \mathcal{N}_{\bm{k}} \mathcal{N}_{-\bm{q}} \sum_{m,n}\kappa_{\bm{k}+\bm{q}}(\bm{r}+\bm{R}_{mn})
\end{split}
\end{equation}

\begin{equation}
\kappa_{\bm{k}+\bm{q}}(\bm{r})\equiv \sum_{a,b} e^{-i ab\pi}e^{-i(\bm{k}+\bm{q})\cdot \bm{r}} \exp(-i\frac{ \bm{r}\times \bm{R}_{ab}}{2l_B^2})
\exp(-\frac{z\bar{z}}{4l_B^2}) \exp(-\frac{(z+R_{ab})(\bar{z}+\bar{R}_{ab})}{4l_B^2})\exp(-i\bm{q}\cdot \bm{R}_{ab}).
\end{equation}
We notice that $\kappa_{\bm{k}+\bm{q}}(\bm{r})$ is localized around the origin due to the $\exp(-\frac{z\bar{z}}{4l_B^2})$ factor.  Next, we apply the Poisson resummation formula \footnote{We notice that in principle $\kappa_{\bm{k}+\bm{q}}(\bm{r})$ should carry an additional subscript $\kappa_{\bm{k}+\bm{q},\bm{q}}(\bm{r})$. However, in most of the cases the momentum of the condensate part $\bm{q}$ is fixed, so we suppress this subscript unless necessary. The same applies to $\mathcal{M}_{\bm{k}+\bm{q}}$ as defined below. }:
\begin{equation}
\sum_{m,n}\kappa_{\bm{k}+\bm{q}}(\bm{r}+\bm{R}_{mn})=\sum_{\bm{g}}\tilde{\kappa}_{\bm{k}+\bm{q}}(\bm{g})\exp(i\bm{g}\cdot \bm{r}),
\end{equation}
where $\bm{g}$ is summed over reciprocal lattice vectors. The expression of $\tilde{\kappa}_{\bm{k}+\bm{q}}(\bm{g})$ is

\begin{equation}
\begin{split}
\tilde{\kappa}_{\bm{k}+\bm{q}}(\bm{g})&=\frac{1}{2\pi l_B^2}\int d\bm{r} \kappa_{\bm{k}+\bm{q}}(\bm{r})\exp(-i\bm{g}\cdot \bm{r})\\
&=\sum_{a,b}e^{-iab\pi}\exp(-\frac{|\bm{k}+\bm{q}+\bm{g}|^2l_B^2}{2})\exp(-\frac{|a\bm{a}_1+b\bm{a}_2|^2}{4l_B^2})\exp(\frac{i}{2}a(\bar{k}+\bar{q}+\bar{g})a_1)\exp(\frac{i}{2}b(\bar{k}+\bar{q}+\bar{g})a_2)\\
&\exp(-\frac{ia}{2}(q\bar{a}_1+\bar{q}a_1)) \exp(-\frac{ib}{2}(q\bar{a}_2+\bar{q}a_2)).
\end{split}
\end{equation}
$\tilde{\kappa}_{\bm{k}+\bm{q}}(\bm{g})$ can represented using the Riemann theta function, defined as \cite{DLMF_riemanntheta}:
\begin{equation}
\Theta(\zeta|\Lambda)\equiv \sum_{n\in \mathbb{Z}^{p} }\exp(2\pi i (\frac{1}{2}n^{T}\Lambda n+n^{T}\zeta)).
\end{equation}
Here, $p$ is a positive integer, $\zeta\in \mathbb{C}^p$, $\Lambda$ is a $p\times p$ symmetric matrix with a positive definite imaginary part. The Riemann theta function satisfies the following quasi-periodicity relation for $m,n\in \mathbb{Z}^p$:
\begin{equation}
\Theta(\zeta+m+\Lambda n|\Lambda)=\exp(-2\pi i (\frac{1}{2} n^{T}\Lambda n+n^T \zeta)) \Theta(\zeta|\Lambda).
\end{equation}
Given these definitions, we rewrite $\tilde{\kappa}_{\bm{k}+\bm{q}}(\bm{g})$ using Riemann theta function:
\begin{equation}
\tilde{\kappa}_{\bm{k}+\bm{q}}(\bm{g})=\exp(-\frac{|\bm{k}+\bm{q}+\bm{g}|^2l_B^2}{2})\Theta(\zeta_{\bm{k}+\bm{q},\bm{g}}|\Lambda_0),
\end{equation}
\begin{equation}
\zeta_{\bm{k}+\bm{q},\bm{g}}=\frac{1}{2\pi}
\begin{pmatrix}
(\bar{k}+\bar{q}+\bar{g})\frac{a_1}{2}-(\bm{q}\cdot\bm{a}_1)\\
(\bar{k}+\bar{q}+\bar{g})\frac{a_2}{2} -(\bm{q}\cdot\bm{a}_2)
\end{pmatrix},
\end{equation}
\begin{equation}
\Lambda_0=\frac{1}{\pi i}
\begin{pmatrix}
-a_1\bar{a}_1/(4l_B^2) & -\frac{i\pi}{2}-\frac{a_1\bar{a}_2+a_2\bar{a}_1}{8l_B^2}\\
 -\frac{i\pi}{2}-\frac{a_1\bar{a}_2+a_2\bar{a}_1}{8l_B^2} & -a_2\bar{a}_2/(4l_B^2)
\end{pmatrix}.
\end{equation}
Here, we have assumed that $\bm{g}=n_1\bm{b}_1+n_2\bm{b}_2$. By using the relation,
\begin{equation}
b_1=-ia_2/l_B^2, \quad b_2 =ia_1/l_B^2, \quad \frac{i}{2}(a_1\bar{a}_2-\bar{a}_1a_2)=2\pi l_B^2,
\end{equation}
we obtain
\begin{equation}
\zeta_{\bm{k}+\bm{q},\bm{g}}=\zeta_{\bm{k}+\bm{q},\bm{0}}+\Lambda_0 \begin{pmatrix}
-n_2\\
n_1
\end{pmatrix}+
\begin{pmatrix}
n_1\\
0
\end{pmatrix},
\end{equation}

\begin{equation}
\tilde{\kappa}_{\bm{k}+\bm{q}}(\bm{g})=\exp(-\frac{|\bm{k}+\bm{q}+\bm{g}|^2l_B^2}{2})\Theta(\zeta_{\bm{k}+\bm{q},\bm{0}}|\Lambda_0)\exp(\frac{l_B^2}{4}(g\bar{g}+2\bar{k}g))\exp(i\pi n_1n_2)\exp(\frac{l_B^2}{2} q\bar{g}).
\end{equation}

Combining the results above, we derive the momentum space representation of the Bloch state
\begin{equation}\label{eq:momentumbloch}
\psi_{\bm{k}+\bm{q},0}(\bm{r})=\mathcal{M}_{\bm{k}+\bm{q}}\sum_{\bm{g}}\exp(i\pi n_1 n_2)\exp(-\frac{l_B^2}{4}(g\bar{g}+2(k+q)\bar{g}))e^{i(\bm{k}+\bm{g}+\bm{q})\cdot (\bm{r}-\bm{R}_0)}
\end{equation}
where all terms independent of $\bm{g}$ are absorbed by the overall factor $\mathcal{M}_{\bm{k}+\bm{q}}$, and $\bm{R}_0=(-q_y,q_x)l_B^2$ is a shift of spatial origin.

If we use the higher Landau level wavefunction in the definition of  the Bloch state,
\begin{equation}
\psi_{\bm{k}+\bm{q},n}(\bm{r})\equiv \phi_{\bm{k}}^{\textrm{LLL}}(\bm{r})\phi^{\textrm{nLL}*}_{-\bm{q}}(\bm{r}),
\end{equation}
where the normalized magnetic Bloch states of higher Landau level is (assuming that $\phi_{\bm{k}}^{\textrm{LLL}}$ is normalized):
\begin{equation}
\phi^{\textrm{nLL}}_{\bm{k}}(\bm{r})=\frac{1}{\sqrt{n!}}(a^{\dagger})^n \phi^{\textrm{LLL}}_{\bm{k}}(\bm{r})=\frac{l^n_B}{\sqrt{n!}}(-i\sqrt{2}\partial_z+\frac{i}{2\sqrt{2}l_B^2}\bar{z})^n \phi^{\textrm{LLL}}_{\bm{k}}(\bm{r}).
\end{equation}
Using Eq.\ref{eq:mangeticBloch}, we can verify the following commutation relation:
\begin{equation}\label{eq:zcommutation}
[\partial_{\bar{z}},\phi^{\textrm{LLL}}_{\bm{k}}(\bm{r})]=-\frac{z}{4 l_B^2} \phi^{\textrm{LLL}}_{\bm{k}}(\bm{r}).
\end{equation}
It then follows that
\begin{equation}
\psi_{\bm{k}+\bm{q},n}(\bm{r})=\phi^{\textrm{LLL}}_{\bm{k}}(\bm{r})\phi^{\textrm{nLL}*}_{-\bm{q}}(\bm{r})= \frac{l_B^n}{\sqrt{n!}} (i\sqrt{2}\partial_{\bar{z}})^n [\phi^{\textrm{LLL}}_{\bm{k}}(\bm{r})\phi^{\textrm{LLL}*}_{-\bm{q}}]=  \frac{l_B^n}{\sqrt{n!}} (i\sqrt{2}\partial_{\bar{z}})^n \psi_{\bm{k}+\bm{q},0}(\bm{r}).
\end{equation}
Using Eq.\ref{eq:momentumbloch}, we conclude that the plane wave expansion of $\psi_{\bm{k}+\bm{q},n}(\bm{r})$ is \footnote{Strictly speaking $\mathcal{M}_{\bm{k}+\bm{q}}$ should carry an additional index $\bm{q}$, $\mathcal{M}^{(\bm{q})}_{\bm{k}+\bm{q}}$. This will only be important in the later TDVP section.}
\begin{equation}\label{eq:planewave}
\psi_{\bm{k}+\bm{q},n}(\bm{r})=\mathcal{M}_{\bm{k}+\bm{q}}\sum_{\bm{g}}\exp(i\pi n_1 n_2)(-\frac{k+g+q}{\sqrt{2}})^n \frac{l_B^n}{\sqrt{n!}} \exp(-\frac{l_B^2}{4}(g\bar{g}+2(k+q)\bar{g}))e^{i(\bm{k}+\bm{q}+\bm{g})\cdot (\bm{r}-\bm{R}_0)},
\end{equation}
where we note that $\mathcal{M}_{\bm{k}+\bm{q}}$ is independent of the Landau level index $n$.

\subsection{Momentum space occupation and kinetic energy of many-body wavefunction}\label{sec:relationtomanybody}

The ansatz many-body wavefunction is

\begin{equation}
\Psi_{\{\ell\}}^{\text{ansatz}}(\{\bm{r}\})=\Phi_{\text{LLL}}^{\nu=1}(\{\bm{r}\})\prod_j \chi_{\ell_j}(\bm{r}_j).\\
\end{equation}
Notice that $\Phi_{\text{LLL}}^{\nu=1}(\{\bm{r}\})$ can be written as
\begin{equation}
\Phi_{\text{LLL}}^{\nu=1}(\{\bm{r}\})=\text{det}_{\bm{k} m}[\phi^{\text{LLL}}_{\bm{k}}(\bm{r}_m)].
\end{equation}
Since $\chi_{\ell_i}(\bm{r}_i)$ does not depend on $\bm{k}$, it follows that
\begin{equation}
\Psi_{\{\ell\}}^{\text{ansatz}}(\{\bm{r}\})=\text{det}_{\bm{k} m}[\phi^{\text{LLL}}_{\bm{k}}(\bm{r}_m)\chi_{\ell_m}(\bm{r}_m)].
\end{equation}
We now identify $w_n(\bm{r})$ used in the main text as
\begin{equation}
w_n(\bm{r})=\phi_{-\bm{q}}^{\textrm{nLL}*}(\bm{r}).
\end{equation}
If we choose
\begin{equation}
\chi_0(\bm{r})=w_0(\bm{r}),
\end{equation}
it follows that
\begin{equation}
\begin{split}
\phi^{\textrm{LLL}}_{\bm{k}}(\bm{r})\chi_{\ell}(\bm{r})=(-2i\gamma\partial_{\bar{z}})^{\ell}[\phi^{\textrm{LLL}}_{\bm{k}}(\bm{r})\chi_{0}(\bm{r})]=\sqrt{\ell!}(-\frac{\sqrt{2}\gamma}{l_B})^{\ell}  \phi^{\textrm{LLL}}_{\bm{k}}(\bm{r}) w_{\ell}(\bm{r})\to \chi_{\ell}(\bm{r})= \sqrt{\ell!}(-\frac{\sqrt{2}\gamma}{l_B})^{\ell} w_{\ell}(\bm{r}).
\end{split}
\end{equation}
This gives Eq.\ref{eq:supp_final_psi_nu1}. In general, if we choose
\begin{equation}
\chi_0(\bm{r})=\sum_{n}c_n w_n(\bm{r}),
\end{equation}
Using the result in the last section, it is straight-forward to derive that

\begin{equation}
\begin{split}\label{eq:ansatzexpansion}
&\psi^{\textrm{ansatz}}_{\bm{k}+\bm{q},l}(\bm{r})\equiv \phi^{\textrm{LLL}}_{\bm{k}}\chi_{l}(\bm{r})\\
&=\sum_n c_n (-2i\gamma \partial_{\bar{z}})^l \phi_{\bm{k}}(\bm{r})\phi^{\textrm{nLL}*}_{-\bm{q}}(\bm{r})\\
&=\mathcal{M}_{\bm{k}+\bm{q}}(-2i\gamma \partial_{\bar{z}})^l[\sum_{\bm{g},n}\exp(i\pi n_1 n_2)(-\frac{k+g+q}{\sqrt{2}})^n \frac{c_n l_B^n}{\sqrt{n!}} \exp(-\frac{l_B^2}{4}(g\bar{g}+2(k+q)\bar{g}))e^{i(\bm{k}+\bm{q}+\bm{g})\cdot (\bm{r}-\bm{R}_0)}\frac{1}{N_{\bm{k}+\bm{g}+\bm{q}}}s_{\bm{k}+\bm{g}+\bm{q},0}]\\
&=\mathcal{M}_{\bm{k}+\bm{q}}\sum_{\bm{g},n}e^{i\pi n_1 n_2}(-\frac{k+g+q}{\sqrt{2}})^n \frac{c_n l_B^n}{\sqrt{n!}} \exp(-\frac{l_B^2}{4}(g\bar{g}+2(k+q)\bar{g}))e^{i(\bm{k}+\bm{q}+\bm{g})\cdot (\bm{r}-\bm{R}_0)}\frac{1}{N_{\bm{k}+\bm{g}+\bm{q}}}s_{\bm{k}+\bm{g}+\bm{q},l}\\
&=\mathcal{M}_{\bm{k}+\bm{q}}e^{\frac{l_B^2|\bm{k}+\bm{q}|^2}{4}}\sum_{\bm{g},n}e^{i\pi n_1 n_2}(-\frac{k+g+q}{\sqrt{2}})^n \frac{c_n l_B^n}{\sqrt{n!}} \exp(-\frac{l_B^2}{4}|\bm{k}+\bm{q}+\bm{g}|^2)\exp(\frac{il_B^2}{2}(\bm{k}+\bm{q})\times\bm{g})\frac{e^{i(\bm{k}+\bm{q}+\bm{g})\cdot (\bm{r}-\bm{R}_0)}}{N_{\bm{k}+\bm{g}+\bm{q}}}s_{\bm{k}+\bm{g}+\bm{q},l},
\end{split}
\end{equation}
where $s_{\bm{k}+\bm{g},l}$ is defined in the main text, and $N_{\bm{k}+\bm{g}+\bm{q}}$ is its normalization factor. The many-body wavefunction is the Slater determinant state formed by filling $\psi^{\textrm{ansatz}}_{\bm{k},l}(\bm{r})$ for all the allowed $\bm{k}$. The momentum space distribution function follows from the expression above:

\begin{equation}
n(\bm{k}+\bm{g})=\langle c^{\dagger}_{\bm{k},\bm{g}} c_{\bm{k},\bm{g}}\rangle =\frac{ |\sum_n (-l_B\frac{k+g}{\sqrt{2}})^n \frac{c_n}{\sqrt{n!}}|^2\exp(-\frac{l_B^2}{2}|\bm{k}+\bm{g}|^2) N^{-2}_{\bm{k}+\bm{g}}}{\sum_{\bm{g}'}|\sum_n (-l_B\frac{k+g+g'}{\sqrt{2}})^n \frac{c_n}{\sqrt{n!}}|^2 \exp(-\frac{l_B^2}{2}|\bm{k}+\bm{g}+\bm{g}'|^2) N^{-2}_{\bm{k}+\bm{g}+\bm{g}'}}.
\end{equation}
Here, we use $c^{\dagger}_{\bm{k},\bm{g}}$ and $c_{\bm{k},\bm{g}}$ to denote the electron creation and annihilation operators corresponding to the state $e^{i(\bm{k}+\bm{g})\cdot\bm{r}}s_{\bm{k}+\bm{g},l}$. We use the convention that $\bm{k}$ is confined with the mBZ, while $\bm{g}$ and $\bm{g}'$ are reciprocal lattice vectors (different from the main text where $\bm{k}$ is not restricted to the mBZ). It follows that when the dispersion of $s_{\bm{k}+\bm{g},l}$ is $\mathcal{E}(\bm{k}+\bm{g})$ (when $\lambda=0.0$), the kinetic energy from fully filling the band $\psi_{\bm{k},l}^{\textrm{ansatz}}(\bm{r})$, is
\begin{equation}
E_{\textrm{kin}}=\sum_{\bm{k},\bm{g}} \mathcal{E}(\bm{k}+\bm{g})n(\bm{k}+\bm{g}).
\end{equation}

\section{Single-particle and Hartree-Fock bandstructures of R5G}\label{sec:HFsingle}
In this section, we describe the technical details of Hartree-Fock calculations on Rhombohedral Multilayer Graphene (RMG). Our approach aligns closely with the convention outlined in Ref. \cite{tan2024parent,dong2024anomalous}.

Sec. \ref{sec:Tight-binding Hamiltonian},   \ref{sec:dipersiveideallimit}, \ref{sec:hBNmoirepotential} lay out the basic details of the sinlge particle Hamiltonian used in the Hartree-Fock calculation; Sec. \ref{sec:Tight-binding Hamiltonian} introduces the realistic tight-binding Hamiltonian, Sec. \ref{sec:dipersiveideallimit} describes the simplifications we make in the ideal limit, and Sec. \ref{sec:hBNmoirepotential} discusses the hBN moir\'e potential.

Sec. \ref{sec:The effect of hBN on the RMG spinor} discusses the local band structure of RMG/hBN for different stacking and alignment. Sec. \ref{sec:Moire potential from the filled valence bands of RMG/hBN} discusses the moir\'e potential from the filled valence band of RMG, which also enters into our Hartree-Fock calculation. Sec. \ref{sec:HFdetails} explains the Hartree-Fock approximation for the interaction. Sec. \ref{sec:Wavefunction overlap between analytic wavefunction and Hartree-Fock wavefunction} explains how we compute the overlap between the analytic ansatz of the Bloch wavefunction and those obtained from the numerics (i.e. how we obtain the 0.997 in the main text Fig. 3). Sec. \ref{sec:Determination of phase boundaries} explains how we determine the phase boundary in the ideal limit, as plotted in the main text Fig. 3 (a). Sec. \ref{sec:Trace condition violation of filled Hartree Fock band} defines the trace condition violation and explains how we compute it in numerics. Sec. \ref{sec:pontragin} defines the skyrmion density and shows that the skyrmion lattice interpretation holds for real R5G/hBN. Sec. \ref{sec:Adiabatic continuity between ideal and realistic models} contains additional data that shows the ideal limit is smoothly connected to the real R5G/hBN.

\subsection{Tight-binding Hamiltonian}\label{sec:Tight-binding Hamiltonian}

\begin{figure}
    \centering
    \includegraphics[width=1.0\linewidth]{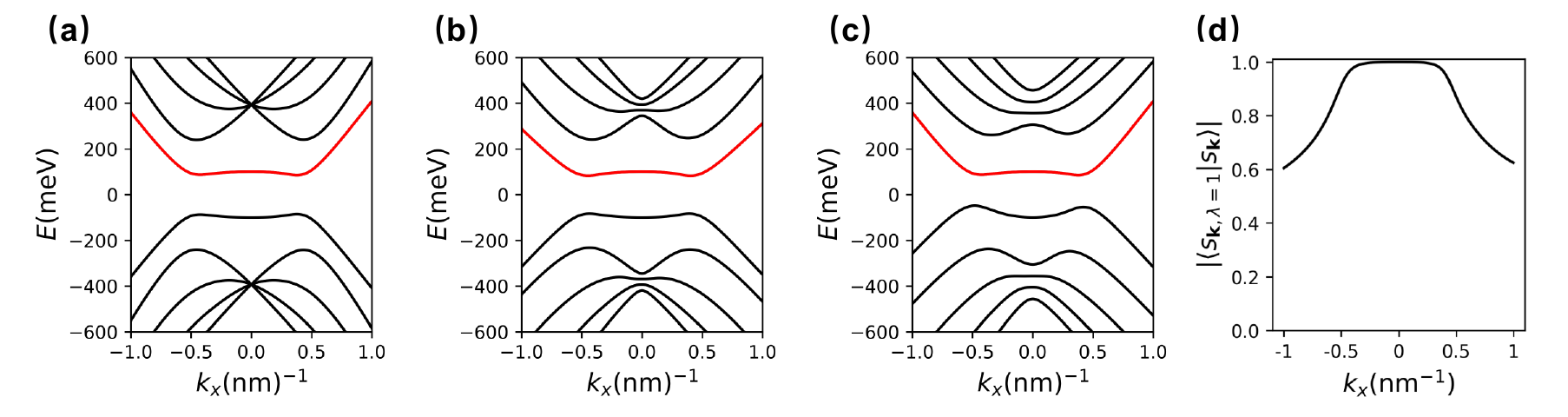}
    \caption{Single particle band structure of $h_{\textrm{inp}}$ for different interpolation ratio $\lambda$. We set $\Delta_B=0\textrm{meV}$, and $\mathcal{E}(\bm{k})$ to be the dispersion of the first conduction band of $h_{\textrm{R5G}}$ at $u_D=50\textrm{meV}$. \textrm{(a)}$\lambda=0$, \textrm{(b)} $\lambda=0.5$, \textrm{(c)}$\lambda=1$. The red band is the band we project to in Hartree-Fock calculation. \textrm{(d)} The overlap between the wavefunction of the first conduction band from (a) and (c) as a function of momentum $\bm{k}$. $\ket{s_{\bm{k},\lambda=1}}$ is the first conduction band wavefunction of $h_{\textrm{RMG}}(\bm{K}+\bm{k})$.}
    \label{Sfig_bandstructure}
\end{figure}

The real-space lattice of RMG is defined by the basis vectors:
\begin{equation}
\bm{R}_1=a_{\text{Gr}}(1,0),\quad \bm{R}_2=a_{\text{Gr}}(\frac{1}{2},\frac{\sqrt{3}}{2}),
\end{equation}
where $a_{\textrm{Gr}}=0.246\textrm{nm}$ is the graphene lattice constant. The reciprocal lattice of RMG is spanned by the basis vectors:
\begin{equation}
\bm{G}_1=\frac{4\pi}{\sqrt{3}a_{\text{Gr}}}(\frac{\sqrt{3}}{2},-\frac{1}{2}),\quad \bm{G}_2=\frac{4\pi}{\sqrt{3}a_{\text{Gr}}}(0,1).
\end{equation}

The tight-binding Hamiltonian of RMG is

\begin{equation}\label{eq:TBHAM}
H_{\textrm{RMG}}=\sum_{\bm{k}\in \text{BZ},\sigma,\sigma',l,l'}c^{\dagger}_{\bm{k},\sigma,l} [h_{\textrm{RMG}}(\bm{k})]_{\sigma,l;\sigma',l'}c_{\bm{k},\sigma',l'}.
\end{equation}
In this expression, $c$ and $c^{\dagger}$ are fermion annihilation and creation operators, respectively. $\sigma={A,B}$ labels the sublattice, and $l=0...N_{L}-1$ labels the layer. Throughout this paper $N_L=5$. BZ denotes Brillouin zone. The intralayer ($l=l'=L$)  Hamiltonian is
\begin{equation}
[h_{\textrm{RMG}}(\bm{k})]_{\sigma,L;\sigma',L}=
\begin{pmatrix}
u_L & -t_0f(\bm{k})\\
-t_0 f^*(\bm{k}) & u_L
\end{pmatrix}_{\sigma\sigma'}.
\end{equation}
The interlayer tunneling Hamiltonians are
\begin{equation}
[h_{\textrm{RMG}}(\bm{k})]_{\sigma,L;\sigma',L+1}=
\begin{pmatrix}
t_4 f(\bm{k}) & t_3 f^*(\bm{k})\\
-t_1 & t_4 f(\bm{k})
\end{pmatrix}_{\sigma\sigma'},
\end{equation}
\begin{equation}
[h_{\textrm{RMG}}(\bm{k})]_{\sigma,L;\sigma',L+2}=
\begin{pmatrix}
0 & t_2/2\\
0 & 0
\end{pmatrix}_{\sigma\sigma'},
\end{equation}
\begin{equation}
f(\bm{k})=\sum_{i=1}^3 \exp(i\bm{k}\cdot \bm{\delta_i})
; \;\;\;\;\;
\bm{\delta}_i=R_{\frac{i 2\pi}{3}}(0,\frac{a_{Gr}}{\sqrt{3}})^T.
\end{equation}
Here, $u_{L}=(-L+\frac{N_L-1}{2})u_D$ is the onsite potential energy on the $L-$th layer. $(t_0,t_1,t_2,t_3,t_4)=(3100,-380,-21,290,141)$meV are hopping parameters of graphene \cite{PhysRevB.82.035409}. $R_{\theta}$ denotes a counter-clockwise rotation by angle $\theta$.

\subsection{Dispersive ideal limit}\label{sec:dipersiveideallimit}
We expand $f(\bm{k})$ near the $\bm{K}=(\frac{4\pi}{3a_{Gr}},0)$ valley in the atomic BZ
\begin{equation}
f(\bm{K}+\bm{q})\approx -\frac{\sqrt{3}}{2}a_{\textrm{Gr}}(q_x-iq_y)\equiv -\frac{v_F}{t_0}(q_x-iq_y).
\end{equation}

The continuum Hamiltonian of RMG in the idealized limit near $\bm{K}$ valley is
\begin{equation}
H_{\textrm{holo}}=\sum_{\bm{k}\in \mathbb{R}^2,\sigma,\sigma',l,l'}c^{\dagger}_{\bm{k}+\bm{K},\sigma,l} [h_{\textrm{flat}}(\bm{k})+h_{\textrm{disp}}(\bm{k})]_{\sigma,l;\sigma',l'}c_{\bm{k}+\bm{K},\sigma',l'}.
\end{equation}
The intralayer hopping matrix is
\begin{equation}
[h_{\textrm{flat}}(\bm{k})]_{\sigma,L;\sigma',L}=
\begin{pmatrix}
0 & v_F(k_x-ik_y)\\
v_F(k_x+ik_y) & 0
\end{pmatrix}_{\sigma\sigma'}\quad L=0...N_{L}-2,
\end{equation}

\begin{equation}
[h_{\textrm{disp}}(\bm{k})]_{\sigma,L;\sigma',L}=
\begin{pmatrix}
\mathcal{E}(\bm{k}) & 0\\
0 & -\mathcal{E}(\bm{k})-\Delta_B
\end{pmatrix}_{\sigma\sigma'}\quad L=0...N_{L}-1.
\end{equation}
The interlayer hopping matrix is
\begin{equation}
[h_{\textrm{flat}}(\bm{k})]_{\sigma,L;\sigma',L+1}=
\begin{pmatrix}
0 & 0\\
-t_1 & 0
\end{pmatrix}_{\sigma\sigma'}.
\end{equation}
$h_{\textrm{flat}}(\bm{k})$ has a zero-energy A-sublattice-polarized eigenstate,
\begin{equation}
h_{\textrm{flat}}(\bm{k})\ket{s_{\bm{k}}}=0.
\end{equation}
This state is also an eigenstate of $h_{\textrm{flat}}(\bm{k})+h_{\textrm{disp}}(\bm{k})$ with energy $\mathcal{E}(\bm{k})$.
The components of $\ket{s_{\bm{k}}}$ in the basis of $(A_0,B_0,A_1..)$ are
\begin{equation}\label{eq:holospinor}
\ket{s_{\bm{k}}}=N_{\bm{k}}(1,0,\frac{v_F(k_x+ik_y)}{t_1},0,(\frac{v_F(k_x+ik_y)}{t_1})^2,0...),\quad N_{\bm{k}}=(\sum^{N_L-1}_{l=0}(\gamma|\bm{k}|)^{2l})^{-\frac{1}{2}},\quad \gamma\equiv \frac{v_F}{t_1}.
\end{equation}
For suitably chosen $\mathcal{E}(\bm{k})$, $\ket{s_{\bm{k}}}$  is the first conduction band of $H_{\textrm{holo}}$ (cf. Fig.\ref{Sfig_bandstructure}). Furthermore, we define an interpolated Hamiltonian,
\begin{equation}
h_{\textrm{inp}}(\bm{k};\lambda)=\lambda h_{\textrm{RMG}}(\bm{K}+\bm{k})+(1-\lambda)(h_{\textrm{flat}}(\bm{k})+h_{\textrm{disp}}(\bm{k})).
\end{equation}

For positive $u_{D}$, the first conduction band of $h_{\textrm{RMG}}$ is mostly localized on the A-sublattice of the first layer, that is, similar to $\ket{s_{\bm{k}}}$.  The  first conduction band of $h_{\textrm{inp}}(\bm{k},\lambda)$ smoothly interpolates between these two limits without going through band crossing as we change $\lambda$. In Fig.\ref{Sfig_bandstructure} (d), we plot the overlap between the first conduction band of $h_{\textrm{RMG}}(\bm{K}+\bm{k})$, $\ket{s_{\bm{k},\lambda=1}}$, and the first conduction band of $h_{\textrm{flat}}(\bm{k})+h_{\textrm{disp}}(\bm{k})$, i.e. $\ket{s_{\bm{k}}}$. It turns out that that the overlap is almost one for small $|\bm{k}|$, and remains close to one for $|\bm{k}|\approx 1 \textrm{nm}^{-1}> \gamma^{-1}=0.58\textrm{nm}^{-1}$. This fact provides justification for ``cutting the bond'' approximation introduced in the main text.

\subsection{hBN moir\'e potential}\label{sec:hBNmoirepotential}

We place Hexagonal Boron Nitride (hBN) in proximity to the $l=N_L-1$ layer of RMG with a twist angle $\theta$. The direct lattice of hBN is spanned by
\begin{equation}
\bm{R}^{\text{hBN}}_i=\frac{a_{\text{hBN}}}{a_{\text{Gr}}}R_{\theta}\bm{R}_i,
\end{equation}
where $a_{\text{hBN}}=0.2504$nm is the lattice constant of hBN. The moir\'e reciprocal lattice basis vectors of RMG+hBN system are

\begin{equation}\label{eq:moirelattice}
\bm{b}_{i}=\left(I-\frac{a_{\text{Gr}}}{a_{\text{hBN}}}R_{\theta}\right)\bm{G}_i.
\end{equation}

The corresponding moir\'e lattice basis vectors $\bm{a}_i$ are defined by
\begin{equation}
\bm{a}_i\cdot \bm{b}_j=2\pi \delta_{ij}
\end{equation}

We model the bare moir\'e potential introduced by the hBN, $H_{\textrm{hBN}}$ by \cite{dong2024anomalous}:

\begin{equation}\label{eq:hBNbarepotential}
H_{\textrm{hBN}}=\sum_{\bm{k}\in \mathbb{R}^2,\sigma} V_0 c^{\dagger}_{\bm{K}+\bm{k},\sigma,N_L-1}c_{\bm{K}+\bm{k},\sigma,N_L-1}+\sum_{\bm{k}\in \mathbb{R}^2,\sigma,\sigma'}\sum^{3}_{i=1} [V_1 \exp(-i\psi) c^{\dagger}_{\bm{K}+\bm{k}+\bm{b}_i,\sigma,N_L-1} N^i_{\sigma,\sigma'} c_{\bm{K}+\bm{k},\sigma',N_L-1}+h.c.]
\end{equation}

\begin{equation}
N^1_{\sigma,\sigma'}=\begin{pmatrix}
1&1\\
\omega&\omega
\end{pmatrix}_{\sigma,\sigma'}\quad
N^2_{\sigma,\sigma'}=\begin{pmatrix}
1&\omega^*\\
\omega^*&\omega
\end{pmatrix}_{\sigma,\sigma'}\quad N^3_{\sigma,\sigma'}=\begin{pmatrix}
1&\omega\\
1&\omega
\end{pmatrix}_{\sigma,\sigma'}.
\end{equation}
We have defined $\bm{b}_3\equiv -\bm{b}_1-\bm{b}_2$, $\omega\equiv \exp(i\frac{2\pi}{3})$ in the equation above. $(V_0,V_1,\psi)=(28.9\text{meV},21\text{meV},-0.29)$ are coupling parameters between hBN and RMG \cite{PhysRevB.90.155406}. While the bare potential only affects the bottom layer, in Sec. \ref{sec:Moire potential from the filled valence bands of RMG/hBN} we will discuss how this induces a potential on the other layers through modifying the charge distribution of the valence bands~\cite{kwan2025moire,moirecapacitor}.

Finally, we comment that there are two possible orientations between hBN and graphene \cite{uzan2025hbn,kwan2025moire,PhysRevB.90.155406}, labelled by $\xi=0$ and $\xi=1$ (defined in the next section). The bare moir\'e potential produced by hBN in these two cases differ by a swap of sublattice index in Eq.\ref{eq:hBNbarepotential}.  The effect of this alignment is disucssed further in the next subsection. Here, we only point out that Eq.\ref{eq:hBNbarepotential} corresponds to $\xi=1$ in our convention.

\subsection{The effect of hBN on the RMG spinor}\label{sec:The effect of hBN on the RMG spinor}

In this section we compute the effect of the hBN alignment on the RMG conduction band spinor and compare the result to the spinor we use for the idealized model.

\begin{figure}
    \centering
    \includegraphics[width=1.0\linewidth]{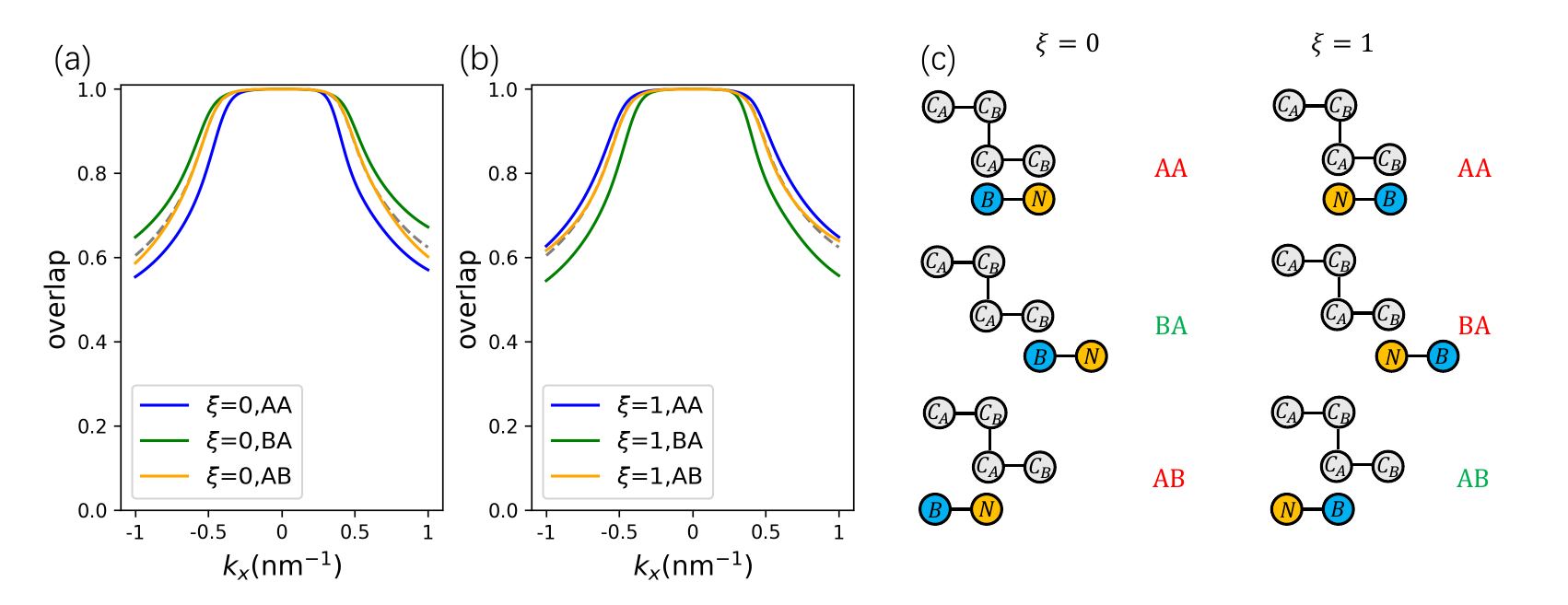}
    \caption{\textbf{(a)(b)} show the overlap of the first conduction band wavefunction between $H_{\textrm{full}}$ and $H_{\textrm{holo}}$, $|\langle s_{\bm{k},\textrm{full}} |s_{\bm{k}} \rangle|$ for different hBN/RMG relative orientation $\xi$ and different local stacking configurations. We use $u_D=50\textrm{meV}$ when calculating $\ket{s_{\bm{k},\textrm{full}}}$. The gray dashed line is the case when there is no hBN, i.e., the same as Fig.\ref{Sfig_bandstructure} (d). \textbf{(c)} shows the local stacking configurations between RMG (gray circles) and hBN (blue and orange circles) for different $\xi$. Green labels the stacking configuration that is energetically favorable for each $\xi$.}
    \label{Sfig_withhBNoverlap}
\end{figure}

We introduced the hBN moir\'e potential in the last section. This potential can be obtained from a perturbative calculation in which hBN is integrated out. Here, we will find it convenient to use the full Hamiltonian, with the hBN degrees of freedom included. Before integrating out the hBN degrees of freedom, the Hamiltonian describing the RMG/hBN system (near K valley) is \cite{kwan2025moire,herzog2024moire,PhysRevB.89.205414,PhysRevB.90.155406,PhysRevB.108.155406}
\begin{equation}
H_{\textrm{full}}(\bm{r})=\begin{pmatrix}h_{\textrm{RMG}}&T(\bm{r})\\T^{\dagger}(\bm{r})& h_{\textrm{hBN}}
\end{pmatrix}.
\end{equation}
 $T(\bm{r})$ is periodic with respect to the RMG/hBN moir\'e period. It describes the tunneling between RMG and hBN. $h_{\textrm{RMG}}$ and $h_{\textrm{hBN}}$ are the Hamiltonians describing isolated RMG and hBN near the K valley, respectively. The low-energy Hamiltonian of hBN is
\begin{equation}
h_{\textrm{hBN}}=\begin{pmatrix}V_B+u_{N_L}&0\\
0&V_N+u_{N_L}
\end{pmatrix}, \textrm{if $\xi$=0}\quad \begin{pmatrix}V_N+u_{N_L}&0\\
0&V_B+u_{N_L}
\end{pmatrix},
\textrm{if $\xi$=1}.
\end{equation}
$\xi$ describes the relative orientation between hBN and RMG as depicted in Fig~\ref{Sfig_withhBNoverlap}c,d. $u_{N_L}$ is the potential energy on the hBN layer. Note that the potential energy on the layers of RMG is $u_{0...N_L-1}$. $(V_B, V_N)=(3352, -1388)\textrm{meV}$ are the onsite potentials of boron and nitrogen atoms, respectively \cite{kwan2025moire,PhysRevB.90.155406}.  These numbers are so large that we can safely ignore the momentum-dependent part of the hBN band energies \cite{PhysRevB.90.155406}. There are three high-symmetry stacking configurations that preserve three-fold rotation symmetry, which are A(graphene)A(hBN), AB, BA. For $\xi=0$ the ``A" (B) sublattice of hBN corrresponds to Boron (Nitrogen), while for $\xi=1$ it corresponds to Nitrogen (Boron). At these three stacking configurations, $T(\bm{r})$, which describes the tunneling between  hBN and the bottom layer of RMG, takes the form \cite{kwan2025moire}:
\begin{equation}
T_{\textrm{Gr},\sigma,N_L-1;\textrm{hBN},\sigma'}=3w\begin{pmatrix}
1&0\\
0&1
\end{pmatrix}_{\sigma,\sigma'},\textrm{for AA },\quad  3w\begin{pmatrix}
0&1\\
0&0
\end{pmatrix}_{\sigma,\sigma'},\textrm{for AB},\quad 3w\begin{pmatrix}
0&0\\
1&0
\end{pmatrix}_{\sigma,\sigma'},\textrm{for BA}.
\end{equation}
Here, $w=152\textrm{meV}$ is the tunneling amplitude \cite{PhysRevB.90.155406}. Locally around those high-symmetry positions, the RMG/hBN system looks like two commensurate honeycomb lattices, as illustrated in Fig.\ref{Sfig_withhBNoverlap} (c),  with a tunneling Hamiltonian given by the expression above. For example, the local Hamiltonian near the AA stacking region is:
\begin{equation}
H_{\textrm{full}}(\bm{k})=\begin{pmatrix}
h_{\textrm{RMG}}(\bm{K}+\bm{k})&3w\begin{pmatrix}
1&0\\
0&1
\end{pmatrix}... \\
3w\begin{pmatrix}
1&0\\
0&1
\end{pmatrix}...&h_{\textrm{hBN}}
\end{pmatrix}.
\end{equation}

In Fig.\ref{Sfig_withhBNoverlap}, we diagonalize $H_{\textrm{full}}(\bm{k})$ for both values of $\xi$ and for each of the high-symmetry stacking configurations. We extract the normalized RMG part from the first conduction band of $H_{\textrm{full}}(\bm{k})$, denoted by $\ket{s_{\textrm{full},\bm{k}}}$. We then calculate its overlap with the ideal spinor, $|\braket{s_{\bm{k},\text{full}}}{s_{\bm{k}}}|$. As shown in Fig.\ref{Sfig_withhBNoverlap}, the overlap only changes slightly for different $\xi$ and stacking. In addition, the difference compared with the no-hBN case is not significant.

Finally, lattice relaxation favors the carbon-on-boron local stacking configuration \cite{uzan2025hbnalignmentorientationcontrols,bocarsly2024haas,krisna2023moire,arrighi2023non,PhysRevB.96.085442}. In our convention, this will favor the BA stacking for $\xi=0$ and AB stakcing for $\xi=1$. These two stacking configurations are labelled with green in Fig.\ref{Sfig_withhBNoverlap} (c). As shown in panel (a) and (b), hBN in these two configurations drive the overlap closer to one slightly compared with the case without hBN, which is more favorable for the ideal limit discussed in this paper.

\subsection{Moir\'e potential from the filled valence bands of RMG/hBN}\label{sec:Moire potential from the filled valence bands of RMG/hBN}

$H_{\textrm{hBN}}$ significantly modifies the charge distribution of the filled valence bands of $H_{\textrm{RMG}}$. This is because when $u_D>0$, the valence band states of RMG are polarized on the $l=N_L-1$ layer, which is in contact with the hBN substrate.
$H_{\textrm{hBN}}$ produces a periodic charge pattern in RMG with respect to the RMG/hBN moir\'e interface~\cite{kwan2025moire}. This moir\'e charge pattern will in turn produce a moir\'e potential and imprint itself onto the conduction band electron via Coulomb interaction.
The primary effect of this can be taken into account by a periodic scalar potential on the conduction band electrons (we thank J. Herzog-Arbeitman and Y. Kwan for sharing unpublished results demonstrating this ``moir\'e capacitor effect''~\cite{moirecapacitor}).
In this section, we explain how we take this effect into account in our Hartree-Fock calculation.

\begin{figure}
    \centering
    \includegraphics[width=1.0\linewidth]{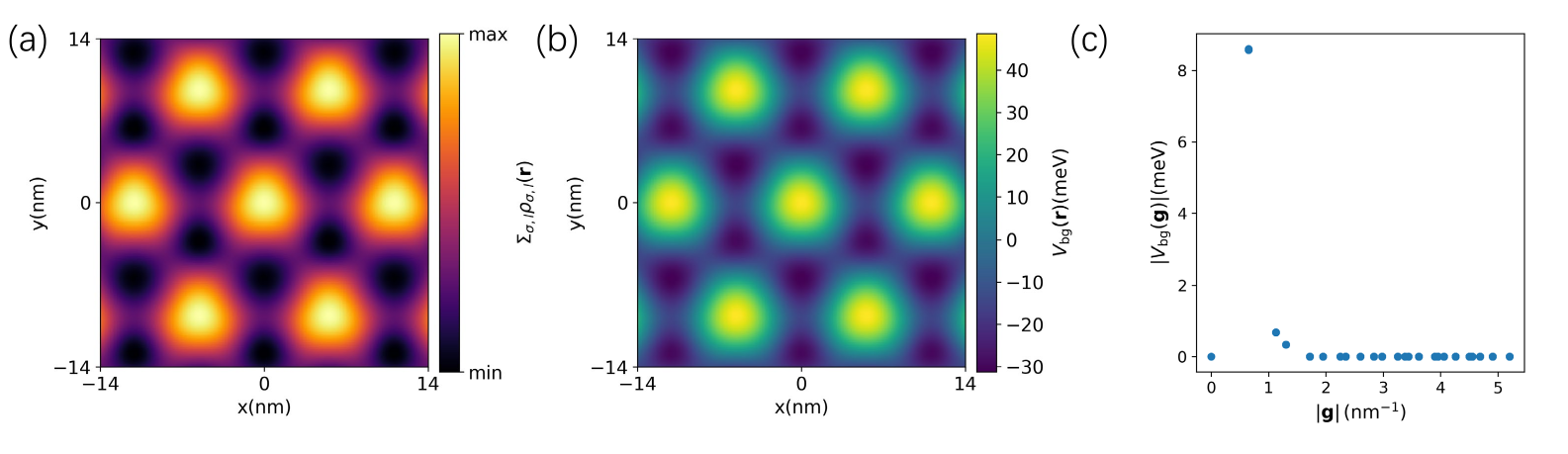}
    \caption{\textbf{(a)} Valence band charge profile for R5G + hBN at $\theta=0.77^{\circ}$, $u_D=50\textrm{meV}$. \textbf{(b)} Coulomb potential produced by the charge profile in (a). \textbf{(c)} Fourier transform of the potential in (b). We have subtracted out the uniform part so that $V_{\text{bg}}(\bm{0})=0$. For reference, we find that for the first harmonics, $\psi_{\text{bg}}(\bm{b}_1)=117.12^{\circ}$. The coordinate system is rotated and translated such that one of the real lattice spanning vector is in the x direction, and the charge density maximum is at the origin. This is the same transformation of the coordinate system as specfied in Sec.\ref{sec:HFdetails}. }
    \label{Sfig_valencecharge}
\end{figure}

We first diagonalize the single-particle moir\'e Hamiltonian $H_{\textrm{RMG}}+H_{\textrm{hBN}}$ and use $f^{\dagger}_{\bm{k},n}$ to denote its eigenstates ($\bm{k} \in \textrm{mBZ}$.):
\begin{equation}
f^{\dagger}_{\bm{k},n}=\sum_{\bm{g} \in \textrm{mRL}} U_{\bm{g}\sigma l,n}(\bm{k})c^{\dagger}_{\bm{K}+\bm{k}+\bm{g},\sigma,l}.
\end{equation}
Here, $n$ is the band index,  $\bm{g}$ is on the moir\'e reciprocal lattice (mRL) spanned by $\bm{b}_i$, and $\bm{k}$ is in the moir\'e Brillouin zone (mBZ), which is the paralleogram spanned by $\bm{b}_i$. We keep all $|\bm{g}|<3.51|\bm{b}_i|$ when diagonalizing the single-particle Hamiltonian. The layer-sublattice-resolved electron density from filling the valence bands states, $\{f^{\dagger}_{\bm{k},n}\}_{\bm{k}\in \text{mBZ},n=1..N_L}$, is:
\begin{equation}
\rho(\bm{r})_{\sigma,l}=4\sum_{\substack{n=1..N_L\\
\bm{g}_1,\bm{g}_2\in \textrm{mRL}\\
\bm{k} \in \textrm{mBZ}}} U_{\bm{g}_1\sigma l,n}(\bm{k})U^*_{\bm{g}_2\sigma l, n}(\bm{k})e^{i(\bm{g}_1-\bm{g}_2)\cdot \bm{r}}\equiv \sum_{\bm{g}\in \textrm{mRL}}\rho_{\sigma,l}(\bm{g})e^{i\bm{g}\cdot \bm{r}}
\end{equation}
Here, the prefactor $4$ is due to the spin-valley degeneracy we assume for the filled valence band. We have also defined the Fourier components of the charge density $\rho_{\sigma,l}(\bm{g})$ in the equation above. In Fig.\ref{Sfig_valencecharge}, we plot the total charge distribution $\sum_{\sigma,l}\rho_{\sigma,l}(\bm{r})$. The maximum of the redistributed charge forms a triangular lattice.

We will later project the Hamiltonian to the conduction band of RMG, i.e. only the conduction band electrons are taken as ``active'' while the valence bands are taken to be ``frozen''. We treat $\rho(\bm{r})_{\textrm{tot}}$ as the background providing a spatially varying Coulomb potential. And the Hamiltonian of electrons moving in this Coulomb potential is:

\begin{equation}\label{eq:valmoirepo}
\begin{split}
H_{\textrm{bg}}&=\sum_{\bm{g}_1,\bm{g}_2\in \textrm{mRL},\bm{k}\in \textrm{mBZ},\sigma,l}\frac{V(\bm{g_2})}{A} [\sum_{\sigma',l'}\rho_{\sigma',l'}(\bm{g}_2)] c^{\dagger}_{\bm{K}+\bm{k}+\bm{g}_1+\bm{g}_2,\sigma,l}c_{\bm{K}+\bm{k}+\bm{g}_1,\sigma,l}\\
&\equiv \sum_{\bm{k},\sigma,l,\bm{g}_1,\bm{g}_2} V_{\textrm{bg}}(\bm{g}_2)\exp(-i\psi_{\textrm{bg}}(\bm{g}_2))\sum^3_{i=1}c^{\dagger}_{\bm{K}+\bm{k}+\bm{g}_1+\bm{g}_2,\sigma,l}c_{\bm{K}+\bm{k}+\bm{g}_1,\sigma,l}\\
&\approx \sum_{\bm{k},\sigma,l,\bm{g}_1} V_{\textrm{bg}}\exp(-i\psi_{\textrm{bg}})\sum^3_{i=1}c^{\dagger}_{\bm{K}+\bm{k}+\bm{g}_1+\bm{b}_i,\sigma,l}c_{\bm{K}+\bm{k}+\bm{g}_1,\sigma,l}+h.c.
\end{split}
\end{equation}
Here, $V(\bm{g})$ is the Fourier transform of the dual-gated Coulomb interaction:
\begin{equation}
V(\bm{q})=\frac{e^2\textrm{tanh}(|\bm{q}|D)}{2\epsilon_r\epsilon_0 |\bm{q}| },\quad D=25\textrm{nm},\quad \epsilon_r=5.
\end{equation}
Going from the second line to the third line of Eq.\ref{eq:valmoirepo}, we made the approximation that the potential is well-approximated by the first harmonics. In principle, $V_{\textrm{bg}}$ and $\psi_{\textrm{bg}}$ can be calculated for each displacement field $u_D$ and twist angle $\theta$. However, when these two parameters are within the experimentally relevant regime ($u_D\sim 30 -50 \text{meV},\theta\sim 0.6^{\circ}-0.9^{\circ} $), we find that their values do not change significantly:
\begin{equation}\label{eq:valencepotentialvalue}
V_{\textrm{bg}}\approx 8\textrm{meV},\quad  \psi_{\textrm{bg}}\approx \frac{2\pi}{3}.
\end{equation}
In Fig.\ref{Sfig_valencecharge} (a), we show $\rho_{\textrm{tot}}(\bm{r})$ at $\theta=0.77^{\circ}$, $u_D=50\textrm{meV}$. The corresponding Coulomb potential,
\begin{equation}
V_{\textrm{bg}}(\bm{r})=\sum_{\bm{g}}V_{\text{bg}}(\bm{g})\exp(-i\psi_{\text{bg}}(\bm{g}))\exp(i\bm{g}\cdot\bm{r}),
\end{equation}
is shown in Fig.\ref{Sfig_valencecharge} (b). The Fourier components of the potential $V_{\textrm{bg}}(\bm{g})$ are shown in Fig.\ref{Sfig_valencecharge} (c). Clearly, the Fourier components are dominated by the first harmonics, justifying our approximation above.

For most of our Hartree-Fock calculation, we will use the parameter in Eq.\ref{eq:valencepotentialvalue}. However, we will allow for $V_{\textrm{bg}}$ to be a free parameter when we explore the effect of this potential on the hole-doped excited states of the system.

Finally, we remark that had we taken into account of all the bands instead of only the first conduction band RMG when doing Hartree-Fock calculation, we would not need to put in this additional moir\'e potential by hand. Hartree-Fock would have included this correction self-consistently.

\subsection{Hartree-Fock}\label{sec:HFdetails}
 We use $c^{\dagger}_{\bm{k},\bm{g}}$   to denote the  operator that creates an electron in the first conduction band of $h_{\textrm{inp}}(\bm{k}+\bm{g};\lambda)$. Thus, at $\lambda=0$,
 \begin{equation}
 c^{\dagger}_{\bm{k},\bm{g}}\ket{0}=e^{i(\bm{k+\bm{g}})\cdot \bm{r}}\ket{s_{\bm{k}+\bm{g}}},
 \end{equation}
 where $\ket{0}$ is the vacuum. The notation is chosen such that $\bm{k}$ is confined within the mBZ spanned by $\bm{b}_i$, and $\bm{g}$ is a  moir\'e reciprocal lattice (mRL) vector.
 We then define the form factor $F(\bm{k}_1+\bm{g}_1,\bm{k}_2+\bm{g}_2)$ of this band to be

 \begin{equation}
F(\bm{k}_1+\bm{g}_1,\bm{k}_2+\bm{g}_2)\equiv \langle  0|c_{\bm{k}_1,\bm{g}_1} e^{-i(\bm{k}_2+\bm{g}_2-\bm{k}_1-\bm{g}_1)\cdot \bm{r}}c^{\dagger}_{\bm{k}_2,\bm{g}_2} |0\rangle=\sum_{\sigma,l}s^{*}_{\bm{k}_1+\bm{g}_1,\sigma,l}s_{\bm{k}_2+\bm{g}_2,\sigma,l}.
\end{equation}
 At $\lambda=0$, the coefficients $s_{\bm{k}+\bm{g},\sigma,l}$ are given by Eq. \ref{eq:holospinor}.  For general $\lambda$, they must be obtained numerically.
 We project the kinetic energy, the moir\'e potential, as well as the interaction to this band. The projected Hamiltonian comprises four parts:
 \begin{equation}
H_{\textrm{total}}=H^{\textrm{proj}}_{\textrm{kin}}+H^{\textrm{proj}}_{\textrm{hBN}}+H^{\textrm{proj}}_{\textrm{bg}}+H^{\textrm{proj}}_{\textrm{int}},
 \end{equation}

\begin{equation}
H^{\textrm{proj}}_{\textrm{kin}}=\sum_{\bm{k}\in \textrm{mBZ}}\sum_{\bm{g}\in \textrm{mRL}} \epsilon(\bm{k}+\bm{g};\lambda) c^{\dagger}_{\bm{k},\bm{g}}c_{\bm{k},\bm{g}},
\end{equation}

\begin{equation}
\begin{split}
H_{\textrm{hBN}}^{\textrm{proj}}=&\sum_{\substack{\bm{k}\in \textrm{mBZ}\\ \bm{g}\in \textrm{mRL}}}\sum_{\sigma} \frac{V_0}{2} s^{*}_{\bm{k}+\bm{g},\sigma,N_L-1} s_{\bm{k}+\bm{g},\sigma,N_L-1} c^{\dagger}_{\bm{k},\bm{g}}c_{\bm{k},\bm{g}}\\
&+\sum_{\substack{\bm{k}\in \textrm{mBZ}\\ \bm{g}\in \textrm{mRL}}}\sum^3_{i=1} \sum_{\sigma,\sigma'} V_1\exp(-i\psi) s^{*}_{\bm{k}+\bm{g}+\bm{b}_i,\sigma,N_L-1} s_{\bm{k}+\bm{g},\sigma',N_L-1}N^{i}_{\sigma,\sigma'} c^{\dagger}_{\bm{k},\bm{g}+\bm{b}_i}c_{\bm{k},\bm{g}}+h.c.,
\end{split}
\end{equation}

\begin{equation}
\begin{split}
H_{\textrm{bg}}^{\textrm{proj}}=
\sum_{\substack{\bm{k}\in \textrm{mBZ}\\ \bm{g}\in \textrm{mRL}}}\sum^3_{i=1} \sum_{\sigma,l} V_{\textrm{bg}}\exp(-i\psi_{\textrm{bg}}) s^{*}_{\bm{k}+\bm{g}+\bm{b}_i,\sigma,l} s_{\bm{k}+\bm{g},\sigma,l} c^{\dagger}_{\bm{k},\bm{g}+\bm{b}_i}c_{\bm{k},\bm{g}}+h.c.,
\end{split}
\end{equation}

\begin{equation}
\begin{split}
H^{\textrm{proj}}_{\textrm{int}}=&\frac{1}{2A} \sum_{\substack{\bm{k}_1-\bm{k}_4 \in \textrm{mBZ}\\ \bm{g}_1-\bm{g}_4 \in \textrm{mRL}}} V(\bm{k}_1+\bm{g}_1-\bm{k}_3-\bm{g}_3)F(\bm{k}_1+\bm{g}_1,\bm{k}_3+\bm{g}_3) F(\bm{k}_2+\bm{g}_2,\bm{k}_4+\bm{g}_4)\\
&c^{\dagger}_{\bm{k}_1,\bm{g}_1}c^{\dagger}_{\bm{k}_2,\bm{g}_2}c_{\bm{k}_4,\bm{g}_4}c_{\bm{k}_3,\bm{g}_3}\delta_{\bm{k}_1+\bm{k}_2+\bm{g}_1+\bm{g}_2, \bm{k}_3+\bm{k}_4+\bm{g}_3+\bm{g}_4}.
\end{split}
\end{equation}
Here, $V(\bm{q})$ is the Fourier transform of the interaction. Hartree-Fock approximation amounts to replacing $H^{\textrm{proj}}_{\textrm{int}}$ by $H^{\textrm{proj}}_{\textrm{mf}}$, which is defined as
\begin{equation}
\begin{split}
H^{\textrm{proj}}_{\textrm{mf}}=&\frac{1}{A} \sum_{\substack{\bm{k}_1,\bm{k}_2 \in \textrm{mBZ}\\ \bm{g}_1-\bm{g}_4 \in \textrm{mRL}}} V(\bm{g}_1-\bm{g}_3)F(\bm{k}_1+\bm{g}_1,\bm{k}_1+\bm{g}_3) F(\bm{k}_2+\bm{g}_2,\bm{k}_2+\bm{g}_4) c^{\dagger}_{\bm{k}_1,\bm{g}_1}c_{\bm{k}_1,\bm{g}_3}P_{\bm{g}_4,\bm{g}_2}(\bm{k}_2) \delta_{\bm{g}_1+\bm{g}_2, \bm{g}_3+\bm{g}_4} \\
-&\frac{1}{A} \sum_{\substack{\bm{k}_1,\bm{k}_2 \in \textrm{mBZ}\\ \bm{g}_1-\bm{g}_4 \in \textrm{mRL}}} V(\bm{k}_1+\bm{g}_1-\bm{k}_2-\bm{g}_3)F(\bm{k}_1+\bm{g}_1,\bm{k}_2+\bm{g}_3) F(\bm{k}_2+\bm{g}_2,\bm{k}_1+\bm{g}_4) c^{\dagger}_{\bm{k}_1,\bm{g}_1}c_{\bm{k}_1,\bm{g}_4}P_{\bm{g}_3,\bm{g}_2}(\bm{k}_2) \delta_{\bm{g}_1+\bm{g}_2, \bm{g}_3+\bm{g}_4}.
\end{split}
\end{equation}
In the equation above, we have defined the projector $P_{\bm{g}_1,\bm{g}_2}(\bm{k})$ as
\begin{equation}\label{eq:projectordef}
P_{\bm{g}_1,\bm{g}_2}(\bm{k})\equiv \langle c^{\dagger}_{\bm{k},\bm{g}_2} c_{\bm{k},\bm{g}_1} \rangle.
\end{equation}
The expectation value is evaluated with respect to the Slater determinant state obtained by diagonalizing the Hartree-Fock Hamiltonian $H^{\textrm{proj}}_{\textrm{kin}}+H^{\textrm{proj}}_{\textrm{hBN}}+H^{\textrm{proj}}_{\textrm{bg}}+H^{\textrm{proj}}_{\textrm{mf}}$. $d^{\dagger}_{n,\bm{k}}$ creates an electron in the eigenstate of the Hartree-Fock Hamiltonian, where $n$ is the band index. We assume they are related to $c^{\dagger}_{\bm{k},\bm{g}}$ by:

\begin{equation}\label{eq:hfbasisdefin}
d^{\dagger}_{n,\bm{k}}=\sum_{\bm{g}\in \textrm{mRL}}\beta_{n,\bm{k},\bm{g}}c^{\dagger}_{\bm{k},\bm{g}}\to P_{\bm{g}_1,\bm{g}_2}(\bm{k})=\sum_{n\in \textrm{filled}}\beta_{n,\bm{k},\bm{g}_1} \beta^*_{n,\bm{k},\bm{g}_2}.
\end{equation}
For calculations in this paper (except the one-hole state, which are discussed in Sec.\ref{sec:enlargedunitcell}), we choose the filling to be one electron per unit cell, i.e. only $n=1$ is filled.

Finally, the permissible values of $\bm{k}$ are
\begin{equation}\label{eq:allowedk}
\bm{k}=\frac{n_1}{N_1}\bm{b}_1+\frac{n_2}{N_2}\bm{b}_2,\quad n_i=0,..,N_i-1.
\end{equation}
This corresponds to a system on a torus with $N_1\times N_2$ moir\'e unit cells.

The calculation in Fig.3 of the main text is done with $N_1=N_2=6$, and keeping all $|\bm{g}|<3.51|\bm{g}_1|$. The calculation is done with fixed $\lambda=0$. We do not include $H^{\textrm{proj}}_{\textrm{hBN}}$ and $H^{\textrm{proj}}_{\textrm{bg}}$ in this calculation (Only for this plot and the corresponding TDHF calculation in the main text. $H^{\textrm{proj}}_{\textrm{hBN}}$  and $H^{\textrm{proj}}_{\textrm{bg}}$ are included for calculation in all other plots). We set $\mathcal{E}(\bm{k})=\alpha|\bm{k}|^2$. $\Delta_B$ is set to some small positive value to split the degeneracy of $h_{\textrm{inp}}$ at $|\bm{k}|=0$. We take the interaction to be density-density contact interaction:
\begin{equation}
V(\bm{q})=V_0
\end{equation}

The calculation in Fig.4 of the main text is done with $N_1=N_2=9$, and keeping all $|\bm{g}|<3.51|\bm{g}_1|$. The rotation angle is set to be $\theta=0.77^{\circ}$. At $\lambda=1$, we set $u_D=50\textrm{meV}$ for $h_{\textrm{RMG}}$. At $\lambda=0$, we take $\mathcal{E}(\bm{k})$ to be the energy of the first conduction band of $h_{\textrm{RMG}}(\bm{K}+\bm{k})$ with $u_D=50\textrm{meV}$ and set $\Delta_B=0$. We take the interaction to be the density-density contact interaction plus gate-screened Coulomb interaction

\begin{equation}\label{eq:contactplusCoulomb}
V(\bm{q})=V_0+\frac{e^2\textrm{tanh}(|\bm{q}|D)}{2\epsilon_r\epsilon_0 |\bm{q}| }\quad D=25\textrm{nm}\quad \epsilon_r=5.
\end{equation}
In Fig. \ref{Sfig_bandstructure}, we plot the band structure of $h_{\textrm{inp}}$ as a function of $\lambda$ for the $\mathcal{E}(\bm{k})$ we use in the main text Fig. 4 calculation. As can be seen, the evolution is smooth without band crossing.

Finally, we note that when we plot the charge density and the magnetic field in the main text and in this supplementary material (cf. Fig.\ref{Sfig_skyrmion}, Fig. \ref{Sfig_effectiveB}), we have rotated and shifted the coordinate system so that (1) one of the moir\'e lattice basis vectors lies on the $x$ axis. (2) The minimum of the charge density on the zeroth layer $\rho_{l=0}(\bm{r})$ (skyrmion core) is at the origin.  The same coordinate transformation is also applied when we plot out the valence band charge density (Fig.\ref{Sfig_valencecharge}), TDHF charge density time evolaution (Fig.\ref{Sfig_timedependence}), and the real-space density profile for one-hole excited states in the presence of a defect (Fig.\ref{Sfig_hole}, Fig.\ref{Sfig_nematic}, Fig.\ref{Sfig_symvac1}).

\subsection{Wavefunction overlap between analytic wavefunction and Hartree-Fock wavefunction}\label{sec:Wavefunction overlap between analytic wavefunction and Hartree-Fock wavefunction}

According to the discussion in the main text and in Sec. \ref{sec:relationtomanybody},
we expect that in the limit of large contact interaction at $\lambda=0$, the (normalized) Hartree-Fock Bloch wavefunction should approach the analytic form, $\psi^{\textrm{ansatz}}_{\bm{k},l}(\bm{r})$, with $c_n=0$ for $n\neq 0$:
\begin{equation}
\begin{split}
\label{eq:psianalytic}
\psi_{\bm{k},l,\textrm{ana}}(\bm{r})&=\braket{\bm{r},l}{\psi_{\bm{k},\textrm{ana}}}\\
&=\psi^{\textrm{ansatz}}_{\bm{k},l}(\bm{r})|_{c_n=0,n>0}\\
&=\mathcal{\tilde{M}}_{\bm{k}}\sum_{\bm{g}}e^{i\pi n_1 n_2} \exp(-\frac{\pi}{2\Omega}(g\bar{g}+2k\bar{g}))e^{i(\bm{k}+\bm{g})\cdot (\bm{r}-\bm{R}_0)}\frac{1}{N_{\bm{k}+\bm{g}}}s_{\bm{k}+\bm{g},l}.
\end{split}
\end{equation}
Here, we change from $\mathcal{M}_{\bm{k}}$ to $\tilde{\mathcal{M}_{\bm{k}}}$ to make $\psi_{\bm{k},l,\textrm{ana}}$ normalized.  $\bm{R}_0$ is an arbitrary\footnote{$\bm{R}_0$ is in principle related to the momentum of the condensate part of the wavefunction, $\chi_l(\bm{r})$. However, that momentum can be chosen arbitrarily in the absence of a pinning moir\'e potential without affecting the energy.} choice of spatial origin, $\Omega=|\bm{b}_1\times \bm{b}_2|$ is the BZ area, $N_{\bm{k}}$ is the normalization factor of the spinor $\ket{s_{\bm{k}}}$, as defined above. To calculate the overlap between the analytic formula and the numerical result, we choose $\bm{R}_0$ such that the real space charge density minimum associated with $\ket{\psi_{\bm{k},\textrm{ana}}}$ is the same as that obtained from Hartree-Fock numerical wavefunction. We define the per-particle overlap between the analytic wavefunction and the Hartree-Fock wavefunction as
\begin{equation}
\textrm{overlap}\equiv (\prod_{\bm{k}\in   \textrm{mBZ}} |\braket{\psi_{\bm{k},\textrm{ana}}}{\psi_{\bm{k},\textrm{HF}}}|^2)^{\frac{1}{N_1N_2}} \quad \ket{\psi_{\bm{k},\textrm{HF}}}\equiv d^{\dagger}_{1,\bm{k}}\ket{0}.
\end{equation}

\begin{figure}
    \centering
    \includegraphics[width=0.4\linewidth]{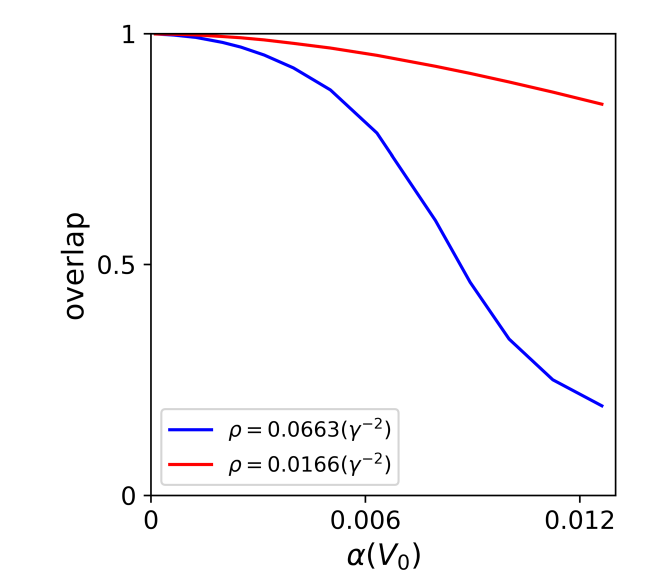}
    \caption{The overlap between the Hartree-Fock wavefunction with ideal R5G ($\lambda=0.0$) and the analytic formula $\ket{\psi_{\bm{k},\textrm{ana}}}$. The dispersion is chosen as $\mathcal{E}(\bm{k})=\alpha|\bm{k}|^2$. $\rho=0.0663 \gamma^{-2}$ is the density used in the main text for plotting $n(\bm{k})$. The interaction is chosen to be density-density contact interaction. hBN moir\'e potential and the moir\'e potential from redistributed valence band charge are not included in this calculation.}
    \label{Sfig_overlap}
\end{figure}

In Fig.\ref{Sfig_overlap}, we plot the overlap between the Hartree-Fock wavefunction and the analytical wavefunction $\ket{\psi_{\bm{k},\textrm{ana}}}$ for different contact interaction strength $V_0$ at the density used in the main text Fig.3. We choose the interaction to be density-density contact interaction, and we choose the dispersion to be $\mathcal{E}(\bm{k})=\alpha|\bm{k}|^2$. We set $\lambda=0.0$ (ideal) for this calculation.

\subsection{Determination of phase boundaries}\label{sec:Determination of phase boundaries}
In the main text Fig. 3, we provide a phase diagram when $\lambda=0.0$, $\mathcal{E}(\bm{k})=\alpha |\bm{k}|^2$, with contact interaction. The boundary between phases with different Chern number is trivial to obtain. In this section, we explain how we obtain the phase boundary among $w_0$, $w_1$ and $w_2$ states. These states are distinguished by their angular momentum, i.e. their eigenvalue under a 3-fold rotation. This section explains how these eigenvalues are obtained.

Operationally, we first shift the spatial origin to a high-symmetry position. Because for the calculation in the main text Fig. 3, we do not impose any pinning moir\'e potential, the total charge density from the Hartree-Fock wavefunction can have its minimum at any position in the real space. If the minimum of charge density corresponding to filling $d^{\dagger}_{1,\bm{k}}$ is at position $\bm{R}_1$ (cf. Eq.\ref{eq:hfbasisdefin}), we can define a new set of Hartree-Fock states $d^{\dagger'}_{1,\bm{k}}$:
\begin{equation}
d^{\dagger'}_{1,\bm{k}}=\sum_{\bm{g}\in \textrm{mRL}}\beta^{'}_{1,\bm{k},\bm{g}}c^{\dagger}_{\bm{k},\bm{g}},\quad \beta^{'}_{1,\bm{k},\bm{g}}\equiv \beta_{1,\bm{k},\bm{g}}\exp(i (\bm{k}+\bm{g})\cdot \bm{R}_1).
\end{equation}
The charge density associated with $d^{\dagger'}_{1,\bm{k}}$ now has its minimum at the origin. Due to the continuous translation symmetry of the Hamiltonian, the energy of the Slater determinant is the same regardless of filling $d^{\dagger}_{1,\bm{k}}$ or $d^{\dagger'}_{1,\bm{k}}$. From now on, we will work with $d^{\dagger'}_{1,\bm{k}}$.

Then, we note that when rotating around the origin, the 3-fold rotation operator $C_3$ acts in the following way:
\begin{equation}
C_3 c^{\dagger}_{\bm{k},\bm{g}} C^{-1}_3= c^{\dagger}_{C_3\bm{k},C_3\bm{g}}.
\end{equation}
Here, $C_3\bm{k}$ rotates the vector in the clockwise direction by $\frac{2\pi}{3}$. This allows us to calculate the rotation eigenvalue of $d^{\dagger'}_{1,\bm{0}}$:
\begin{equation}
\omega'\equiv \langle 0|d^{'}_{1,\bm{0}}C_3 d^{\dagger'}_{1,\bm{0}}|0 \rangle=\sum_{\bm{g}}\beta^{*}_{1,C_3\bm{k},C_3\bm{g}}\beta_{1,\bm{k},\bm{g}}.
\end{equation}

A few comments are in place. First, in principle, there is no guarantee
 that the  system is  $C_3$ symmetric around total charge density minimum. However, if it is not, it will show up  as $\omega'$ being values other than $\{1,\exp(\pm i\frac{2\pi}{3})\}$. This is a self-consistency check we perform.

 Second, the rotation eigenvalue of $d^{\dagger'}_{1,\bm{0}}$ is the same for different rotation centers. Since for $d^{\dagger'}_{1,\bm{0}}$, one of the $C_3$ rotation symmetry centers is the origin with rotation eignevalue $\omega'$, it implies that
 \begin{equation}
 \beta'_{1,\bm{0},\bm{g}}=\omega'\beta'_{1,\bm{0},C_3\bm{g}}.
 \end{equation}
 For a triangular lattice, the additional rotation centers can only be at $\frac{\bm{a}_1+\bm{a}_2}{3}$ or $\frac{2(\bm{a}_2+\bm{a}_2)}{3}$.  We note that $\bm{b}_2$ is obtained by rotating $\bm{b}_1$ counter-clockwise by $\frac{2\pi}{3}$, while $\bm{a}_2$ is obtained by rotating $\bm{a}_1$ counter-clockwise by $\frac{\pi}{3}$ (cf. Eq .\ref{eq:moirelattice}). This is because if the additional rotation center (B) is at $\bm{R}_2=x\bm{a}_1+y\bm{a}_2$, while the original rotation center (A) is the origin, we can always shift our coordinate system to make B the origin. The Hartree-Fock wavefunction coefficients in this new coordinate system is
 \begin{equation}
\beta''_{1,\bm{k},\bm{g}}=\beta'_{1,\bm{k},\bm{g}}\exp(i(\bm{k}+\bm{g})\cdot \bm{R}_2).
 \end{equation}
 Since in this new coordinate system, the wavefunction should still be 3-fold rotation symmetric around the origin, it implies that
 \begin{equation}
 \beta''_{1,\bm{0},\bm{g}'}=\omega'' \beta''_{1,\bm{0},C_3\bm{g}'},\quad \omega''\in \{1,\exp(\pm i\frac{2\pi}{3})\}.
 \end{equation}
Making use of the two equations above, one can show that this is only possible when $x=y=\frac{1}{3}$ or $x=y=\frac{2}{3}$. In both cases $\omega''=\omega'$. Thus, the rotation eigenvalue of the $\bm{k}=\bm{0}$ wavefunction is independent of the choice of rotation center (This is not the case for the $\bm{k}=\bm{\kappa}$ and $\bm{k}=\bm{\kappa}'$ wavefunction).

Third, by making use of the results in  Sec.\ref{sec:planewave}, we can show that $w_{3n}$, $w_{3n+1}$, $w_{3n+2}$, $n\in \mathcal{Z}$ carries angular momentum $0$, $1$, $2$ (mod 3), respectively. For the $w_n$ state in the ideal limit, with appropriate choice of origin, the wavefunction coefficients are of the form (cf. Eq.\ref{eq:ansatzexpansion}):
\begin{equation}
\beta'_{1,\bm{0},\bm{g}}\propto e^{i\pi (n_1-1)(n_2-1)} e^{-\frac{|\bm{g}|^2 l_B^2}{4}}(gl_B)^n \frac{1}{N_{\bm{g}}},\quad \bm{g}=n_1\bm{b}_1+n_2\bm{g}_2.
\end{equation}
Here, the proportionality constant is independent of $\bm{g}$. This trivially implies that
\begin{equation}
\omega'=\exp(i\frac{2n\pi}{3}).
\end{equation}

\subsection{Trace condition violation of filled Hartree Fock band}\label{sec:Trace condition violation of filled Hartree Fock band}

For a general Bloch wavefunction $\ket{\psi_{\bm{k}}}=e^{i\bm{k}\cdot \bm{r}}\ket{u_{\bm{k}}}$, trace condition violation (TCV) is defined as \footnote{Notice that we have a sign difference in the definition of $\Omega(\bk)$ compared with Ref.\cite{tan2024parent}. Under this convention, $\ket{s_{\bm{k}}}$ has positive Berry curvature.}

\begin{equation}
\begin{split}
\Omega(\bm{k})&\equiv \Im\sum_{\mu,\nu}[\epsilon_{\mu\nu}\braket{\partial_{k_\mu}u_{\bm{k}}}{\partial_{k_\nu}u_{\bm{k}}}],
\end{split}
\end{equation}

\begin{equation}
g^{\mathrm{FS}}_{\mu\nu}(\bm{k})=\Re\left[\braket{\partial_{k_\mu} u_{\bm{k}}}{\partial_{k_\nu} u_{\bm{k}}} -
\braket{\partial_{k_\mu} u_{\bm{k}}}{u_{\bm{k}}}
\braket{u_{\bm{k}}}{\partial_{k_\nu}u_{\bm{k}}}
\right],
\end{equation}

\begin{equation}
\textrm{TCV}\equiv \frac{1}{2\pi}\int_{\textrm{BZ}} d\bm{k} (\textrm{Tr}[g^{\textrm{FS}}_{\mu\nu}(\bm{k})]-|\Omega(\bm{k})|),
\end{equation}
where $g^{\mathrm{FS}}_{\mu\nu}(\bm{k})$ is the Fubini-Study metric, and $\Omega(\bm{k})$ is the Berry curvature. Numerically, Hartree-Fock wavefunction and mean-field Hamiltonian are only computed on a sparse momentum grid, as given by Eq.\ref{eq:allowedk}. To calculate TCV accurately, we interpolate $H^{\textrm{proj}}_{\textrm{mf}}(\bm{k})$ over the entire Brillouin zone. We diagonalize the interpolated Hamiltonian to obtain the Bloch wavefunction over the entire Brillouin zone, which we can use to calculate the TCV accurately. In practice, we find that interpolating $H^{\textrm{proj}}_{\textrm{mf}}(\bm{k})$ on a $\sim 30\times 30$ grid of the mBZ is sufficient for convergence.

\subsection{skyrmion density and the existence of skyrmion lattice away from the ideal limit}\label{sec:pontragin}
The real space density matrix (projector)  in the microscopic layer/sublattice basis can be obtained from the projector $P_{\bm{g}_1,\bm{g}_2}(\bm{k})$ (cf. Eq.\ref{eq:projectordef}) via:
\begin{equation}
\begin{split}
P_{\sigma_1,l_1;\sigma_2,l_2}(\bm{r})&\equiv\sum_{\bm{k},\bm{g}_1,\bm{g}_2}\langle c^{\dagger}_{\bm{K}+\bm{k}+\bm{g}_2,\sigma_2,l_2} c_{\bm{K}+\bm{k}+\bm{g}_1,\sigma_1,l_1}\rangle e^{i(\bm{g}_1-\bm{g}_2)\cdot \bm{r}}\\
&=\sum_{\bm{k},\bm{g}_1,\bm{g}_2} P_{\bm{g}_1,\bm{g}_2}(\bm{k})  e^{i(\bm{g}_1-\bm{g}_2)\cdot \bm{r}} s^{*}_{\bm{k}+\bm{g}_2,\sigma_2,l_2}s_{\bm{k}+\bm{g}_1,\sigma_1,l_1}.
\end{split}
\end{equation}
Here, we use $s_{\bm{k}+\bm{g},\sigma,l}$ to represent the sublattice/layer component of Bloch wavefunction of the first conduction band of $h_{\textrm{inp}}$. When $\lambda=0$, it reduces to $s_{\bm{k}+\bm{g}}$, as defined in Eq.\ref{eq:holospinor}. We diagonalize the real space density matrix $P_{\sigma_1,l_1;\sigma_2,l_2}(\bm{r})$ to obtain the eigenvector $\ket{\zeta_{\bm{r}}}$ corresponding to the largest eigenvalue $\tau_{2N_L}(\bm{r})$ ($\tau_1\leq  \tau_2 \leq... \tau_{2N_L}$).  The skyrmion density $\rho_{sk}(\bm{r})$ associated with $\ket{\zeta_{\bm{r}}}$ is defined as

\begin{equation}
\rho_{sk}(\bm{r})\equiv \frac{1}{2\pi} \textrm{Im}\sum_{\mu,\nu}[\epsilon_{\mu\nu}\langle \partial_{r_{\mu}}\zeta_{\bm{r}}|\partial_{r_{\nu}} \zeta_{\bm{r}} \rangle].
\end{equation}

We first comment that in the ideal limit $(\lambda=0.0,  V_0\to\infty)$, the real space density matrix has only one eigenvector with non-zero eigenvalue, $\bm{\chi}(\bm{r})$. This is because that in that limit, the (normalized) Bloch wavefunction should approach the form
\begin{equation}
\psi^{\textrm{ansatz}}_{\bm{k},l}(\bm{r})=\tilde{\mathcal{M}}_{\bm{k}}\phi^{\textrm{LLL}}_{\bm{k}-\bm{q}}(\bm{r})\chi_l(\bm{r}),\quad \chi_0(\bm{r})=\sum_n c_n\phi^{\textrm{nLL}*}_{-\bm{q}}(\bm{r}).
\end{equation}
Here $\tilde{\mathcal{M}}_{\bm{k}}$ is an overall normalization factor that normalize $\psi_{\bm{k},l,\textrm{ana}}(\bm{r})$.
Then the real space density matrix is (the sublattice index is dropped in this limit)
\begin{equation}
P_{l_1,l_2}(\bm{r})=(\sum_{\bm{k}}|\mathcal{\tilde{M}}_{\bm{k}}|^2 |\phi^{\textrm{LLL}}_{\bm{k}-\bm{q}}(\bm{r})|^2) \chi_{l_1}(\bm{r})\chi^{*}_{l_2}(\bm{r}).
\end{equation}
Thus, the normalized largest eigenvalue $\tau_{2N_L}(\bm{r})/\textrm{Tr}[P(\bm{r})]$ is 1 everywhere in the ideal limit.  The integrated skyrmion density associated with $\bm{\chi}(\bm{r})$ is equal to an integer $N_{\phi}$. A non-zero integrated skyrmion density is the signature of skyrmion.

\begin{figure}
    \centering
    \includegraphics[width=1.0\linewidth]{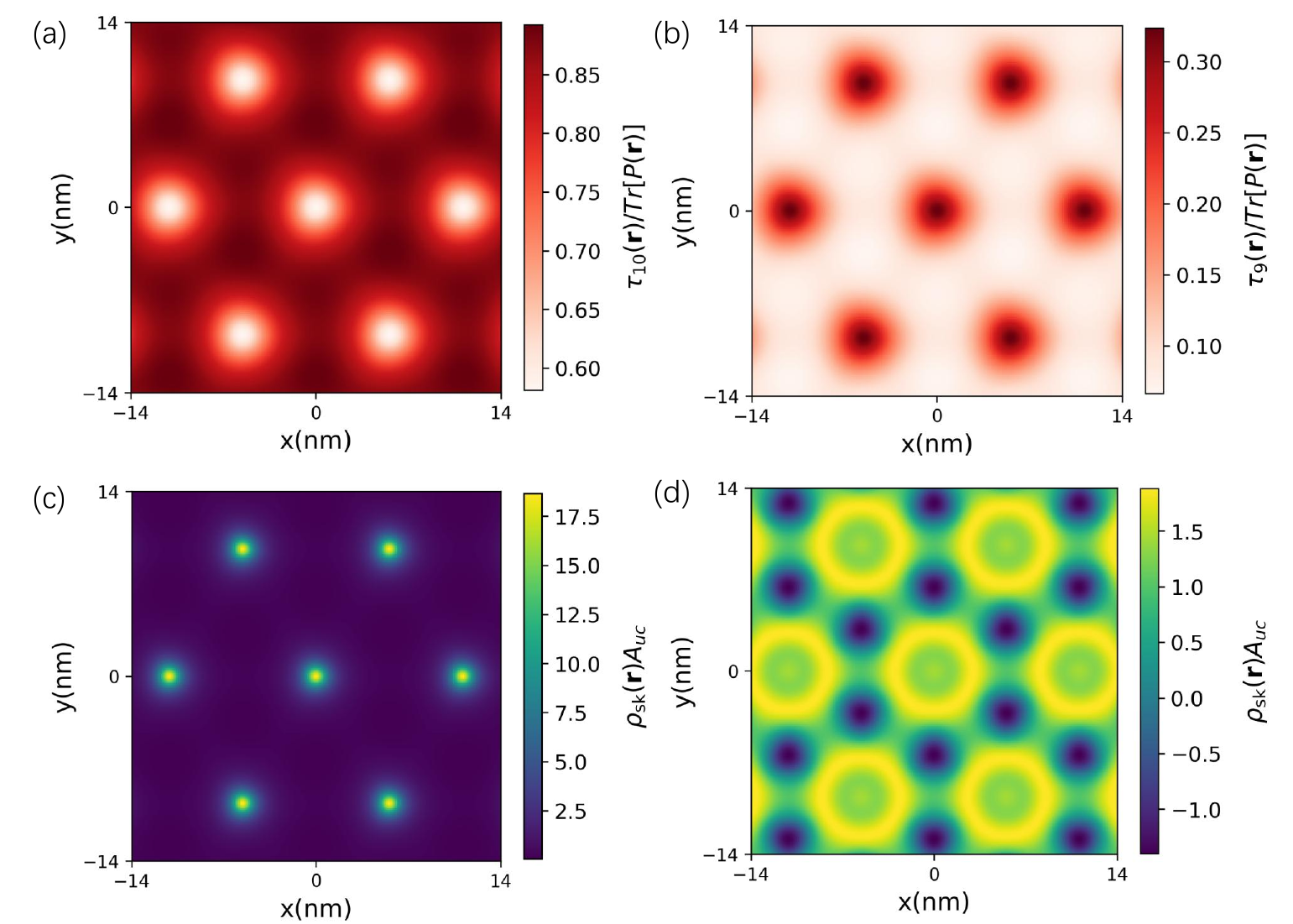}
    \caption{\textbf{(a-c)} Hartree-Fock calculation at twist angle $\theta=0.77^{\circ}$, $u_D=50\textrm{meV}$,$\lambda=1$,$V_{\text{bg}}=8\textrm{meV}$, $\psi_{\text{bg}}=\frac{2\pi}{3}$ . The interaction is taken to be gate-screened Coulomb interaction. This is the same set of parameters as those used in the main text plot. (a) (b) are the largest and second- largest eigenvalue of the real space density matrix $P_{\sigma_1,l_1;\sigma_2,l_2}(\bm{r})$, normalized by the trace of the density matrix.  (c) is the skyrmion density $\rho_{\textrm{sk}}(\bm{r})$. $A_{\text{uc}}$ is the  moir\'e unit cell area. The integrated skyrmion density in one unit cell is 1. \textbf{(d)} The skyrmion density from the ansatz wave function (cf. Eq.\ref{eq:psianalytic}) at $\theta=0.77^{\circ}$ and $\lambda=0$.}
    \label{Sfig_skyrmion}
\end{figure}

In Fig.\ref{Sfig_skyrmion} (a) and (b), we plot the normalized largest eigenvalue $\tau_{10}(\bm{r})/\textrm{Tr}[P(\bm{r})]$ and the normalized second largest eigenvalue $\tau_{9}(\bm{r})/\textrm{Tr}[P(\bm{r})]$for the Hartree-Fock, which is carried out with gate-screened Coulomb interaction, $\lambda=1$, $\theta=0.77^\circ$, $u_D=50\textrm{meV}$. In the entire unit cell, the normalized largest eigenvalue is larger than 0.55, while the normalized second largest eigenvalue  is at most 0.35 in the , showing that the real space density matrix is dominated by $\ket{\zeta_{\bm{r}}}$. In Fig. \ref{Sfig_skyrmion} (c), we plot the skyrmion density $\rho_{\textrm{sk}}(\bm{r})$ corresponding to $\ket{\zeta_{\bm{r}}}$. The curvature integrates to $2\pi$ in the unit cell. Combing these two results, we have shown that the skyrmion in the ideal limit survives in the realistic calculation. We have also included in Fig.\ref{Sfig_skyrmion} (d) the skyrmion density from the ansatz wavefunction (cf. Eq.\ref{eq:psianalytic}) at $\theta=0.77^{\circ}$, $\lambda=0.0$ for completeness. The skyrmion density integrates to 1 in one moir\'e unit cell.

Finally, we comment that the non-trivial integrated skyrmion density $\rho_{\textrm{sk}}$ of the higest eigenvector of $P(\bm{r})$  and a non-trivial Chern number is not necessarily in one-to-one correspondence. In fact, we find that the gap between $\tau_9$ and $\tau_{10}$ closes at around $u_D=40\textrm{meV}$ for $\theta=0.77^{\circ}$. The integrated $\rho_{\textrm{sk}}(\bm{r})$ becomes trivial below that displacement field. However, the Chern number remains to be 1 until much lower $u_D$. The real-space electron pseudospin texture completely follows that of the highest eigenvector of $P(\bm{r})$ only in the ideal limit. Once we move away from that limit, the $P(\bm{r})$ starts to develop other non-vanishing eigenvalue, then the real-space electron pseudospin texture is not completely determined by the highest eigenvector. However, as long as $\tau_{2N_L}$ and $\tau_{2N_L-1}$ remains well gapped, the integrated skyrmion density for the highest eigenvector is a good proxy for the winding of real electron texture.

\subsection{Adiabatic continuity between ideal and realistic models}\label{sec:Adiabatic continuity between ideal and realistic models}

\begin{figure}
    \centering
    \includegraphics[width=1.0\linewidth]{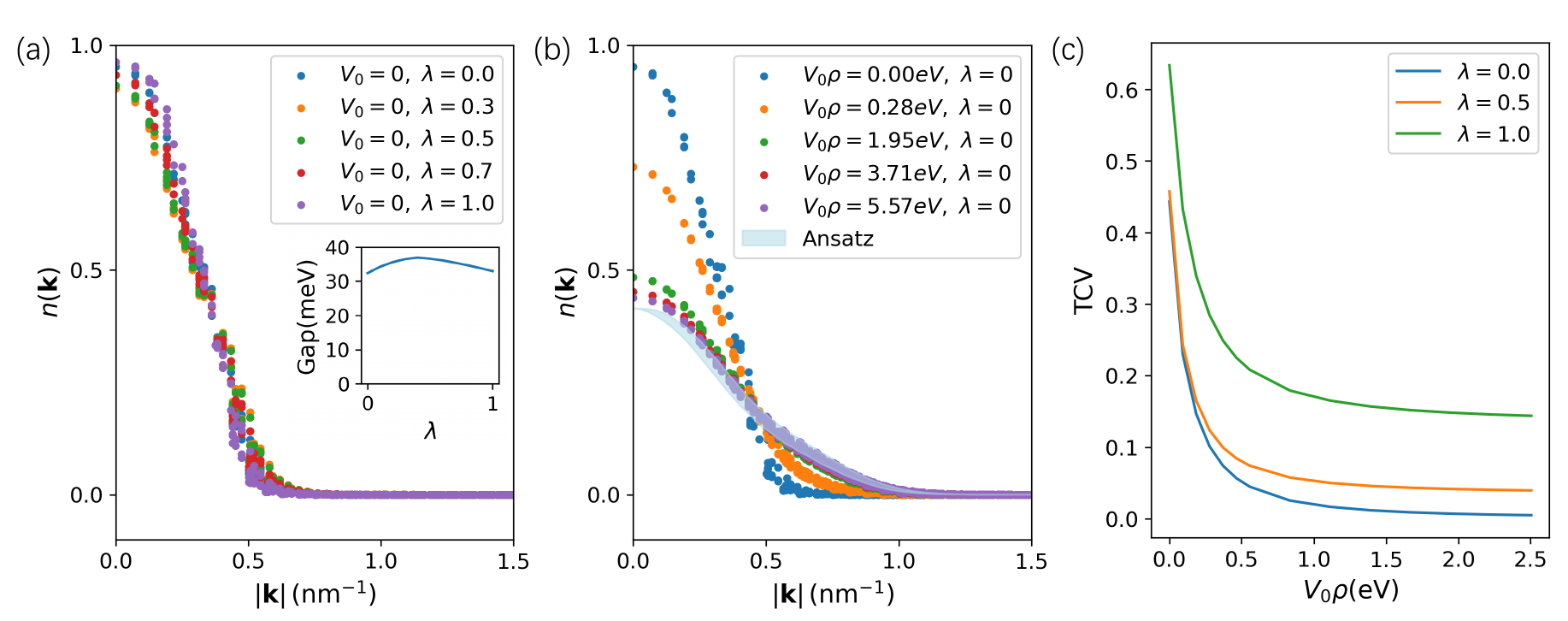}
    \caption{Hartree-Fock calculation at $u_D=50\textrm{meV}$ and $\theta=0.77^{\circ}$, $V_{\textrm{bg}}=8\textrm{meV}$, $\psi_{\text{bg}}=\frac{2\pi}{3}$. This corresponds to a real space moir\'e period of 11.16 nm. We choose $\mathcal{E}(\bm{k})$ to be the real dispersion of the first conduction band $h_{\textrm{R5G}}$ at $u_D=50\textrm{meV}$. The interaction is chosen to be density-density contact interaction plus gate-screened Coulomb interaction. \textbf{(a)} Momentum space occupation number of Hartree-Fock wavefunction for different $\lambda$ when contact interaction strength $V_0=0$. Inset is the Hartree-Fock indirect band gap as a funtion of $\lambda$ when $V_0=0$. \textbf{(b)} Momentum space occupation number of Hartree-Fock wavefunction for different contact interaction strength when $\lambda=0$. \textbf{(c)} Trace condition violation (TCV) as a function of contact interaction strength for different interpolation ratio $\lambda$.
    }
    \label{Sfig_TCV_theta0.77}
\end{figure}

In Fig.\ref{Sfig_TCV_theta0.77}, we show additional information to the results presented in Fig.4 of the main text. These calculation is done at $\theta^{\circ}=0.77^{\circ}$, $u_D=50\textrm{meV}$, $V_{\textrm{bg}}=8\textrm{meV}$, $\psi_{\textrm{bg}}=\frac{2\pi}{3}$. We choose $\mathcal{E}(\bm{k})$ to be the real dispersion of the first conduction band $h_{\textrm{R5G}}$ at $u_D=50\textrm{meV}$.

Fig.\ref{Sfig_TCV_theta0.77} (a) (b) shows the momentum space occupation number for different interpolation ratio $\lambda$ and contact interaction strength $V_0$. These data are also presented in the main text, we reproduce them here for completeness. Fig.\ref{Sfig_TCV_theta0.77} (c) and the inset of (a) shows the TCV and indirect Hartree-Fock band gap, respectively. All quantities evolve smoothly, showing the adiabatic continuity between the fully realistic model ($\lambda=1$, $V_0=0$) and the ideal limit of R5G.

\section{Hartree-Fock with enlarged-unit cell and one-hole-doped excited states}\label{sec:enlargedunitcell}
The goal of this section is to study what happens when one hole is doped to the $\nu=1$ skyrmion lattice in the presence of a point defect. We will present a phase diagram for this one-hole state, and show the corresponding charge density profile. We first describe the numerical setup (Sec. A and Sec. B). We then present the one-hole phase diagram (Sec. C) and quantify the role of band mixing (Sec. D). In Sec. E, we project to the lowest $\nu=1$ Hartree-Fock band and study the one-hole state under such band projection. We show that the one-hole state obtained in this manner is qualitatively different from those obtained without band projection (Sec. C).

\subsection{Set Up}\label{sec:setupenlargedHF}
In the main text, we present the electron density profile for the one-hole-doped excited states of RMG (relative to the skyrmion lattice ground state at moir\'e filling factor $\nu=1$). This calculation is also Hartree-Fock, but it is done by relaxing the assumption of translational invariance and using an enlarged unit cell.  Specifically, the discussion in Sec.\ref{sec:HFdetails} is under the assumption that the RMG system is placed on a torus spanned by $\bm{L}_i=N_i \bm{a}_i$ and we assume translational invariance with respect to $\bm{a}_i$. Hence, the mBZ is spanned by $\bm{b}_i$ and the Bloch momentum $\bm{k}$ can take $N_1N_2$ distinct values inside the mBZ (cf.\ref{eq:allowedk}).

For the calculation in this section, we assume translational invariance with respect to $\bm{L}_i$, i.e. the whole torus is the unit cell. Hence, the new reduced mBZ is spanned by $\bm{b}_i/N_i$, and the Bloch momentum $\bm{k}$ can only take the unique value of $\bm{0}$. In addition, $\bm{g}$ now takes values on the new reduced mRL:
\begin{equation}\label{eq:reducedRL}
\bm{g}=n_1 \frac{\bm{b}_1}{N_1}+n_2 \frac{\bm{b}_2}{N_2}, \quad n_i \in \mathbb{Z}.
\end{equation}
With this understanding, the technical details described in Sec.\ref{sec:HFdetails} are still valid. In addition, moir\'e filling factor $\nu=1$ in Sec.\ref{sec:HFdetails} will translate to filling the lowest $N=N_1N_2$ states in this enlarged unit cell set up.

We caution the readers that the computation efforts increase dramatically in the current set up. For example, suppose we take a cutoff of $|\bm{g}|<3.51 |\bm{b}_1|$. There will be around 40 reciprocal vectors $\bm{g}$ if we assume translational invariance with respect to $\bm{a}_i$. In the current set up, there will be around $N_1N_2 \times 40$ $\bm{g}$. The construction of Hartree-Fock Hamiltonian and diagonalization of it will take significantly more time. In addition, there will be more competing states with the relaxed translational symmetry, making the convergence even harder. To deal with this difficulty, we initially start with  random projectors $P$. After a few sweeps of the phase diagram, we use the converged projector as seed to carry out finer sweeps of the phase diagram.

\subsection{Defect Potential}
We now consider a negatively charged defect in the hBN spacer layer.
The effect of this defect is modeled as a scalar potential acting on RMG of the form

\begin{equation}
H_{\textrm{defect}}=\sum_{\sigma, l}\sum_{\bm{g},\bm{g}'\in \textrm{reduced mRL}} \frac{V_{\textrm{defect}}(\bm{g})}{A}c^{\dagger}_{\bm{K}+\bm{g}+\bm{g}',\sigma,l} c_{\bm{K}+\bm{g},\sigma,l}.
\end{equation}

\begin{equation}
V_{\textrm{defect}}(\bm{g})=\exp(-i\bm{g}\cdot \bm{R}_{\textrm{defect}}) \frac{\sinh((D-d_{\textrm{defect}})|\bm{g}|)}{\cosh(D|\bm{g}|)}\frac{e^2}{2\epsilon_0\epsilon_r |\bm{g}|},\quad \epsilon_r=5, \quad D=25\textrm{nm}.
\end{equation}
Here, $V_{\textrm{defect}}(\bm{g})$ is the in-plane Fourier transform at $z=0$ plane of the Coulomb potential produced by a defect charge placed at $z=d_{\textrm{defect}}$,  derived in the appendix of Ref.\cite{kwan2025moire}.  The width and the strength of the defect potential is controlled by the parameter $d_{\textrm{defect}}$, which becomes sharper with smaller $d_{\textrm{defect}}$. This form of $V_{\textrm{defect}}(\bm{g})$ ignores the thickness of the RMG. When we carry out Hartree-Fock, $H_{\textrm{defect}}$ is projected to the first conduction band of RMG, similar to what is done to $H_{\textrm{bg}}$. $\bm{R}_{\textrm{defect}}$ is the in-plane position of the defect charge. For the purpose of simplifying our discussion, we set
\begin{equation}
\bm{R}_{\textrm{defect}}=\frac{2}{3}(2\bm{a}_2-\bm{a}_1).
\end{equation}
This choice is such that the defect (as well as the potential maximum it produces) is right on top of the maximum of the background moir\'e potential $V_{\textrm{bg}}(\bm{r})$. With this choice, the Hamiltonian remains $C_3$ symmetric in its presence. One can also use other $\bm{R}_{\textrm{defect}}$, and it may favor some of the phases that break $C_3$ as we demonstrate below. However, we expect that the presence of a skyrmion vacancy is robust with respect to this choice.

\subsection{Result}\label{sec:enlargedresult}
\begin{figure}
    \centering
    \includegraphics[width=0.8\linewidth]{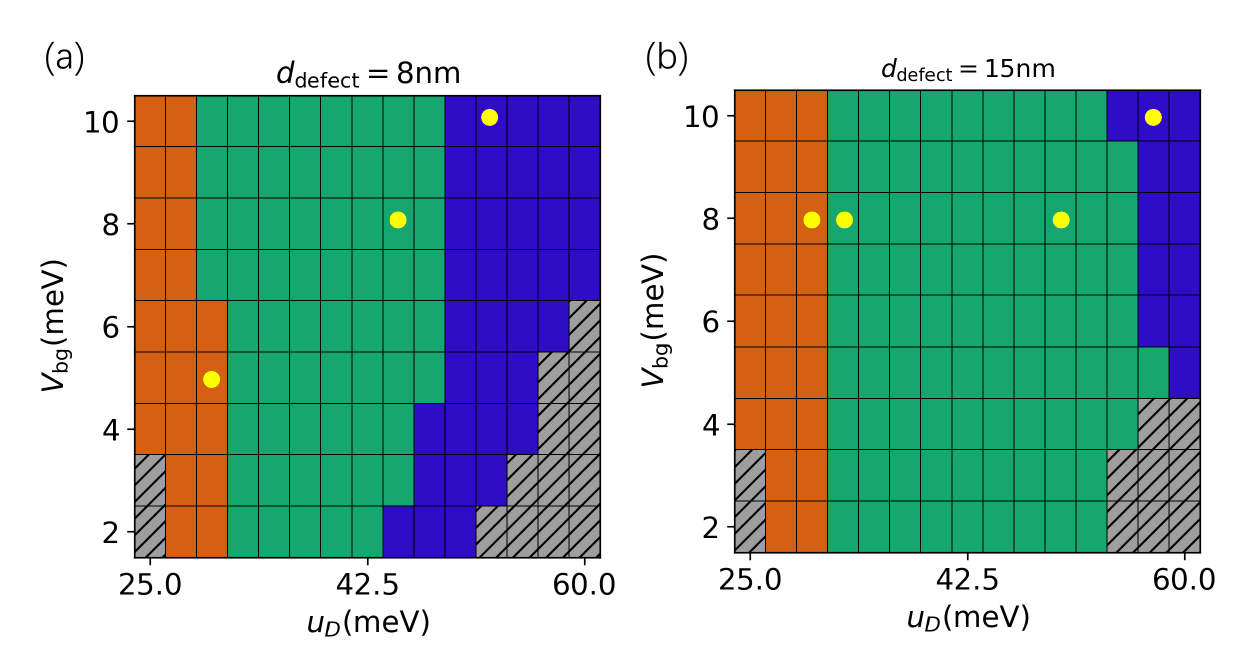}
    \caption{Phase diagram when one hole is doped to the $\nu=1$ state for two different defect-RMG distances, $d_{\textrm{defect}}$. The phase diagram is obtained with $\lambda=1$, $\psi_{\textrm{bg}}=\frac{2\pi}{3}$, $\theta=0.77^{\circ}$. The shaded gray region is where we are not certain about the lowest-energy one-hole state. The orange region is where the doped hole is delocalized on the torus. The green and blue region are both skyrmion-vacancy phases. In the green region, $C_3$ symmetry is preserved, while $C_3$ is spontaneously broken in the blue region. In Fig.\ref{Sfig_hole}, \ref{Sfig_nematic}, and \ref{Sfig_symvac1}, we show the $l=0$ charge density profile for representative parameters, corresponding to the yellow dots on the phase diagram.
    }
    \label{Sfig_Nminusonephase}
\end{figure}

The calculations presented in this section are performed with $N_1=N_2=6$, $\theta=0.77^{\circ}$, $\lambda=1$, $\psi_{\textrm{bg}}=\frac{2\pi}{3}$, $\bm{R}_{\textrm{defect}}=\frac{2}{3}(2\bm{a}_2-\bm{a}_1)$. We keep all $|\bm{g}|<3.51 |\bm{b}_1|$.
In Fig.\ref{Sfig_Nminusonephase} we present the phase diagram when one hole is doped into the $\nu=1$ state for two different defect distances, $d_{\textrm{defect}}$. The two resulting phase diagrams are qualitatively similar, with slight shifts in the phase boundary.

The shaded gray region is either where the $\nu=1$ skyrmion lattice ground state is unstable to this relaxed translational symmetry (enlarged unit cell), or where many one-hole states compete closely in energy such that we are not certain about the lowest-energy state. Outside this area, the phase diagram can be divided into three regions. We show the density profiles for representative points in the phase diagram.

\begin{figure}
    \centering
    \includegraphics[width=1.0\linewidth]{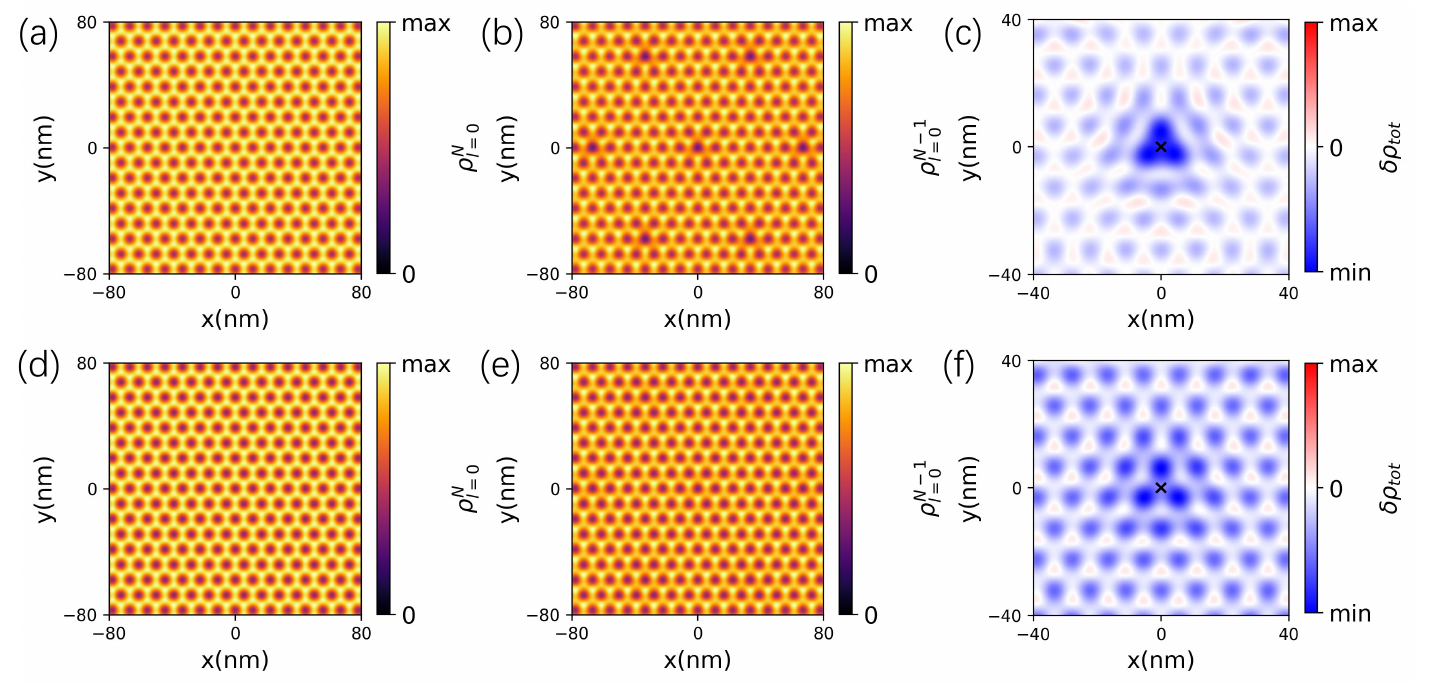}
    \caption{Charge density profile of delocalized hole state (cf. the orange region in Fig. \ref{Sfig_Nminusonephase}) from Hartree-Fock calculation with enlarged unit cell at $\lambda=1$, $\psi_{\textrm{bg}}=\frac{2\pi}{3}$, $\theta=0.77^{\circ}$. All calculations are done on a $N_1=N_2=6$ torus. \textbf{(a,b,c)} use $u_D=30.0\textrm{meV}$, $V_{\textrm{bg}}=5\textrm{meV}$. \textbf{(d,e,f)} use  $u_D=30.0\textrm{meV}$, $V_{\textrm{bg}}=8\textrm{meV}$.  \textbf{(a,d)} are calculated without defect potential. \textbf{(b)} uses $d_{\textrm{defect}}=8\textrm{nm}$. \textbf{(e)} uses $d_{\textrm{defect}}=15\textrm{nm}$. There are  36 electrons per enlarged unit cell for \textbf{(a,d)}, and 35 electrons per enlarged unit cell for $\textbf{(b,e)}$. \textbf{(c,f)} show the change in the total density in a zoomed-in window, computed as $\delta\rho_{\text{tot}}=\sum_l (\rho^{N-1}_{l}-\rho^N_l$). The defect position is marked with a black cross.}
    \label{Sfig_hole}
\end{figure}

For small $u_D$ (orange in the phase diagram), the doped hole is delocalized across the entire torus. We show the layer-resolved charge density of the one-hole state, as well as the change $\delta \rho$ compared to the $\nu=1$ skyrmion lattice state in Fig.\ref{Sfig_hole}. These plots show that the doped hole is not bound to the defect. Rather, it is spread over the entire enlarged unit cell (cf. Fig.\ref{Sfig_hole} (c)(f)). In the one-hole state, there are still $N$ pronounced minima in  the $l=0$ charge density within the enlarged unit cell. The charge density of the one-hole state (cf. Fig.\ref{Sfig_hole} (b)(e)) is qualitatively similar to the $\nu=1$ skyrmion-lattice state (cf. Fig.\ref{Sfig_hole} (a)(d)). This differentiates this phase from the two phases discussed below.

\begin{figure}
    \centering  \includegraphics[width=1.0\linewidth]{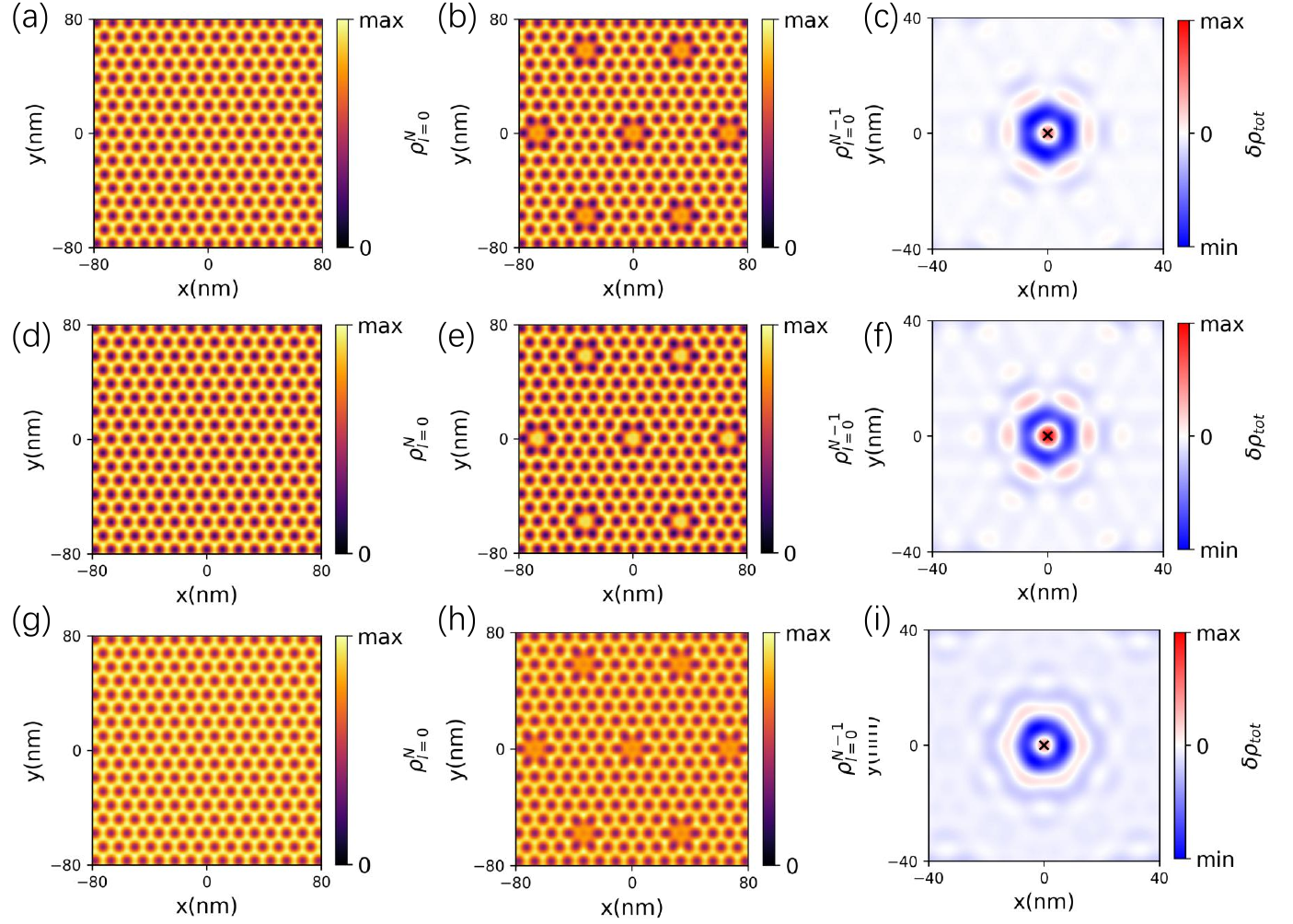}
    \caption{Charge density profile of skyrmion vacancy state (cf. the green region in Fig.\ref{Sfig_Nminusonephase}) from Hartree-Fock calculation with enlarged unit cell at $\lambda=1$ $\psi_{\textrm{bg}}=\frac{2\pi}{3}$, $\theta=0.77^{\circ}$.  All calculations are done on a $N_1=N_2=6$ torus. \textbf{(a,b,c)} use $u_D=45.0\textrm{meV}$, $V_{\textrm{bg}}=8\textrm{meV}$. \textbf{(d,e,f)} uses  $u_D=50.0\textrm{meV}$, $V_{\textrm{bg}}=8\textrm{meV}$. \textbf{(g,h,i)} use $u_D=32.5\textrm{meV}$, $V_{\textrm{bg}}=8\textrm{meV}$.\textbf{(a,d,g)} are calculated without defect potential. \textbf{(b)} uses $d_{\textrm{defect}}=8\textrm{nm}$. \textbf{(e,h)} uses $d_{\textrm{defect}}=15\textrm{nm}$. There are  36 electrons per enlarged unit cell for \textbf{(a,d,g)}, and 35 electrons per enlarged unit cell for $\textbf{(b,e,h)}$. \textbf{(c,f,i)} show the change in the density in a zoomed-in window. The defect position is marked with a black cross.
    }
    \label{Sfig_symvac1}
\end{figure}

As $u_D$ increases, a large portion of the phase diagram (green) corresponds to the regime where the one-hole state is a charged skyrmion vacancy. We show three examples in Fig. \ref{Sfig_symvac1}, where we plot the charge density profile for the corresponding one-hole state (cf. panel (b)(e)(h)). Compared with the $\nu=1$ skyrmion lattice state (cf. panel (a)(d)(g)), we clearly observe that there is one fewer minimum in the $l=0$ charge density. The change in the charge density is also localized in space (cf. panel (c)(f)(i)), showing that the skyrmion vacancy is tightly bound to the defect.

\begin{figure}
    \centering
    \includegraphics[width=0.8\linewidth]{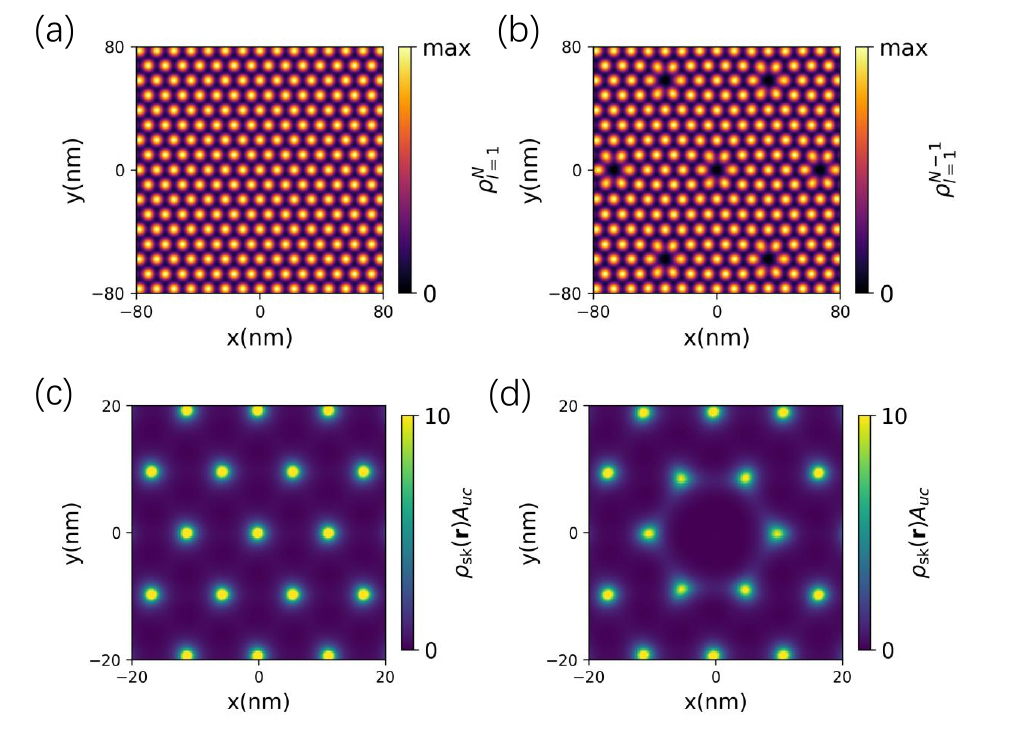}
    \caption{Additional Hartree-Fock results with enlarged unit cell at $\lambda=1$, $\psi_{\textrm{bg}}=\frac{2\pi}{3}$, $u_D=50\textrm{meV}$, $d_{\textrm{defect}}=15\textrm{nm}$, $\theta=0.77^{\circ}$, $V_{\textrm{bg}}=8\textrm{meV}$. All calculations are done on a $N_1=N_2=6$ torus.  (\textbf{a}) shows the $l=1$ charge density for $N_e=36$ electron per unit cell state.  (\textbf{b}) shows the $l=1$ charge density for $N_e=35$ electron per unit cell state. (\textbf{c},\textbf{d}) show the skyrmion density $\rho_{\textrm{sk}}(\bm{r})$ corresponding to (\textbf{a},\textbf{b}). The skyrmion density integrates of $36$ (\textbf{c}) and $35$ (\textbf{d}) over the torus enclosed by $\bm{L}_i$, respectively.
    }
    \label{Sfig_symvac1_l=1}
\end{figure}

In Fig.\ref{Sfig_symvac1_l=1}, we present additional data for the skyrmion vacancy state at $d_{\textrm{defect}}=15\textrm{nm}$, $V_{\textrm{bg}}=8\textrm{meV}$. Panel (\textbf{b}) shows the $l=1$ charge density for $N_e=35$ per unit cell skyrmion-vacancy state. Compared with the $l=1$ charge density of $N_e=36$ per unit cell skyrmion-lattice state (panel (\textbf{a})), one peak is clearly missing. The same feature is also visible in the corresponding skyrmion density $\rho_{\textrm{sk}}(\bm{r})$ (panel \textbf{c},\textbf{d}). $\rho_{\textrm{sk}}(\bm{r})$ integrates to 36 and 35 over the torus enclosed by $\bm{L}_i$ for the skyrmion-lattice state and the skyrmion-vacancy state, respectively, showing that there is indeed one skyrmion missing per enlarged unit cell.

\begin{figure}
    \centering
    \includegraphics[width=1.0\linewidth]{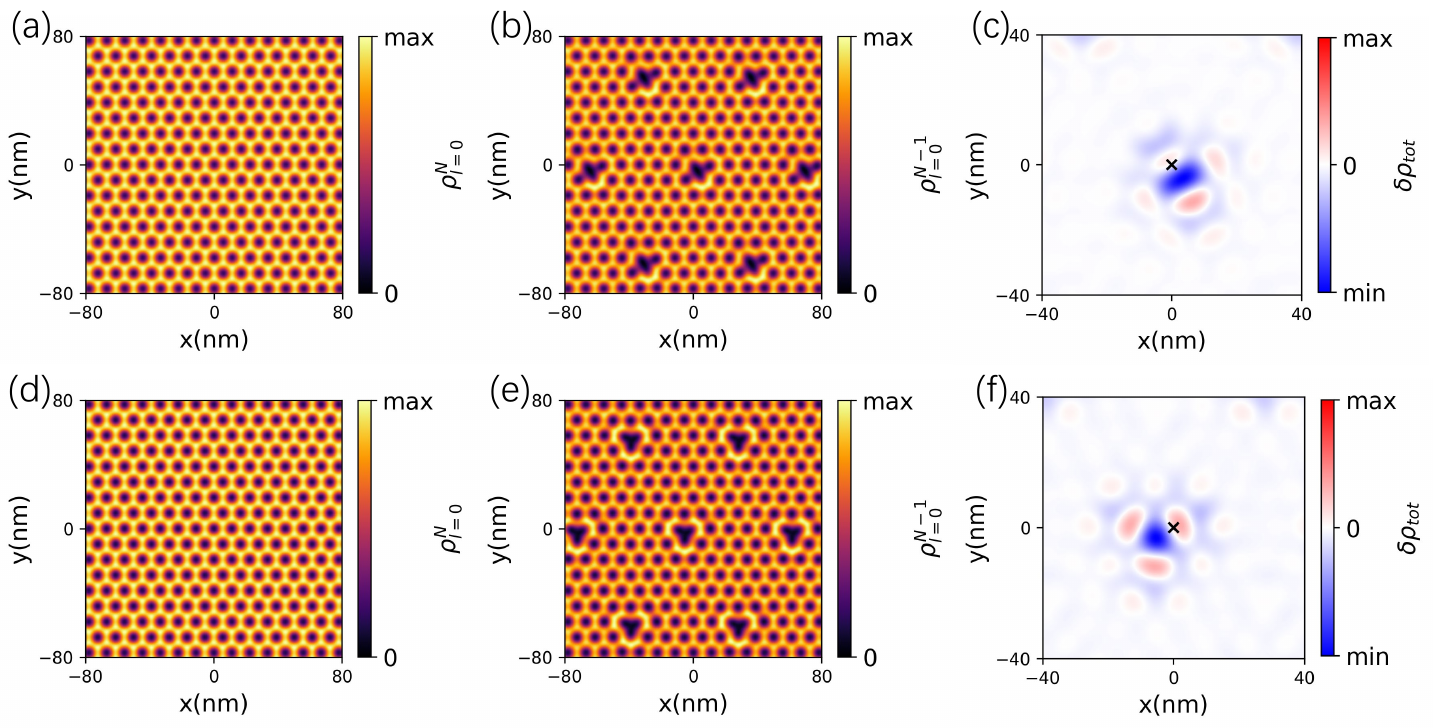}
    \caption{Charge density profile of $C_3$-breaking state (cf. the blue region in Fig.\ref{Sfig_Nminusonephase}) from Hartree-Fock calculation with enlarged unit cell at $\lambda=1$, $\psi_{\textrm{bg}}=\frac{2\pi}{3}$, $\theta=0.77^{\circ}$.  All calculations are done on a $N_1=N_2=6$ torus. \textbf{(a,b,c)} use $u_D=52.5\textrm{meV}$, $V_{\textrm{bg}}=10\textrm{meV}$. \textbf{(d,e,f)} use  $u_D=57.5\textrm{meV}$, $V_{\textrm{bg}}=10\textrm{meV}$. \textbf{(a,d)} are calculated without defect potential. \textbf{(b)} uses $d_{\textrm{defect}}=8\textrm{nm}$. \textbf{(e)} uses $d_{\textrm{defect}}=15\textrm{nm}$.  There are  36 electrons per enlarged unit cell for \textbf{(a,d)}, and 35 electrons per enlarged unit cell for $\textbf{(b,e)}$. \textbf{(c,f)} show the change in the density in a zoomed-in window. The defect position is marked with a black cross.
    }
    \label{Sfig_nematic}
\end{figure}

For even larger $u_D$ (blue), we find phases in which the one-hole state spontaneously breaks the $C_3$ symmetry of the Hamiltonian. Two examples are shown in Fig.\ref{Sfig_nematic}. We clearly see that the change in the charge density $\delta \rho$ (cf. panel (c)(f)) is not $C_3$ symmetric around the origin (the defect position, marked with a black cross).  We compute the skyrmion density using the Fourier-transformed density matrix following the procedure in Sec. \ref{sec:pontragin}. The integrated skyrmion density is $N-1$, suggesting that this phase can still be understood as a charged skyrmion vacancy. However, the doped hole spontaneously breaks the three-fold rotation symmetry of the system.

We rotate and translate the coordinate system so that the defect lies at the origin (marked with a black cross in Fig.\ref{Sfig_hole},  \ref{Sfig_symvac1}, and \ref{Sfig_nematic}), and $\bm{a}_1$ lies along the x-axis. We define the change in  density, $\delta \rho$, as the difference between the density profile of the one-hole state obtained with the defect potential and the $N$-electron state ($\nu=1$) obtained without the defect potential (but with enlarged unit cell), $\delta\rho=\rho^{N-1}-\rho^{N}$.

\subsection{Band mixing is necessary to describe the skyrmion-vacancy state}\label{sec:bandmixHF}
In the main text, we noted that Hartree-Fock band mixing is necessary to capture the skyrmion-vacancy excitation. We quantify this statement in this section.  All calculations presented in this section are done with $N_1=N_2=6$, $\theta=0.77^{\circ}$, $\bm{R}_{\textrm{defect}}=\frac{2}{3}(2\bm{a}_2-\bm{a}_1)$ unless otherwise specified. We keep all $|\bm{g}|<3.51 |\bm{b}_1|$.

Let us denote by  $\ket{\Psi_{N_e}}$ the many-body state obtained from the Hartree-Fock calculation with enlarged unit cell, where $N_e$ is the number of electrons on the torus. We would like to know how much weight  this many-body wavefunction carries in the Hilbert space spanned by the bands obtained from the Hartree-Fock calculation with the normal unit cell, $d^{\dagger}_{n,\bm{k}}$ (cf. Eq.\ref{eq:hfbasisdefin}). We define the many-body projector into this Hilbert space as
\begin{equation}
\hat{P}=\prod^{\infty}_{n= N_{c}+1}\prod_{\bm{k}}(1-d^{\dagger}_{n,\bm{k}}d_{n,\bm{k}}).
\end{equation}
Here, $N_c$ is the cutoff of the one-particle Hilbert space  spanned by $d^{\dagger}_{1..N_c,\bm{k}}$.
The weight of a normalized Slater determinant in the corresponding Hilbert space is defined to be
\begin{equation}
\textrm{weight}\equiv \langle \Psi_{N_e}|\hat{P}|\Psi_{N_e} \rangle =\det(C^{\dagger}D D^{\dagger}C).
\end{equation}
Here, $C$ is a rectangular matrix with columns being the orthonormal orbitals of $\ket{\Psi_{N_e}}$, written in the $c^{\dagger}_{\bm{k=0},\bm{g}}$ basis. $\bm{g}$ is on the reduced reciprocal lattice (Eq.\ref{eq:reducedRL}). $D$ is also a rectangular matrix whose columns are the orbitals spanning the one-particle Hilbert space, i.e. $d^{\dagger}_{1..N_c,\bm{k}\in \text{mBZ}}$, written in the same $c^{\dagger}_{\bm{k=0},\bm{g}}$ basis. We note that, to compute the weight, we need to carry out two Hartree-Fock calculations, one to obtain the many-body states with enlarged unit cell $\ket{\Psi_{N_e}}$, the other to obtain $d^{\dagger}_{n,\bm{k}}$.

For all the results presented in this section, we check that at those parameters, when the defect is absent, enlarging the unit cell does not change the ground state at $\nu=1$, i.e., $\ket{\Psi_{N_e=36}}$ has weight 1 for $N_c=1$ when the defect is absent.

If the weight is small, this means that the truncated one-particle Hilbert space cannot adequately describe the corresponding many-body state. In that case, numerical methods involving projecting onto the Hartree-Fock bands, such as band-projected exact diagonalization, become less trustworthy.

We note that, with this definition, if even one orbital of $\ket{\Psi_{N_e}}$ is completely orthogonal to the one-particle Hilbert space spanned by $d^{\dagger}_{1..N_c,\bm{k}}$, the weight is zero. This is a desirable feature of this definition,  since for a skyrmion-vacancy state, the wavefunction differs from the $\nu=1$ skyrmion-lattice state only locally. Other possible definitions of the weight, e.g. $\langle \Psi_{N_e}|\sum_{n,\bm{k}}d^{\dagger}_{n,\bm{k}} d_{n,\bm{k}}|  \Psi_{N_e} \rangle$, include a bulk contribution that overwhelms the localized contribution from the skyrmion vacancy.

\begin{figure}
\includegraphics[width=0.8\linewidth]{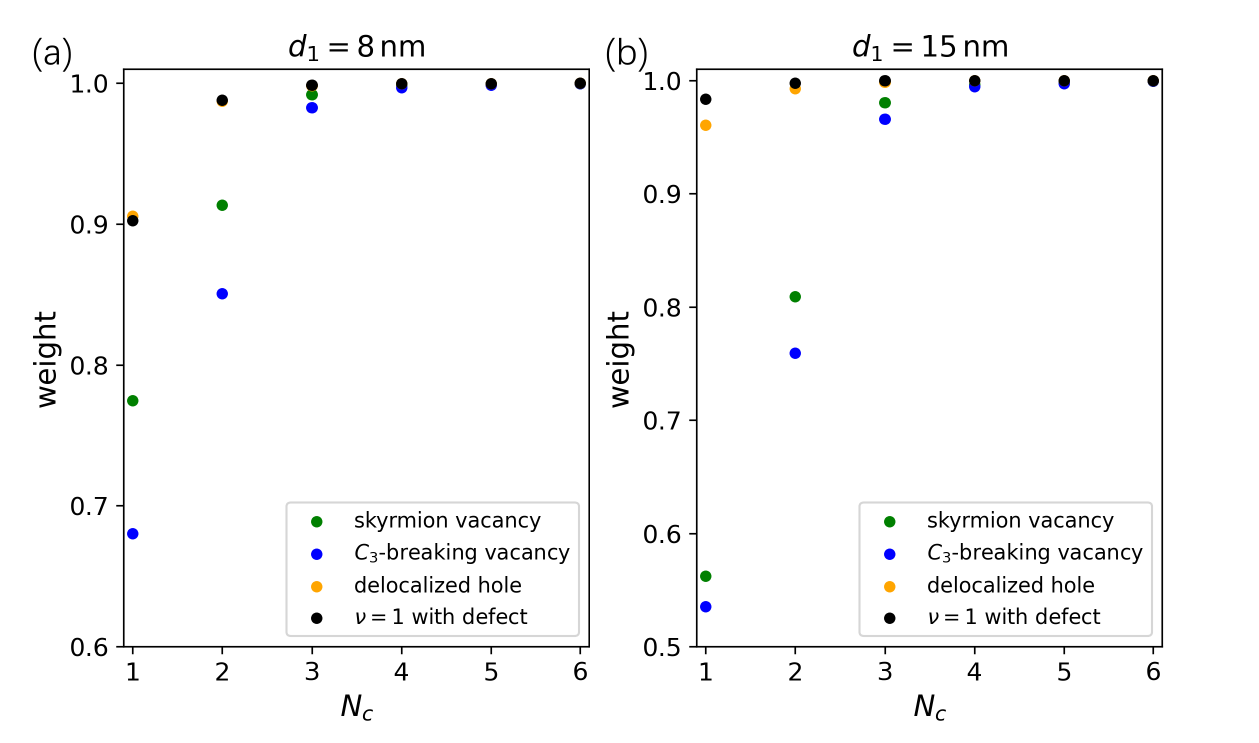}
    \caption{Weight of various many-body states in the  Hilbert space spanned by Hartree-Fock orbitals $d^{\dagger}_{n,\bm{k}}$ as a function of the cutoff $N_c$. The many-body states are obtained from Hartree-Fock calculation with enlarged unit cell. Both panels are calculated using $\lambda=1$, $\psi_{\textrm{bg}}=\frac{2\pi}{3}$, with dual-gate screened Coulomb interaction. The left panel uses $d_{\textrm{defect}}=8\textrm{nm}$. The right panel uses $d_{\textrm{defect}}=15\textrm{nm}$. Other parameters used include \textbf{(a)} $u_D=45\textrm{meV}$, $V_{\textrm{bg}}=8\text{meV}$, $N_e=35$ for the skyrmion vacancy state; $u_D=52.5\textrm{meV}$, $V_{\textrm{bg}}=10\text{meV}$, $N_e=35$ for the $C_3$ breaking vacancy state; $u_D=30\textrm{meV}$, $V_{\textrm{bg}}=5\text{meV}$, $N_e=35$ for the delocalized hole state;  $u_D=45\textrm{meV}$, $V_{\textrm{bg}}=8\text{meV}$, $N_e=36$ for the $\nu=1$  with defect state.
    \textbf{(b)} $u_D=50\textrm{meV}$, $V_{\textrm{bg}}=8\text{meV}$, $N_e=35$ for the skyrmion vacancy state; $u_D=57.5\textrm{meV}$, $V_{\textrm{bg}}=10\text{meV}$, $N_e=35$ for the $C_3$-breaking vacancy state; $u_D=30\textrm{meV}$, $V_{\textrm{bg}}=10\text{meV}$, $N_e=35$ for the delocalized hole state;  $u_D=50\textrm{meV}$, $V_{\textrm{bg}}=8\text{meV}$, $N_e=36$ for the $\nu=1$ with defect state. These parameters are also marked with yellow in Fig.\ref{Sfig_Nminusonephase}. }
    \label{Sfig_weight}
\end{figure}

In Fig.\ref{Sfig_weight}, we show the weight of the one-hole state in the Hartree-Fock Hilbert space as a function of the cutoff $N_c$ for various phases of the one-hole state ($N_e=35$).  For the skyrmion-vacancy state (green and blue in Fig.\ref{Sfig_Nminusonephase}), when $N_c=1$, the weight is small, showing that the skyrmion-lattice band $d_{1,\bm{k}}$ is insufficient to capture this one-hole state. Including additional Hartree-Fock bands is necessary to capture the skyrmion-vacancy excitation. In contrast, for the delocalized hole state (orange in Fig.\ref{Sfig_Nminusonephase}), the weight is near unity even for $N_c=1$, showing that band mixing is less important for describing this type of excitation.

In both panels of Fig.\ref{Sfig_weight}, we have included the weight of the  ``$\nu=1$ with defect" state. This state (black dots) is obtained using the same parameters ($u_D,\theta,V_{\textrm{bg}}, d_{\textrm{defect}}...$.) as the skyrmion-vacancy state (green dots), but with $N_e=36$ electrons per enlarged unit cell. The presence of a defect makes the weight ``$\nu=1$ with defect" state different from one when $N_c=1$. However, that weight is still much larger than the weight of skyrmion-vacancy state in $N_c=1$ Hilbert space. Thus, the defect cannot explain why the skyrmion-vacancy state has small weight
in the $N_c=1$ Hilbert space.
Taken together, these results show that Hartree-Fock band mixing is important for capturing the skyrmion vacancy excitation.

\begin{figure}
\includegraphics[width=0.5\linewidth]{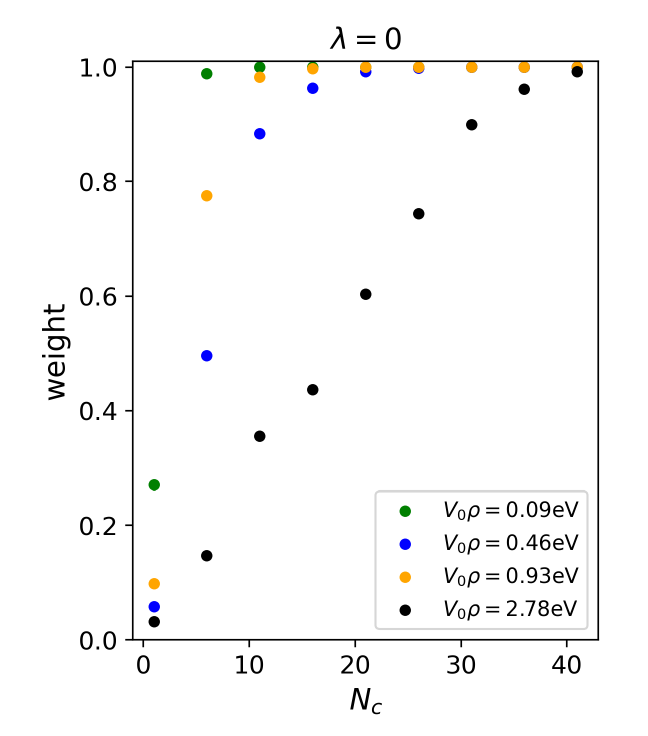}
    \caption{Weight of $N_e=35$ many-body states in the  Hilbert space spanned by Hartree-Fock orbitals $d^{\dagger}_{n,\bm{k}}$ as a function of the cutoff $N_c$. The many-body states are obtained by Hartree-Fock calculation with enlarged unit cell. We use $\lambda=0$, $\psi_{\textrm{bg}}=\frac{2\pi}{3}$, $u_D=50\textrm{meV}$, $V_{\textrm{bg}}=8\text{meV}$, $N_e=35$, $\theta=0.77^{\circ}$.  The interaction takes the form of Eq. \ref{eq:contactplusCoulomb}. We do not include the defect potential in this calculation.  }
    \label{Sfig_ideal_weight}
\end{figure}

Finally, for completeness, we have also calculated the weight of the skyrmion vacancy states when $\lambda=0$, and for large contact interaction $V_0$. The results are shown in Fig.\ref{Sfig_ideal_weight}. All states shown are $N_e=35$ skyrmion-vacancy states. We do not include the defect potential in this calculation. Therefore, the ground state is degenerate with respect to the  skyrmion vacancy position.  When $V_0$ increases, the weight quickly decreases. At $V_0\rho=2.78\textrm{eV}$, one must keep more than $N_c=30$ bands to achieve a weight of 0.9.

\subsection{Evidence from band-projected Hartree-Fock}\label{sec:bandprojection}
In the last section, we computed the weight of the many-body state $\ket{\Psi_{N_e=35}}$ in the single-particle Hilbert space spanned by $d^{\dagger}_{n,\bm{k}}$. In particular, we note that when $N_c=1$, the weight is very small, which implies that the numerical approaches based on projection onto the Hartree-Fock bands can be unreliable. Below, we directly shows that if we project to the $d^{\dagger}_{1,\bm{k}}$ band to study the one-hole state, the results are qualitatively different.

We first carry out a standard Hartree-Fock calculation with translation symmetry imposed with respect to $\bm{a}_i$, as outlined in Sec.\ref{sec:HFdetails}. We then project onto a set of minibands obtained from this calculation, $d^{\dagger}_{1..N_c,\bm{k}}$ and carry out a second Hartree-Fock calculation with translation symmetry relaxed to the torus spanning vectors $\bm{L}_i$. We then seek an $N_e$-electron state within this restricted Hilbert space. We will show that, when $N_c=1$, the ground state obtained in this manner is fundamentally different from the skyrmion-vacancy state discussed in the previous section.

We note that, when $N_c\to \infty$, it is guaranteed that the ground state obtained in this band-projected Hartree-Fock procedure will coincide with the states presented in Sec.\ref{sec:setupenlargedHF} and Sec.\ref{sec:enlargedresult}. This is because, when $N_c\to \infty$, the variational space of this two calculations are the same. However, the goal of this section is to show that there are states in the variational space of the latter calculation that are fundamentally unreachable by this band-projected Hartree-Fock when $N_c$ is small.

We use $\ket{n,\bm{k}}$ to denote the Bloch wavefunction corresponding to $d^{\dagger}_{n,\bm{k}}=\sum_{\bm{g}}\beta_{n,\bm{k},\bm{g}}c^{\dagger}_{\bm{k},\bm{g}}$. We first project all single-particle Hamiltonian into this basis. We remind the reader that this is a second projection step. We have already projected our Hamiltonian onto the first conduction band of RMG. We now further project onto the first miniband obtained from Hartree-Fock calculations of Sec.\ref{sec:HFdetails}. For example, the twice-projected kinetic Hamiltonian has the form:
\begin{equation}
H^{\textrm{2nd-proj}}_{\textrm{kin}}=\sum^{N_c}_{n_1,n_2=1}\sum_{\bm{k}\in \textrm{mBZ}}\sum_{\bm{g}\in\textrm{mRL}}\beta^{*}_{n_1,\bm{k},\bm{g}}\beta_{n_2,\bm{k},\bm{g}}\epsilon(\bm{k}+\bm{g};\lambda)d^{\dagger}_{n_1,\bm{k}}d_{n_2,\bm{k}}.
\end{equation}
All the single-particle Hamiltonians, $H_{\textrm{kin}}$,  $H_{\textrm{defect}}$, $H_{\textrm{bg}}$, $H_{\textrm{hBN}}$, are projected similarly. We omit the explicit forms here to avoid cluttering the presentation with equations. We only note that $H_{\textrm{defect}}$ is not diagonal in $\bm{k}$. This is expected, since the defect potential is only periodic with respect to $\bm{L}_i$. The others projected single-particle Hamiltonians are all diagonal in $\bm{k}$.

We also project the interaction into this subset of Hartree-Fock bands:
\begin{equation}
H^{\textrm{2nd-proj}}_{\textrm{int}}=\frac{1}{2}\sum_{\bm{k}_1-\bm{k}_4\in \mathrm{mBZ}}^{\alpha,\beta,\gamma,\delta}V_{1234} d^{\dagger}_{\alpha,\bm{k}_1}d^{\dagger}_{\beta,\bm{k}_2}d_{\delta,\bm{k}_4}d_{\gamma,\bm{k}_3}.
\end{equation}
Here, $V_{1234}$ is the Coulomb matrix element:

\begin{equation}
V_{1234}=\bra{\alpha \bm{k}_1,\beta\bm{k}_2}\hat{V}\ket{\gamma\bm{k}_3,\delta\bm{k}_4}=\sum_{\bm{q}}\frac{V(\bm{q})}{A}\braket{\alpha\bm{k}_1}{\exp(i\bm{q}\cdot \bm{r})|\gamma\bm{k}_3}\braket{\beta\bm{k}_2}{\exp(-i\bm{q}\cdot \bm{r})|\delta\bm{k}_4}.
\end{equation}
In the equation above, $\bm{q}$ loops through all vectors such that $\exp(i\bm{q}\cdot\bm{L}_i)=1$. $V(\bm{q})$ is given by Eq.\ref{eq:contactplusCoulomb}. In Hartree-Fock approximation, we replace the 4-fermion operator in the interaction Hamiltonian by various two-fermion contractions:
\begin{equation}
\begin{split}
d^{\dagger}_{\alpha,\bm{k}_1}d^{\dagger}_{\beta,\bm{k}_2}d_{\delta,\bm{k}_4}d_{\gamma,\bm{k}_3}\to&
(\langle d^{\dagger}_{\alpha,\bm{k}_1} d_{\gamma,\bm{k}_3}\rangle d^{\dagger}_{\beta,\bm{k}_2} d_{\delta,\bm{k}_4}+ d^{\dagger}_{\alpha,\bm{k}_1} d_{\gamma,\bm{k}_3} \langle d^{\dagger}_{\beta,\bm{k}_2} d_{\delta,\bm{k}_4}\rangle)\\
&-(\langle d^{\dagger}_{\alpha,\bm{k}_1} d_{\delta,\bm{k}_4}\rangle d^{\dagger}_{\beta,\bm{k}_2} d_{\gamma,\bm{k}_3}+d^{\dagger}_{\alpha,\bm{k}_1} d_{\delta,\bm{k}_4} \langle d^{\dagger}_{\beta,\bm{k}_2} d_{\gamma,\bm{k}_3}\rangle).
\end{split}
\end{equation}
We further define the projector $P$ and the form factors $\Lambda$
\begin{equation}
P_{\alpha,\bm{k}_1;\beta,\bm{k}_2}\equiv \langle d^{\dagger}_{\beta,\bm{k}_2} d_{\alpha,\bm{k}_1} \rangle,
\end{equation}

\begin{equation}
\Lambda_{\bm{q}}(\bm{k}_1)_{\alpha,\beta}=\braket{\alpha\bm{k}_1}{e^{-i\bm{q}\cdot \bm{r}}|\beta\bm{k}_2}\delta_{\bm{k}_1,\lceil \bm{k}_2-\bm{q}\lceil}.
\end{equation}
Here, $\lceil ... \rceil$ sends the argument back to the mBZ enclosed by $\bm{b}_i$. Since we have relaxed the translation symmetry to $\bm{L}_i$, it is not required that $\bm{k}_1=\bm{k}_2$ in the projector $P$. After this replacement, the interaction Hamiltonian is replaced by its mean-field counterpart:
\begin{equation}
\begin{split}
&H^{\textrm{2nd-proj}}_{\textrm{int}}\to H^{\textrm{2nd-proj}}_{\textrm{mf}}\\
&=\sum_{\substack{\bm{k}_2,\bm{k}_4,\beta,\delta\\\bm{g}\in \textrm{mRL}}} \frac{V(\bm{g}+\lceil \bm{k}_4-\bm{k}_2\rceil)}{A} \left(\sum_{\bm{k}_1,\alpha,\gamma}\Lambda^{*}_{\bm{\bm{g}+\lceil \bm{k}_4-\bm{k}_2 \rceil }}(\lceil \bm{k}_1+\bm{k}_2-\bm{k}_4 \rceil)_{\gamma,\alpha}P_{\gamma\lceil \bm{k}_1+\bm{k}_2-\bm{k}_4 \rceil,\alpha\bm{k}_1}\right)\Lambda_{\bm{g}+\lceil \bm{k}_4-\bm{k}_2 \rceil}(\bm{k}_2)_{\beta,\delta}d^{\dagger}_{\beta,\bm{k}_2}d_{\delta,\bm{k}_4}\\
&-\sum_{\substack{\bm{k}_1,\bm{k}_2,\bm{k}_3,\beta,\gamma\\\bm{g}\in \textrm{mRL}}}\frac{V(\bm{g}+\lceil \bm{k}_1-\bm{k}_3 \rceil)}{A}\left(\sum_{\alpha,\delta}\Lambda_{\bm{g}+\lceil \bm{k}_1-\bm{k}_3 \rceil}(\bm{k}_2)_{\beta,\delta}P_{\delta,\lceil \bm{k}_1+\bm{k}_2-\bm{k}_3 \rceil;\alpha,\bm{k}_1}\Lambda^{*}_{\bm{g}+\lceil \bm{k}_1-\bm{k}_3 \rceil}(\bm{k}_3)_{\gamma,\alpha}\right) d^{\dagger}_{\beta,\bm{k}_2}d_{\gamma,\bm{k}_3}.
\end{split}
\end{equation}
We again obtain an effective single-particle Hamiltonian that depends on the projector $P$. The projector $P$ is obtained from diagonalizing the effective single-particle Hamiltonian, similar to Eq.\ref{eq:hfbasisdefin}.  This procedure is iterated until self-consistency.
We use a random Hermitian matrix as the initial guess for the projector in order to find the lowest-energy state.

\begin{figure}
    \centering
    \includegraphics[width=0.7\linewidth]{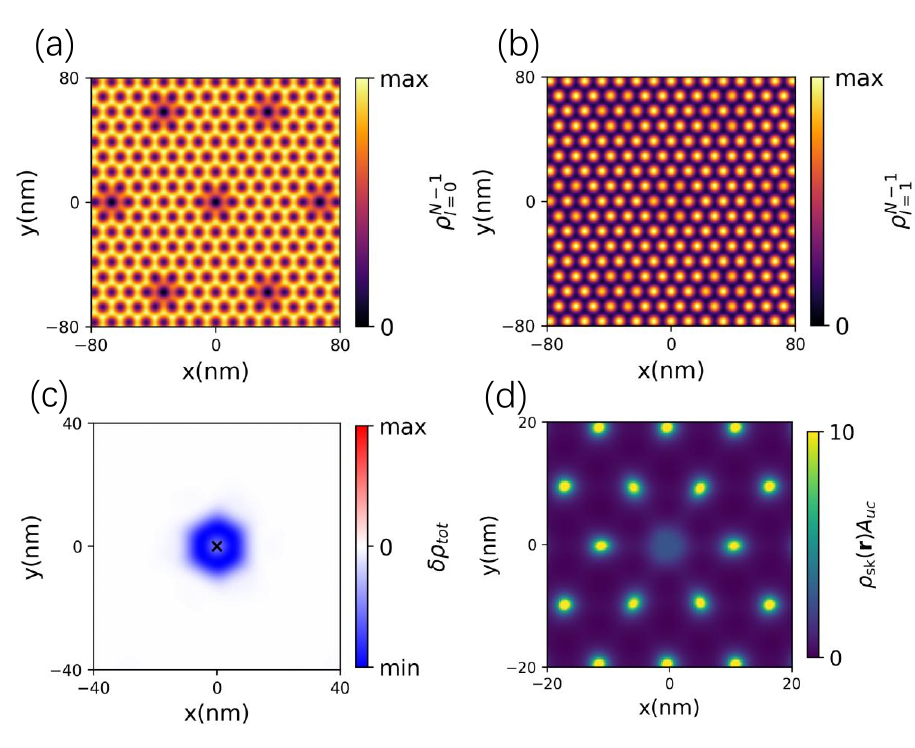}
    \caption{Results for $N_e=35$ electron per enlarged unit cell state obtained from band-projected Hartree-Fock when translation symmetry is only assumed with respect to the torus spanning vectors $\bm{L}_i$. The parameters used are $N_1=N_2=6$ (torus size), $N_c=1$ (band projection cutoff), $\lambda=1$,$V_0=0$, $V_{\textrm{bg}}=8\textrm{meV}$, $\psi_{\textrm{bg}}=\frac{2\pi}{3}$, $d_{\textrm{defect}}=15\textrm{nm}$, $u_D=50\textrm{meV}$. (\textbf{a}) and (\textbf{b}) show the $l=0,1$ charge density, respectively. (\textbf{c}) shows the change in total charge density, computed as the difference between the state in (\textbf{a},\textbf{b}) and $N_e=36$ skyrmion lattice state (shown in Fig.\ref{Sfig_symvac1_l=1} (a) and Fig.\ref{Sfig_symvac1} (d)) . (\textbf{d}) shows the skyrmion density, $\rho_{\textrm{sk}}(\bm{r})$, which integrates to 36 over the torus enclosed by $\bm{L}_i$.   $A_{uc}$ in the legend denotes the moir\'e unit cell area.
    }
    \label{Sfig_bandprojectedHF}
\end{figure}

Fig.\ref{Sfig_bandprojectedHF}  shows the $N_e=35$ results for the same set of parameters as Fig.\ref{Sfig_symvac1_l=1} ($u_D=50\textrm{meV}$, $d_{\text{defect}}=15\textrm{nm}$, $V_{\textrm{bg}}=8\textrm{meV}$, $\lambda=1$...), but calculated using $N_c=1$ band-projected Hartree-Fock. The most important distinction is that there are still 36 maxima in the $l=1$ charge density, and 36 minima in the $l=0$ charge density within the torus, as shown in panel (a) and (b). We also show the change in charge density, $\delta \rho$, compared with the $\nu=1$ skyrmion lattice state (no defect) in panel (c).  The doped hole is localized around the defect (black cross in (c)). However, the electron density does not increase near the skyrmion core, which is different from that shown in Fig.\ref{Sfig_symvac1} (f). Finally, the integrated skyrmion density remains 36 for this 35-electron state. This evidence combined shows that the one-hole state we obtain from band-projected Hartree-Fock describes a localized hole on top of the $\nu=1$ skyrmion lattice state, which is  qualitatively different from the skyrmion-vacancy state shown in Fig.\ref{Sfig_symvac1}.

\section{Effective Magnetic Field}\label{sec:effectivemagnetic}

In this section, we will derive the expression of the effective magnetic field that we quoted in in the main text (Sec. A). We then relate its average flux to the winding number of the skyrmion texture (Sec. B). Finally in Sec. C we discuss how the effective field and associated wavefunctions can be interpreted in terms of a parton construction.

\subsection{Calculation of the field profile}

We start with a Slater determinant, whose constituent single-particle orbitals are of the form
\begin{equation}
\psi_{m,l}(\bm{r})=f_m(z)\bm{\xi}_l(\bm{r})\equiv \phi_{\textrm{AC}}(\bm{r})\hat{\bm{\xi}}_l(\bm{r}).
\end{equation}
Here $m$ labels single-particle orbitals, and $l$ labels layers, $\hat{\bm{\xi}}(\bm{r})\equiv\bm{\xi}(\bm{r})/|\bm{\xi}(\bm{r})|$, $\phi_{\textrm{AC}}\equiv f_m(z)|\bm{\xi}(\bm{r})|$. Notice that by writing down the wavefunction in this form, we assume that the spinor $\bm{\xi}(\bm{r})$ is A-sublattice-polarized, hence carrying only layer index.

According to the discussion in the main text, this form of the wavefunction guarantees that the Slater determinant is a zero-energy eigenstate of  density-density contact interaction.  We further require that $\psi_{m,l}(\bm{r})$ to be zero energy eigenstate of the single-particle Hamiltonian $h_{\textrm{flat}}$, the equations to be satisfied are
\begin{equation}\label{eq:setofNeq}
-2i\gamma \partial_{\bar{z}}[\phi_{\textrm{AC}}(\bm{r})\hat{\bm{\xi}}_l(\bm{r})]-\phi_{\textrm{AC}}(\bm{r})\hat{\bm{\xi}}_{l+1}(\bm{r})=0\quad l=0..N-2.
\end{equation}

This is a set of $N-1$ equations. For an arbitrarily chosen $\hat{\bm{\xi}}(\bm{r})$, these equations can not be satisfied simultaneously in general. However, if we impose the constraint that
\begin{equation}
\hat{\bm{\xi}}_{l+1}(\bm{r})=-2i\gamma\partial_{\bar{z}}\hat{\bm{\xi}_l}(\bm{r})-g(\bm{r})\hat{\bm{\xi}_l}(\bm{r}),
\end{equation}
where $g(\bm{r})$ is independent of $l$. Then the $N-1$ equations Eq.\ref{eq:setofNeq} are organized into a single equation
\begin{equation}\label{eq:ACband}
[-2i\gamma \partial_{\bar{z}}+g(\bm{r})]\phi_{\textrm{AC}}(\bm{r})=0.
\end{equation}
We further note that Eq.\ref{eq:setofNeq} can be equivalently written as
\begin{equation}
-2i\gamma \partial_{\bar{z}}[\bm{\xi}_l(\bm{r})]-\bm{\xi}_{l+1}(\bm{r})=0\quad l=0..N-2.
\end{equation}
These provide the expression for $g(\bm{r})$
\begin{equation}
g(\bm{r})=2i\gamma\partial_{\bar{z}}\textrm{log}(|\bm{\xi}(\bm{r})|).
\end{equation}
We can then identify Eq.\ref{eq:ACband} as Aharonov-Casher equation for the zero mode of a massless Dirac fermion in a magnetic field \cite{PhysRevA.19.2461}

\begin{equation}
(p_x+ip_y+A_x(\bm{r})+iA_y(\bm{r}))\phi_{\textrm{AC}}(\bm{r})=0,
\end{equation}
where $A_x$ and $A_y$ are components of the gauge field. Hence, we can identify the expression for the gauge field

\begin{equation}
A_x(\bm{r})+iA_y(\bm{r})=\frac{1}{\gamma}g(\bm{r}).
\end{equation}
This provides the final expression for the magnetic field

\begin{equation}
\bm{B}=\nabla\times \bm{A}=-B_{\textrm{eff}}\hat{z}, B_{\textrm{eff}}=-\nabla^2 \textrm{log}(|\bm{\xi}(\bm{r})|)
\end{equation}

\begin{figure}
    \centering
    \includegraphics[width=1.0\linewidth]{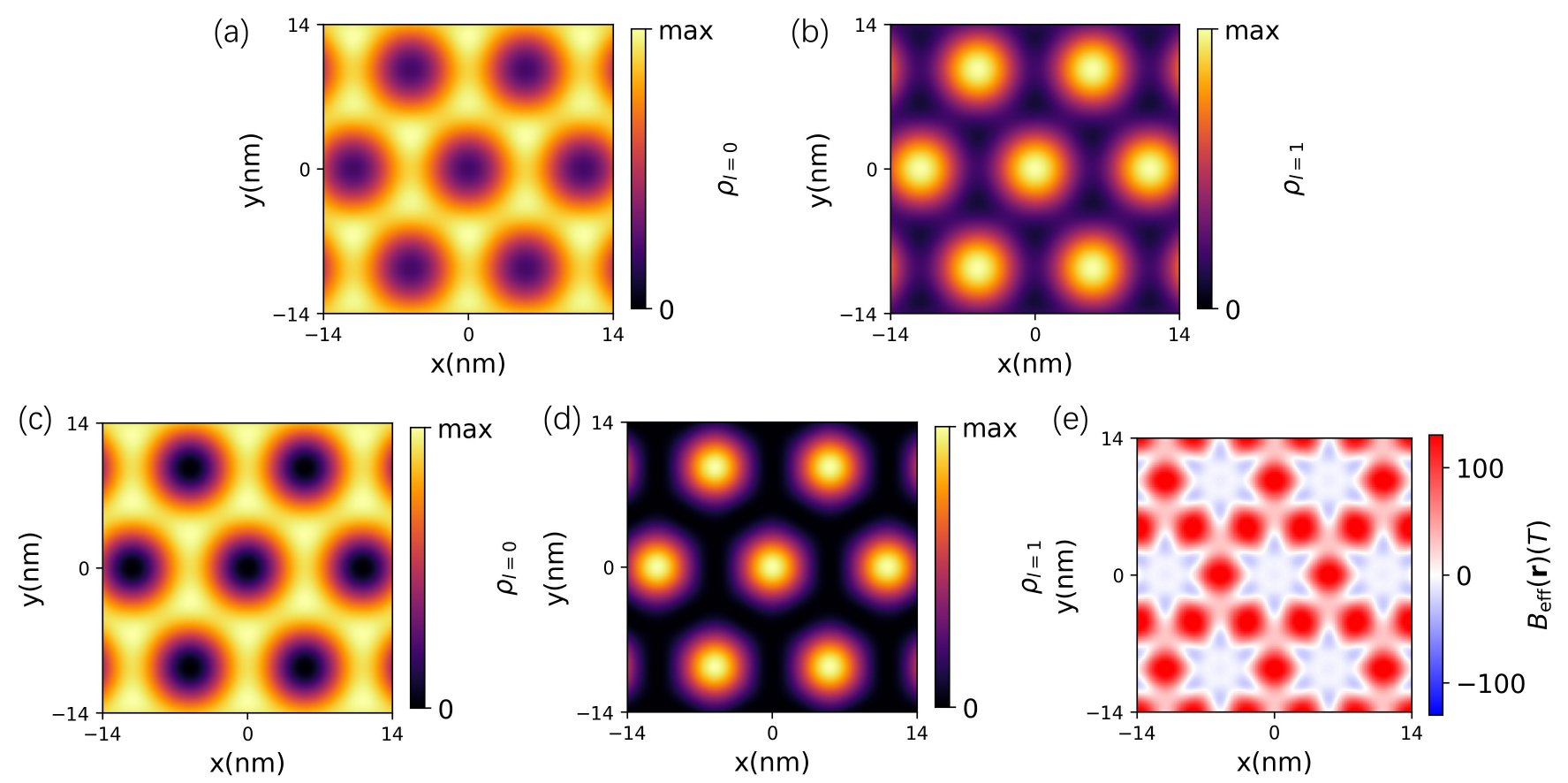}
    \caption{\textbf{(a)(b)} Charge density profile for $l=0,1$ from  Hartree-Fock calculation at $u_D=50\textrm{meV}$, $\theta=0.77^{\circ}$, $V_{\text{bg}}=8\textrm{meV}$, $\psi_{\text{bg}}=\frac{2\pi}{3}$ and $\lambda=1$. This corresponds to a real space moir\'e period of 11.16 nm. The interaction is chosen to be gate-screened Coulomb interaction. \textbf{(c)(d)} Charge density profile for $l=0,1$ from the ansatz wavefunction (cf. Eq.\ref{eq:psianalytic}). \textbf{(e)} Effective magnetic field calculated from the charge density profile in (c)(d).
    }
    \label{Sfig_effectiveB}
\end{figure}

Numerically,  we can use the analytic ansatz wavefunction, $\ket{\psi_{\bm{k},\textrm{ana}}}$, as defined in Eq.\ref{eq:psianalytic}. The real space amplitude of this wavefunction is
\begin{equation}
\psi_{\bm{k},l,\textrm{ana}}(\bm{r})=\braket{\bm{r},l}{\psi_{\bm{k},\textrm{ana}}}.
\end{equation}
The effective magnetic field is numerically evaluated as

\begin{equation}
B_{\textrm{eff}}(\bm{r})=-\frac{\nabla^2}{2} \textrm{log}(\sum_{l} |\psi_{\bm{k},l,\textrm{ana}}(\bm{r})|^2)
\end{equation}
Notice that we can take any $\bm{k}$, and there is no $\bm{k}$ dependence for this procedure. This is because according to the discussion above, in real space, the ansatz wavefunction is related to $\bm{\xi}(\bm{r})$ by

\begin{equation}
\psi_{\bm{k},l,\textrm{ana}}(\bm{r})=f_{\bm{k}}(z)\bm{\xi}_l(\bm{r}),
\end{equation}
where $f_{\bm{k}}(z)$ is some holomorphic function in $z$. Using the identity that $\nabla^2 \textrm{log}(f_{\bm{k}}(z)f_{\bm{k}}^{*}(z))=0$, we arrive at the expression of the effective magnetic field given above. We notice that the argument above holds for all $\bm{r}$ except where $f(z)$ vanishes, and hence $\textrm{log}(\sum_{l} |\psi_{\bm{k},l,\textrm{ana}}(\bm{r})|^2)$ becomes ill-defined. For the main text and supplementary plot, we patch together the magnetic field calculated using two different $\bm{k}$. The final results are free from singularity. The average magnetic flux per moir\'e unit cell is $2\pi$. In Fig.\ref{Sfig_effectiveB} (e), we show the profile of the effective magnetic field following the procedure above.

\subsection{Skyrmion winding and net flux}

In this section we relate the skyrmion winding number, $N_w$, to the number of flux quanta of the effective magnetic field, $N_\phi$,
\begin{equation}
    N_w = \frac{1}{2\pi i} \int_{\text{unit cell}}\epsilon^{\mu\nu} \Tr P \partial_\mu P \partial_\nu P d^2 \br = N_{\phi}
    \label{eq:pontryaginisflux}
\end{equation}
as was claimed in the main text (see also Refs. \cite{dongCompositeFermiLiquid2023,duranTMDAdiabaticMagic,shi2024adiabatic,guerci2025layer}). Here we used the antisymmetric symbol $\varepsilon^{xy}=-\varepsilon^{yx}=1$ and the repeated index summation convention. Indices are placed for typographic convenience (they are raised or lowered by $\delta^{\mu \nu}$).

We recall
\begin{equation}
    \xi_\ell(\br)=e^{-\frac{1}{4}l_B^{-2}|\br|^2}\chi_\ell(\br), \qquad \hat{\chi}_\ell(\br) = \chi_\ell(\br)/\norm{\chi(\br)} = \hat{\xi}_\ell(\br), \qquad P = \hat{\chi} \hat{\chi}^\dag,
\end{equation}
and $\chi_\ell(\br)$ satisifes the magnetic Bloch boundary conditions
\begin{equation}
    \chi_\ell(\br+\bm{R})= e^{-i \varphi_{\bm{R}}(\br)} \chi_\ell(\br), \qquad e^{-i \varphi_{\bm{R}}(\br)} = e^{i\bm{q}\cdot\bm{a}}\chi_\ell(\br)e^{-\frac{i}{2}l_B^{-2}\bm{a}\times\br},
    \label{eq:suppchibcs}
\end{equation}
where $l_B^{-2} = e B$ and $B$ is the average of $B_{\rm eff}(\br) = B + \delta B$, and in this section we use $\bm{R}$ to denote lattice vectors.

We begin by proving
\begin{equation}
    -i \epsilon^{\mu\nu} \Tr P \partial_\mu P \partial_\nu P = \nabla \times a = \epsilon^{\mu \nu} (\partial_\mu \hat{\chi})^\dag (\partial_\nu \hat{\chi})^\dag, \qquad a_\mu = -i \hat{\chi}^\dag \partial_\mu \hat{\chi}.
    \label{eq:windingisflux}
\end{equation}
Afterwards, we will show that the flux associated with $\nabla \times a$ is equivalent to that of $B_{\rm eff}$ through the boundary conditions \eqref{eq:suppchibcs}.

To show \eqref{eq:windingisflux}, we first write $\Tr P \partial_\mu P \partial_\nu P = \hat{\chi}^\dag \partial_\mu P \partial_\nu P \hat{\chi}$, using the cyclic property of the trace, and compute
\begin{equation}
    \partial_\nu P \hat{\chi} = \left[\hat{\chi}^\dag (\partial_\nu \hat{\chi}) + (\partial_\nu \hat{\chi}^\dag) \hat{\chi}  \right] \hat{\chi} = \left[\hat{\chi}^\dag (\partial_\nu \hat{\chi}) -\hat{\chi}^\dag \partial_\nu \hat{\chi}  \right] \hat{\chi} = (1-P)\partial_\nu \hat{\chi}
    \label{eq:projectortoflux}
\end{equation}
where we used $(\partial_\nu \hat{\chi})^\dag \hat{\chi} = -\hat{\chi}^\dag \partial_\nu \hat{\chi}$ (from differentiating $\hat{\chi}^\dag \hat{\chi} = 1$). These computations, and the Hermitian conjugate of \eqref{eq:projectortoflux}, lead to
\begin{equation}
    -i \epsilon^{\mu \nu} \Tr P \partial_\mu P \partial_\nu P = -i \epsilon^{\mu \nu} \hat{\chi}^\dag \partial_\mu P \partial_\nu P \hat{\chi} = i \epsilon^{\mu \nu} (\partial_\mu \hat{\chi})^\dag (1-P) \partial_\nu \hat{\chi} = \hat{z} \cdot \bm{\nabla} \times \bm{a} - i \epsilon^{\mu \nu} (\partial_\mu \hat{\chi})^\dag P \partial_\nu \hat{\chi} = \hat{z} \cdot \bm{\nabla} \times \bm{a}.
\end{equation}
In the last step we used that the second term on the RHS vanishes due to antisymmetry: $\epsilon^{\mu \nu} (\partial_\mu \hat{\chi})^\dag P \partial_\nu \hat{\chi} = -\epsilon^{\mu \nu} a_\mu a_\nu = 0  $.

We now use the boundary conditions \eqref{eq:suppchibcs} to compute $N_w$. We use Stokes theorem to write the flux of $\nabla \times \bm{a}$ as a line integral around the unit cell parallelogram defined by the points $\br, \br+\bm{R}_1, \br + \bm{R}_2, \br + \bm{R}_1 + \bm{R}_2$. We can then group paths that are related by the primitive lattice vectors, $\bm{R}_{1,2}$, and use $a_\mu(\br+\bR) = a_\mu(\br) - \partial_\mu \varphi_\bR(\br)$. We obtain
\begin{equation}
\begin{aligned}
    2\pi N_w & =  \int_{\text{unit cell}} \hat{z} \cdot \bm{\nabla} \times \bm{a} = \int_{\text{uc-boundary}} \bm{a} \cdot d \br\\
    & = \frac{1}{2\pi}\int_{\br}^{\br+\bm{R}_1} (\bm{a}(\br) - \bm{a}(\br + \bm{R}_1))\cdot d \br + \int_{\br}^{\br+\bm{R}_2} (\bm{a}(\br+\bm{R}_2) - \bm{a}(\br))(\br)\cdot d \br \\
    & = -\int_{\br}^{\br+\bm{R}_1} \bm{\nabla} \varphi_{\bR_2}(\br)\cdot d \br + \int_{\br}^{\br+\bm{R}_2} \bm{\nabla}\varphi_{\bR_1}(\br)\cdot d \br \\
    & = \varphi_{\bR_1}(\br + \bR_2) - \varphi_{\bR_1}(\br) + \varphi_{\bR_2}(\br) - \varphi_{\bR_2}(\br + \bR_1) \\
    & = l_B^2 \bR_1 \times \bR_2 = l_B^2 \abs{\bR_1 \times \bR_2} = 2\pi,
\end{aligned}
\end{equation}
implying $N_w = N_\phi = 1$, as claimed in the main text.

\subsection{Parton construction}

We now discuss how we can obtain the main text wavefunctions exactly through a parton description, following Ref. \cite{dongCompositeFermiLiquid2023} Appendix D. We begin by fractionalizing the electron on layer $l$ as $\psi_l(\br) = \phi(\br) \chi_l(\br)$. We regard $\phi$ as a fermionic chargon and $\chi$ as a bosonic ``layeron" (a layer-space analog of a spinon). There is a $U(1)$ gauge redundancy associated with this decomposition: replacing $\phi \to e^{-i \alpha(\br)} \phi$ and $\chi_l(\br) \to e^{i \alpha(\br)} \chi_l(\br)$ leaves the physical electron $\psi_l$ invariant. There is an associated gauge field $a_\mu$, under which $\phi$ has charge $-1$ and $\chi$ has charge $+1$.

We assume a simple ansatz within the expanded parton Hilbert space, in which $\hat{z} \cdot \nabla \times \bm{a}$ has a nonzero average and the partons are unentangled: the chargons are in a many-body state $\Phi$ whereas the layerons are in an independent many-body state $X$. One then obtains the physical electron wavefunction through Gutzwiller projecting to the physical subspace. This amounts to evaluating $\Phi$ and $X$ at the same set of positions $\{ \br_i \}$ \cite{luSymmetryprotectedFractionalChern2012a}:
\begin{equation}
    \Psi_{\{l_i\}}(\{\br_i\}) = \Phi(\{\br_i\}) X_{\{l_i\}}(\{\br_i\}).
    \label{eq:generalparton}
\end{equation}
These are precisely the form of the wavefunctions we discussed in the main text.

We largely focused on the simplest case where the chargons form an integer quantum Hall state, $\Phi = \prod_{i<j} (z_i - z_j) \prod_i e^{-\frac{\abs{\br}^2}{4l_B^2} }$, and the layerons condense into a vortex lattice, $\prod_i \chi_{l_i}(\br_i)$. Other states can be written down from this point of view as well. For example, if we expand to higher powers in the interaction range as $V_d(\bq) = V_0 + V_1 (\bq d)^2 + V_2 (\bq d)^2 + \cdots$\cite{TrugmanKivelson1985,pokrovskySimpleModelFractional1985}, then the linear order chiral zeros only suffice for the zeroth order contact interaction $V_0$. If $V_1, V_2$ are appreciable, then it makes sense to add more zeros through putting $\Phi$ in a fractional quantum Hall state $\prod_{i<j} (z_i - z_j)^3 e^{-\frac{\abs{\br}^2}{4l_B^2} }$. In this case, $\Psi$ pays zero energy to $V_{0,1,2}$ and describes a fractional Chern insulating (FCI) state\cite{TrugmanKivelson1985,pokrovskySimpleModelFractional1985,ledwith2023vortexability}.

This ability to describe the integer and fractional Chern insulator within this same parton framework is directly tied to ideal quantum geometry of the filled band of the integer state\cite{dongCompositeFermiLiquid2023,guerci2025layer}. In general, if a band has ideal quantum geometry then fractional Chern insulating states within the band of interest can be obtained through attaching vortices to the integer state, $\ket{\Psi_0} \to \prod_{i<j} (z_i-z_j)^2 \ket{\Psi_0}$\cite{ledwith2023vortexability}. An ideal $C=1$ band always has wavefunctions of the form $\psi_{\bk l}(\br) = \phi_\bk(\br) \chi_l(\br)$\cite{wang2021exact}. Any many-body wavefunction within such a band therefore has the layeron-condensate form $\Psi = \Phi \prod_i \chi_{\{l_i\}}(\br_i)$, and vortex attachment simply turns the integer quantum Hall state of chargons into a fractional quantum Hall state. We therefore see that, in the $C=1$ case, ideal geometry implies the exactness of the above parton description and relates different states within it through vortex attachment. A similar, though more intricate, situation occurs in $C>1$ ideal bands\cite{dong2023many,wang2023origin} in which there are $C$ species of layerons, each with $1/C$ average winding per unit cell\cite{guerci2025layer}.

\section{Time-dependent Hartree-Fock}
This section discuss the details of time-dependent Hartree-Fock (TDHF), and present additional data that is omitted in the main text. Sec.\ref{sec:gf_tdvp_v2} and Sec. \ref{sec:coupling_gf} give a pedagogical review on the general formalism of TDHF, starting from the time-dependent variational principle (TDVP). The rest of this section focuses specifically on TDHF in RMG, presents additional data on the rotation eigenvalue and coupling strength to the external perturbation for various excitation modes in RMG.

\subsection{General Formalism}\label{sec:gf_tdvp_v2}
TDHF is an application of TDVP\cite{khalaf2020softmodesmagicangle,dong2025phonons,BultinckMagLabWinterSchool2023}\footnote{In addition, the author acknowledges useful private discussion with N. Bultinck and Y. Kwan on this topic}. TDHF is used to determine the excited state within some given manifold of states. The goal of this section is to derive this relation. We set $\hbar=1$. Consider the following Lagrangian $L$ and action $S$, where $L$ is a functional of the unnormalized state $\psi(x,t)$:
\begin{equation}
S=\int dt L(\psi(x,t),\psi(x,t)^*),
\end{equation}

\begin{equation}
L=\frac{i}{2}\frac{\braket{\psi}{\dot{\psi}}-\braket{\dot{\psi}}{\psi}}{\braket{\psi}{\psi}}-\frac{\langle \psi |H|\psi \rangle}{\braket{\psi}{\psi}}.
\end{equation}
Here, $H$ is the Hamiltonian of the system under consideration, and we use $x$ to refer to general spatial coordinates. For simplicity, we will assume $H$ to be time-independent. We first notice that if we attach a position-independent, time-dependent scalar function $f(t)$ to $\ket{\psi}$, the corresponding change in the Lagrangian is a total time derivative:
\begin{equation}
\delta L=\frac{i}{2} \frac{d(\ln(f(t))-\ln(f^*(t)))}{dt}.
\end{equation}
Consequently, the equation of motion (EOM) derived from such action can only determine the state up to such scalar function, which reflects the projective nature of TDVP. The EOM obtained by requiring $\delta S=0$ with respect to arbitrary small changes in $\ket{\psi}$ is:
\begin{equation}
(i\frac{d}{dt}-H)\psi(x,t)=(i\frac{\braket{\psi}{\dot{\psi}}}{\braket{\psi}{\psi}}-\frac{\langle \psi|H|\psi\rangle}{\braket{\psi}{\psi}})\psi(x,t).
\end{equation}
We can define the scalar function $f(t)$:
\begin{equation}
f(t)=\exp (i\int^{t} dt' \frac{\bra{\psi} (i \frac{d}{dt'}-H)\ket{\psi}}{\langle \psi|\psi\rangle}),
\end{equation}
then $f(t)\psi(x,t)$ will satisfy the standard Schr\"{o}dinger equation:
\begin{equation}
i\frac{d}{dt}(f(t)\psi(x,t))=H(f(t)\psi(x,t)).
\end{equation}
The formulation above is for a state in an unrestricted Hilbert space. TDHF is applying TDVP in a restricted Hilbert space. Let us assume that the time-dependence of the state is parametrized by a set of complex parameters $\{z_i(t)\}$, $\psi(x,\{z_i(t)\})$. Then the Lagrangian written in terms of these parameters is:
\begin{equation}
L(\{z_i\},\{\dot{z}_i\},\{z^*_i\},\{\dot{z^*_i}\})=\frac{i}{2}(\sum_i \dot{z}_i \partial_{z_i}-\dot{z^*_i}\partial_{z_i^*})\ln \mathcal{N}(\{z_i\},\{z^*_i\})-\mathcal{H}(\{z_i\},\{z^*_i\}),
\end{equation}

\begin{equation}
\mathcal{N}(\{z_i\},\{z^*_i\})=\braket{\psi(\{z_i^*\})}{\psi(\{z_i\})},
\end{equation}

\begin{equation}
\mathcal{H}(\{z_i\},\{z^*_i\})=\frac{\langle \psi(\{z_i^*\})|H| \psi(\{z_i\}) \rangle}{\braket{\psi(\{z_i^*\})}{\psi(\{z_i\})}}.
\end{equation}

The EOMs obtained by varying $L$ with respect to small changes $\delta z$ and $\delta z^*$ are
\begin{equation}
i\sum_i C_{ij}\dot{z}^*_i=-\frac{\partial\mathcal{H}}{\partial z_j},\quad i\sum_{j} C_{ij}\dot{z}_j=\frac{\partial\mathcal{H}}{\partial z^*_i},
\end{equation}
where we have defined
\begin{equation}
C_{ij}=\frac{\partial^2\ln\mathcal{N}}{\partial z^{*}_i \partial z_j}.
\end{equation}
We will now focus on a particular parametrization of states. We denote the Hartree-Fock ground state (Slater determinant) by $\ket{\psi_0}$. According to Thouless theorem \cite{thouless1960stability}, any Slater determinant that is not orthogonal to $\ket{\psi_0}$ can be parametrized as
\begin{equation}
\ket{z}=\ket{\psi(\{z_{im}\})}=\exp(\sum_{i,m}z_{im}c^{\dagger}_mc_i)\ket{\psi_0}.
\end{equation}
Here $z_{im}$ is a complex number, and we have defined a shorthand $\ket{z}$ for the state. We will use $h,i,j,k,l$ for the occupied orbitals (hole) and $m,n,o,p$ for the unoccupied orbitals (particle). Therefore,
\begin{equation}
\ket{\psi_0}=\prod_i c^{\dagger}_{i}\ket{0}.
\end{equation}
Here, $\ket{0}$ is the vacuum. By making use of the Baker--Campbell--Hausdorff formula, we have
\begin{equation}\label{eq:thoulessslater}
\ket{z}=\prod_i (c^{\dagger}_i+\sum_m z_{im}c^{\dagger}_m)\ket{0}.
\end{equation}
This shows that $\ket{z}$ is indeed a Slater determinant. Using the expression above, we can easily obtain the overlap between two Slater determinants parametrized in this manner:
\begin{equation}
\braket{z'}{z}=\det(I+ZZ'^{\dagger}), \quad Z_{i,m}\equiv z_{im},
\end{equation}
where we have defined the matrix $Z$ in the equation above. We further notice that for a square matrix $A$, $\ln (\det(A))=\textrm{Tr}(\ln(A))$. Therefore, we can calculate the derivative in the EOM as
\begin{equation}
\frac{\partial \ln(\braket{z}{z})}{\partial z_{im}}=(Z^{\dagger}(I+ZZ^{\dagger})^{-1})_{mi}.
\end{equation}
This further allows us to calculate the coefficient matrix $C$ appearing in EOMs:
\begin{equation}
C_{jn,im}=\frac{\partial^2 \ln(\braket{z}{z})}{\partial z^{*}_{jn} \partial z_{im}}=\sum_k \partial_{z^*_{jn}}(Z^{\dagger}_{mk}(I+ZZ^{\dagger})^{-1}_{ki})=(I+Z^{\dagger}Z)^{-1}_{mn}(I+ZZ^{\dagger})^{-1}_{ji}.
\end{equation}
In deriving the equation above, we make use the following properties: If a square matrix depends on some parameter $t$, then the derivative of its inverse is
\begin{equation}
\frac{dA^{-1}(t)}{dt}=-A^{-1}\frac{dA}{dt}A^{-1}.
\end{equation}
To summarize, the EOMs written in the Thousless parametrization $\{z_{im}\}$ are
\begin{equation}
i\sum_{im}C_{im,jn}\dot{z}^*_{im}=-\frac{\partial \mathcal{H}}{\partial z_{jn}},\quad i\sum_{jn}C_{im,jn}\dot{z}_{jn}=\frac{\partial \mathcal{H}}{\partial z^*_{im}}.
\end{equation}
We now proceed to calculate the RHS of the EOMs above. The denominator of $\mathcal{H}$ is
\begin{equation}
\braket{z}{z}=1+\sum_{im}|z_{im}|^2+O(z^4).
\end{equation}
The numerator is more tedious to calculate. We first notice that the Hamiltonian written in the Hartree-Fock orbital basis is
\begin{equation}
H=\sum_{ab}T_{ab}c^{\dagger}_a c_b+\frac{1}{2}\sum_{abcd}V_{abcd}c^{\dagger}_a c^{\dagger}_b c_d c_c.
\end{equation}
Here, $a,b,c,d$ loop through all the orbitals. $T$ is the single-body term and  $V$ is the two-body term. Self-consistency of Hartree-Fock equation (explained in Sec.\ref{sec:HFsingle}) implies that
\begin{equation}
T_{ab}+\sum_{i\in \textrm{occupied}} V_{aibi}-V_{aiib}=E_{a}\delta_{ab}, \quad \forall a,b,
\end{equation}
where $E_a$ is the Hartree-Fock energy obtained in diagonalizing the Hartree-Fock Hamiltonian \footnote{Note that the sum of $E_a$ of the occupied orbital is not the energy of the system. The interaction part is off by a factor of 2. }. Using the expression above and Eq. \ref{eq:thoulessslater}, we can  calculate the numerator of $\mathcal{H}$ to quadratic order:
\begin{equation}
\begin{split}
\langle z|H|z\rangle&=\langle \psi_0|H|\psi_0 \rangle(1+\sum_{im}|z_{im}|^2)+\sum_{im}|z_{im}|^2 (E_{m}-E_{i})+\sum_{ijmn}z_{im}z^*_{jn}(V_{inmj}-V_{nimj})\\
&\quad+\sum_{ijmn}\frac{z^{*}_{im}z^{*}_{jn}}{2}(V_{mnij}-V_{nmij})+\sum_{ijmn}\frac{z_{im}z_{jn}}{2}(V_{ijmn}-V_{jimn})+O(z^3)\\
&=\langle \psi_0|H|\psi_0 \rangle(1+\sum_{im}|z_{im}|^2)+\frac{1}{2}(z^* z)
\begin{pmatrix}
A&B\\
B^*&A^*
\end{pmatrix}
\begin{pmatrix}
z\\
z^*
\end{pmatrix}+O(z^3),
\end{split}
\end{equation}
where we have defined the $A$ matrix and $B$ matrix,
\begin{equation}
A_{im,jn}=\delta_{ij}\delta_{mn}(E_m-E_i)+V_{jmni}-V_{mjni},
\end{equation}

\begin{equation}
B_{im,jn}=V_{nmji}-V_{mnji}.
\end{equation}
We note that $A$ is Hermitian and B is symmetric. To proceed further, we must now linearize the EOMs. To leading order, $C$ is just the identity matrix \footnote{In later section where we carry out $\chi-\textrm{TDVP}$, this is not the case}. Thus, the linearized EOMs are
\begin{equation}
i\begin{pmatrix}
\dot{z}\\
\dot{z}^*
\end{pmatrix}
=\begin{pmatrix}
A & B\\
-B^* & -A^*
\end{pmatrix}
\begin{pmatrix}
z\\
z^*
\end{pmatrix}\equiv
K\begin{pmatrix}
z\\
z^*
\end{pmatrix}.
\end{equation}
This is a set of coupled first order differential equations, and the general solution is of the form
\begin{equation}\label{eq:generalsolution}
z_{im}=X_{im}\exp(-i\omega t)+Y^*_{im}\exp(i\omega t).
\end{equation}
In terms of $X$ and $Y$, the linearized EOMs are

\begin{equation}\label{eq:TDVP_XYequation}
\omega\begin{pmatrix}
X\\
Y
\end{pmatrix}
=\begin{pmatrix}
A & B\\
-B^* & -A^*
\end{pmatrix}
\begin{pmatrix}
X\\
Y
\end{pmatrix}= K \begin{pmatrix}
X\\
Y
\end{pmatrix}.
\end{equation}
This is an eigenvalue problem for the matrix $K$. We note that since $K$ is in general non-Hermitian, an imaginary eigenvalue $\omega$ may appear. If that occurs, it implies the invalidity of the linearized EOMs, and typically an instability of the system. We remark that in the TDVP language, we can see that $\omega$ is the excitation energy since TDHF is the Schr\"{o}dinger equation projected to a manifold, and $\omega$ is the frequency of the evolution of the states in that manifold.

If $v^{\mu}=(X^{\mu},Y^{\mu})$ is an eigenvector of $K$ with eigenvalue $\omega^{\mu}$, it is straightforward to show that $v^{\tilde{\mu}}=(Y^{\mu*},X^{\mu*})$ is an eigenvector with eigenvalue $-\omega^{\mu}$. This is the reason why the general solution of $\{z_{im}\}$ must be of the form Eq. \ref{eq:generalsolution}. The normalization of the eigenvector is chosen to be \footnote{In the case of a zero-energy Goldstone mode, it is guaranteed that $X^{\dagger}X-Y^{\dagger}Y=0$ for that mode \cite{ring2004nuclear} .}
\begin{equation}
X^{\dagger}X-Y^{\dagger}Y=\text{sgn}(\omega).
\end{equation}
This convention is chosen because if we have two modes with different energy $\omega^{\mu}$ and $\omega^{\nu}$,
\begin{equation}\label{eq:orthoTDHF}
v^{\mu\dagger}
\begin{pmatrix}
A&B\\
B^{*}&A^{*}
\end{pmatrix}v^{\nu}=\omega^{\nu}(X^{\mu\dagger}X^{\nu}-Y^{\mu\dagger}Y^{\nu})=\omega^{\mu} (X^{\mu\dagger}X^{\nu}-Y^{\mu\dagger}Y^{\nu})\geq 0 \to X^{\mu\dagger}X^{\nu}-Y^{\nu\dagger}Y^{\nu}=0.
\end{equation}
The inequality in the derivation above arises because for a stable Hartree-Fock ground state, the matrix between $v^{\mu\dagger}$ and $v^{\nu}$ is semi-positive definite.

We now proceed to connect the solution of TDHF to the perturbation to density matrix. The density matrix $\rho_{ab}(\{z_{im}\})$ is defined to be
\begin{equation}
\rho_{ab}(\{z_{im}\})\equiv \langle z|c^{\dagger}_b c_a|z \rangle.
\end{equation}
We expand $\ket{z}$ to linear order in $z$,
\begin{equation}\label{eq:LOexpand}
\ket{z}=\ket{\psi_0}+\sum_{im}z_{im}c^{\dagger}_m c_i\ket{\psi_0}+O(z^2).
\end{equation}
Thus, it is straightforward to calculate the particle-particle, hole-hole, particle-hole part of the density matrix to the same order:
\begin{equation}
\rho_{ij}=\delta_{ij}+O(z^2)\quad  \rho_{mn}=O(z^2) \quad \rho_{im}=z_{im}^*+O(z^2)\quad \rho_{mi}=z_{im}+O(z^2)
\end{equation}

\subsection{Coupling to the external perturbation}\label{sec:coupling_gf}
In this section, we study the effect of introducing a time-dependent perturbation to the formalism derived above. Let us introduce a time-dependent perturbation to the original Hamiltonian:
\begin{equation}
\delta H(t)=\delta u(t) \hat{F}.
\end{equation}
Here, $F$ is some single-body operator representing the perturbing `` field''. For example, if we want to describe the effect of a time-dependent displacement field, $F$ will be the layer-z operator. For now, we will keep it general. In its presence, the linearized EOMs are modified to be \footnote{$\delta u$ is assumed to be small and of the same order as z.}
\begin{equation}
i\begin{pmatrix}
\dot{z}\\
\dot{z}^*
\end{pmatrix}
=K
\begin{pmatrix}
z\\
z^*
\end{pmatrix}+ \delta u(t)
\begin{pmatrix}
f\\
-f^{*}
\end{pmatrix},
\end{equation}
where we have defined
\begin{equation}
f^{*}_{im}\equiv \frac{\partial \mathcal{F}}{\partial z_{im}}|_{z=0},\quad f_{im}\equiv \frac{\partial \mathcal{F}}{\partial z^*_{im}}|_{z=0},\quad \mathcal{F}\equiv \frac{\langle z |\hat{F}| z \rangle}{\braket{z}{z}}.
\end{equation}
This is a set of inhomogeneous ODEs, with the external field providing the source term. To proceed, we expand $(z,z^*)^T$ in the eigenbasis of $K$ matrix :
\begin{equation}
\begin{pmatrix}
z\\
z^*
\end{pmatrix}=\sum_{\nu}\eta^{\nu}(t)
\begin{pmatrix}
X^{\nu}\\
Y^{\nu}
\end{pmatrix}=\sum_{\nu} \eta^{\nu}(t) v^{\nu},
\end{equation}
where the $\nu$ labels different eigenvectors. By using the condition that $z^{*}$ is complex conjugate of $z$, we can show that the coefficients of the two modes of opposite energy are complex conjugate of each other ($\omega^{\tilde{\nu}}=-\omega^{\nu}$):
\begin{equation}
\eta^{\nu}=\eta^{\tilde{\nu}*}.
\end{equation}
After expanding $z$ and $z^{*}$ in the eigenbasis, we can use Eq.\ref{eq:orthoTDHF} to show that $\eta$ obeys the following equation:
\begin{equation}
i\dot{\eta^{\nu}}(t)=\omega^{\nu}\eta^{\nu}(t)+\delta u(t)[X^{\nu\dagger}f+Y^{\nu\dagger}f^*].
\end{equation}
On the RHS of the equation above, $X^{\nu\dagger}f+Y^{\nu\dagger}f^*$ describes the coupling strength between the excited mode and the external perturbation. We will later plot this quantity for different modes we obtained in the RMG system. In its absence, the solution is $\eta^{\nu}=\exp(-i\omega^{\nu}t)$, consistent with Eq.\ref{eq:generalsolution}.

Finally, we proceed to give an explicit formula for the matrix element $f$. The denominator of $\mathcal{F}$ is not important since the leading non-trivial term starts at $O(z^2)$. For the numerator, we can expand $\hat{F}$ in the Hartree-Fock orbital basis and use Eq.\ref{eq:LOexpand}, which leads to:
\begin{equation}
\hat{F}=\sum_{ab}F_{ab}c^{\dagger}_a c_b\to f^{*}_{im}=F_{im},\quad f_{im}=F_{mi}.
\end{equation}

\subsection{TDHF on RMG}
We now apply the formalism derived above to obtain the excited modes in RMG. According to the discussion above, the most general Thouless parametrization for the excited mode is $z_{h,\bm{k}_1;p,\bm{k}_2}$. We remind the readers that h refers to hole (occupied Hartree-Fock orbitals) and p refers to particle (unoccupied Hartree-Fock orbitals).
However, due to momentum conservation of the interaction matrix element, the matrix equation Eq.\ref{eq:TDVP_XYequation}
is decoupled into different blocks labeled by momentum transfer $\bm{q}$. For a fixed $\bm{q}$, the set of coefficients $\{X_{p,\bm{k}+\bm{q},h,\bm{k}}\}_{\bm{k},p,h}$ is only coupled with the set of coefficients $\{Y_{p,\bm{k}-\bm{q},h,\bm{k}}\}_{\bm{k},p,h}$. Concretely, the matrix equation for that block is

\begin{equation}
\begin{pmatrix}
A^{\bm{q}}&  B^{\bm{q}}\\
-(B^{\bm{q}})^{\dagger}& - (A^{-\bm{q}})^*
\end{pmatrix}
\begin{pmatrix}
X^{a,\bm{q}}\\Y^{a,\bm{q}}
\end{pmatrix}
=\omega^{a,\bm{q}} \begin{pmatrix}
X^{a,\bm{q}}\\Y^{a,\bm{q}}
\end{pmatrix},
\end{equation}

\begin{equation}
\begin{split}\label{eq:TDHF}
A^{\bm{q}}_{(p_1,h_1),\bm{k}_1;(p_2,h_2),\bm{k}_2}=&(\epsilon_{p_1,\bm{k}_1+\bm{q}}-\epsilon_{h_1,\bm{k}_1})\delta_{\bm{k}_1,\bm{k}_2}\delta_{p_1,p_2}\delta_{h_1,h_2}\\
&+V_{p_1,\bm{k}_1+\bm{q};h_2,\bm{k}_2;h_1,\bm{k}_1;p_2,\bm{k}_2+\bm{q}}-V_{p_1,\bm{k}_1+\bm{q};\bm{k}_2,h_2;\bm{k}_2+\bm{q},p_2;\bm{k}_1,h_1},
\end{split}
\end{equation}

\begin{equation}
B^{\bm{q}}_{(p_1,h_1),\bm{k}_1;(p_2,h_2),\bm{k}_2}=V_{p_1,\bm{k}_1+\bm{q};p_2,\bm{k}_2-\bm{q};h_1,\bm{k}_1;h_2,\bm{k}_2}-V_{p_1,\bm{k}_1+\bm{q};p_2,\bm{k}_2-\bm{q};h_2,\bm{k}_2;h_1,\bm{k}_1}.
\end{equation}
In the equation above, we have adopted the notation that $\bm{q}$ labels different blocks corresponding to different momentum transfer, and $a$ labels different modes within that sector. In addition, we have changed the notation, in which the momentum transfer $\bm{q}$ becomes a `` flavor'' index.  $X^{a,\bm{q}}$ and $Y^{a,\bm{q}}$ are vectors with components
\begin{equation}
X^{a,\bm{q}}_{(p,h),\bm{k}}=X_{h,\bm{k},p,\bm{k}+\bm{q}}\quad Y^{a,\bm{q}}_{(p,h),\bm{k}}=Y_{h,\bm{k},p,\bm{k}-\bm{q}}.
\end{equation}
This is the TDHF equation commonly seen in the literature, e.g. in Ref.\cite{kwan2025moire, PhysRevB.104.035119}, though this set of equations is derived using a different approach there.

In the main text, we plot the time evolution of the charge density of the excited mode. We use $P_0$ to denote the density matrix (projector) of the Hartree-Fock ground state . According to the discussion in Sec.\ref{sec:gf_tdvp_v2}, the excitation corresponds to a time-dependent perturbation $P_1$ on the ground state density matrix

\begin{equation}
P(t)=P_0+\eta P_1 e^{-i\omega t}+\eta P_1^{\dagger} e^{i\omega t},
\end{equation}
where $\eta$ is an arbitrary small number since the EOMs is linear. Since the ground state is a Slater determinant, the matrix element of $P_0$ is
\begin{equation}
P_{0 n_1,\bm{k}_1;n_2,\bm{k}_2}=\langle d^{\dagger}_{n_2,\bm{k}_2} d_{n_1,\bm{k}_1} \rangle
\begin{cases}
1,\quad \textrm{if}\quad \bm{k}_1=\bm{k}_2,\quad n_1=n_2 \in \textrm{occupied}\\
0,\quad \textrm{otherwise},
\end{cases}
\end{equation}
where the expectation value is evaluated with respect to the Hartree-Fock ground state. According to the discussion in Sec.\ref{sec:gf_tdvp_v2}, the matrix element of the perturbation $P_1$ is related to $X$ and $Y$ by \cite{ring2004nuclear}
\begin{equation}
P_{1p,\bm{k}+\bm{q};h,\bm{k}}=X^{a,\bm{q}}_{(p,h),\bm{k}},\quad P_{1h,\bm{k};p,\bm{k}-\bm{q}}=Y^{a,\bm{q}}_{(p,h),\bm{k}}.
\end{equation}

The density matrix $P_0$ and $P_1$ are currently written in the Hartree-Fock band basis. They can be transformed into the microscopic layer/sublattice basis by

\begin{equation}
\begin{split}
P_{\bm{k}_1,\bm{g}_1,\sigma_1,l_1;\bm{k}_2,\bm{g}_2,\sigma_2,l_2}&\equiv \langle c^{\dagger}_{\bm{K}+\bm{k}_2+\bm{g}_2,\sigma_2,l_2}c_{\bm{K}+\bm{k}_1+\bm{g}_1,\sigma_1,l_1}\rangle\\
&=s_{\bm{k}_1+\bm{g}_1,\sigma_1,l_1}s^*_{\bm{k}_2+\bm{g}_2,\sigma_2,l_2}\sum_{n_1,n_2}\beta_{n_1,\bm{k}_1,\bm{g}_1} \beta^{*}_{n_2,\bm{k}_2,\bm{g}_2}P_{n_1,\bm{k}_1;n_2,\bm{k}_2},
\end{split}
\end{equation}
where $\beta$ is defined by Eq.\ref{eq:hfbasisdefin}. In the main text, we plot the time-dependent total charge density associated with the excited mode. The charge density on layer $l$ and sublattice $\sigma$, $\rho_{\sigma,l}$, is related to the projector by
\begin{equation}
\begin{split}
\rho_{\sigma,l}(\bm{r})&=\sum_{\bm{k}_1,\bm{k}_2,\bm{g}_1,\bm{g}_2}\langle c^{\dagger}_{\bm{K}+\bm{k}_1+\bm{g}_1,\sigma,l} c_{\bm{K}+\bm{k}_2+\bm{g}_2,\sigma,l}\rangle e^{i(\bm{k}_2+\bm{g}_2-\bm{k}_1-\bm{g}_1)\cdot \bm{r}}\\
&=\sum_{\bm{k}_1,\bm{k}_2,\bm{g}_1,\bm{g}_2}P_{\bm{k}_2,\bm{g}_2,\sigma,l;\bm{k}_1,\bm{g}_1,\sigma,l} e^{i(\bm{k}_2+\bm{g}_2-\bm{k}_1-\bm{g}_1)\cdot \bm{r}}.
\end{split}
\end{equation}

\subsection{Rotation Eigenvalue}
In the main text, we categorize different modes at $\bm{q}=\bm{0}$ by their angular momentum. We explain how that quantity is defined in this section. The presence of hBN moir\'e potential breaks the continuous rotation symmetry down to 3-fold rotation symmetry, which means that the angular momentum in this problem is only defined modulo 3. We note that for the case where there is no hBN, the angular momentum is defined modulo 6 since we assume a triangular lattice in Hartree-Fock, but we can still characterize it using 3-fold rotation. One can also calculate the $C_6$ eigenvalue in that case, and the procedure will be essentially the same as what we do below.

We use Eq. \ref{eq:LOexpand} to carry out linear order expansion of the state, then the variation on top of the original Hartree-Fock ground state is
\begin{equation}
 \ket{\delta \psi}=\sum_{a}\eta^{a} (t) \sum_{p,h,\bm{k}}X^{a,\bm{0}}_{(p,h),\bm{k}}d^{\dagger}_{p,\bm{k}}  d_{h,\bm{k}}\ket{\psi_0}.
\end{equation}
Here, $\ket{\psi_0}=\prod_{h,\bm{k}}d^{\dagger}_{h,\bm{k}}\ket{0}$ is the Hartree-Fock ground state, and we have only included the $\bm{q}=0$ mode in our expansion. We also ignore the $Y$ component of expansion since a particle operator acting on $\ket{\psi_0}$ annihilates it. We now make only one of the $\eta^a$ non-zero.
\begin{equation}
 \ket{\delta\psi}\propto  \sum_{p,h,\bm{k}}X^{a,\bm{0}}_{(p,h),\bm{k}}d^{\dagger}_{p,\bm{k}}  d_{h,\bm{k}}\ket{\psi_0}.
\end{equation}
Though the physical constraint requires the expansion coefficient for the pair with opposite energy to be complex conjugate of each other, it is legitimate to have only one of the $\eta^a$ be non-zero now since we are interested in assigning a rotation eigenvalue to that particular mode. Thus, $\ket{\delta \psi}$ is a linear combination of the possibility of replacing the originally occupied orbital $d^{\dagger}_{1,\bm{k}}$ by $\sum_{p}X^{a,\bm{0}}_{(p,1),\bm{k}}d^{\dagger}_{p,\bm{k}}$. Under $C_3$ rotation, the electron operators transform as
\begin{equation}
C_3 c^{\dagger}_{\bm{k},\bm{g}}C_3^{-1}=c^{\dagger}_{C_3\bm{k},C_3\bm{g}}.
\end{equation}
Using this relation, we compute the $C_3$ rotation eigenvalue of the state $d^{\dagger}_{1,\bm{0}}\ket{0}$ and  $\sum_{p}X^{a,\bm{0}}_{(p,1),\bm{0}}d^{\dagger}_{p,\bm{0}}\ket{0}$. The difference between these two eigenvalues is the angular momentum of the corresponding mode.

\subsection{Breathing Mode}
\begin{figure}
    \centering
    \includegraphics[width=0.6\linewidth]{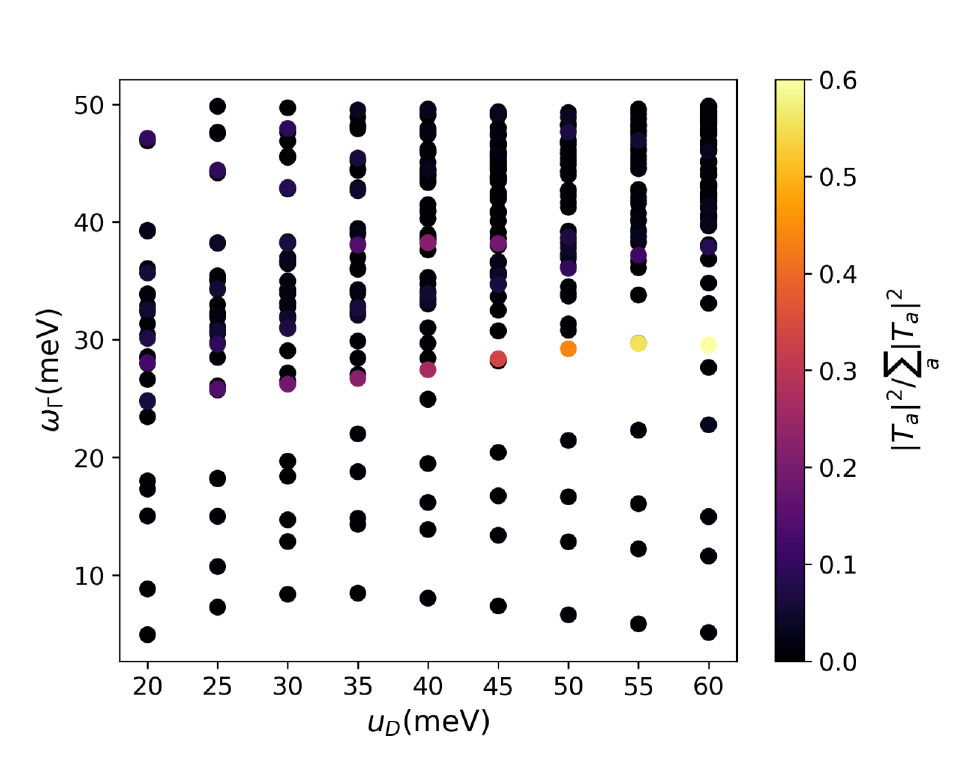}
    \caption{TDHF spectrum of $\bm{q}=\bm{0}$ mode for different displacement field $u_D$. Other parameters are $\theta=0.77^{\circ}$, $\lambda=1$, $V_{\textrm{bg}}=8\textrm{meV}$, $\psi_{\textrm{bg}}=\frac{2\pi}{3}$. The interaction is take to be gate-screened Coulomb interaction, $V(\bm{q})=\frac{e^2\textrm{tanh}(|\bm{q}|D)}{2\epsilon_r\epsilon_0 |\bm{q}| }, D=25\textrm{nm}, \epsilon_r=5$. Color of the points is proportional to the magnitude of transition matrix elements $|T_a|^2$.}
    \label{Sfig_breathingmode}
\end{figure}

On top of the fixed displacement field $u_D$, we can add another oscillating time-dependent displacement field as perturbation.  The transition matrix element $T_{a}$ measures the transition probability between the ground state and the excited state $(a,\bm{q}=\bm{0})$. The perturbation $\hat{F}$ is described by the layer-z operator:
\begin{equation}
\begin{split}
\hat{F}=\sum_{\bm{k},\bm{g},l_1,l_2,\sigma} (\tau_z)_{l_1l_2} c^{\dagger}_{\bm{K}+\bm{k}+\bm{g},\sigma,l_1}c_{\bm{K}+\bm{k}+\bm{g},\sigma,l_2},\\
(\tau_z)_{ll}=l-\frac{N_l-1}{2},\quad l=0...N_l-1.
\end{split}
\end{equation}
For convenience, we can expand the operator in the Hartree-Fock basis:
\begin{equation}
\hat{F}=\sum_{\bm{k}\in \textrm{mBZ},p,h} F_{p,\bm{k};h,\bm{k}}d^{\dagger}_{p,\bm{k}}d_{h,\bm{k}}+F_{h,\bm{k};p,\bm{k}}d^{\dagger}_{h,\bm{k}}d_{p,\bm{k}}+...
\end{equation}
According to the discussion in Sec.\ref{sec:coupling_gf}, the transition matrix element $T_a$ is
\begin{equation}
T_{a}=\sum_{\bm{k},p,h}F_{h,\bm{k};p,\bm{k}}X^{a,\bm{0}}_{(p,h),\bm{k}}+F_{p,\bm{k};h,\bm{k}}Y^{a,\bm{0}}_{(p,h),\bm{k}}.
\end{equation}
In Fig.\ref{Sfig_breathingmode}, we plot the magnitude of the transition matrix elements for different excitation modes. It is obvious that in the low energy spectrum, there is one mode (e.g. the fifth mode at $u_D=50$meV) that couples most strongly to the displacement field.  In addition, this mode carries an angular momentum $0$ (mod 3). This mode is called breathing mode because if we plot out the the corresponding charge density perturbation, it corresponds to the skyrmion core expanding and retracting periodically as a function of time, hence ``breathing'' (cf. Fig.\ref{Sfig_timedependence}). More intuitively, the reason why a breathing mode couples strongly to a displacement field is that the breathing of the layer skyrmion core induces a charge oscillation between layers. When the oscillating field has the matching frequency, the coupling is strong.

\subsection{Extended Data}
\begin{figure}
    \centering
    \includegraphics[width=1.0\linewidth]{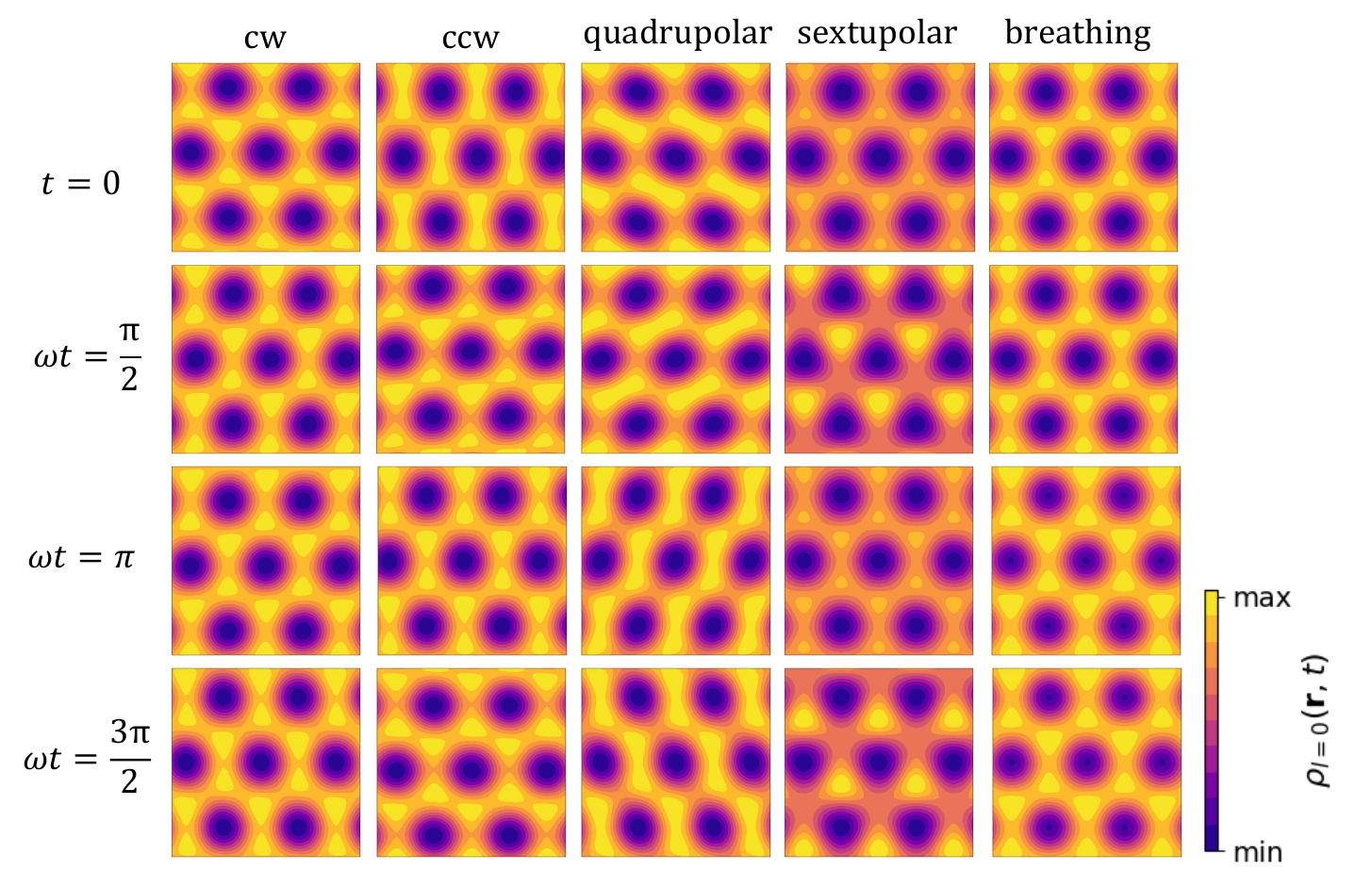}
    \caption{The top-layer $l=0$ charge density for different excitation modes at different instance of time. The calculation is done at $u_D=50\textrm{meV}$, $\theta=0.77^{\circ}$, $\lambda=1$, $V_{\text{bg}}=8\textrm{meV}$, $\psi_{\text{bg}}=\frac{2\pi}{3}$. We choose $\mathcal{E}(\bm{k})$ to be the real dispersion of the first conduction band $h_{\textrm{R5G}}$ at $u_D=50\textrm{meV}$. The interaction is taken to be gate-screened Coulomb interaction $V(\bm{q})=\frac{e^2\textrm{tanh}(|\bm{q}|D)}{2\epsilon_r\epsilon_0 |\bm{q}| },D=25\textrm{nm}, \epsilon_r=5$. For clockwise and counter-clockwise  gyrotropic mode, we use $\eta=2.5$. For other modes, we use $\eta=1$. Videos of the time-dependence these mode are also available in the supplementary material of this paper.}
    \label{Sfig_timedependence}
\end{figure}

In Fig.\ref{Sfig_timedependence}, we show the time dependence of different excitation modes  at $\theta=0.77^{\circ}$ and $u_D=50\textrm{meV}$, i.e. those modes  highlighted in the main text Fig. 6. The Hartree-Fock and time-dependent Hartree-Fock is done with $\lambda=1$ and gate screened Coulomb interaction.  For Fig.\ref{Sfig_timedependence}, we plot the charge density using an exaggerated $\eta=1$ and $\eta=2.5$ to show the time-dependence of the modes. We have also included in the supplementary material the videos of the time-varying $l=0$ density profile. The motion is more obvious in those videos.

\begin{figure}
    \centering
    \includegraphics[width=1.0\linewidth]{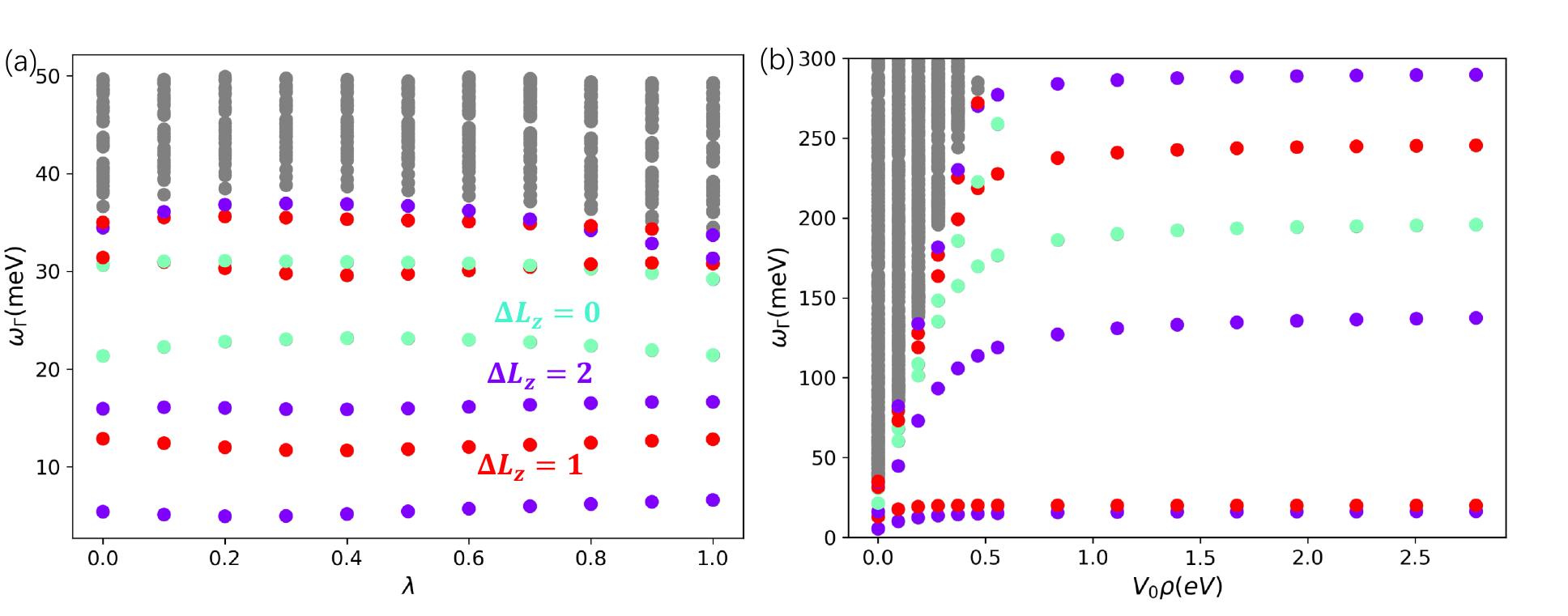}
    \caption{TDHF spectrum of $\bm{q}=\bm{0}$ mode at $\theta=0.77^{\circ}$, $u_D=50\textrm{meV}$, $V_{\text{va}}=8\textrm{meV}$, $\psi_{\text{bg}}=\frac{2\pi}{3}$ as a function \textbf{(a)}  interpolation ratio $\lambda$ ($V_0=0$). \textbf{(b)} contact interaction strength $V_0$ ($\lambda=0.0$). The color encodes the angular momentum of the low-lying excitation modes. Notice that due to the presence of hBN moir\'e potential, angular momentum is only defined mod 3.  We choose $\mathcal{E}(\bm{k})$ to be the real dispersion of the first conduction band $h_{\textrm{R5G}}$ at $u_D=50\textrm{meV}$}
    \label{Sfig_TDHF_theta0.77}
\end{figure}

In Fig.\ref{Sfig_TDHF_theta0.77}, we show the $\bm{q}=\bm{0}$ TDHF spectrum at $\theta=0.77^{\circ}$, both as a function of $\lambda$ at $V_0=0.0$, and as a function of $V_0$ at $\lambda=0$.  We set $\mathcal{E}(\bm{k})$ to be the dispersion of the first conduction band of $h_{\textrm{RMG}}$ with $u_D=50.0$meV. The interaction is taken to be

\begin{equation}
V(\bm{q})=V_0+\frac{e^2\textrm{tanh}(|\bm{q}|D)}{2\epsilon_r\epsilon_0 |\bm{q}| }\quad D=25\textrm{nm}\quad \epsilon_r=5.
\end{equation}

We use different colors to encode the angular momentum carried by different excitation modes. From Fig.\ref{Sfig_TDHF_theta0.77} (a), we observe that the third and the fourth excitation mode in the realistic scenario ($\lambda=1$) is smoothly connected to that of $\lambda=0$. From Fig.\ref{Sfig_TDHF_theta0.77} (b), we further observe that the third and fourth excitation mode at $\lambda=0$, $V_0=0$ is adiabatically connected to the corresponding modes in the ideal limit $\lambda=0$,$V_0\to \infty$. These combined shows that the excitation modes in the real RMG/hBN can be adiabatically traced back to the ideal limit.

\section{$\chi$-TDVP for Skyrmion Shape Modes}

In the main text we reported on a TDVP calculation corresponding to the time-dependent ansatz
\begin{equation}
\chi_0^{\text{exc}}(\br,t) = w_0^{(\bm{q}_0)}(\br) + \sum_{n \bq} \eta_{n\bm{q}}(t) w_n^{(\bm{q}_0+\bm{q})}(\br).
\label{eq:tdansatz_supp}
\end{equation}
In this supplementary section we derive the TDVP equations of motion corresponding to this variational space, linearize them and derive a generalized eigenvalue problem, and discuss how to implement the eigenvalue problem using the momentum-space computational basis.

TDVP generalizes TDHF in that it corresponds to Hilbert space dynamics restricted to a variational manifold\cite{kramerGeometryTimeDependentVariational1981}. While TDHF corresponds to TDVP applied to the full set of Slater determinants, here we apply it to the much more restricted ansatz \eqref{eq:tdansatz_supp} in order to confirm that the tower of states can indeed be identified as skyrmion shape modes of the conjectured form.

The equations of motion follow from varying the action
\begin{equation}
    S = \int \bra{\Psi(\{\eta_{n\bq}\})} (i \partial_t - H) \ket{\Psi(\{ \eta_{n \bq} \})} dt ,
    \label{eq:tdvp_action}
\end{equation}
where $\ket{\Psi(\{ \eta_{n \bq} \})}$ is the normalized many-body Slater determinant corresponding to \eqref{eq:tdansatz_supp}.
The action \eqref{eq:tdvp_action} reproduces the full many-body  Schr\"{o}dinger equation if one allows for fully general variations of $\bra{\Psi}$. Here we only allow variations of $\overline{\eta_{n\bq}}$, which will project the time evolution to the variational manifold \eqref{eq:tdansatz_supp}. It is convenient to parameterize these variations as
\begin{equation}
    \ket{\Psi(\{ \eta_{n \bq} \})} = \frac{\ket{\tilde{\Psi}_{\eta}}}{\norm{\tilde{\Psi}_\eta}}, \qquad \ket{\tilde{\Psi}_\eta} = e^{\sum_{n \bq} \eta_{n \bq}  \hat{M}_{n \bq}} \ket{\Psi(0)}
\end{equation}
Here $\hat{M}_{n \bq} = c^\dag M_{n \bq} c$, where $c^\dag$ and $c$ are regarded as row and column vectors in the space of single particle states and $M_{n \bq}$ is the corresponding first-quantized operator that acts on the space of single particle states. This parameterization will be convenient because $\bra{\tilde{\Psi}_\eta}$ depends only on $\{\overline{\eta_{n \bq} }\}$ and because we will ultimately be able to express everything in terms of Wick-contractible correlators of $M_{n \bq}$ and the kinetic energy in the ground state $\ket{\Psi(0)}$.

We choose $M_{n \bq}$ so that
\begin{equation}
    M_{n \bq} \ket{\phi_\bk w_{m \bq_0 + \bq'}} = \delta_{m,0} \delta_{\bq',0} \ket{\phi_{\bk} w_{n \bq_0 + \bq}}, \quad \implies M_{n \bq} M_{n' \bq'} = 0.
    \label{eq:propertyofM}
\end{equation}
In this section, we label states by their $l=0$ component, so
$\braket{\br,l}{\phi_\bk w_{n\bq}} = \phi_\bk(\br) \chi_l(\br)$ where $\chi_0=w_{n\bq}$ and $\chi_{l>0}$ is determined from the zero-mode condition,
and $\phi_{\bk}(\br)$ is a fixed LLL Bloch state (independent of $\chi$).

Notably we do not include a $\chi$-dependent prefactor that normalizes the wavefunction. This choice to forgo normalization is made, alongside \eqref{eq:propertyofM}, such that
\begin{equation}
    e^{\eta_{n \bq} M_{n \bq}} \ket{\phi_\bk w_{0 \bq_0}} = \ket{\phi_\bk w_{0 \bq_0}} + \eta_{n \bq} \ket{\phi_\bk w_{n \bq_0 + \bq})} = \ket{\phi_\bk(w_{0 \bq_0}  + \eta_{n \bq} w_{n \bq_0 + \bq})}.
\end{equation}
This, in turn, ensures
\begin{equation}
\begin{aligned}
    \ket{\tilde{\Psi}_\eta} = e^{\sum_i \eta_{n \bq}M^{(i)}_{n \bq}} \ket{\Psi(0)} & = \prod_i (1+\eta_{n \bq}M^{(i)}_{n \bq})  \mathcal{N}(0)\bigwedge_{i} \ket{\phi_{\bk_i} w_{0,\bq_0}}_i \\
    & = \mathcal{N}(0)\bigwedge_{i} (1+\eta_{n \bq}M^{(i)}_{n \bq})\ket{\phi_{\bk_i} w_{0,\bq_0}}_i \\
    & = \mathcal{N}(0)\bigwedge_{i} \ket{\phi_{\bk_i} (w_{0 \bq_0}  + \eta_{n \bq} w_{n \bq_0 + \bq})}_i  \\
    & \propto \ket{\Psi(\eta_{n \bq})},
    \end{aligned}
\end{equation}
as desired. Above, we worked in first quantization for convenience, used $M^{(i)}_{n \bq}$ to denote the action of $\hat{M}_{n \bq}$ on particle $i=1,\ldots,N_e$, and used $\bigwedge$ to denote the antisymmetric tensor product appropriate for fermionic wavefunctions. A factor $\mathcal{N}(0)$ is included such that $\ket{\Psi(0)}$ is normalized.

We pause to comment that \eqref{eq:propertyofM} functions as a definition of $M_{n \bq}$ as an abstract linear operator within our space of interest. We will use these properties going forward and discuss concrete aspects associated with numerical implementation at the end, where we will be able to justify some convenient simplifications.

The action may be rewritten in terms of the Berry connections in the $\eta_{n \bq}$ complex plane as
\begin{equation}
    S  = \int dt  \left(\sum_{n \bq} \left[ A_{n\bq}(\{\eta_{n \bq} \}) \partial_t \eta_{n \bq} + \overline{A}(\eta_{n \bq}) \partial_t \overline{\eta}_{n \bq}\right] - \langle H \rangle_\eta \right)
\end{equation}
where
\begin{equation}
    \langle H \rangle_\eta  = \frac{\bra{\tilde{\Psi}_\eta} H \ket{\tilde{\Psi}_\eta}}{\braket{\tilde{\Psi}_\eta}}, \qquad A_{n\bq}(\{\eta_{n \bq}\})  = i \braket{\Psi(\eta_{n \bq})}{\partial_{\eta_{n \bq}} \Psi(\eta_{n \bq})} = i \partial_{\eta_{n \bq}} \log \norm{\tilde{\Psi}_\eta},
\end{equation}
In the evaluation of the holomorphic Berry connection $A$ we used that $\partial_{\eta_{n \bq}} \bra{\tilde{\Psi}} = 0$. The antiholomorphic Berry connection $\overline{A}$ is the complex conjugate of $A$. We now vary the action with respect to $\overline{\eta}_{n \bq} \to \overline{\eta}_{n \bq} + \delta \overline{\eta}_{n \bq}$. The corresponding Euler-Lagrange equations are
\begin{equation}
    \sum_{n\bq'} G_{m\bq n\bq'}(\{ \eta_{n \bq} \}) i \partial_t \eta_{n \bq'} = \overline{\partial}_{\eta_{m \bq}} \langle H \rangle_\eta
    \label{eq:fullEOM_supp}
\end{equation}
where $G$ is ($1/4$ of) the non-Abelian Berry curvature in this parameter space,
\begin{equation}
    G_{m \bq n \bq'}(\{ \eta_{n \bq} \}) = -i \overline{\partial}_{\eta_{m \bq}} A_{n \bq'} = \overline{\partial}_{\eta_{m \bq}} \partial_{\eta_{n \bq}} \log \norm{\tilde{\Psi}_\eta} = -\frac{i}{2} (\overline{\partial}_{\eta_{m \bq}} A_{n \bq'} - \partial_{\eta_n \bq} \overline{A}_{m \bq'}).
\end{equation}
The equations of motion can be interpreted as effective classical dynamics on the variational manifold where $G$ plays the role of a symplectic form $\sum_{m\bq n \bq'} G_{m \bq n \bq'} d \overline{\eta}_{m\bq} \wedge \eta_{n\bq}$.

We now expand around the ground state ansatz $\eta_{m \bq} = 0$ to obtain linearized equations of motion. We only need $G$ to zeroth order in $\eta$, so we expand $\log \norm{\Psi} = \frac{1}{2} \log \braket{\tilde{\Psi}_\eta} $ and act with $\overline{\partial}_{\eta_{m \bq}} \partial_{\eta_{n \bq}}$. We need to keep the linear terms in the expansion of $\braket{\tilde{\Psi}_\eta}$ as well as the $\overline{\eta} \eta$ quadratic term in order to get the $\overline{\eta} \eta$ terms in the log. We obtain
\begin{equation}
    G_{m \bq n \bq'}(0) = G_{mn}(\bq) \delta_{\bq \bq'} = \langle M^\dag_{m \bq} \hat{M}_{n \bq} \rangle_{0,c} = \langle \hat{M}^\dag_{m \bq} \hat{M}_{n \bq} \rangle_{0} - \langle \hat{M}^\dag_{m \bq} \rangle_0 \langle \hat{M}_{n \bq} \rangle_0.
\end{equation}
where we use the subscript ``$c$" to denote connected correlators
\begin{equation}
    \langle A B \rangle_c = \langle A B \rangle - \langle A \rangle \langle B \rangle, \quad \langle ABC\rangle_c = \langle ABC \rangle - \langle A \rangle \langle BC \rangle_c - \langle B\rangle \langle AC \rangle_c - \langle A B \rangle_c \langle C \rangle -  \langle A \rangle \langle B \rangle \langle C \rangle
\end{equation}
and the subscript ``$0$" to denote expectation values in the normalized ground state $\ket{\Psi(0)}$. We used crystalline translation symmetry to set $\bq = \bq'$. The disconnected part is nonzero only for $\bq = 0$ here.

We must now expand the right hand side using
\begin{equation}
    \langle H \rangle_\eta = \frac{\left\langle \exp\left(\sum_{m \bq} \overline{\eta}_{m \bq} \hat{M}^\dag_{m \bq}\right) H  \exp\left(\sum_{n \bq'} \eta_{m \bq'} \hat{M}_{n \bq'} \right) \right\rangle_0}{\left\langle \exp\left(\sum_{m \bq} \overline{\eta}_{m \bq} \hat{M}^\dag_{m \bq}\right)  \exp\left(\sum_{n \bq'} \eta_{m \bq'} \hat{M}_{n \bq'} \right) \right\rangle_0}
\end{equation}
The above expression can be interpreted as $\langle H \rangle_{\eta}$ with respect to an effective partition function, the denominator, with source fields $\overline{\eta}_{m \bq}$ and $\eta_{n \bq'}$.
Accordingly, we observe that differentiation with respect to $\overline{\eta}_{m \bq}$ and $\eta_{n \bq}$ produces connected correlation functions, e.g. $[(\overline{\partial}_{\eta_{m \bq}})^k (\partial_{\eta_{n \bq'}})^l \langle H \rangle_{\eta}]_{\eta=0} = \langle (\hat{M}^\dag_{m \bq})^k H (\hat{M}_{n \bq})^l\rangle_c$, where as usual the numerator generates the full correlator $\langle (\hat{M}^\dag_{m \bq})^k H (\hat{M}_{n \bq})^l\rangle$ and the denominator cancels off the disconnected parts. We will find it convenient to use this formal property, but the calculations below can also be verified through more direct computations. We pause to comment that differentiating each exponential is straightforward here because second quantization preserves commutators, $[\hat{M}_{n \bq}, \hat{M}_{n' \bq'}] = \widehat{[M_{n \bq}, M_{n' \bq'}]}$, and the first quantized operators commute due to \eqref{eq:propertyofM}. However, $\hat{M}_{n \bq} \hat{M}_{n' \bq} \neq 0$, unlike the first quantized version \eqref{eq:propertyofM}, essentially because the operators can act on different particles.

The zeroth order contribution to the right hand side of \eqref{eq:fullEOM_supp} is
\begin{equation}
\begin{aligned}
   [\overline{\partial}_{\eta_{m \bq}} \langle H \rangle_\eta]_{\eta=0} = \langle \hat{M}^\dag_{m \bq} H \rangle_{0,c}
\end{aligned}
  \label{eq:firstorderterm}
\end{equation}
where the subtraction of the disconnected part comes from expanding the denominator as usual.
The connected correlator in \eqref{eq:firstorderterm}, as well as the one obtained by differentiating with respect to $\eta_{n \bq'}$, must vanish
since $\{\eta \} = 0$ is the minimum energy state:
\begin{equation}
    \langle \hat{M}^\dag_{m \bq} H \rangle_{0,c} = \langle H \hat{M}_{n \bq'} \rangle_{0,c} = 0.
    \label{eq:firstorderconds}
\end{equation}
This is consistent with the fact that the variational ground state should be a steady state.  We note that, in practice, $\chi=w_0$ isn't \emph{exactly} the ground state as there is a very small mixing with $w_6$ in practice. One should then expand around the true, slightly mixed, steady state. For simplicity we assume $w_0$ is the exact steady state and restrict $n,m \leq 5$.

To obtain the linearized equation of motion we calculate the RHS of \eqref{eq:fullEOM_supp} to linear order in $\eta,\overline{\eta}$. These linear order terms are
\begin{equation}
\begin{aligned}
     & \sum_{n \bq'}\eta_{n \bq'}[\partial_{\eta_{n \bq'}} \overline{\partial}_{\eta_{m \bq}} \langle H \rangle_{\eta}]  + \overline{\eta}_{n \bq'}[\overline{\partial}_{\eta_{n \bq'}} \overline{\partial}_{\eta_{m \bq}} \langle H \rangle_{\eta}] \\
     & = \sum_{n \bq'}\eta_{n \bq'} \langle \hat{M}^\dag_{m \bq} H \hat{M}_{n \bq'} \rangle_{0,c}  -  \overline{\eta}_{n \bq'} \langle \hat{M}^\dag_{m \bq'}  \hat{M}^\dag_{n \bq} H \rangle_{0,c}, \\
     & = \sum_{n \bq'}\eta_{n \bq'} \big[ \langle \hat{M}^\dag_{m \bq} H \hat{M}_{n \bq'} \rangle_{0} - \langle \hat{M}^\dag_{m \bq}\rangle_0 \langle H \hat{M}_{n \bq'}\rangle_{0,c} - \langle \hat{M}^\dag_{m \bq} H \rangle_{0,c} \langle \hat{M}_{n \bq'} \rangle_0 - E_0 \langle \hat{M}^\dag_{m \bq} \hat{M}_{n \bq'}\rangle_{0,c}  - E_0 \langle \hat{M}^\dag_{m \bq}\rangle_0 \langle \hat{M}_{n \bq'} \rangle_0 \big]\\
     & + \overline{\eta}_{n \bq'}\big[ \langle \hat{M}^\dag_{m \bq} \hat{M}^\dag_{n \bq'} H  \rangle_{0} - \langle \hat{M}^\dag_{m \bq}\rangle_0 \langle \hat{M}^\dag_{n \bq'} H \rangle_{0,c} - \langle \hat{M}^\dag_{m \bq} H \rangle_{0,c} \langle \hat{M}^\dag_{n \bq'} \rangle_0 - E_0 \langle \hat{M}^\dag_{m \bq} \hat{M}^\dag_{n \bq'}\rangle_{0,c}  - E_0 \langle \hat{M}^\dag_{m \bq}\rangle_0 \langle \hat{M}^\dag_{n \bq'} \rangle_0 \big]\\
     & =  \sum_{n}\eta_{n \bq'} \langle \hat{M}^\dag_{m \bq} (H-E_0) \hat{M}_{n \bq} \rangle_{0} + \overline{\eta}_{n,-\bq} \langle \hat{M}^\dag_{m \bq} \hat{M}^\dag_{n,-\bq}(H-E_0)  \rangle_{0}
\end{aligned}
\end{equation}
To obtain the last line we used \eqref{eq:firstorderconds} repeatedly and also used translation symmetry to set $\bq' = +\bq$ in the first term and $\bq'=-\bq$ in the second term.

Our linearized equations of motion are therefore given by
\begin{equation}
\begin{aligned}
    i \sum_n G_{m n}(\bq) \partial_t \eta_{n \bq} & = \sum_n A_{mn}(\bq) \eta_{n \bq} + \sum_{n} B_{mn}(\bq) \overline{\eta}_{n,-\bq} \\
    G_{mn}(\bq) &  = \langle \hat{M}^\dag_{m \bq} \hat{M}_{n \bq} \rangle_{0,c} = \langle \hat{M}^\dag_{m \bq} \hat{M}_{n \bq} \rangle_{0} - \langle \hat{M}^\dag_{m \bq} \rangle_0 \langle \hat{M}_{n \bq} \rangle_{0} \\
    A_{mn}(\bq) & = \langle \hat{M}^\dag_{m \bq} H \hat{M}_{n \bq} \rangle_{0,c} = \langle \hat{M}^\dag_{m \bq} (H-E_0) \hat{M}_{n \bq} \rangle_0 \\
    B_{mn}(\bq) & = \langle \hat{M}^\dag_{m \bq}  \hat{M}^\dag_{n -\bq} H\rangle_{0,c} = \langle \hat{M}^\dag_{m \bq} \hat{M}^\dag_{n -\bq}  (H-E_0) \rangle_0
    \end{aligned}
    \label{eq:linearizedEOM}
\end{equation}
where we Fourier transformed $\eta$ in the time domain and used translation symmetry to set $\bq'\to \pm \bq$ as appropriate. Solutions are of the form $\eta_{n \bq} = x_{n \bq} e^{i \omega t} + \overline{y}_{n,-\bq} e^{-i \omega t} $ where
\begin{equation}
    \begin{pmatrix} A_\bq & B_\bq \\ B^*_{-\bq} & A^*_{-\bq}  \end{pmatrix}
    \begin{pmatrix} x_\bq \\ y_\bq \end{pmatrix}
    = \omega \begin{pmatrix} G_\bq & 0 \\ 0 & -G^*_{-\bq} \end{pmatrix}
    \begin{pmatrix} x_\bq \\ y_\bq \end{pmatrix}
    \label{eq:evalproblem}
\end{equation}
This is a generalized eigenvalue problem that we can solve to obtain the excitation frequencies $\omega$.

The single mode approximation can be obtained straightforwardly by replacing the $G, A, B$ matrices with the diagonal elements $G_{nn}, A_{nn}, B_{nn}$. We note that $G,A,B$ can be evaluated using Wick's theorem as
\begin{equation}
\begin{aligned}
    G_{mn}(\bq) & = \Tr P_0 M^\dag_{m \bq} (1-P_0) M_{n \bq},    \\
    A_{mn}(\bq) & = \Tr P_0 M^\dag_{m \bq} (1-P_0) h (1-P_0) M_{n \bq} + \Tr P_0 M^\dag_{m \bq} (1-P_0) M_{n \bq} (1-P_0) h, \\
    B_{mn}(\bq) & = \Tr P_0 M^\dag_{m \bq} (1-P_0) M^{\dag}_{n -\bq} (1-P_0) H + \Tr P_0 M^\dag_{m \bq} (1-P_0) H (1-P_0) M^{\dag}_{n -\bq}.
\end{aligned}
\end{equation}
where we used that $\hat{H}= c^\dag H c$ can be taken to be a single particle kinetic energy operator because  $\langle \hat{V}_0 \rangle$ vanishes for all of these states. We note that $\langle c c^\dag \rangle$ contractions produce $1-P_0$, where $P_0$ is the Hartree Fock projector corresponding to $\ket{\Psi(0)}$.

\subsection{Numerical Procedure for TDVP}
In this section, we outline the numerical procedure for implementing TDVP, which is best illustrated in the second-quantized language. In our discussion, we implicitly assume that $\bm{q}\neq \bm{0}$. However, it turns out that the final expression of $G_{mn}(\bm{q})$, $A_{mn}(\bm{q})$, $B_{mn}(\bm{q})$ are all applicable to the $\bm{q}\neq\bm{0}$ case.  The (unnormalized) ground state is
\begin{equation}
\ket{\tilde{\Psi}_0}=\prod_{\bm{k}\in BZ}c^{\dagger}_{\bm{k},0\bm{q}_0}\ket{0}.
\end{equation}
We note that the expectation value of any operator $\hat{O}$ is evaluated with respect to the normalized ground state
\begin{equation}
\langle \hat{O} \rangle_0 \equiv \frac{\langle \tilde{\Psi}_0|\hat{O}|  \tilde{\Psi}_0\rangle}{\braket{\tilde{\Psi}_0}{\tilde{\Psi}_0}}
\end{equation}
The notation is chosen such that
\begin{equation}
c^{\dagger}_{\bm{k},n\bm{q}_0}\ket{0}=\ket{\phi_{\bm{k}}w_{n\bm{q}_0}}.
\end{equation}
We notice that $\ket{\phi_{\bm{k}}w_{n\bm{q}_0}}$ is unnormalized, and it stands for a wavefunction whose 0th-layer component is the product of two normalized Landau level wavefunction
$\phi^{\textrm{LLL}}_{\bm{k}}(\bm{r})\phi^{\textrm{nLL}*}_{-\bm{q_0}}(\bm{r})(\bm{r})$. The commutation relation is thus
\begin{equation}
\{c_{\bm{k}_1,n_1\bm{q}_1},c^{\dagger}_{\bm{k}_2,n_2\bm{q}_2}\}=\braket{\phi_{\bm{k}_1}w_{n_1\bm{q}_1}}{\phi_{\bm{k}_2}w_{n_2\bm{q}_2}}.
\end{equation}

According to Eq.\ref{eq:propertyofM}, the action of the second-quantized opertor $\hat{M}_{n\bm{q}}$ on the ground state is to produce a coherent superposition of excited states:
\begin{equation}\label{eq:actionofM}
\hat{M}_{n\bm{q}}\ket{\tilde{\Psi}_0}=\sum^{N_e}_{i=1}c^{\dagger}_{\bm{k}_1,0\bm{q}_0}... c^{\dagger}_{\bm{k}_i,n\bm{q}_0+\bm{q}}...c^{\dagger}_{\bm{k}_{N_e},0\bm{q}_0}\ket{0}.
\end{equation}
It follows that $G_{mn}(\bm{q})$ is
\begin{equation}
\begin{split}
G_{mn}(\bm{q})&=\langle \hat{M}^{\dagger}_{m\bm{q}} M_{n\bm{q}}\rangle_0-\langle \hat{M}^{\dagger}_{m\bm{q}}\rangle_0 \langle \hat{M}_{n\bm{q}}\rangle_0\\
&=\sum_{\bm{k}}
\text{det}
\begin{pmatrix}
\braket{\phi_{\bm{k}}w_{m\bm{q}_0+\bm{q}}}{\phi_{\bm{k}}w_{n\bm{q}_0+\bm{q}}} & \braket{\phi_{\bm{k}}w_{m\bm{q}_0+\bm{q}}}{\phi_{\bm{k}+\bm{q}}w_{0\bm{q}_0}}\\
\braket{\phi_{\bm{k}+\bm{q}}w_{0\bm{q}_0}}{\phi_{\bm{k}}w_{n\bm{q}_0+\bm{q}}}&\braket{\phi_{\bm{k}+\bm{q}}w_{0\bm{q}_0}}{\phi_{\bm{k}+\bm{q}}w_{0\bm{q}_0}}
\end{pmatrix}
/(\braket{\phi_{\bm{k}+\bm{q}}w_{0\bm{q}_0}}{\phi_{\bm{k}+\bm{q}}w_{0\bm{q}_0}}\braket{\phi_{\bm{k}}w_{0\bm{q}_0}}{\phi_{\bm{k}}w_{0\bm{q}_0}}).
\end{split}
\end{equation}
There is a simple way to get the second line from the first. One can view $\hat{M}_{n\bm{q}}\ket{\tilde{\Psi}_0}$ as a coherent  superposition of the possibility of sending an electron from $(\bm{k},0\bm{q}_0)$ (i.e. $\ket{\phi_{\bm{k}w_{n\bm{q}_0}}}$) to $(\bm{k},n\bm{q}_0+\bm{q})$. For each of these terms, it only has non-zero overlap with the corresponding term in $\bra{\tilde{\Psi}_0}\hat{M}^{\dagger}_{m\bm{q}}$
that sends an electron from $(\bm{k},0\bm{q}_0)$ to $(\bm{k},m\bm{q}_0+\bm{q})$ due to momentum conservation. The overlap then produces the determinant in the equation above.

To evaluate $A_{mn}(\bm{q})$ and $B_{mn}(\bm{q})$, we will need the matrix element of the Hamiltonian $\langle \phi_{\bm{k}_1}w_{n_1\bm{q}_1}|H|\phi_{\bm{k}_2}w_{n_2\bm{q}_2} \rangle$, which includes only the kinetic energy contribution.  If the plane wave expansion of the state $\ket{\phi_{\bm{k}}w_{n\bm{q}}}$ is
\begin{equation}\label{eq:expansioncoeff}
\ket{\phi_{\bm{k}}w_{n\bm{q}}}=\sum_{\bm{g}}\psi_{\lceil \bm{k}+\bm{q}\rceil,\bm{q},n,\bm{g}}e^{i(\lceil \bm{k}+\bm{q} \rceil +\bm{g})\cdot \bm{r}}\ket{s_{\lceil \bm{k}+\bm{q}\rceil+\bm{g}}}\Leftrightarrow c^{\dagger}_{\bm{k},n\bm{q}}=\sum_{\bm{g}}\psi_{\lceil \bm{k}+\bm{q}\rceil,\bm{q},n,\bm{g}}c^{\dagger}_{\bm{k},\bm{g}},
\end{equation}
then the kinetic energy matrix element can be expressed in terms of these expansion coefficients as

\begin{equation}
\langle \phi_{\bm{k}_1}w_{n_1\bm{q}_1}|H|\phi_{\bm{k}_2}w_{n_2\bm{q}_2} \rangle=
\sum_{\bm{g}}\psi^{*}_{\lceil\bm{k}_1+\bm{q}_1 \rceil,\bm{q}_1 ,n_1,\bm{g}} \psi_{\lceil\bm{k}_2+\bm{q}_2 \rceil,\bm{q}_2,n_2,\bm{g}}\mathcal{E}(\lceil \bm{k}_2+\bm{q}_2 \rceil+\bm{g}_2)\delta_{\lceil \bm{k}_1+\bm{q}_1\rceil, \lceil \bm{k}_2+\bm{q}_2\rceil}.
\end{equation}
Here, $\lceil ... \rceil$ sends its argument back to the mBZ, and $\ket{s_{\bm{k}}}$ was defined in Eq.\ref{eq:holospinor}.  We note the first subscript $\lceil \bm{k}+\bm{q} \rceil$ is the Bloch momentum of the wavefunction, while the second index $\bm{q}$ denotes the momentum of the condensate part. Since we will later deal with states with different condensate momentum $\bm{q}$, it is the best to keep this index explicit. This expression above is obtained by making use of the second-quantized form of the kinetic energy operator:
\begin{equation}\label{eq:secondquantizedKinetic}
H=\sum_{\bm{k},\bm{g}}\mathcal{E}(\bm{k}+\bm{g})c^{\dagger}_{\bm{k},\bm{g}}c_{\bm{k},\bm{g}},\quad c^{\dagger}_{\bm{k},\bm{g}}\ket{0}=e^{i(\bm{k}+\bm{g})\cdot \bm{r}}\ket{s_{\bm{k}+\bm{g}}}.
\end{equation}
In addition,  the overlap between two states can also be expressed using these expansion coefficients:
\begin{equation}
\braket{\phi_{\bm{k}_1}w_{n_1\bm{q}_1}}{\phi_{\bm{k}_2}w_{n_2\bm{q}_2}}=\sum_{\bm{g}}\psi^{*}_{\lceil\bm{k}_1+\bm{q}_1 \rceil,\bm{q}_1 ,n_1,\bm{g}} \psi_{\lceil\bm{k}_2+\bm{q}_2 \rceil ,\bm{q}_2,n_2,\bm{g}}\delta_{\lceil \bm{k}_1+\bm{q}_1\rceil, \lceil \bm{k}_2+\bm{q}_2\rceil}.
\end{equation}
In principle, Eq.\ref{eq:actionofM}, Eq.\ref{eq:expansioncoeff}, Eq.\ref{eq:secondquantizedKinetic} are sufficient to compute $A_{mn}(\bm{q})$ and $B_{mn}(\bm{q})$ by brute force. However, there is again a simpler way to get the final expression without going through lengthy algebra.

To compute $A_{mn}(\bm{q})$, we need  $\langle \hat{M}^{\dagger}_{m\bm{q}} H \hat{M}_{n\bm{q}}\rangle_0$ and $\langle \hat{M}^{\dagger}_{m\bm{q}} E_0 \hat{M}_{n\bm{q}}\rangle_0$. The ground state energy $E_0$ is
\begin{equation}
E_0=\langle H \rangle=\sum_{\bm{p}} \frac{\langle \phi_{\bm{p}}w_{0\bm{q}_0}|H| \phi_{\bm{p}}w_{0\bm{q}_0} \rangle}{ \langle \phi_{\bm{p}}w_{0\bm{q}_0}|\phi_{\bm{p}}w_{0\bm{q}_0} \rangle}.
\end{equation}
Thus,  $\langle \hat{M}^{\dagger}_{m\bm{q}} E_0 \hat{M}_{n\bm{q}}\rangle_0$ is
\begin{equation}
\langle \hat{M}^{\dagger}_{m\bm{q}} E_0 \hat{M}_{n\bm{q}}\rangle_0
=\sum_{\bm{k},\bm{p}} \frac{\langle \phi_{\bm{p}}w_{0\bm{q}_0}|\hat{K}| \phi_{\bm{p}}w_{0\bm{q}_0} \rangle}{ \langle \phi_{\bm{p}}w_{0\bm{q}_0}|\phi_{\bm{p}}w_{0\bm{q}_0} \rangle}
\frac{\text{det}
\begin{pmatrix}
\braket{\phi_{\bm{k}}w_{m\bm{q}_0+\bm{q}}}{\phi_{\bm{k}}w_{n\bm{q}_0+\bm{q}}} & \braket{\phi_{\bm{k}}w_{m\bm{q}_0+\bm{q}}}{\phi_{\bm{k}+\bm{q}}w_{0\bm{q}_0}}\\
\braket{\phi_{\bm{k}+\bm{q}}w_{0\bm{q}_0}}{\phi_{\bm{k}}w_{n\bm{q}_0+\bm{q}}}&\braket{\phi_{\bm{k}+\bm{q}}w_{0\bm{q}_0}}{\phi_{\bm{k}+\bm{q}}w_{0\bm{q}_0}}
\end{pmatrix}}{
\braket{\phi_{\bm{k}+\bm{q}}w_{0\bm{q}_0}}{\phi_{\bm{k}+\bm{q}}w_{0\bm{q}_0}}\braket{\phi_{\bm{k}}w_{0\bm{q}_0}}{\phi_{\bm{k}}w_{0\bm{q}_0}}}.
\end{equation}
We further note that $\bra{\tilde{\Psi}_0} \hat{M}^{\dagger}_{m\bm{q}}$ is a coherent superposition of sending a particle from $(\bm{k}_1,0\bm{q}_0)$ to $(\bm{k}_1,0\bm{q}_0+\bm{q})$, and $M_{n\bm{q}}\ket{\tilde{\Psi}_0}$  is a coherent superposition of sending a particle from $(\bm{k}_2,0\bm{q}_0)$ to $(\bm{k}_2,0\bm{q}_0+\bm{q})$. By momentum conservation, only when $\bm{k}_1=\bm{k}_2$, the corresponding term has a non-zero contribution to $\langle \hat{M}^{\dagger}_{m\bm{q}} H \hat{M}_{n\bm{q}}\rangle_0$. This immediately gives:

\begin{equation}
\begin{split}
\langle \hat{M}^{\dagger}_{m\bm{q}} H \hat{M}_{n\bm{q}}\rangle_0=
&\sum_{\bm{k}} \left(\sum_{\bm{p}\neq \bm{k},\bm{k}+\bm{q}}\frac{\langle \phi_{\bm{p}}w_{0\bm{q}_0}|H| \phi_{\bm{p}}w_{0\bm{q}_0} \rangle}{ \langle \phi_{\bm{p}}w_{0\bm{q}_0}|\phi_{\bm{p}}w_{0\bm{q}_0} \rangle}
\frac{\text{det}
\begin{pmatrix}
\braket{\phi_{\bm{k}}w_{m\bm{q}_0+\bm{q}}}{\phi_{\bm{k}}w_{n\bm{q}_0+\bm{q}}} & \braket{\phi_{\bm{k}}w_{m\bm{q}_0+\bm{q}}}{\phi_{\bm{k}+\bm{q}}w_{0\bm{q}_0}}\\
\braket{\phi_{\bm{k}+\bm{q}}w_{0\bm{q}_0}}{\phi_{\bm{k}}w_{n\bm{q}_0+\bm{q}}}&\braket{\phi_{\bm{k}+\bm{q}}w_{0\bm{q}_0}}{\phi_{\bm{k}+\bm{q}}w_{0\bm{q}_0}}
\end{pmatrix}}{
\braket{\phi_{\bm{k}+\bm{q}}w_{0\bm{q}_0}}{\phi_{\bm{k}+\bm{q}}w_{0\bm{q}_0}}\braket{\phi_{\bm{k}}w_{0\bm{q}_0}}{\phi_{\bm{k}}w_{0\bm{q}_0}}}
\right)\\
&+\sum_{\bm{k}}
\frac{\text{det}
\begin{pmatrix}
\langle\phi_{\bm{k}}w_{m\bm{q}_0+\bm{q}}|H|\phi_{\bm{k}}w_{n\bm{q}_0+\bm{q}}\rangle & \braket{\phi_{\bm{k}}w_{m\bm{q}_0+\bm{q}}}{\phi_{\bm{k}+\bm{q}}w_{0\bm{q}_0}}\\
\langle \phi_{\bm{k}+\bm{q}}w_{0\bm{q}_0}|H|\phi_{\bm{k}}w_{n\bm{q}_0+\bm{q}}\rangle&\braket{\phi_{\bm{k}+\bm{q}}w_{0\bm{q}_0}}{\phi_{\bm{k}+\bm{q}}w_{0\bm{q}_0}}
\end{pmatrix}}{
\braket{\phi_{\bm{k}+\bm{q}}w_{0\bm{q}_0}}{\phi_{\bm{k}+\bm{q}}w_{0\bm{q}_0}}\braket{\phi_{\bm{k}}w_{0\bm{q}_0}}{\phi_{\bm{k}}w_{0\bm{q}_0}}}
\\
&+\sum_{\bm{k}}
\frac{\text{det}
\begin{pmatrix}
\langle\phi_{\bm{k}}w_{m\bm{q}_0+\bm{q}}|\phi_{\bm{k}}w_{n\bm{q}_0+\bm{q}}\rangle & \langle \phi_{\bm{k}}w_{m\bm{q}_0+\bm{q}}|H|\phi_{\bm{k}+\bm{q}}w_{0\bm{q}_0} \rangle\\
\langle \phi_{\bm{k}+\bm{q}}w_{0\bm{q}_0}|\phi_{\bm{k}}w_{n\bm{q}_0+\bm{q}}\rangle&\langle \phi_{\bm{k}+\bm{q}}w_{0\bm{q}_0}|H|\phi_{\bm{k}+\bm{q}}w_{0\bm{q}_0}\rangle
\end{pmatrix}}{
\braket{\phi_{\bm{k}+\bm{q}}w_{0\bm{q}_0}}{\phi_{\bm{k}+\bm{q}}w_{0\bm{q}_0}}\braket{\phi_{\bm{k}}w_{0\bm{q}_0}}{\phi_{\bm{k}}w_{0\bm{q}_0}}}.
\end{split}
\end{equation}
Combining these two terms, we arrive at the final expression for $A_{mn}(\bm{q})$:
\begin{equation}
\begin{split}
A_{mn}(\bm{q})&=\langle \hat{M}^{\dagger}_{m\bm{q}} (H-E_0) \hat{M}_{n\bm{q}}\rangle_0\\
&=-\sum_{\bm{k}} \frac{\langle \phi_{\bm{k}}w_{0\bm{q}_0}|H| \phi_{\bm{k}}w_{0\bm{q}_0} \rangle}{ \langle \phi_{\bm{k}}w_{0\bm{q}_0}|\phi_{\bm{k}}w_{0\bm{q}_0} \rangle}
\frac{\text{det}
\begin{pmatrix}
\braket{\phi_{\bm{k}}w_{m\bm{q}_0+\bm{q}}}{\phi_{\bm{k}}w_{n\bm{q}_0+\bm{q}}} & \braket{\phi_{\bm{k}}w_{m\bm{q}_0+\bm{q}}}{\phi_{\bm{k}+\bm{q}}w_{0\bm{q}_0}}\\
\braket{\phi_{\bm{k}+\bm{q}}w_{0\bm{q}_0}}{\phi_{\bm{k}}w_{n\bm{q}_0+\bm{q}}}&\braket{\phi_{\bm{k}+\bm{q}}w_{0\bm{q}_0}}{\phi_{\bm{k}+\bm{q}}w_{0\bm{q}_0}}
\end{pmatrix}}{
\braket{\phi_{\bm{k}+\bm{q}}w_{0\bm{q}_0}}{\phi_{\bm{k}+\bm{q}}w_{0\bm{q}_0}}\braket{\phi_{\bm{k}}w_{0\bm{q}_0}}{\phi_{\bm{k}}w_{0\bm{q}_0}}}
\\
&-\sum_{\bm{k}} \frac{\langle \phi_{\bm{k}+\bm{q}}w_{0\bm{q}_0}|H| \phi_{\bm{k}+\bm{q}}w_{0\bm{q}_0} \rangle}{ \langle \phi_{\bm{k}+\bm{q}}w_{0\bm{q}_0}|\phi_{\bm{k}+\bm{q}}w_{0\bm{q}_0} \rangle}
\frac{\text{det}
\begin{pmatrix}
\braket{\phi_{\bm{k}}w_{m\bm{q}_0+\bm{q}}}{\phi_{\bm{k}}w_{n\bm{q}_0+\bm{q}}} & \braket{\phi_{\bm{k}}w_{m\bm{q}_0+\bm{q}}}{\phi_{\bm{k}+\bm{q}}w_{0\bm{q}_0}}\\
\braket{\phi_{\bm{k}+\bm{q}}w_{0\bm{q}_0}}{\phi_{\bm{k}}w_{n\bm{q}_0+\bm{q}}}&\braket{\phi_{\bm{k}+\bm{q}}w_{0\bm{q}_0}}{\phi_{\bm{k}+\bm{q}}w_{0\bm{q}_0}}
\end{pmatrix}}{
\braket{\phi_{\bm{k}+\bm{q}}w_{0\bm{q}_0}}{\phi_{\bm{k}+\bm{q}}w_{0\bm{q}_0}}\braket{\phi_{\bm{k}}w_{0\bm{q}_0}}{\phi_{\bm{k}}w_{0\bm{q}_0}}}
\\
&+\sum_{\bm{k}}
\frac{\text{det}
\begin{pmatrix}
\langle\phi_{\bm{k}}w_{m\bm{q}_0+\bm{q}}|H|\phi_{\bm{k}}w_{n\bm{q}_0+\bm{q}}\rangle & \braket{\phi_{\bm{k}}w_{m\bm{q}_0+\bm{q}}}{\phi_{\bm{k}+\bm{q}}w_{0\bm{q}_0}}\\
\langle \phi_{\bm{k}+\bm{q}}w_{0\bm{q}_0}|H|\phi_{\bm{k}}w_{n\bm{q}_0+\bm{q}}\rangle&\braket{\phi_{\bm{k}+\bm{q}}w_{0\bm{q}_0}}{\phi_{\bm{k}+\bm{q}}w_{0\bm{q}_0}}
\end{pmatrix}}{
\braket{\phi_{\bm{k}+\bm{q}}w_{0\bm{q}_0}}{\phi_{\bm{k}+\bm{q}}w_{0\bm{q}_0}}\braket{\phi_{\bm{k}}w_{0\bm{q}_0}}{\phi_{\bm{k}}w_{0\bm{q}_0}}}
\\
&+\sum_{\bm{k}}
\frac{\text{det}
\begin{pmatrix}
\langle\phi_{\bm{k}}w_{m\bm{q}_0+\bm{q}}|\phi_{\bm{k}}w_{n\bm{q}_0+\bm{q}}\rangle & \langle \phi_{\bm{k}}w_{m\bm{q}_0+\bm{q}}|H|\phi_{\bm{k}+\bm{q}}w_{0\bm{q}_0} \rangle\\
\langle \phi_{\bm{k}+\bm{q}}w_{0\bm{q}_0}|\phi_{\bm{k}}w_{n\bm{q}_0+\bm{q}}\rangle&\langle \phi_{\bm{k}+\bm{q}}w_{0\bm{q}_0}|H|\phi_{\bm{k}+\bm{q}}w_{0\bm{q}_0}\rangle
\end{pmatrix}}{
\braket{\phi_{\bm{k}+\bm{q}}w_{0\bm{q}_0}}{\phi_{\bm{k}+\bm{q}}w_{0\bm{q}_0}}\braket{\phi_{\bm{k}}w_{0\bm{q}_0}}{\phi_{\bm{k}}w_{0\bm{q}_0}}}.
\end{split}
\end{equation}

To calculate $B_{mn}(\bm{q})$, we must obtain an expression of $\langle \hat{M}^{\dagger}_{m\bm{q}} \hat{M}^{\dagger}_{n\bm{q}} H \rangle$. Again, $\bra{\tilde{\Psi}_0}\hat{M}^{\dagger}_{m\bm{q}}$ creates a coherent superposition of sending a particle from $(\bm{k}_1,0\bm{q}_0)$ to $(\bm{k}_1+\bm{q},m\bm{q}_0)$. When $M^{\dagger}_{n-\bm{q}}$ further acts on it, it can in principle send any particle from $(\bm{k}_2,0\bm{q}_0)$ to $(\bm{k}_2-\bm{q},n\bm{q}_0)$. However, due to momentum conservation, only when $\lceil \bm{k}_2-\bm{q} \rceil =\bm{k}_1$, the corresponding term will have non-zero overlap with $H\ket{\tilde{\Psi}_0}$. Thus, the expression of $B_{mn}(\bm{q})$ is:
\begin{equation}
\begin{split}
B_{mn}(\bm{q})&=\langle \hat{M}^{\dagger}_{m\bm{q}}  \hat{M}^{\dagger}_{n-\bm{q}} (H-E_0)\rangle_0\\
&=\sum_{\bm{k}}\frac{\langle \phi_{\bm{k}}w_{0\bm{q}_0}|H|\phi_{\bm{k}} w_{0\bm{q}_0}\rangle}{ \langle \phi_{\bm{k}}w_{0\bm{q}_0}|\phi_{\bm{k}} w_{0\bm{q}_0}\rangle} \frac{\langle \phi_{\bm{k}}w_{m\bm{q}_0+\bm{q}}|\phi_{\bm{k}+\bm{q}} w_{0\bm{q}_0}\rangle}{ \langle \phi_{\bm{k}+\bm{q}}w_{0\bm{q}_0}|\phi_{\bm{k}+\bm{q}} w_{0\bm{q}_0}\rangle}
\frac{\langle \phi_{\bm{k}+\bm{q}}w_{n\bm{q}_0-\bm{q}}|\phi_{\bm{k}} w_{0\bm{q}_0}\rangle}{ \langle \phi_{\bm{k}}w_{0\bm{q}_0}|\phi_{\bm{k}} w_{0\bm{q}_0}\rangle}\\
&+\sum_{\bm{k}}\frac{\langle \phi_{\bm{k}+\bm{q}}w_{0\bm{q}_0}|H|\phi_{\bm{k}+\bm{q}} w_{0\bm{q}_0}\rangle}{ \langle \phi_{\bm{k}+\bm{q}}w_{0\bm{q}_0}|\phi_{\bm{k}+\bm{q}} w_{0\bm{q}_0}\rangle} \frac{\langle \phi_{\bm{k}}w_{m\bm{q}_0+\bm{q}}|\phi_{\bm{k}+\bm{q}} w_{0\bm{q}_0}\rangle}{ \langle \phi_{\bm{k}+\bm{q}}w_{0\bm{q}_0}|\phi_{\bm{k}+\bm{q}} w_{0\bm{q}_0}\rangle}
\frac{\langle \phi_{\bm{k}+\bm{q}}w_{n\bm{q}_0-\bm{q}}|\phi_{\bm{k}} w_{0\bm{q}_0}\rangle}{ \langle \phi_{\bm{k}}w_{0\bm{q}_0}|\phi_{\bm{k}} w_{0\bm{q}_0}\rangle}\\
&-\sum_{\bm{k}}\frac{\langle \phi_{\bm{k}}w_{m\bm{q}_0+\bm{q}}|H|\phi_{\bm{k}+\bm{q}} w_{0\bm{q}_0}\rangle}{ \langle \phi_{\bm{k}+\bm{q}}w_{0\bm{q}_0}|\phi_{\bm{k}+\bm{q}} w_{0\bm{q}_0}\rangle}
\frac{\langle \phi_{\bm{k}+\bm{q}}w_{n\bm{q}_0-\bm{q}}|\phi_{\bm{k}} w_{0\bm{q}_0}\rangle}{ \langle \phi_{\bm{k}}w_{0\bm{q}_0}|\phi_{\bm{k}} w_{0\bm{q}_0}\rangle}\\
&-\sum_{\bm{k}}\frac{\langle \phi_{\bm{k}}w_{m\bm{q}_0+\bm{q}}|\phi_{\bm{k}+\bm{q}} w_{0\bm{q}_0}\rangle}{ \langle \phi_{\bm{k}+\bm{q}}w_{0\bm{q}_0}|\phi_{\bm{k}+\bm{q}} w_{0\bm{q}_0}\rangle}
\frac{\langle \phi_{\bm{k}+\bm{q}}w_{n\bm{q}_0-\bm{q}}|H|\phi_{\bm{k}} w_{0\bm{q}_0}\rangle}{ \langle \phi_{\bm{k}}w_{0\bm{q}_0}|\phi_{\bm{k}} w_{0\bm{q}_0}\rangle}.
\end{split}
\end{equation}

Finally, we comment on how to obtain the expansion coefficients $\psi_{\lceil \bm{k}+\bm{q} \rceil,\bm{q},n,\bm{g}}$ .  We first construct the normalized Landau level wavefunction $\phi^{\textrm{nLL}}_{\bm{k}}(\bm{r})$. According to Eq.\ref{eq:planewave}, $\psi_{\lceil \bm{k}+\bm{q}\rceil,\bm{q},n,\bm{g}}$ is known up to an $\bm{g}$-independent, $n$-independent factor $\mathcal{M}^{(\bm{q})}_{\lceil \bm{k}+\bm{q}\rceil} $:
\begin{equation}
\psi_{\lceil \bm{k}+\bm{q}\rceil,\bm{q},n,\bm{g}}=\mathcal{M}^{(\bm{q})}_{\lceil \bm{k}+\bm{q} \rceil}\frac{\exp(i\pi n_1 n_2)}{N_{\bm{\lceil \bm{k}+\bm{q} \rceil+\bm{g}}}}(-\frac{\lceil k+q \rceil +g}{\sqrt{2}})^n \frac{l_B^n}{\sqrt{n!}} \exp(-\frac{l_B^2}{4}(g\bar{g}+2\lceil k+q \rceil\bar{g})) e^{-i(\lceil\bm{k}+\bm{q}\rceil+\bm{g})\cdot\bm{R}_0}.
\end{equation}
 We compare the $0$-th layer component of $\ket{\phi_{\bm{k}}w_{n\bm{q}}}$ obtained by using the expansion coefficient $\psi_{\lceil \bm{k}+\bm{q}\rceil,\bm{q},n,\bm{g}}$, with $\phi^{\textrm{LLL}}_{\bm{k}}(\bm{r}) \phi^{\textrm{nLL}*}_{-\bm{q}}(\bm{r})$, and require them to be the same. This will fix $\mathcal{M}^{(\bm{q})}_{\lceil \bm{k}+\bm{q} \rceil}$, which is sufficient for the numerics.

Finally, for the $\chi-\textrm{TDVP}$ calculation, we allow $m,n\in [0,5]$. The reason of excluding $n\geq 6$ modes is that $\ket{\tilde{\Psi}_0}$ is not the true ground state, but only a very good approximation to the ground state, which have a tiny mixture from $c^{\dagger}_{\bm{k},6\bm{q}_0}$ and $c^{\dagger}_{\bm{k},12\bm{q}_0}$..... Including $n\geq 6$ modes will result in instability, i.e. the excitation energy becomes imaginary. However, the fact that the TDVP spectrum agrees extremely well with the full TDHF calculation, as shown in the main text, confirms that $\ket{\tilde{\Psi}_0}$ is indeed a very good approximation to the ground state.

\subsection{Phonon modes}

We now discuss how the small $\bm{q}$ phonon modes emerge from $\chi$-TDVP. To do so, we must expand $G,A,B$ in small $\bm{q}$. We focus on $m, n \in \{0,1\}$ since these are the branches with appropriate angular momentum around $\bq = 0$ to form the phonons. A pattern becomes clear upon rewriting $G_{mn}(\bm{q})$ as
\begin{equation}
\begin{split}
G_{mn}(\bm{q})&=\langle \hat{M}^{\dagger}_{m\bm{q}} M_{n\bm{q}}\rangle_0-\langle \hat{M}^{\dagger}_{m\bm{q}}\rangle_0 \langle \hat{M}_{n\bm{q}}\rangle_0\\
&=\sum_{\bm{k}} \frac{\bra{\phi_{\bm{k}}w_{m\bm{q}_0+\bm{q}}}Q_{0}(\bk + \bq)\ket{\phi_{\bm{k}}w_{n\bm{q}_0+\bm{q}}}}{\braket{\phi_{\bm{k}}w_{0\bm{q}_0}}{\phi_{\bm{k}}w_{0\bm{q}_0}}}
\end{split}
\end{equation}
where
\begin{equation}
    Q_0(\bk + \bq) = 1-\frac{\ket{\phi_{\bm{k}+\bm{q}}w_{0\bm{q}_0}}\bra{\phi_{\bm{k}+\bm{q}}w_{0\bm{q}_0}}}{\braket{\phi_{\bm{k}+\bm{q}}w_{0\bm{q}_0}}}
\end{equation}
projects out the parts of the excited state that are parallel to the existing state. To expand in $\bq$ then, we will need to compute the difference between states like $\ket{\phi_{\bk+\bq} w_{0 \bq_0 + \bq}}$ and $\ket{\phi_\bk w_{0 \bq_0 + \bq}}$. We will use that they can be related by (non-magnetic) continuous translations. Indeed, the momentum operator $\bm{P} = -i \bm{\nabla}_\br$ acts as
\begin{equation}
    \bm{P}(\phi_\bk w_{n\bq_0}) = (\bm{K}_+\phi_\bk)w_{n\bq_0} + \phi_\bk (\bm{K}_- w_{n\bq_0})
\label{eq:momentum_onbasisstates}
\end{equation}
where $\bm{K}_+ = \bm{\pi} + l_B^{-2} \hat{z} \times \bm{r} = \bm{P} + e \bm{A} + e B \hat{z} \times \bm{r}$ generates magnetic translations and $\bm{K}_- = \bm{P} - e \bm{A} - e B \hat{z} \times \bm{r}$ generates conjugate magnetic translations (opposite effective field). Since $[\bm{K}_+,\bm{K}_-] = 0$, we can exponentiate \eqref{eq:momentum_onbasisstates} to obtain
\begin{equation}
    e^{i \hat{\bm{P}} \cdot \bm{d}_\bq} (\phi_\bk w_{n\bq_0}) = (e^{i \bm{K}_+ \cdot \bm{d}_\bq} \phi_\bk)(e^{i \bm{K}_- \cdot \bm{d}_\bq} w_{n\bq_0}) = \phi_{\bk+\bq} w_{n \bk - \bq}, \qquad \bm{d}_\bq = l_B^2 \hat{z} \times \bm{q}.
\end{equation}
In the last step we used that $t(\bm{d}) \phi_\bk = e^{i \bm{K}_+ \cdot \bm{d}} \phi_\bk$ has magnetic Bloch momentum $\bk - l_B^{-2} \hat{z} \times \bm{q}$; this can be verified by acting with the magnetic translation operator $t(\bm{R})$, where $\bm{R}$ is in the lattice, and using the algebra $t(\bm{d}_1) t(\bm{d_2})=t(\bm{d}_2) t(\bm{d_1})e^{i l_B^{-2} \bm{d}_1 \times \bm{d}_2}$. We can therefore expand
\begin{equation}
    Q_{0,\bk+\bq}\ket{\phi_\bk w_{0 \bq_0 + \bq}}  = Q_{0,\bk+\bq}e^{i \bm{P} \cdot (-\bm{d}_{\bm q})} \ket{\phi_{\bk+\bq} w_{0 \bq_0}} \approx Q_{0,\bk+\bq}(1+i \bm{P} \cdot (-\bm{d}_{\bm{q}}))\ket{\phi_{\bk+\bq} w_{0 \bq_0}} = -i d_{\bm q}^\nu Q_{0,\bk+\bq}P_\nu \ket{\phi_{\bk+\bq} w_{0 \bq_0}}
\end{equation}
where summation over $\nu$ is implied and we do not distinguish between upper and lower indices.

We also expand
\begin{equation}
    \ket{\phi_\bk w_{1 \bq_0 + \bq}}  \approx \ket{\phi_{\bk + \bq} w_{1 \bq_0}} = i\sqrt{2}l_B\partial_{\bar{z}} \ket{\phi_{\bk + \bq} w_{0 \bq_0}} = -l_B \omega^\mu P_\mu \ket{\phi_{\bk + \bq} w_{0 \bq_0}}
\end{equation}
to leading order in $\bq$ and wrote $\partial_{\overline{z}} = \frac{1}{\sqrt{2}}\omega^\mu \partial_\mu$ with $\bm{\omega} = \frac{1}{\sqrt{2}} \begin{pmatrix} 1 & i \end{pmatrix}^T$ for imminent convenience. We note that we expand the $n=0$ mode to linear order in $\bq$ whereas we expand the $n=1$ mode to constant order in $\bq$. That is, $\eta_0 q$ is at the ``same order" as $\eta_1$. At the end of the section we discuss why this is the appropriate power counting, though we note here that it is necessary to expand to this order in order to obtain a non-singular $G(\bq)$.

Substituing into $G(\bq)$, we obtain
\begin{equation}
\begin{aligned}
    G(\bq) = \begin{pmatrix} l_B^{-2} d_\bq^\mu d_\bq^\nu  &  -i l_B^{-1}\omega^\mu d^\nu_\bq \\  il_B^{-1}\overline{\omega}^\mu d^\nu_\bq & \overline{\omega}^\mu \omega^\nu \end{pmatrix} l_B^2 \sum_\bk \frac{\bra{\phi_{\bm{k+q}}w_{0\bm{q}_0}}P_\mu Q_{0}(\bk + \bq)P_\nu \ket{\phi_{\bm{k+q}}w_{0\bm{q}_0}}}{\braket{\phi_{\bm{k}}w_{0\bm{q}_0}}{\phi_{\bm{k}}w_{0\bm{q}_0}}} \\
\end{aligned}.
\end{equation}
We can furthermore ignore the difference between the $\bk + \bq$ argument in the numerator and $\bk$ in the denominator. The result of the above manipulations therefore justify the replacements
\begin{equation}
    M_{0 \bq} \to - i d^\mu P_\mu, \qquad M_{1 \bq} \to -l_B \omega^\mu P_\mu
\end{equation}
to leading order in $\bq$. We additionally use the commutativity of momentum operators, $\langle P_\mu P_\nu \rangle_c = \langle P_\nu P_\mu \rangle_c$
as well as $C_{3z}$ symmetry, to conclude $l_B^2\langle P_\mu P_\nu \rangle_c = g_0 \delta_{\mu \nu}$. We then obtain
\begin{equation}
    G(\bq) = g_0 \begin{pmatrix} l_B^2 \abs{q}^2 & \frac{1}{\sqrt{2}} l_B q \\ \frac{1}{\sqrt{2}} l_B \overline{q} & 1 \end{pmatrix}
\end{equation}
where $q = q_x + i q_y=-i\omega_\mu d^\mu$.
We can justify the same substitution in $A$. This leads to
\begin{equation}
    A(\bq) = \omega_c g_0 \begin{pmatrix} l_B^2 \abs{q}^2 & \frac{1}{\sqrt{2}} l_B q \\ \frac{1}{\sqrt{2}} l_B \overline{q} & 1 \end{pmatrix}, \qquad \omega_c\delta_{\mu \nu} = \frac{\langle P_\mu H P_\nu \rangle_c}{\langle P_\mu  P_\nu \rangle_c} = \delta_{\mu \nu} \frac{\langle M_{1,0}^\dag H M_{1,0} \rangle_c}{\langle M^\dag_{1,0} M_{1,0} \rangle_c},
\end{equation}
where we identified the SMA energy, $\omega_c$, for the $w_1$ mode.

Similarly
\begin{equation}
    B(\bq) = \omega_c g_0 \begin{pmatrix} l_B^2 \abs{q}^2 & -\frac{1}{\sqrt{2}} l_B \overline{q}  \\ \frac{1}{\sqrt{2}} l_B \overline{q} & 0  \end{pmatrix},
\end{equation}
where we used $\overline{\omega}^\mu \overline{\omega}_\mu = 0$ for the $B_{11}$ entry.

We see that $G,A,B$ all have very similar forms. We thus multiply the linearized equations of motion \eqref{eq:linearizedEOM} by inverting $G$ and computing
\begin{equation}
    G^{-1}(\bq) A(\bq) = \omega_c, \qquad G^{-1}(\bq) B(\bq) = \omega_c \begin{pmatrix} 1 & -(\tfrac{1}{\sqrt{2}}l_B q)^{-1} \\ 0 & \frac{\overline{q}}{q} \end{pmatrix},
\end{equation}
to obtain
\begin{equation}
\begin{aligned}
    i \partial_t \eta_{0\bq} & = \omega_c\left(\eta_{0 \bq} + \overline{\eta}_{0 -\bq} - \frac{\sqrt{2}}{l_B q} \overline{\eta}_{1 -\bq}\right) \\
    i \partial_t \eta_{1 \bq} & = \omega_c\left(\eta_{1 \bq} + \frac{\overline{q}}{q }\overline{\eta}_{1 -\bq}\right)
\end{aligned}
\label{eq:supp_etaEOM}
\end{equation}
In matrix form, with $\bm{\eta}\equiv (\eta_{0,\bq},\bar{\eta}_{0,-\bq},\eta_{1,\bq},\bar{\eta}_{1,-\bq})^{\mathsf{T}}$, we have
\begin{equation}\label{eq:supp_etamatrixEOM}
i\partial_t\bm{\eta}=\omega_c\begin{pmatrix}
1 & 1 & 0 & -\sqrt{2}/(l_B q) \\
-1 & -1 & -\sqrt{2}/(l_B \bar{q}) & 0\\
0 & 0 & 1 & \bar{q}/q \\
0 & 0 & -q/\bar{q} & -1
\end{pmatrix}
\bm{\eta}
\end{equation}
which is what we quoted in the main text. We state the solutions here for convenience: $\bm{\eta}^{\text{shear}} = (1,-1,0,0)^{\mathsf{T}}$ and $\bm{\eta}^{\text{long}} = i(1,1,-\sqrt{2}l_B \bar{q}, \sqrt{2}l_B q)^{\mathsf{T}}$.

We note that we have only expanded out the leading non-singular $\bq$-dependence. To obtain phonon velocities with this method one must compute $O(\abs{q}^2)$ corrections to $G,A,B$ in order to obtain the nonzero restoring force associated with inhomogeneous displacements. The above $\bq$ dependence captures the fact that displacements associated with $\eta_1$ are comparable to $\abs{q}$ times the displacements associated with $\eta_0$, but not the spacial inhomogeneity of these displacements.

To understand the role of $\bq$ in generating displacements in the $n=0$ sector, consider the shift in the skyrmion core position
\begin{equation}
    \chi_0^{\rm phonon}(\br) \approx \delta \overline{z} + i l_B^2 \overline{q}(\eta_{0 \bq} - \eta_{0 -\bq}) + i \sqrt{2} \ell_B(\eta_{1 \bq} + \eta_{1 -\bq}),
    \label{eq:mixphonons}
\end{equation}
where we used $w_{0}^{(\bq_0 \pm \bq)}(\br) \approx \pm i l_B^2 \overline{q}$ and $w_1^{(\bq_0 \pm \bq)}(\br) \approx i \sqrt{2} l_B$ for small $\bq$ near the core $\delta \overline{z} \approx0$ for concreteness. Near other cores, such as $\br \approx \bm{R}$, the $\eta_{m \pm \bq}$ terms should be multiplied by $e^{\pm i \bq \cdot \bm{R}}$.

 We can read off two independent $\omega= 0 + O(\abs{\bq})$ ``steady-state" solutions of \eqref{eq:supp_etaEOM}
 (i) $0=\eta_{0\bq}+\overline{\eta}_{0,-\bq}=\eta_{1,\bq}$, which corresponds to a shear displacement $\chi_0 \approx \delta \overline{z} + 2i \overline{q}l_B^2\Re \eta_{0\bq} $ in Eq. \eqref{eq:mixphonons}. This is the shear mode $\bm{\eta}^{\text{shear}} = (1,-1,0,0)^{\mathsf{T}}$. We also have
 (ii) $\eta_{0,\bq}=\overline{\eta}_{0,-\bq}=(\sqrt{2} l_B q)^{-1} \overline{\eta}_{1 -\bq}=-(\sqrt{2} l_B \overline{q})^{-1} \eta_{1 \bq}$,
 which corresponds to a longitudinal displacement $\chi_0(\br) \approx \delta \overline{z} +  2\overline{q} \Im \eta_{0 \bq}$ and can be identified with the longitudinal mode $\bm{\eta}^{\text{long}} = i(1,1,-\sqrt{2}l_B \bar{q}, \sqrt{2}l_B q)^{\mathsf{T}}$.

 The expansion \eqref{eq:mixphonons} also makes clear why the power counting $\eta_0 q \sim \eta_1$ is appropriate: the $n=0$ mode displacements are accompanied by factors of $q$ relative to the $n=1$ displacements.

\end{document}